\documentclass[a4paper,12pt]{article}
\pdfoutput=1
\usepackage{jheppub}
\usepackage[utf8]{inputenc}
\usepackage{url}
\usepackage{mathrsfs}
\usepackage{pdflscape}
\usepackage{refcount}
\usepackage{booktabs}
\usepackage{todonotes}
\usepackage{enumitem}
\usepackage{adjustbox}
\usepackage{afterpage}
\usepackage{mathdots}
\usepackage{ytableau, youngtab}
\usepackage{amsfonts}
\usepackage{bbm}
\usepackage{pdfpages}
\usepackage{caption}
\usepackage{subcaption}
\usepackage{tikz}
\usetikzlibrary{positioning}
\usetikzlibrary{arrows}
\usetikzlibrary{decorations.pathreplacing}
\usetikzlibrary{shapes.geometric}
\usetikzlibrary{calc}
\usetikzlibrary{decorations.pathmorphing}
\usetikzlibrary{shapes.misc}
\usetikzlibrary{positioning,trees,decorations.pathmorphing,decorations.markings,decorations.pathreplacing,calc,shapes,patterns,arrows,chains,arrows.meta,fit,fadings,decorations.markings,graphs,graphs.standard,quotes,plotmarks,calligraphy,decorations.pathreplacing}
\usetikzlibrary{intersections, calc}
\usetikzlibrary{shapes.geometric}
\usetikzlibrary{matrix, arrows.meta}
\usetikzlibrary{shapes.geometric}
\usetikzlibrary{decorations.pathmorphing}
\usetikzlibrary{shapes.misc}
\usetikzlibrary{arrows.meta}
\usetikzlibrary{angles, quotes}

\definecolor{goodgreen}{RGB}{55,169,49}
\definecolor{darkyellow}{RGB}{230,170,10}
\definecolor{brightyellow}{RGB}{255,240,190}

\tikzset{flavour/.style={draw=none,minimum size=0.3mm,fill=white, regular polygon,regular polygon sides=4,draw}}
\tikzset{gaugeBig/.style={inner sep=1mm,draw=none,fill=white,minimum size=2mm,circle, draw}}
\tikzset{bd/.style={circle, draw=black, inner sep=0pt, fill=black, minimum size=2mm}}
\tikzset{wd/.style={circle, draw=black, inner sep=0pt, fill=white, minimum size=2mm}}
\tikzset{Dynkin/.style={circle, draw=black, inner sep=0pt, fill=white, minimum size=2mm}}
\tikzstyle{ligne}=[draw, very thick]
\tikzstyle{gridline}=[draw, gray]
\tikzset{gauge/.style={circle, draw,inner sep=2.5pt}}
\tikzset{gaugeo/.style={circle, draw,inner sep=2.5pt,fill=orange}}
\tikzset{gaugec/.style={circle, draw,inner sep=2.5pt,fill=cyan}}
\tikzset{gauger/.style={circle, draw,inner sep=2.5pt,fill=red}}
\tikzset{gaugeb/.style={circle, draw,inner sep=2.5pt,fill=blue}}
\tikzset{gaugeg/.style={circle, draw,inner sep=2.5pt,fill=green}}
\tikzset{gaugem/.style={circle, draw,inner sep=2.5pt,fill=magenta}}
\tikzset{hasse/.style={circle, fill,inner sep=2pt}}
\tikzset{shrinky/.style={circle, fill,inner sep=1pt}}
\tikzset{sized/.style={circle, draw, inner sep=1.5pt}}
\tikzset{seven/.style={circle, draw,inner sep=3pt}}
\tikzset{dotto/.style={circle, orange, draw,inner sep=1.5pt,fill=orange}}
\tikzset{dottp/.style={circle, purple, draw,inner sep=1.5pt,fill=purple}}
\tikzset{dottc/.style={circle, cyan, draw,inner sep=1.5pt,fill=cyan}}
\tikzset{dottr/.style={circle, red, draw,inner sep=1.5pt,fill=red}}
\tikzset{dottb/.style={circle, blue, draw,inner sep=1.5pt,fill=blue}}
\tikzset{dottg/.style={circle, green, draw,inner sep=1.5pt,fill=green}}
\tikzset{dottm/.style={circle, magenta, draw,inner sep=1.5pt,fill=magenta}}

\tikzset{redgauge/.style={draw=none,minimum size=0.4cm,fill=red,circle, draw}}
\tikzset{gauge3/.style={draw=none,minimum size=0.4cm,fill=white,circle, draw}}
\tikzset{bluegauge/.style={draw=none,minimum size=0.4cm,fill=blue,circle, draw}}
\tikzset{redflavor/.style={draw=none,minimum size=0.6cm,fill=red, regular polygon,regular polygon sides=4,draw}}
\tikzset{blueflavor/.style={draw=none,minimum size=0.6cm,fill=blue, regular polygon,regular polygon sides=4,draw}}
\tikzset{flavour2/.style={draw=none,minimum size=0.6cm,fill=white, regular polygon,regular polygon sides=4,draw}}
\tikzset{rede/.style={line width=0.5mm,red}}
\tikzset{bluee/.style={line width=0.5mm,blue}}
\tikzset{pinkline/.style={line width=0.5mm,purple}}
\pgfdeclarelayer{edgelayer}
\pgfdeclarelayer{nodelayer}

\DeclareMathOperator{\U}{U}
\DeclareMathOperator{\SU}{SU}
\DeclareMathOperator{\SO}{SO}

\DeclareMathOperator{\USp}{USp}

\DeclareMathOperator{\rank}{rank}

\usepackage{epsfig}
\usepackage{amsmath}
\usepackage{amssymb}
\usepackage{amsfonts}
\usepackage{amsxtra}
\usepackage{amsthm}
\usepackage{mathrsfs}
\usepackage{makeidx}
\usepackage{graphics}
\usepackage{dsfont}
\usepackage{mathtools}
\usepackage{graphicx}
\usepackage{placeins}
\usepackage{bm}
\usepackage[capitalise]{cleveref}
\usepackage{empheq}
\usepackage{colortbl}
\usepackage{xcolor}
\usepackage{titlesec}
\usepackage{longtable}
\usepackage{hhline}
\usepackage{float}
\usepackage{color}
\usepackage{tikz}
\usepackage{xfrac}
\usepackage{footnote}
\usepackage{rotating}
\usepackage{lscape}
\usepackage{makecell}
\usepackage{environ}
\usepackage{tabularx}
\usepackage{subfiles}
\usepackage{ytableau}
\usepackage{tikz-3dplot}
\usepackage{slashed}
\usepackage{pifont}
\usepackage{multirow}
\usepackage{mdframed}
\usepackage{bbm}
\usepackage[normalem]{ulem}
\usepackage{arydshln}
\usepackage{enumitem}
\usepackage{fancybox}
\usetikzlibrary{positioning,trees,decorations.pathmorphing,decorations.markings,decorations.pathreplacing,calc,shapes,patterns,arrows,chains,arrows.meta,fit,fadings,decorations.markings,graphs,graphs.standard,quotes,plotmarks,calligraphy,decorations.pathreplacing}

\usetikzlibrary{positioning}
\usetikzlibrary{chains}
\usetikzlibrary{arrows, arrows.meta ,fit,decorations.pathreplacing}
\tikzstyle{every picture}+=[remember picture]
\tikzstyle{na} = [baseline]
\tikzstyle{ligne}=[draw, thick]

\usetikzlibrary{arrows, decorations.markings, calc, fadings, decorations.pathreplacing, patterns, decorations.pathmorphing, positioning}
\tikzset{>={Latex[width=1.5mm,length=1.5mm]}}
\tikzset{bd/.style={circle, draw=black, inner sep=0pt, fill=black, minimum size=1.2mm}}
\tikzset{bld/.style={circle, draw=blue, inner sep=0pt, fill=blue, minimum size=1.2mm}}
\tikzset{wd/.style={circle, draw=black, inner sep=0pt, fill=white, minimum size=1.2mm}}
\tikzset{rd/.style={circle, draw=red, inner sep=0pt, fill=red, minimum size=.9mm}}
\tikzset{wrd/.style={circle, draw=red, inner sep=0pt, fill=white, minimum size=.9mm}}
\usetikzlibrary{graphs,graphs.standard,quotes}

\tikzstyle{every picture}+=[remember picture]
\tikzstyle{na} = [baseline=-.5ex]

\providecommand{\abs}[1]{\lvert#1\rvert}

\newcommand{\eg}{e.g. }

\newcommand{\ie}{i.e. }

\numberwithin{equation}{section}
\newcommand{\bes}[1]{\begin{equation} \begin{split} #1\end{split} \end{equation}}

\newcommand{\be}{\begin{equation}} \newcommand{\ee}{\end{equation}}
\newcommand{\bea}{\begin{equation} \begin{aligned}} \newcommand{\eea}{\end{aligned} \end{equation}}

\def\tilde{\widetilde}

\def\hat{\widehat}

\def\rt2{\sqrt{2}}

\def\mod{{\rm mod}}
\def\det{\mathop{\rm det}}

\def\coker{{\mathop {\rm coker}}}

\def\Tr{\mathrm{Tr}}

\def\abs#1{\left|#1\right|}

\def\CI{{\cal I}}

\def\CN{{\cal N}}

\def\1{{\ds 1}}

\def\Z{\hbox{$\bb Z$}}

\newcommand{\fg}{\mathfrak{g}}
\newcommand{\fm}{\mathfrak{m}}

\newcommand{\fu}{\mathfrak{u}}

\def\SO{\mathrm{SO}}

\def\SU{\mathrm{SU}}

\def\Spin{\mathrm{Spin}}

\def\su{\mathfrak{su}}

\def\so{\mathfrak{so}}

\def\usp{\mathfrak{usp}}

\def\fr{\mathfrak{r}}

\def\repa{\raise4pt\hbox{$\square$}\mkern-14mu\raise-4pt\hbox{$\square$}}
\def\repab{\overline{\raise4pt\hbox{$\square$}\mkern-14mu\raise-4pt\hbox{$\square$}\mkern-1mu}}
\def\smileface{\ensuremath{\hbox{\large$\bigcirc$}\mkern-15mu\raise-1pt\hbox{\scriptsize$\smallsmile$}%
\mkern-10mu\raise4pt\hbox{..}\mkern4mu}}
\def\frownface{\ensuremath{\hbox{\large$\bigcirc$}\mkern-15mu\raise-1pt\hbox{\scriptsize$\smallfrown$}%
\mkern-10mu\raise4pt\hbox{..}\mkern4mu}}

\newcommand{\ba}{\begin{array}}
\newcommand{\ea}{\end{array}}
\newcommand{\bi}{\begin{itemize}}
\newcommand{\ei}{\end{itemize}}
\def\vec#1{\bm{#1}}
\def\bea#1\eea{\allowdisplaybreaks \begin{align}#1\end{align}}
 \newcommand{\ben}{\begin{enumerate}}
\newcommand{\een}{\end{enumerate}}
\newcommand{\bean}{\begin{eqnarray*}}
\newcommand{\eean}{\end{eqnarray*}}
\newcommand{\eref}[1]{(\ref{#1})}
\newcommand{\sref}[1]{\S\ref{#1}}

\newcommand{\PE}{\mathop{\rm PE}}
\newcommand{\PL}{\mathop{\rm PL}}

\newcommand{\BC}{\mathbb{C}}

\newcommand{\BP}{\mathbb{P}}

\newcommand{\BZ}{\mathbb{Z}}

\definecolor{light-gray}{gray}{0.5}
\definecolor{new-green}{rgb}{0,0.7,0.3}
\definecolor{cerulean}{rgb}{0.0, 0.48, 0.65}
\definecolor{claret}{rgb}{0.50, 0.09, 0.20}
\definecolor{darkred}{rgb}{0.7, 0.11, 0.11}
\definecolor{scarlet}{rgb}{1.0, 0.13, 0.0}
\definecolor{orange-red}{rgb}{1.0, 0.27, 0.0}
\definecolor{blue-green}{rgb}{0.0, 0.5, 0.65}
\definecolor{green-red}{rgb}{0.5, 0.65, 0.0}

\def\aup#1 {\overset{#1}{\uparrow} \, \overset{\tilde{#1}}{\downarrow}}

\tikzset{snake it/.style={decorate, decoration={snake, amplitude=.4mm, segment length=2mm,
                       post length=0mm,pre length=0mm}}}

\DeclareMathAlphabet{\mymathds}{U}{BOONDOX-ds}{m}{n}
\tikzstyle{double_border} = [draw, double, double distance=1pt]

\colorlet{verdescuro}{green!50!black}

\theoremstyle{plain}

\def\Z{\mathbb{Z}}

\newcommand{\cw}{\mathrm{cw}}
\newcommand{\crt}{\mathrm{cr}}   
\def\PSU{\mathrm{PSU}}
\newcommand{\kCS}{k_{\mathrm{CS}}}

\title{Generalised global symmetries in 5d $\mathcal{N}=1$ theories from the blow-up equations}
\author[a,b]{William Harding}
\author[b,c]{and Noppadol Mekareeya}
\affiliation[a]{Dipartimento di Fisica, Universit\`a di Milano-Bicocca, Piazza della Scienza 3, I-20126 Milano, Italy}
\affiliation[b]{INFN, sezione di Milano-Bicocca, Piazza della Scienza 3, I-20126 Milano, Italy}
\affiliation[c]{The Institute for Fundamental Study ``The Tah Poe Academia Institute'', Naresuan University, Phitsanulok 65000, Thailand}
\emailAdd{w.harding@campus.unimib.it}
\emailAdd{n.mekareeya@gmail.com}
\abstract{
Five-dimensional $\mathcal{N}=1$ superconformal field theories admit a rich variety of generalised global symmetries, including higher-form and 2-group symmetries and their 't~Hooft anomalies. We show that this data can be extracted directly from the blow-up equations governing the instanton partition functions of such theories on the $\Omega$-background. The central object is the classical prefactor $\exp(-V_n)$ weighting each magnetic flux on the blown-up geometry: evaluated on a background for the electric 1-form symmetry, the fractional parts of its exponents encode the cubic self-anomaly of the 1-form symmetry and its mixed anomalies with the instanton, flavour, gravitational, and $\mathrm{SU}(2)_R$ symmetries. Combined with the faithful continuous global symmetry of the ultraviolet fixed point, determined from the superconformal index, the same data decides whether the theory possesses a 2-group symmetry or a mixed 't~Hooft anomaly. We illustrate the method in gauge theories, including $\mathrm{SU}(4)$ and $\mathrm{USp}(4)$ with antisymmetric hypermultiplets and $\mathrm{Spin}(7)$ and $\mathrm{Spin}(8)$ with vector hypermultiplets, as well as in several families of non-Lagrangian theories. New results include the effective prepotentials of the $B_N$ and $B_N^{(1,2,3)}$ families, the cubic 1-form anomalies of the rank-two theories $\mathbb{P}^2\cup\mathbb{F}_3$ and $\mathbb{P}^2\cup\mathbb{F}_6$, and several mixed 1-form--flavour and 1-form--$\mathrm{SU}(2)_R$ anomalies.
}

\begin{document}
\maketitle

\section{Introduction}
\label{sec:intro}

The modern viewpoint on global symmetries, initiated in \cite{Gaiotto:2014kfa}, identifies a symmetry with the existence of topological operators of arbitrary codimension in a quantum field theory. This perspective has led to far-reaching generalisations of the familiar notion of a group-like symmetry acting on local operators, including higher-form symmetries acting on extended operators, higher-group symmetries \cite{Cordova:2018cvg,Benini:2018reh,Cordova:2020tij}, and non-invertible symmetries; see \cite{Cordova:2022ruw,Schafer-Nameki:2023jdn,Brennan:2023mmt,Bhardwaj:2023kri,Kaidi:2026urc} for reviews. Like ordinary symmetries, generalised global symmetries and their 't~Hooft anomalies are preserved along renormalisation group flows, and therefore provide powerful non-perturbative handles on the dynamics of quantum field theories.

Five-dimensional superconformal field theories (SCFTs) constitute an especially fertile arena for these ideas. Although gauge theories in five dimensions are infrared free, a large class of them arises as relevant deformations of strongly coupled ultraviolet fixed points \cite{Seiberg:1996bd,Morrison:1996xf,Intriligator:1997pq}, which can be engineered in M-theory on canonical Calabi--Yau threefold singularities or via five-brane webs \cite{Aharony:1997ju}; see \cite{Jefferson:2018irk,Bhardwaj:2018yhy,Bhardwaj:2018vuu,Apruzzi:2019opn,Bhardwaj:2019jia,Bhardwaj:2020gyu} for progress on the systematic classification of 5d SCFTs. The 1-form symmetries of these theories, together with their 't~Hooft anomalies and 2-group structures, have been analysed extensively from the geometry of the Calabi--Yau threefold and the associated symmetry topological field theory \cite{Morrison:2020ool,Albertini:2020mdx,Bhardwaj:2020phs,BenettiGenolini:2020doj,Gukov:2020btk,Apruzzi:2021nmk,Apruzzi:2021vcu,DelZotto:2022fnw,Apruzzi:2022dlm,DelZotto:2022joo,Cvetic:2022imb,Genolini:2022mpi,Cvetic:2025lat,Chakrabhavi:2026iku}. More recently, strong field-theoretic evidence for 2-group symmetries in 5d SCFTs was found via the dimensional reduction of 6d theories \cite{Zafrir:2025xca}, and the interplay between the mixed instanton--1-form anomaly of 5d Yang--Mills theories and the choice of global variant was investigated in \cite{Bertolini:2025wyj}.

In parallel, the \emph{blow-up equations} have emerged as one of the most powerful computational tools for the partition functions of theories with eight supercharges. Originally introduced by Nakajima and Yoshioka \cite{Nakajima:2003pg,Nakajima:2005fg,Gottsche:2006bm} for the Nekrasov partition function \cite{Nekrasov:2002qd,Nekrasov:2003rj} of supersymmetric Yang--Mills theory, they have been extended to arbitrary 5d $\CN=1$ gauge theories, and even to non-Lagrangian theories admitting no gauge-theory deformation \cite{Huang:2017mis,Kim:2019uqw,Kim:2020hhh,Kim:2021gyj,Kim:2023qwh,Kim:2025qaf}. The equations determine the partition function on the $\Omega$-background in terms of partition functions on the blown-up geometry, summed over magnetic fluxes threading the exceptional $\BP^1$; each flux is weighted by a classical prefactor $\exp(-V_n)$, fixed by the effective prepotential of the theory.

The purpose of this paper is to point out that the blow-up equations know about the generalised global symmetries of the theory, and to turn this observation into a practical, purely field-theoretic diagnostic. The starting point is the remark that the magnetic fluxes summed over in the blow-up equations take values in the coweight lattice of the gauge group, and that fluxes in a non-trivial coset of the coweight lattice modulo the coroot lattice are precisely the 't~Hooft fluxes of the electric 1-form symmetry. Evaluating $\exp(-V_n)$ on such a background, the \emph{fractional parts} of its exponents encode the 't~Hooft anomalies involving the 1-form symmetry: the exponent of the instanton fugacity measures the fractional instanton number, i.e.\ the mixed 1-form--$\U(1)_I$ anomaly; the constant part of the $(p_1 p_2)$-exponent packages the cubic self-anomaly of the 1-form symmetry together with its mixed gravitational and $\SU(2)_R$ couplings; and the exponents of the flavour fugacities yield the mixed 1-form--flavour anomalies. In this way, we reproduce, from an elementary evaluation, anomalies previously derived via geometry and the symmetry topological field theory \cite{BenettiGenolini:2020doj,Gukov:2020btk,Apruzzi:2021nmk,Apruzzi:2021vcu,Apruzzi:2022dlm}, and we obtain new ones.

Beyond anomalies, the same evaluation furnishes a criterion for 2-group symmetry. Demanding integrality of the gauge-fugacity exponents of $\exp(-V_n)$ at the generator of the genuine 1-form symmetry fixes the admissible background fluxes and leaves behind a definite fractional flux for the partner instanton and flavour symmetries --- the \emph{forced flux}. Whether the forced flux signals a genuine 2-group, a mixed 't~Hooft anomaly, or a split (trivial) extension is decided by the faithful continuous global symmetry of the ultraviolet fixed point, which we determine from the superconformal index \cite{Bhattacharya:2008zy,Kim:2012gu}. The use of superconformal indices as a probe of generalised global symmetries and of the faithful symmetry group is in the same spirit as the three-dimensional analyses of \cite{Beratto:2021xmn,Mekareeya:2022spm,Comi:2023lfm,Sacchi:2023omn,Grimminger:2024mks,Harding:2025vov,Garavaglia:2025cgz}. This interplay is illustrated most sharply by the $\SU(4)_{\kCS}$ theories with antisymmetric hypermultiplets: the blow-up returns the \emph{same} forced flux for $\kCS = 0, 2, 4$, and it is solely the faithful symmetry --- which varies with $\kCS$ --- that separates non-split 2-groups from mixed anomalies and split extensions. Our findings are summarised in Tables~\ref{tab:mixed} and~\ref{tab:2groupsummary}.

Our analysis covers both Lagrangian and non-Lagrangian theories: on the Lagrangian side, pure $\SU(N)_{\kCS}$ gauge theories, $\SU(4)_{\kCS}$ and $\USp(4)$ with antisymmetric hypermultiplets, and $\Spin(7)$ and $\Spin(8)$ with vector hypermultiplets; on the non-Lagrangian side, the toric families $B_N$ and $B_N^{(1,2,3)}$ of \cite{Eckhard:2020jyr, Morrison:2020ool} and the rank-two theories $\mathbb{P}^2\cup\mathbb{F}_3$ and $\mathbb{P}^2\cup\mathbb{F}_6$ of \cite{Jefferson:2018irk,Apruzzi:2019opn} (models 67 and 68 in the classification of \cite{Apruzzi:2019opn}). Along the way, we obtain a number of results that, to our knowledge, are new.
\begin{itemize}[leftmargin=*]
    \item Closed-form expressions for the blow-up function $\exp(-V_n)$ for a general gauge group with hypermultiplet matter, together with a simple two-step recipe fixing the quantisation of all background magnetic fluxes, as well as the effective prepotentials of several non-Lagrangian theories, including those of the $B_N$ and $B_N^{(1,2,3)}$ families, most of which have not appeared in the literature.
    \item For the non-Lagrangian families, the quantisation of the magnetic fluxes, including their spin$^c$ offsets, which must be carefully distinguished from the bare 1-form symmetry backgrounds on which the anomalies are evaluated.
    \item The cubic 1-form anomalies of the rank-two theories $\mathbb{P}^2\cup\mathbb{F}_3$ and $\mathbb{P}^2\cup\mathbb{F}_6$, together with the mixed 1-form--flavour anomalies of the $B_N^{(2)}$ and $B_N^{(3)}$ families. The anomaly of $\mathbb{P}^2\cup\mathbb{F}_3$ first arises as a mixed 1-form--gravitational anomaly; a mod $24$ counterterm ambiguity allows us to recast it as a purely cubic one. This theory arises from the isolated orbifold $\mathbb{C}^3/\Z_5(1,1,3)$, and its anomaly can also be extracted by the complementary methods of \cite{Cvetic:2025lat,Chakrabhavi:2026iku}, based on boundary $\eta$-invariants and on quiver Hilbert series, respectively. Our blow-up derivation is independent of these; moreover, the mod $24$ ambiguity we exploit has the same index-theoretic origin as the $\Z_6$ redundancy discussed in \cite{Cvetic:2025lat}. To the best of our knowledge, the remaining results have not appeared in the literature.
    \item The mixed 1-form--$\mathrm{SU}(2)_R$ anomalies of all of the above theories. For the $E_0$ theory, this furnishes an independent, intrinsically five-dimensional derivation of the mixed anomaly recently proposed in \cite{Sacchi:2023omn}.
    \item The superconformal index of $\USp(4)_{\theta=0} + 2\Lambda^2$ from the blow-up equations, including the subtleties associated with the $\Lambda$ factors and the decoupled free hypermultiplet. This computation is carried out in parallel with that of $\SU(3)_{\kCS=6}$ --- many steps of which were presented in \cite{Kim:2020hhh,Kim:2023qwh}, and for which we spell out further details.
\end{itemize}

The rest of the paper is organised as follows. In \S\ref{sec:blow-up}, we review the blow-up equations for 5d $\CN=1$ theories on the $\Omega$-background. In \S\ref{sec:expV}, we present the blow-up function $\exp(-V_n)$ in closed form for the theories of interest, and derive the quantisation of the gauge and background magnetic fluxes. In \S\ref{sec:anomalies}, we extract the 't~Hooft anomalies involving the 1-form symmetry from the fractional parts of the exponents of $\exp(-V_n)$. In \S\ref{sec:faithfulsym}, we determine the faithful continuous global symmetries of the ultraviolet fixed points from their superconformal indices. In \S\ref{sec:twogroup}, we formulate the flux criterion for 2-group symmetries and apply it to all of the theories under consideration. Appendices~\ref{app:BN} and~\ref{app:PF} contain the detailed toric analysis of the non-Lagrangian theories, and appendix~\ref{app:blow-upSU36USp4} collects the blow-up computation of the superconformal indices of $\SU(3)_{\kCS=6}$ and $\USp(4)_{\theta=0} + 2\Lambda^2$.

\section{Review of the blow-up equations}
\label{sec:blow-up}
In this section, we review the blow-up approach to compute instanton partition functions of 5d $\CN=1$ gauge theories. The main protagonist of this strategy is a set of equations, known as {\it blow-up equations}, first introduced in \cite{Nakajima:2003pg,Nakajima:2005fg,Gottsche:2006bm} to study the Nekrasov partition function \cite{Nekrasov:2002qd,Nekrasov:2003rj} of 4d and 5d supersymmetric Yang--Mills theory for any gauge group \cite{Keller:2012da} (see also \cite{Benvenuti:2010pq,Keller:2011ek,Hanany:2012dm,Cremonesi:2014xha} for complementary methods), and later extended to the case of generic 5d $\CN=1$ gauge theories in \cite{Huang:2017mis,Kim:2019uqw,Kim:2020hhh} (see also \cite{Kim:2021gyj,Kim:2025qaf} for partition functions with the insertion of defects, as well as \cite{Gu:2018gmy,Gu:2019dan,Gu:2019pqj,Gu:2020fem,Kim:2023glm} for 6d theories).

Before starting our discussion on the blow-up equations, we review some preliminary concepts regarding 5d $\CN=1$ gauge theories, namely the $\Omega$--deformed partition function and the effective prepotential, which are necessary tools for understanding the subsequent parts of this section.

\subsection{The $\Omega$--deformed partition function}
The starting point is the $\Omega$--deformed partition function of a given 5d theory on $\BC^2\times S^1$ with $\Omega$--deformed parameters $\epsilon_{1,2}$, given by the following Witten index:
\bes{ \label{Ztrace}
Z = \Tr (-1)^F p_1^{\left(j_1 + j_R\right)} p_2^{\left(j_2 + j_R\right)} \prod_i z_i^{G_i} \prod_j w_j^{H_j}\,,
}
where $(-1)^F$ is the fermion number operator, and we introduce $p_{1,2}$, which are related to the $\Omega$--deformed parameters by $p_i = e^{-\epsilon_i}$. Moreover, the Cartan generators of the $\SO(4)$ Lorentz rotations and the $\SU(2)_R$ R-symmetry are denoted as $j_{1,2}$ and $j_R$, respectively. Finally, $z_i$ and $w_j$ stand for the gauge and global symmetry fugacities, respectively, with $G_i$ and $H_j$ the corresponding charges. The partition function \eref{Ztrace} receives classical, one-loop and instanton contributions, meaning it can be factorised as
\bes{ \label{Zclass-oneloop-inst}
Z = Z_{\text{class}} Z_{\text{one-loop}} Z_{\text{inst}}\,,
}
where, for a theory with gauge group $G$ of rank $r$ and $N_{\mathbf{R}}$ hypermultiplets transforming in the representation $\mathbf{R}$ of $G$, the one-loop contribution is given by
\bes{ \label{Zoneloop}
Z_{\text{one-loop}} = \PE\Bigg[&-\frac{1 + p_1 p_2}{\left(1-p_1\right) \left(1-p_2\right)} \sum_{\alpha \in \Delta^+} \prod_{i=1}^r z^{\alpha_i}_i \\ &+ \frac{\sqrt{p_1 p_2}}{\left(1-p_1\right) \left(1-p_2\right)} \sum_{h=1}^{N_{\mathbf{R}}} \sum_{\lambda \in \mathbf{R}_h} e^{-\abs{\langle \lambda, f\rangle + m_h}}\Bigg]\,,
}
with $\Delta^+$ being the set of positive roots of the Lie algebra $\fg$ associated with $G$, and $\lambda \in \mathbf{R}_h$ denoting the weights of the $h$--th hypermultiplet, where the gauge and flavour fugacities are given by $z_i = e^{f_i}$, $w_h = e^{-m_h}$. We also recall that the plethystic exponential is defined as $\PE\left[f(a)\right]= \exp\left[\sum_{n=1}^{\infty} \frac{1}{n} f(a^n)\right]$. The explicit expressions for the set of positive roots and the weights for the various theories considered in this paper are collected in appendix~\ref{app:grouptheory} and in the table \eref{eq:letters}, respectively.

The classical contribution in \eref{Zclass-oneloop-inst} can instead be directly derived as 
\bes{
Z_{\text{class}} = e^{\mathcal{E}\left(\epsilon_1, \epsilon_2, f_i, m_j\right)}\,,
}
where $\mathcal{E}\left(\epsilon_1, \epsilon_2, f_i, m_j\right)$ is the {\it effective prepotential}, which we discuss in \S\ref{sec:EffPrep}, whereas $f_i$ and $m_j$ are the chemical potentials for the gauge and global symmetry, respectively, explicitly related to the corresponding fugacities as $z_i = e^{f_i}$, with $i = 1, \ldots, r$, and $w_j = e^{-m_j}$, with $j = 0, 1, \ldots, N_{\mathbf{R}}$. In particular, $w_0 = e^{-m_0}\equiv q$ is identified with the fugacity associated with the instantonic symmetry.

Finally, the instanton contribution in \eref{Zclass-oneloop-inst} is expressed as an infinite power series in the instanton fugacity $q = e^{-m_0}$:
\bes{ \label{Zinst}
Z_{\text{inst}}(p_1,p_2; q; z_i; w_h) &= 1 + \sum_{k=1}^{\infty} q^k Z_{\text{inst}, \, (k)}(p_1,p_2; z_i; w_h)\,,
}
with $Z_{\text{inst}, \, (k)}$ denoting the partition function for $k$ $G$ instantons, which can be computed either using the standard ADHM construction \cite{Nekrasov:2002qd,Nekrasov:2003rj,Nekrasov:2004vw,Fucito:2004gi,Shadchin:2005mx,Kim:2012gu,Hwang:2014uwa,Bergman:2015dpa} if the UV completion of the corresponding instanton moduli space is known, or by solving the blow-up equations. Since the ADHM construction for the most general instanton moduli space of a theory with gauge group $G$ and matter is not known, we mostly rely on the blow-up equations approach, which is presented in \S\ref{sec:blow-upEq}.

\subsection{The effective prepotential of 5d $\CN=1$ gauge theories} \label{sec:EffPrep}
The effective prepotential evaluated on the $\Omega$--background is given in \cite[(2.12)]{Kim:2020hhh} to be
\bes{ \label{EffectivePrepotential}
\mathcal{E}\left(\epsilon_1, \epsilon_2, f_i, m_j\right) = \frac{1}{\epsilon_1 \epsilon_2} \left[\mathcal{F} + \frac{1}{48} \sum_{i=1}^r C^G_i f_i \left(\epsilon^2_1 + \epsilon^2_2\right) + \frac{1}{2} \sum_{i=1}^r C^R_i f_i \epsilon^2_+\right]\,,
}
where we discuss each term appearing in the expression above separately.

The first term appearing in the square bracket in \eref{EffectivePrepotential} is the cubic {\it prepotential} on the Coulomb branch, which, for gauge group $G$ with Chern--Simons level $\kCS$ and matter consisting of $N_{\mathbf{R}}$ hypermultiplets in the representation $\mathbf{R}$ of $G$, reads \cite{Seiberg:1996bd,Intriligator:1997pq}
\bes{ \scalebox{1.01}{$
\mathcal{F}= \frac{m_0}{2} \sum\limits_{i=1}^{\dim \mathbf{F}} \tilde{\phi}^2_i + \frac{\kCS}{6} \sum\limits_{i=1}^{\dim \mathbf{F}} \tilde{\phi}^3_i + \frac{1}{12} \left(\sum\limits_{\alpha \in \Delta} \abs{\langle \alpha, f \rangle}^3 - \sum\limits_{h=1}^{N_{\mathbf{R}}} \sum\limits_{\lambda \in \mathbf{R}_h} \abs{\langle \lambda, f \rangle + m_h}^3\right)\,,
$} \label{Prepotential}
}
with the mass parameters $m_j = \left\{m_0, m_1, \ldots, m_{N_{\mathbf{R}}}\right\}$, where $m_0 = \frac{1}{g^2_0}$ is the square of the inverse gauge coupling. In \eref{Prepotential}, $\alpha \in \Delta$ and $\lambda \in \mathbf{R}_h$ are the set of roots and the set of weights for the $h$--th hypermultiplet in the representation $\mathbf{R}$ of $G$, respectively, which, for the gauge theories considered in this paper, are reported explicitly in \S\ref{sec:expV}. Furthermore, we denote with $\mathbf{F}$ the fundamental representation of $G$, whereas $\tilde{\phi}_i$, with the index $i$ running over the fundamental representation, are the orthonormal Coulomb branch VEVs. Explicitly, for the gauge groups of interest in this paper, they are given in terms of the chemical potentials $f_1, \ldots, f_r$ for the gauge fugacities $z_i = e^{f_i}$ as follows:
\begin{subequations} 
\begin{align} \label{phitofSUN}
\SU(N) \, :& \, \tilde{\phi}_i = \left\{f_1, f_2 - f_1, \ldots, f_{N-1}-f_{N-2}, - f_{N-1}\right\}\,,
\\ \USp(2N) \, :& \, \tilde{\phi}_i = \pm \left\{f_1, f_2 - f_1, \ldots, f_{N-1}-f_{N-2}, f_N - f_{N-1}\right\}\,,
\\ \Spin(2N+1) \, :& \, \tilde{\phi}_i = \{0\} \cup \pm \left\{f_1, f_2 - f_1, \ldots, f_{N-1}-f_{N-2}, 2 f_N - f_{N-1}\right\} \,,
\\ \Spin(2N) \, :& \, \scalebox{0.87}{$ \displaystyle \tilde{\phi}_i = \pm \left\{f_1, f_2 - f_1, \ldots, f_{N-2}-f_{N-3}, f_N + f_{N-1} - f_{N-2}, f_N - f_{N-1}\right\} $} \,. \label{phitofSpin2N}
\end{align} 
\end{subequations}
Note that the cubic term in \eref{Prepotential} containing the Chern--Simons level, \ie $\sum_{i=1}^{\dim \mathbf{F}} \tilde{\phi}^3_i$, is non-zero only for $\SU(N)$ gauge group, with $N \ge 3$.\footnote{For a $G_1 \times \cdots \times G_n$ gauge theory with $n$ non-Abelian subgroups of the whole gauge group, the prepotential can be obtained from \eref{Prepotential} by summing the first two terms (the ones outside the round bracket), as well as the W-boson contribution, namely the first term inside the round bracket, over all the possible groups $G_a$, with $a = 1, \ldots, n$. Moreover, one has to appropriately couple the matter contributions to all of the various subgroups $G_1, \ldots, G_n$ of the total gauge group.}

Next, let us move on to the second term appearing inside the square bracket in the effective prepotential \eref{EffectivePrepotential}. The coefficients $C^G_i$ are the levels of the mixed gauge/gravitational Chern--Simons terms \cite[(4.27)]{Bonetti:2011mw}
\bes{ \label{MixedCSCG}
S_{\text{gauge/grav}} = -\frac{1}{48} \int \sum_{i=1}^r C^G_i A_i \wedge p_1\left(T\right)\,,
}
where $A$ is the 5d one-form vector field and $p_1\left(T\right)$ is the Pontryagin class of the tangent bundle on the 5d spacetime. In detail, the coefficients $C^G_i$ are given by \cite[(2.5)]{Kim:2020hhh}
\bes{ \label{CiGexpression}
C^G_i = -\frac{\partial}{\partial f_i} \left(\sum_{\alpha \in \Delta} \abs{\langle \alpha, f \rangle} - \sum\limits_{h=1}^{N_{\mathbf{R}}} \sum\limits_{\lambda \in \mathbf{R}_h} \abs{\langle \lambda, f \rangle + m_h}\right)\,,
}
and the second term inside the square bracket in \eref{EffectivePrepotential} can be interpreted as an equivariant integral of the mixed Chern--Simons terms \eref{MixedCSCG} upon performing the replacement $p_1\left(T\right) \rightarrow -\left(\epsilon^2_1 + \epsilon^2_2\right)$.

Similarly, the last term appearing in the effective prepotential \eref{EffectivePrepotential} can be thought of as an equivariant integral of the mixed gauge/$\SU(2)_R^2$ Chern--Simons terms \cite[(2.1)]{Bonetti:2013ela}
\bes{ \label{MixedCSCR}
S_{\text{gauge/$\SU(2)_R^2$}} = \frac{1}{2} \int \sum_{i=1}^r C^R_i A_i \wedge c_2\left(R\right)\,,
}
where $c_2(R)$ is the second Chern class of the $\SU(2)_R$ bundle, upon performing the substitution $c_2(R) \rightarrow \epsilon^2_+$, where the latter is defined as $\epsilon_+ = \frac{1}{2} \left(\epsilon_1 + \epsilon_2\right)$.\footnote{Note that, upon redefining $p_1 = x y$ and $p_2 = \frac{x}{y}$, with $x$ the R-symmetry fugacity and $y$ the fugacity associated with spacetime $\SU(2)$ rotations, then $x = e^{-\epsilon_+}$, \ie $\epsilon_+$ is exactly the chemical potential for the R-symmetry fugacity. It is thus clear that the last term in \eref{EffectivePrepotential} is associated with mixed gauge/$\SU(2)_R^2$ Chern--Simons factors. \label{foot:p1p2tox}} Explicitly, the coefficients $C^R_i$ are defined as \cite[(2.7)]{Kim:2020hhh}
\bes{
C^R_i = \frac{1}{2} \frac{\partial}{\partial f_i} \sum_{\alpha \in \Delta} \abs{\langle \alpha, f \rangle}\,,
}
which, using the conventions detailed in \eref{phitofSUN}--\eref{phitofSpin2N}, turn out to be $C^R_i = 2$ for all $i = 1, \ldots, r$, see \eg \cite[(3.10)]{BenettiGenolini:2019zth}.

\subsection{The blow-up equations} \label{sec:blow-upEq}
The main idea of the blow-up equations approach is to replace $\BC^2 \times S^1$, on which the partition function \eref{Ztrace} is defined, with $\hat{\BC}^2 \times S^1$, where, in the blow-up geometry $\hat{\BC}^2$, a two-sphere $\BP^1$ is inserted at the origin of the complex plane $\BC^2$. The blow-up space $\hat{\BC}^2$ can be parametrised by coordinates $\left(z_0, z_1, z_2\right) \sim \left(\lambda^{-1} z_0, \lambda z_1, \lambda z_2\right)$, with $\lambda \in \BC\setminus\{0\}$, where $\BP^1$ is inserted at the locus $z_0 = 0$. The Cartan generators $j_{1,2}$ of the Lorentz rotations induce a $\U(1)^2$ action $T$ on the aforementioned coordinates $T \, : \,\left(z_0, z_1, z_2\right) \rightarrow \left(z_0, e^{\epsilon_1} z_1, e^{\epsilon_2} z_2\right)$, with instantons located at the fixed points of such action, namely the north pole $(0,1,0)$ and the south pole $(0,0,1)$. The local $\lambda$--invariant coordinates around the north and south pole are given by $(z_0 z_1,\frac{z_2}{z_1})$ and $(z_0 z_2,\frac{z_1}{z_2})$, respectively, which are acted upon by $T$ as follows:
\begin{subequations} 
\begin{align} \label{TactionNpole}
\text{North pole} \, :& \, \left(z_0 z_1,\frac{z_2}{z_1}\right) \rightarrow \left(e^{\epsilon_1} z_0 z_1, e^{\epsilon_2 - \epsilon_1} \frac{z_2}{z_1}\right) \,,
\\ \text{South pole} \, :& \, \left(z_0 z_2,\frac{z_1}{z_2}\right) \rightarrow \left(e^{\epsilon_2} z_0 z_2,e^{\epsilon_1 - \epsilon_2} \frac{z_1}{z_2}\right) \,. \label{TactionSpole}
\end{align} 
\end{subequations}
The partition function $\hat{Z}$ defined on $\hat{\BC}^2 \times S^1$ can then be defined by summing over all possible instanton configurations located at the two poles, which are labelled by magnetic fluxes $n_i$ on the two-sphere $\BP^1$, with $i = 1, \ldots, r$ associated with the gauge group $G$ of rank $r$. Importantly, the definition of the blow-up partition function $\hat{Z}$ differs from the one for the partition function $Z$ given in \eref{Ztrace} since, in the presence of a vector of magnetic fluxes $n$ on $\BP^1$, the fermion number operator gets shifted by the quantity $\langle e, n\rangle$, where $e$ is the vector of electric charges of a state. Thus, the fermion number operator is mapped into $(-1)^F \rightarrow (-1)^{F + \langle e, n \rangle} = (-1)^{2 j_R}$, and the partition function $\hat{Z}$ can be defined by just replacing $(-1)^F$ with $(-1)^{2 j_R}$ in \eref{Ztrace} \cite[(3.3)]{Kim:2020hhh} (see also \cite[(2.6)]{Dimofte:2011py} and \cite[(A.2)]{Aharony:2013dha} for analogous considerations in 3d). In particular, since the fermion number operator in \eref{Ztrace} can also be defined as $(-1)^{2 j_1}$, such phase shift $(-1)^{\langle e, n \rangle}$ relating $Z$ to $\hat{Z}$ can simply be implemented by shifting the chemical potential $\epsilon_1$ as
\bes{ \label{ZtoZhatshift}
&\epsilon_1 \rightarrow \epsilon_1 + 2 \pi i \,, \quad \text{\ie} \quad p_1^\fr \rightarrow p_1^\fr e^{-2 \pi i \fr}\,, \quad \text{with $\fr \in \mathbb{Q}$}\, , \\
& \hat{Z} \left(p_1,p_2; q; z_i; w_h\right) = Z \left(p_1 e^{-2 \pi i},p_2; q; z_i; w_h\right) \,.
}
In addition to the gauge fluxes, one can also turn on magnetic fluxes $B_{m_0}$ and $B_{m_h}$ for the instantonic and flavour symmetries. Hence, putting everything together, the celebrated blow-up equations can finally be recovered from the observation that the partition function $\hat{Z}$ on $\hat{\BC}^2 \times S^1$ can be expressed by summing over all possible partition function configurations located at the north and at the south pole, denoted by $\hat{Z}^{(N)}$ and $\hat{Z}^{(S)}$, respectively, as \cite[(6.9) and (6.10)]{Nakajima:2003pg}
\bes{ \label{hatZeqhatZNhatZS}
& \Lambda \left(p_1, p_2; q; m_h\right) \hat{Z} \left(p_1,p_2; q, B_{m_0}; z_i; w_h, B_{m_h}\right) \\ &= \sum_{\left(n_1, \ldots, n_r\right)} (-1)^{\sum_{i=1}^r n_i} \hat{Z}^{(N)} \left(p_1 , \frac{p_2}{p_1}; q p^{B_{m_0}}_1; z_i p^{-n_i}_1; w_h p^{B_{m_h}}_1\right) \\ & \qquad \qquad \qquad \qquad \, \times \hat{Z}^{(S)} \left(\frac{p_1}{p_2} , p_2; q p^{B_{m_0}}_2; z_i p^{-n_i}_2; w_h p^{B_{m_h}}_2\right)\,,
}
where the parameter $\Lambda$, whose role will be remarked below, does not depend on the gauge fugacities $z_i$, and the allowed values of the magnetic fluxes can be deduced either from geometric considerations, as discussed in \cite{Kim:2020hhh}, or by requiring consistency in solving the blow-up equations, as we discuss explicitly in \S\ref{sec:expV}. In \eref{hatZeqhatZNhatZS}, the superscripts $(N)$ and $(S)$ mean that $\hat{Z}^{(N)}$ and $\hat{Z}^{(S)}$ are given by $\hat{Z}$, with a redefinition of the fugacities, as indicated by the arguments in round brackets, due to the presence of magnetic fluxes, as well as according to \eref{TactionNpole} and \eref{TactionSpole}. Explicitly, for the partition function at the north pole, \eref{TactionNpole} tells us that the fugacity $p_1 = e^{-\epsilon_1}$ remains untouched, whereas the fugacity $p_2 = e^{-\epsilon_2}$ is redefined as $p_2 \rightarrow \frac{p_2}{p_1} = e^{-(\epsilon_2 - \epsilon_1)}$ with respect to $\hat{Z}$; moreover, the gauge fluxes induce a shift in the chemical potentials, which translates into the redefined fugacities $z_i = e^{f_i} \rightarrow z_i p^{-n_i}_1 = e^{f_i - (-n_i \epsilon_1)}$ and $w_j = e^{-m_j} \rightarrow w_j p^{B_{m_j}}_1 = e^{-(m_j + B_{m_j} \epsilon_1)}$, where $w_0 = q$, with respect to $\hat{Z}$. Analogously, for the partition function at the south pole, the fugacity $p_2 = e^{-\epsilon_2}$ remains untouched, whereas the fugacity $p_1 = e^{-\epsilon_1}$ is redefined as $p_1 \rightarrow \frac{p_1}{p_2} = e^{-(\epsilon_1 - \epsilon_2)}$ with respect to $\hat{Z}$ due to \eref{TactionSpole}; also in this case, the chemical potentials get shifted by the gauge fluxes, such that the corresponding fugacities are redefined as $z_i = e^{f_i} \rightarrow z_i p^{-n_i}_2 = e^{f_i - (-n_i \epsilon_2)}$ and $w_j = e^{-m_j} \rightarrow w_j p^{B_{m_j}}_2 = e^{-(m_j + B_{m_j} \epsilon_2)}$ with respect to $\hat{Z}$.

Using the factorisation \eref{Zclass-oneloop-inst}, the blow-up equations \eref{hatZeqhatZNhatZS} can be recast as
\bes{ \label{blow-upEq}
& \Lambda \left(p_1, p_2; q; m_h\right) \hat{Z}_{\text{inst}} \left(p_1,p_2; q, B_{m_0}; z_i; w_h, B_{m_h}\right) \\ &= \sum_{\left(n_1, \ldots, n_r\right)} (-1)^{\sum_{i=1}^r n_i} e^{-V_n} \frac{\hat{Z}^{(N)}_{\text{one-loop}} \hat{Z}^{(S)}_{\text{one-loop}}}{\hat{Z}_{\text{one-loop}}} \\ & \qquad \qquad \qquad \qquad \, \times \hat{Z}^{(N)}_{\text{inst}} \left(p_1 , \frac{p_2}{p_1}; q p^{B_{m_0}}_1; z_i p^{-n_i}_1; w_h p^{B_{m_h}}_1\right) \\ & \qquad \qquad \qquad \qquad \, \times \hat{Z}^{(S)}_{\text{inst}} \left(\frac{p_1}{p_2} , p_2; q p^{B_{m_0}}_2; z_i p^{-n_i}_2; w_h p^{B_{m_h}}_2\right)\,,
}
where the function $V_n$\footnote{We emphasise the dependence of such function on the gauge fluxes with the subscript $n$.} is the following linear combination of the effective prepotential \eref{EffectivePrepotential} and the ones evaluated at the north and south poles:
\bes{
V_n =& \mathcal{E}^{(N)}\left(\epsilon_1, \epsilon_2 - \epsilon_1, f_i + n_i \epsilon_1, m_j + B_{m_j} \epsilon_1\right) \\ + &\, \mathcal{E}^{(S)}\left(\epsilon_1 - \epsilon_2, \epsilon_2, f_i + n_i \epsilon_2, m_j + B_{m_j} \epsilon_2\right) - \mathcal{E}\left(\epsilon_1, \epsilon_2, f_i, m_j\right)\,,
}
with the same fugacity redefinitions at the north and south poles as discussed below \eref{hatZeqhatZNhatZS}. Throughout this paper, we refer to the quantity $\exp(-V_n)$ as the {\it blow-up function}. Observe that, using \eref{ZtoZhatshift}, the blow-up one-loop contribution ${\hat{Z}_{\text{one-loop}}}$ is given by \eref{Zoneloop} with a minus sign in front of $\sqrt{p_1 p_2}$ in the second line. As detailed below \eref{hatZeqhatZNhatZS}, $\hat{Z}^{(N)}_{\text{one-loop}}$ is given by ${\hat{Z}_{\text{one-loop}}}$, with the redefinitions $(p_1, p_2, z_i, w_h) \rightarrow (p_1, \frac{p_2}{p_1}, z_i p^{-n_i}_1, w_h p^{B_{m_h}}_1)$, whereas $\hat{Z}^{(S)}_{\text{one-loop}}$ is obtained by sending $(p_1, p_2, z_i, w_h) \rightarrow (\frac{p_1}{p_2}, p_2, z_i p^{-n_i}_2, w_h p^{B_{m_h}}_2)$.

Practically, the equations \eref{blow-upEq} can be solved for the unknown functions $\hat{Z}_{\text{inst}}$, $\hat{Z}^{(N)}_{\text{inst}}$ and $\hat{Z}^{(S)}_{\text{inst}}$ by expanding them as power series in the instantonic fugacity $q$ as in \eref{Zinst}. Additionally, we also expand $\Lambda$ in powers of $q$ as
\bes{ \label{LambdaPowerq}
\Lambda\left(p_1, p_2; q; m_h\right) =\sum_{k=0}^{\infty} q^k \Lambda_{(k)}\left(p_1, p_2; m_h\right) \,.
}
Plugging in \eref{Zinst} and \eref{LambdaPowerq} in \eref{blow-upEq}, we finally obtain
\bes{ \label{blow-upEqPowerq}
& \sum_{k_1=0}^{\infty} q^{k_1} \Lambda_{(k_1)}\left(p_1, p_2; m_h\right) \left[1 + \sum_{k_2=1}^{\infty} q^{k_2} \hat{Z}_{\text{inst}, \, (k_2)}(p_1,p_2; B_{m_0}; z_i; w_h, B_{m_h})\right] \\ &= \sum_{\left(n_1, \ldots, n_r\right)} (-1)^{\sum_{i=1}^r n_i} e^{-V_n} \frac{\hat{Z}^{(N)}_{\text{one-loop}} \hat{Z}^{(S)}_{\text{one-loop}}}{\hat{Z}_{\text{one-loop}}} \\ & \qquad \qquad \, \times \left[1 + \sum_{k_3=1}^{\infty} \left(q p^{B_{m_0}}_1\right)^{k_3} \hat{Z}^{(N)}_{\text{inst}, \, (k_3)}\left(p_1,\frac{p_2}{p_1}; z_i p^{-n_i}_1;w_h p^{B_{m_h}}_1\right)\right] \\ & \qquad \qquad \, \times \left[1 + \sum_{k_4=1}^{\infty} \left(q p^{B_{m_0}}_2\right)^{k_4} \hat{Z}^{(S)}_{\text{inst}, \, (k_4)}\left(\frac{p_1}{p_2},p_2; z_i p^{-n_i}_2;w_h p^{B_{m_h}}_2\right)\right]\,,
}
which can be solved iteratively order by order in $q$ for the three unknown functions $\hat{Z}_{\text{inst}, \, (k)}$, $\hat{Z}^{(N)}_{\text{inst}, \, (k)}$ and $\hat{Z}^{(S)}_{\text{inst}, \, (k)}$. Since \eref{blow-upEqPowerq} are the blow-up equations we actually want to solve to compute instanton partition functions, let us provide the following comments to understand better how the solutions can be worked out.
\ben
\item Since, at each order $k$ in $q$, there are three unknown functions $\hat{Z}_{\text{inst}, \, (k)}$, $\hat{Z}^{(N)}_{\text{inst}, \, (k)}$ and $\hat{Z}^{(S)}_{\text{inst}, \, (k)}$, we need (at least) three different values of the background fluxes $(B_{m_0}, B_{m_h})$ to find a solution of \eref{blow-upEqPowerq}. The possible admissible values of such fluxes are discussed extensively in \S\ref{sec:expV}. Once the equations are solved at order $k$ and the explicit expressions of $\hat{Z}_{\text{inst}, \, (k)}$, $\hat{Z}^{(N)}_{\text{inst}, \, (k)}$ and $\hat{Z}^{(S)}_{\text{inst}, \, (k)}$ are figured out, such $k$--instanton solutions are plugged in to solve iteratively the equations \eref{blow-upEqPowerq} at the next order in $q$, \ie $k+1$.
\item The parameter $\Lambda$ does not depend on the gauge fugacities $z_i$ and can be fixed as follows. At each order $k$ in $q$, the right-hand side of \eref{blow-upEqPowerq} can be evaluated as a function of the unknown functions $\hat{Z}_{\text{inst}, \, (k)}$, $\hat{Z}^{(N)}_{\text{inst}, \, (k)}$ and $\hat{Z}^{(S)}_{\text{inst}, \, (k)}$ by plugging in the solutions for the $p$--instanton partition functions found at order $p < k$. We thus just focus on the $\hat{Z}_{\text{inst}, \, (k)}$--independent (as well as $\hat{Z}^{(N), (S)}_{\text{inst}, \, (k)}$--independent) terms of the right-hand side of \eref{blow-upEqPowerq}, and extract its $z_i$--independent part. We then define
\bes{ \label{eq:defLambda}
\Lambda_{(k)}\left(p_1, p_2; m_h\right) = \left[\text{RHS of \eref{blow-upEqPowerq}} \lvert_{q^k}\right]_{\text{$z_i$ independent}} \,,
}
where $\lvert_{q^k}$ denotes the coefficient of $q^k$ in the power expansion in $q$.\footnote{In practice, in order to compute $\Lambda_{(k)}$, it is useful to expand the coefficient of $q^k$ of the right-hand side of \eref{blow-upEqPowerq} as a power series in $z_i$ around $z_i = \infty$, and subsequently identify $\Lambda_{(k)}$ with the zeroth order coefficient of such series expansion.}
\item We remark again that the solutions $\hat{Z}^{(N)}_{\text{inst}, \, (k)}$ and $\hat{Z}^{(S)}_{\text{inst}, \, (k)}$ are equal to the solution $\hat{Z}_{\text{inst}, \, (k)}$, with the fugacity redefinitions as explained below \eref{hatZeqhatZNhatZS}. Explicitly:
\bes{
& \hat{Z}^{(N)}_{\text{inst}, \, (k)} = \hat{Z}_{\text{inst}, \, (k)}\Big\lvert_{\left(p_1, p_2, z_i, w_h\right) \rightarrow \left(p_1, \frac{p_2}{p_1}, z_i p^{-n_i}_1, w_h p^{B_{m_h}}_1\right)}\, , \\
& \hat{Z}^{(S)}_{\text{inst}, \, (k)} = \hat{Z}_{\text{inst}, \, (k)}\Big\lvert_{\left(p_1, p_2, z_i, w_h\right) \rightarrow \left(\frac{p_1}{p_2}, p_2, z_i p^{-n_i}_2, w_h p^{B_{m_h}}_2\right)}\, .
}
\item Once the solution to the blow-up equations $\hat{Z}_{\text{inst}, \, (k)}$ is recovered, the $k$--instanton partition function ${Z}_{\text{inst}, \, (k)}$ on $\BC^2 \times S^1$ can be recovered by applying the inverse of the map \eref{ZtoZhatshift}, \ie
\bes{ \label{MapZfromZhat}
Z_{\text{inst}, \, (k)} \left(p_1,p_2; z_i; w_h\right) = \hat{Z}_{\text{inst}, \, (k)} \left(p_1 e^{2 \pi i},p_2; z_i; w_h\right) \,.
}
As anticipated in Footnote \ref{foot:p1p2tox}, it is convenient to introduce the variables $x$ and $y$, defined by
\bes{ \label{eq:xydef}
&p_1 = x y\,, \qquad p_2 = \frac{x}{y}\,, \\ \text{\ie} \qquad & x = \left(p_1 p_2\right)^{1/2} = e^{-\epsilon_+}\,, \qquad y = \left(p_1/p_2\right)^{1/2}\,.
}
It follows from \eref{Ztrace} that $x$ is the fugacity associated with the combination $j_1 + j_2 + 2 j_R$, which involves the Cartan generator $j_R$ of the $\SU(2)_R$ R-symmetry; we will loosely refer to $x$ as the $\SU(2)_R$ fugacity. Similarly, $y$ is the fugacity associated with the Cartan generator $j_1 - j_2$ of another $\SU(2)$, related to the spacetime rotations, which we denote by $\SU(2)_y$. Since $y$ is an $\SU(2)$ fugacity, the instanton partition function is expected to be invariant under the Weyl reflection $y \rightarrow y^{-1}$. In terms of these variables, the map \eref{MapZfromZhat} amounts to changing $x \rightarrow -x$ and $y \rightarrow -y$ simultaneously, since both $x$ and $y$ acquire a phase $e^{i \pi}$; equivalently, thanks to the invariance under $y \rightarrow y^{-1}$, it can also be implemented as $x \rightarrow -x$ together with $y \rightarrow -y^{-1}$.
\item An additional consistency check to ensure the correctness of the solution $\hat{Z}_{\text{inst}, \, (k)}$ recovered by solving the blow-up equations \eref{blow-upEqPowerq} proceeds as follows: writing the solution in the $(x, y)$ variables introduced above, and recalling that $x$ is the $\SU(2)_R$ fugacity, one verifies that the instanton partition function is invariant under $x \rightarrow \frac{1}{x}$ (see \cite{Rodriguez-Gomez:2013dpa} and \cite[below (11)]{Bergman:2013aca}), \ie $\hat{Z}_{\text{inst}, \, (k)} (x) = \hat{Z}_{\text{inst}, \, (k)} (\frac{1}{x})$. A further non-trivial consistency check, which also validates the choice of the gauge and background magnetic fluxes, is that every coefficient of the power series in $x$ of the instanton partition function, at each instanton level, must be expressible in terms of characters of representations of the gauge symmetry, of the global symmetries, and of the $\SU(2)_y$ symmetry. Let us also point out an important advantage of the blow-up approach: the instanton partition function at each instanton level obtained from the blow-up equations is \emph{automatically} invariant under $x \rightarrow x^{-1}$, whereas, in the ADHM construction, extra contributions typically have to be supplemented by hand in order to restore this property \cite{Bergman:2013ala,Bergman:2013aca,Hwang:2014uwa,Zafrir:2014ywa,Gaiotto:2015una,Bergman:2015dpa,Zafrir:2015uaa}.
\een
For the sake of clarity, we explicitly solve the blow-up equations in the case of the $\SU(2)$ gauge theory with one fundamental hypermultiplet in \S\ref{sec:exSU2w1F}.\footnote{The explicit computation for the pure $\SU(2)$ gauge theory is described in \cite[Section 3.2.1]{Kim:2020hhh}.} Other examples, such as the $\SU(3)_{\kCS=6}$ gauge theory and the $\USp(4)_{\theta = 0}$ gauge theory with two antisymmetric hypermultiplets, will be carried out explicitly in appendix~\ref{app:blow-upSU36USp4}.

\subsubsection{Superconformal indices} \label{sec:index}
The instanton partition functions derived from the above computation are useful for computing the superconformal index. Being a Witten index, the superconformal index is invariant under supersymmetry-preserving continuous deformations and, in particular, along the renormalisation group flow. It therefore captures the protected multiplets --- most notably the conserved-current multiplets --- of the SCFT at the ultraviolet fixed point, even when computed from the infrared gauge theory description. This is the reason why the faithful continuous global symmetry of the UV fixed point can be read off from the index of the corresponding gauge theory, as we will do in \S\ref{sec:exSU2w1F} and, systematically, in \S\ref{sec:faithfulsym}. Let us give some general details regarding the 5d superconformal index \cite{Bhattacharya:2008zy}, which is the supersymmetric partition function on $S^4 \times S^1$. Thanks to localisation, the superconformal index for a theory with gauge group $G$ of rank $r$ and $N_{\mathbf{R}}$ hypermultiplets transforming in the representation $\mathbf{R}$ of $G$ can be expressed as an integral over the gauge fugacities $z_i$, with $i=1,\ldots,r$, in the following fashion \cite{Kim:2012gu} (see also \cite{Rodriguez-Gomez:2013dpa,Bergman:2013koa,Bergman:2013ala,Bergman:2013aca,Hwang:2014uwa,Zafrir:2014ywa,Bergman:2014kza}):
\bes{ \label{Indexgeneralformula}
\oint \left[d\mu_{G}(z_i)\right] \CI_{\text{pert}}(x,y; z_i; w_h) \CI_{\text{inst}}(x,y; q; z_i; w_h)\,,
}
where the instanton information is collected in the factor
\begin{subequations}
\begin{align} \label{instcontr}
\CI_{\text{inst}}(x,y; q; z_i; w_h) &= Z_{\text{inst}}(x,y; q; z_i; w_h) Z_{\text{inst}}(x,y; q^{-1}; z^{-1}_i; w^{-1}_h)\,,
\\ Z_{\text{inst}}(x,y; q; z_i; w_h) &= 1 + \sum_{k=1}^{\infty} q^k Z_{\text{inst}, (k)}(x,y; z_i; w_h)\,,
\end{align}
\end{subequations}
with $Z_{\text{inst}, (k)}$ denoting the partition function for $k$ $G$ instantons, which can be computed either using the standard ADHM construction \cite{Nekrasov:2004vw,Benvenuti:2010pq,Hanany:2012dm,Kim:2012gu,Hwang:2014uwa,Bergman:2015dpa}, or by solving the blow-up equations, as described in \S\ref{sec:blow-upEq} and appendix~\ref{app:blow-upSU36USp4}. The Haar measure for the gauge group $G$, denoted by $\left[d\mu_{G}(z_i)\right]$, can be built from the set of positive roots $\alpha \in \Delta^+$ of the Lie algebra $\fg$ associated with $G$ as
\bes{ \label{Haarmeasure}
\oint \left[d\mu_{G}(z_j)\right] = \oint \prod_{j = 1}^r \frac{d z_j}{2 \pi i z_j} \prod_{\alpha \in \Delta^+} \left(1-\prod_{j=1}^r z^{\alpha_j}_j\right)\,,
}
whereas the perturbative contribution is given by
\bes{ \label{Indexpert}
\CI_{\text{pert}}(x,y; z_i; w_h) = \PE\Bigg[&-\frac{x \left(y+\frac{1}{y}\right)}{\left(1-x y\right) \left(1-\frac{x}{y}\right)} \chi^{\fg}_{\text{adj}}(z_i) \\ &+ \frac{x}{\left(1-x y\right) \left(1-\frac{x}{y}\right)} \chi^{\fg}_{\mathbf{R}}(z_i) \sum_{h=1}^{N_{\mathbf{R}}} \left(w_h + \frac{1}{w_h}\right)\Bigg]\,,
}
where the plethystic exponential is defined as $\PE\left[f(a)\right]= \exp\left[\sum_{n=1}^{\infty} \frac{1}{n} f(a^n)\right]$. We collect the explicit expressions for the Haar measures and the characters in appendix~\ref{app:grouptheory}. In the following subsection and in \S\ref{sec:faithfulsym}, we will compute the superconformal indices of the various theories of our interest.

\subsubsection{Example: $\SU(2) + 1 \mathbf{F}$}  \label{sec:exSU2w1F}
Upon denoting with $f_1$ the Coulomb branch VEV, with $m_0 = \frac{1}{g^2_0}$ the square of the inverse gauge coupling, and with $m_1$ the mass parameter associated with the fundamental hypermultiplet, the prepotential \eref{Prepotential} in the chamber $f_1 > m_1 > 0$ reads
\bes{
\mathcal{F} = m_0 f_{1}^{2} + \frac{4}{3} f_{1}^{3} - \frac{1}{12} \Big[\left(f_{1} + m_{1}\right)^{3} + \left(f_{1} - m_{1}\right)^{3} \Big] \,,
}
from which the effective prepotential \eref{EffectivePrepotential} can be obtained as
\bes{
\mathcal{E} = \frac{1}{\epsilon_1 \epsilon_2} \left[ \mathcal{F} - \frac{1}{24} f_{1} (\epsilon_{1}^2 + \epsilon_{2}^2) + \frac{1}{4} f_{1} (\epsilon_{1} + \epsilon_{2})^2 \right]\,.
}
This leads to the blow-up function
\bes{ \label{eq:expVsu2f}
\scalebox{0.92}{$
    \exp(-V_n) \;=\; q^{\,n^2} \, (p_1 p_2)^{\frac{n}{24} \left( -7 - 12 B_{m_1}^2 + 24 B_{m_0} n + 28 n^2 \right)} \, w_1^{\,-B_{m_1} n} \, z_1^{\frac{1}{8} \left( -1 + 4 B_{m_1}^2 - 16 B_{m_0} n - 28 n^2 \right)}\,,   
    $}
}
where $n$ is the $\SU(2)$ gauge flux, $B_{m_1}$ is the background flux for the flavour symmetry and $B_{m_0}$ the one for the topological symmetry.

The admissible values of the background fluxes are derived in \S\ref{sec:quantisationwithmatter} (see also \cite[(4.3)--(4.4)]{Kim:2025qaf}), based on the triviality of the 1-form symmetry and the integrality of the exponent of $z_1$: the gauge flux takes integer values, the flavour flux is fixed to $B_{m_1} = \pm\frac{1}{2}$, while $B_{m_0} \in \Z \pm \frac{1}{4}$, and the two inequivalent combinations of $(B_{m_0}, B_{m_1})$ correspond to the two discrete theta angles $\theta = \{0, \pi\}$ of the $\SU(2)$ gauge theory:
\bes{ \label{eq:thetaSU2quant}
\theta=0: &\qquad B_{m_0} \in  \Z+\frac{1}{4}\,, \, B_{m_1} = +\frac{1}{2} \quad \text{or} \quad B_{m_0} \in  \Z-\frac{1}{4}\,, \, B_{m_1} = -\frac{1}{2}\,, \\
\theta=\pi: &\qquad B_{m_0} \in  \Z-\frac{1}{4}\,, \, B_{m_1} = +\frac{1}{2} \quad \text{or} \quad B_{m_0} \in  \Z+\frac{1}{4}\,, \, B_{m_1} = -\frac{1}{2}\, .
}
It is important to note that the concept of the discrete theta angle is {\it unphysical} in the presence of a massless fermion in the fundamental representation (see also \cite[Section 4.2]{Bergman:2013ala}); we will see below that the different choices are simply related by inverting the flavour fugacity in the instanton partition functions and in the superconformal index. Here, we use the labels $\theta = 0$ and $\pi$ merely as a convenient way of distinguishing the different choices of the background fluxes.

We are now ready to solve the blow-up equations \eqref{blow-upEqPowerq}, where the one-loop contribution on $\hat{\BC}^2 \times S^1$ is given explicitly by
\bes{
\hat{Z}_{\text{one-loop}} = \PE\left[-\frac{1 + p_1 p_2}{\left(1-p_1\right) \left(1-p_2\right) z^2_1} - \frac{\sqrt{p_1 p_2}}{\left(1-p_1\right) \left(1-p_2\right)} \frac{1}{z_1} \left(w_1 + \frac{1}{w_1}\right)\right]\,.
}
Note that, at order zero in $q$, where only the gauge flux $n=0$ contributes, the blow-up equations \eqref{blow-upEqPowerq} with the choices of $B_{m_1}$ described above fix $\Lambda_{(0)} = 1$.

As remarked below \eqref{blow-upEqPowerq}, solving the blow-up equations at each instanton order requires three distinct values of the background fluxes. A subtlety arises in the computation of the corresponding parameters $\Lambda_{(1)}$ from \eqref{eq:defLambda}, which depend on the choice of fluxes as follows:
\bes{ \label{eq:LambdaSU2F}
\theta = 0: &\qquad
\begin{array}{ll}
B_{m_0} = \bigl(-\tfrac{3}{4},\, \tfrac{1}{4},\, \tfrac{5}{4}\bigr)\,, \quad B_{m_1} = +\tfrac{1}{2}: & \Lambda_{(1)} = \bigl(0,\, 0,\, 0\bigr)~, \\[4pt]
B_{m_0} = \bigl(-\tfrac{1}{4},\, \tfrac{3}{4},\, \tfrac{7}{4}\bigr)\,, \quad B_{m_1} = -\tfrac{1}{2}: & \Lambda_{(1)} = \bigl(0,\, 0,\, -p_1 p_2\, w_1^{-1/2}\bigr)~,
\end{array} \\[6pt]
\theta = \pi: &\qquad
\begin{array}{ll}
B_{m_0} = \bigl(-\tfrac{3}{4},\, \tfrac{1}{4},\, \tfrac{5}{4}\bigr)\,, \quad B_{m_1} = -\tfrac{1}{2}: & \Lambda_{(1)} = \bigl(0,\, 0,\, 0\bigr)~, \\[4pt]
B_{m_0} = \bigl(-\tfrac{1}{4},\, \tfrac{3}{4},\, \tfrac{7}{4}\bigr)\,, \quad B_{m_1} = +\tfrac{1}{2}: & \Lambda_{(1)} = \bigl(0,\, 0,\, -p_1 p_2\, w_1^{1/2}\bigr)~,
\end{array}
}
where the entries of $\Lambda_{(1)}$ are ordered as the corresponding values of $B_{m_0}$. In other words, for either theta angle, the parameters $\Lambda_{(1)}$ vanish for all three fluxes with $B_{m_0} \in \Z + \frac{1}{4}$, whereas, for the opposite quantisation, the flux with the largest $|B_{m_0}|$ acquires a non-trivial $\Lambda_{(1)}$. Furthermore, at the two-instanton level, $\Lambda_{(2)}=(0,0,0)$ for all choices of the fluxes above.

Solving the blow-up equations \eqref{blow-upEqPowerq} with the fluxes corresponding to $\theta = 0$, the one-instanton partition function takes the closed form
\bes{ \label{eq:Zhat1SU2F}
\hat{Z}^{\theta=0}_{\text{inst}, (1)} = \frac{p_1 p_2\, z_1 \Bigl[ \left(1 + p_1 p_2\right) z_1 + \sqrt{p_1 p_2}\, w_1 \left(1 + z_1^2\right) \Bigr]}{\sqrt{w_1}\, (1-p_1)(1-p_2)\left(p_1 p_2 - z_1^2\right)\left(-1 + p_1 p_2\, z_1^2\right)}~.
}
In terms of the variables $x$ and $y$ of \eref{eq:xydef}, its power series in $x$ reads
\bes{ \label{eq:Zhat1SU2Fexp}
\hat{Z}^{\theta=0}_{\text{inst}, (1)} = \frac{x^2}{\sqrt{w_1}} &+ \left[ \sqrt{w_1}\, \chi_{[1]}(z_1) + \frac{\chi_{[1]}(y)}{\sqrt{w_1}} \right] x^3 \\
&+ \left[ \sqrt{w_1}\, \chi_{[1]}(y)\, \chi_{[1]}(z_1) + \frac{\chi_{[2]}(y) + \chi_{[2]}(z_1)}{\sqrt{w_1}} \right] x^4 \\
&+ \left[ \sqrt{w_1}\, \chi_{[1]}(z_1) + \frac{\chi_{[1]}(y)}{\sqrt{w_1}} \right] \Bigl[ \chi_{[2]}(y) + \chi_{[2]}(z_1) - 1 \Bigr]\, x^5 + \ldots~,
}
where $\chi_{[n]}$ denotes the character of the $(n+1)$-dimensional representation of $\su(2)$, applied to the spacetime fugacity $y$ and the gauge fugacity $z_1$. Observe that the flavour fugacity $w_1$ appears with half-integer powers only, reflecting the half-integral flavour flux $B_{m_1} = \pm\frac{1}{2}$; moreover, the gauge fugacity enters through complete $\su(2)$ characters, as required by Weyl invariance. The two-instanton partition function admits a similar, albeit much lengthier, closed form, whose expansion reads
\bes{ \label{eq:Zhat2SU2Fexp}
\hat{Z}^{\theta=0}_{\text{inst}, (2)} &= \frac{x^4}{w_1} + \left[ \chi_{[1]}(z_1) + \frac{\chi_{[1]}(y)}{w_1} \right] x^5 \\
&+ \left[ w_1 + 2\, \chi_{[1]}(y)\, \chi_{[1]}(z_1) + \frac{2\, \chi_{[2]}(y) + \chi_{[2]}(z_1)}{w_1} \right] x^6 \\
&+ \biggl[ w_1\, \chi_{[1]}(y) \Bigl( \chi_{[2]}(z_1) + 1 \Bigr) + 3\, \chi_{[2]}(y)\, \chi_{[1]}(z_1) + \chi_{[3]}(z_1) + 2\, \chi_{[1]}(z_1) \\
&\qquad \, \, \, + \frac{2\, \chi_{[3]}(y) + \chi_{[1]}(y) + 2\, \chi_{[1]}(y)\, \chi_{[2]}(z_1)}{w_1} \biggr]\, x^7 + \ldots~.
}
For $\theta = \pi$, the solution of the blow-up equations is obtained from the $\theta = 0$ one simply by inverting the flavour fugacity:
\bes{ \label{eq:thetapiw1inv}
\hat{Z}^{\theta=\pi}_{\text{inst}, (k)}\left(p_1, p_2; z_1; w_1\right) = \hat{Z}^{\theta=0}_{\text{inst}, (k)}\left(p_1, p_2; z_1; w_1^{-1}\right)~, \qquad k = 1, 2\,.
}
This is a direct consequence of \eqref{eq:thetaSU2quant}: the two theta angles are exchanged by flipping the sign of $B_{m_1}$ at fixed $B_{m_0}$, and $B_{m_1} \rightarrow -B_{m_1}$ amounts to $w_1 \rightarrow w_1^{-1}$ in the blow-up function \eqref{eq:expVsu2f}. In particular, at one instanton:
\bes{ \label{eq:Zhat1SU2Fpi}
\hat{Z}^{\theta=\pi}_{\text{inst}, (1)} = \frac{p_1 p_2 \Bigl[ \sqrt{p_1 p_2}\, z_1 \left(1 + z_1^2\right) + \left(1 + p_1 p_2\right) w_1\, z_1^2 \Bigr]}{\sqrt{w_1}\, (1-p_1)(1-p_2)\left(p_1 p_2 - z_1^2\right)\left(-1 + p_1 p_2\, z_1^2\right)}~.
}
The one-instanton partition functions \eqref{eq:Zhat1SU2F} and \eqref{eq:Zhat1SU2Fpi} are in perfect agreement with \cite[(19), (20)]{Bergman:2013ala}: their unhatted counterparts, obtained through the shift \eqref{ZtoZhatshift}, coincide exactly with the one-instanton functions computed there for $\theta = 0$ and $\theta = \pi$ respectively, upon identifying the flavour fugacity as $e^{i m} = w_1^{-1}$.

As an application, let us use these results to compute the superconformal index (whose definition we reviewed in \S\ref{sec:index}, with the group theory conventions collected in appendix~\ref{app:grouptheory}). The ultraviolet completion of the $\SU(2) + 1\mathbf{F}$ gauge theory is the $E_2$ SCFT of \cite{Seiberg:1996bd}, whose flavour symmetry algebra is $\mathfrak{e}_2 \cong \su(2) \oplus \fu(1)$, enhancing the topological and flavour symmetries $\fu(1)_I \oplus \fu(1)_{w_1}$ of the gauge theory. The enhancement is manifest in the index (see \cite{Kim:2012gu} for the superconformal indices of the $E_n$ theories): for $\theta = 0$, we find
\bes{ \label{eq:indexSU2F}
\scalebox{0.85}{$
\begin{split}
& \CI^{\theta=0}_{\SU(2)+1\mathbf{F}} = 1 + \Bigl[ 1 + \chi_{[2]}(v) \Bigr]\, x^2 + \chi_{[1]}(y) \Bigl[ 2 + \chi_{[2]}(v) \Bigr]\, x^3 \\
& \quad + \Bigl\{ \chi_{[2]}(y) \Bigl[ 2 + \chi_{[2]}(v) \Bigr] + 1 + \chi_{[4]}(v) - \bigl( u + u^{-1} \bigr)\, \chi_{[1]}(v) \Bigr\}\, x^4 \\
& \quad + \Bigl\{ \chi_{[3]}(y) \Bigl[ 2 + \chi_{[2]}(v) \Bigr] + \chi_{[1]}(y) \Bigl[ 1 + 2\, \chi_{[2]}(v) + \chi_{[4]}(v) - \bigl( u + u^{-1} \bigr)\, \chi_{[1]}(v) \Bigr] \Bigr\}\, x^5 + \ldots~,
\end{split}
$}
}
where
\bes{ \label{eq:E2fugacities}
v = \bigl( q\, w_1^{-1/2} \bigr)^{1/2}~, \qquad u = \bigl( q^{-1} w_1^{-7/2} \bigr)^{1/2}
}
are the fugacities associated with the Cartan of the enhanced $\su(2)$ and with the commuting $\fu(1)$, respectively. In particular, the four conserved currents at order $x^2$, namely $1 + \chi_{[2]}(v)$, fill the adjoint representation of $\mathfrak{e}_2 = \su(2) \oplus \fu(1)$. In accordance with \eqref{eq:thetapiw1inv}, the index for $\theta = \pi$ takes exactly the same form \eqref{eq:indexSU2F} upon inverting the flavour fugacity in \eqref{eq:E2fugacities}, i.e.\ with $v = \bigl( q\, w_1^{1/2} \bigr)^{1/2}$ and $u = \bigl( q^{-1} w_1^{7/2} \bigr)^{1/2}$. The indices for the two theta angles are therefore related by inverting the flavour fugacity, $w_1 \rightarrow w_1^{-1}$. This is as expected, since the discrete theta angle becomes unphysical in the presence of a massless fermion in the fundamental representation.

\section{Quantisation rules from the blow-up function}
\label{sec:expV}
The central object of the blow-up method~\cite{Kim:2020hhh} is the classical prefactor $\exp(-V_n)$, dubbed the {\it blow-up function}, accompanying each magnetic flux assignment on the blown-up geometry. In this section, we present it in closed form: first for pure $\SU(N)_{\kCS}$, then for a general gauge group with hypermultiplet matter, and, finally, explicitly for $\SU(4)_{\kCS}$ and $\USp(4)$ with antisymmetric hypermultiplets, for $\Spin(7)$ and $\Spin(8)$ with vector hypermultiplets, and for the non-Lagrangian theories of appendices~\ref{app:BN} and~\ref{app:PF}. Crucially, we will discuss the allowed values of gauge magnetic fluxes and background magnetic fluxes for global symmetries from the perspective of the blow-up function.

\subsection{Pure $\SU(N)_{\kCS}$ gauge theory}
\label{sec:expVSUN}

We parametrise the gauge fluxes by $n = (n_1, \ldots, n_{N-1})$ in the coroot basis, $n = \sum_i n_i \alpha_i^\vee$, so that the pairing with the simple roots is $\langle n, \alpha_i \rangle = (Cn)_i$ with $C$ the Cartan matrix. It is convenient to introduce
\begin{equation}
    \tilde{n}_i \;\equiv\; n_i - n_{i-1}\,, \qquad i = 1, \ldots, N\,, \qquad n_0 = n_N \equiv 0\,,
\end{equation}
i.e.~the fluxes in the orthonormal basis of the Cartan of $\mathfrak{u}(N)$, satisfying $\sum_{i=1}^N \tilde{n}_i = 0$. The pairing with the positive roots $\alpha = e_i - e_j$ ($1 \leq i < j \leq N$), whose set we denote by $\Delta^+$, is then $\langle n, \alpha \rangle = \tilde{n}_i - \tilde{n}_j$. We further denote by $B_{m_0}$ the background flux for the topological symmetry $\U(1)_I$, by $q$ the instanton fugacity, by $z_i$ ($i = 1, \ldots, N-1$) the gauge fugacities, and by $p_{1,2} = e^{-\epsilon_{1,2}}$ the $\Omega$-background fugacities.

For pure $\SU(N)_{\kCS}$, the blow-up function takes the form
\begin{equation}
    \exp(-V_n) \;=\; q^{\,Q(n)} \, (p_1 p_2)^{\,P(n)} \, \prod_{i=1}^{N-1} z_i^{\,Z_i(n)}\,,
    \label{eq:expVmaster}
\end{equation}
with
\begin{subequations}
\label{eq:expVexponents}
\begin{align}
    Q(n) \;&=\; \frac{1}{2} \sum_{i,j=1}^{N-1} C_{ij}\, n_i n_j \;=\; \sum_{i=1}^{N-1} n_i^2 - \sum_{i=1}^{N-2} n_i n_{i+1} \;=\; \frac{1}{2} \sum_{i=1}^{N} \tilde{n}_i^2\,, \label{eq:Qgen} \\
    P(n) \;&=\; B_{m_0}\, Q(n) \;+\; \frac{1}{6} \sum_{\alpha \in \Delta^+} \Bigl( \langle n, \alpha \rangle^3 - \langle n, \alpha \rangle \Bigr) \;+\; \frac{\kCS}{6} \sum_{i=1}^{N} \tilde{n}_i^3\,, \label{eq:Pgen} \\
    Z_i(n) \;&=\; -\Bigl( \partial_{n_i} P(n) - \partial_{n_i} P(n)\big|_{n=0} \Bigr) \nonumber \\
    \;&=\; - B_{m_0} \bigl( \tilde{n}_i - \tilde{n}_{i+1} \bigr) \;-\; \frac{1}{2} \sum_{\alpha \in \Delta^+} \langle n, \alpha \rangle^2 \, (\alpha, \alpha_i) \;-\; \frac{\kCS}{2} \bigl( \tilde{n}_i^2 - \tilde{n}_{i+1}^2 \bigr)\,. \label{eq:Zgen}
\end{align}
\end{subequations}
The $\kCS$-dependent term in $P(n)$ can equivalently be written directly in terms of the $n_i$ as
\begin{equation}
    \frac{\kCS}{6} \sum_{i=1}^{N} \tilde{n}_i^3 \;=\; \frac{\kCS}{2} \sum_{i=1}^{N-2} \Bigl( n_i^2\, n_{i+1} - n_i\, n_{i+1}^2 \Bigr)\,.
    \label{eq:CStermalt}
\end{equation}
The three terms in $P(n)$ originate, respectively, from the mixed instanton--gauge coupling, from the one-loop prepotential of the W-bosons, and from the classical Chern--Simons prepotential $\frac{\kCS}{6} \sum_i \tilde{\phi}_i^3$ evaluated on the fluxes; the $z_i$-exponents follow from $P(n)$ by differentiation as indicated. 

Note that both the quadratic form $Q(n)$ and the Chern--Simons term admit a uniform derivation from the character of the fundamental representation. Writing
\begin{equation}
    \chi_{\mathbf{N}}(z) \;=\; \sum_{a=1}^{N} e^{\langle \lambda_a, f \rangle} \;=\; z_1 + \frac{z_2}{z_1} + \cdots + \frac{z_{N-1}}{z_{N-2}} + \frac{1}{z_{N-1}}\,,
\end{equation}
with the Coulomb parameters in the $z$-fugacity basis, $z_i = e^{f_i}$, the weights of $\mathbf{N}$, read off from the character, pair with the fluxes exactly as $\langle \lambda_a, n \rangle = n_a - n_{a-1} = \tilde{n}_a$ ($a = 1, \ldots, N$, with $n_0 = n_N \equiv 0$). Consequently,
\begin{equation}
    Q(n) \;=\; \frac{1}{2} \sum_{a=1}^{N} \langle \lambda_a, n \rangle^2 \;=\; \frac{1}{2} \Tr_{\mathbf{N}}\, n^2\,,
    \qquad
    \frac{\kCS}{6} \sum_{a=1}^{N} \langle \lambda_a, n \rangle^3 \;=\; \frac{\kCS}{6} \Tr_{\mathbf{N}}\, n^3\,,
\end{equation}
i.e.\ the $q$-exponent is governed by the quadratic form in the fundamental normalisation, $K_{ij} = \frac{1}{2}\delta_{ij}$, and the $\kCS$-term by the cubic invariant $d_{ijk} = \frac{1}{2}\Tr_{\mathbf{N}}\, T_i \{T_j, T_k\}$ contracted with the fluxes. This is the form of the result that generalises directly to other gauge groups and representations.

\paragraph{Example: $\SU(2)$.} For $N = 2$ the formulae reduce to 
\bes{ \label{eq:exp-VSU2pure}
\exp(-V_n) = q^{n^2} z_1^{-2n(B_{m_0} + 2n)} (p_1 p_2)^{B_{m_0} n^2 + \frac{1}{3}(4n^3 - n)}~.
}

\paragraph{Example: $\SU(3)_{\kCS}$.} For $N = 3$ one finds
\begin{subequations} \label{eq:exp-VSU3pure}
\begin{align}
    P(n) \;=&\; -\frac{1}{3}(n_1 + n_2) + B_{m_0}\, Q(n) + \frac{4}{3}\bigl(n_1^3 + n_2^3\bigr) - \frac{1}{2}\, n_1 n_2 (n_1 + n_2) \nonumber \\
    +&\; \frac{\kCS}{2}\, n_1 n_2 (n_1 - n_2)\,, \\
    Z_1(n) \;=&\; -\frac{1}{2} (2n_1 - n_2) \bigl[ 2 B_{m_0} + 4 n_1 + (1 + \kCS)\, n_2 \bigr]\,, \\
    Z_2(n) \;=&\; -\frac{1}{2} (2n_2 - n_1) \bigl[ 2 B_{m_0} + 4 n_2 + (1 - \kCS)\, n_1 \bigr]\,,
\end{align}
\end{subequations}
with $Q(n) = n_1^2 - n_1 n_2 + n_2^2$. 

\paragraph{Example: $\SU(4)_{\kCS}$.} Writing out \eqref{eq:expVexponents} for $N = 4$, we find
\begin{subequations}
\begin{align}
    P(n) \;=&\; -\frac{1}{3}(n_1 + n_2 + n_3) + B_{m_0}\, Q(n) + \frac{4}{3}(n_1^3 + n_2^3 + n_3^3) - n_2^2 (n_1 + n_3) \nonumber \\
     +&\; \frac{\kCS}{2} \Bigl[ n_1 n_2 (n_1 - n_2) + n_2 n_3 (n_2 - n_3) \Bigr]\,, \\
    Z_1(n) \;=&\; -B_{m_0}(2n_1 - n_2) - 4n_1^2 + n_2^2 - \frac{\kCS}{2}\, n_2 (2n_1 - n_2)\,, \\
    Z_2(n) \;=&\; -B_{m_0}(2n_2 - n_1 - n_3) + 2n_2(n_1 + n_3) - 4n_2^2 \\ +&\; \frac{\kCS}{2} (n_1 - n_3)(2n_2 - n_1 - n_3)\,, \\
    Z_3(n) \;=&\; -B_{m_0}(2n_3 - n_2) - 4n_3^2 + n_2^2 + \frac{\kCS}{2}\, n_2 (2n_3 - n_2)\,,
\end{align}
\end{subequations}
with $Q(n)$ as in \eqref{eq:Qgen}. Charge conjugation $\mathcal{C}$ acts on $\SU(4)$ as the outer automorphism of the Dynkin diagram, $n_1 \leftrightarrow n_3$ with $n_2$ fixed (and $z_1 \leftrightarrow z_3$), together with a sign flip $\kCS \to -\kCS$ of the ($\mathcal{C}$-odd) Chern--Simons term. The exponents above are covariant under this combined operation --- it leaves $Q$ and $P$ invariant and exchanges $Z_1 \leftrightarrow Z_3$ --- but for $\kCS \neq 0$ it is not a symmetry at fixed $\kCS$, i.e.\ charge conjugation is explicitly broken. 

\subsubsection*{Gauge magnetic fluxes} Let us first discuss the case of $\kCS=0$. The gauge magnetic fluxes $n$ take values in the coweight lattice $\Lambda_{\cw}$ (the lattice dual to the root lattice, generated by the fundamental coweights), as required for an integral pairing with all the weights of the gauge representations. Two fluxes differing by an element of the coroot lattice $\Lambda_{\crt} = \mathrm{span}_{\Z}\{\alpha_i^\vee\}$ describe the same genuine $\SU(N)$ bundle and are gauge-equivalent. The inequivalent 't~Hooft fluxes --- equivalently, the backgrounds for the electric $\Z_N^{(1)}$ 1-form symmetry --- are therefore classified by the quotient
\begin{equation}
    \Lambda_{\cw} / \Lambda_{\crt} \;\cong\; \mathcal{Z}\bigl( \SU(N) \bigr) \;=\; \Z_N\,,
\end{equation}
that is the centre of the simply connected gauge group. We label its $N$ elements by $\omega_j$, $j = 0, \ldots, N-1$, with $\omega_0$ trivial and $\omega_1$ a generator; the coset $\omega_j = j\, \omega_1$ admits the representative $n_i = \frac{ij}{N} \pmod 1$ ($i = 1, \ldots, N-1$). As an example, for $\SU(4)$, convenient representatives are
\begin{equation}
\vcenter{\hbox{%
\begin{tabular}{cccc}
\toprule
Coset & $(n_1, n_2, n_3)$ & Generator & Group-theoretic meaning \\
\midrule
$\omega_0$ (trivial) & $(0, 0, 0)$ & $0$ & identity element of $\Z_4$ \\
$\omega_1$           & $\bigl(\tfrac{1}{4}, \tfrac{1}{2}, \tfrac{3}{4}\bigr)$ & $\omega$ & generator of $\Z_4$ \\
$\omega_2$           & $\bigl(\tfrac{1}{2}, 0, \tfrac{1}{2}\bigr)$           & $\omega^2$ & order-2 element \\
$\omega_3$           & $\bigl(\tfrac{3}{4}, \tfrac{1}{2}, \tfrac{1}{4}\bigr)$ & $\omega^3 = \omega^{-1}$ & inverse of $\omega_1$ \\
\bottomrule
\end{tabular}}}
\label{eq:su4cosets}
\end{equation}
where $\omega = \exp(2\pi i/4)$. Evaluating the quadratic form \eqref{eq:Qgen} on these representatives gives
\begin{equation}
    Q(\omega_j) \;=\; \frac{j(N-j)}{2N} \;\;\overset{N=4}{=}\;\; \begin{cases} 0 & j = 0 \\[2pt] \tfrac{3}{8} & j = 1, 3 \\[2pt] \tfrac{1}{2} & j = 2 \end{cases} \,,
    \label{eq:Qcosets}
\end{equation}
where $\omega_1, \omega_2, \omega_3$ give the fractional Pontryagin numbers (well defined modulo $1$): turning on such a non-trivial 1-form symmetry background amounts to considering $\PSU(N) = \SU(N)/\Z_N$ bundles with non-trivial 't~Hooft flux, i.e.\ bundles that do not lift to $\SU(N)$.

In the case of $\kCS \neq 0$, the 1-form symmetry is explicitly broken to $\Z^{(1)}_{\gcd(N, \kCS)}$. The valid gauge fluxes then reside in the sublattice of $\Lambda_{\cw}/\Lambda_{\crt}$ corresponding to $\Z^{(1)}_{\gcd(N, \kCS)}$. For example:
\begin{equation}
\scalebox{0.89}{$
\vcenter{\hbox{%
\begin{tabular}{lll} 
\toprule
Theory & 1-form symmetry & Allowed gauge fluxes $n$ \\
\midrule
$\SU(N)_{\kCS}$ pure & $\Z_m^{(1)}$, $m = \gcd(N,\kCS)$ & multiples of $\omega_{N/m}$ \\
\quad $\SU(4)_{\kCS},\ \kCS \equiv 0 \!\!\pmod 4$ & $\Z_4^{(1)}$ & $\omega_0, \omega_1, \omega_2, \omega_3$ \\
\quad $\SU(4)_{\kCS},\ \kCS \equiv 2 \!\!\pmod 4$ & $\Z_2^{(1)}$ & $\omega_0,\ \omega_2$ \\
\quad $\SU(4)_{\kCS},\ \kCS \equiv 1, 3 \!\!\pmod 4$ & trivial & $\omega_0$ \\
\bottomrule
\end{tabular}}}
$}
\label{eq:su4allowedflux}
\end{equation}

\subsubsection*{Comments on integrality} Since $x^3 - x \equiv 0 \pmod 6$ for any $x \in \Z$, and since $\sum_i \tilde{n}_i^3 \equiv \sum_i \tilde{n}_i = 0 \pmod 6$ for integer fluxes, all exponents in \eqref{eq:expVmaster} are manifestly integers (including the overall $q^{Q(n)}$, since $Q(n) \in \Z$ there) when $n$ lies in the coroot lattice $\Lambda_{\crt}$ and $B_{m_0} \in \Z$, for \emph{any} integer $\kCS$. Fractional values of the exponents can therefore only arise when $n$ is taken in a non-trivial coset of $\Lambda_{\cw}/\Lambda_{\crt}$ --- and precisely these fractional shifts furnish the diagnostics of the global structure of the theory exploited in the rest of the paper.

\subsubsection*{Quantisation of $B_{m_0}$}\label{sec:Bm0quant} The allowed values of the background flux $B_{m_0}$ are determined by the following requirements that the blow-up equations consistently reproduce the partition function.
\begin{enumerate}
\item The blow-up function $\exp(-V_n)$ evaluated at vanishing gauge flux, $n_i = 0$, must be \emph{independent of the gauge fugacities} $z_i$. This follows from the requirement that the instanton partition function at the $0$-th instanton order must be equal to unity. For pure $\SU(N)_{\kCS}$, this is automatic in the normalisation of \eqref{eq:expVmaster}, while in the presence of matter it is a non-trivial condition that fixes the flavour background fluxes (see \S\ref{sec:expVgeneralsub}). 
\item Beyond the zero-flux sector, consistency with integrality of the weights of gauge group representations requires the gauge fugacities to appear with \emph{integer} powers, $Z_i(n) \in \Z$ for all allowed fluxes $n$. This constraint fixes the allowed values of $B_{m_0}$. 
\end{enumerate}
Note that the $q$-exponent $Q(n)$ and the constant part of the $p$-exponent $P(n)$ may well be fractional: these are manifestations of 't~Hooft anomalies and of the presence of 2-groups, which we will discuss below.

Crucially, $B_{m_0}$ must be read off \emph{only at the allowed cosets} --- those that are genuine
backgrounds of the $\Z_\fm^{(1)}$ 1-form symmetry, $\fm=\gcd(N,\kCS)$. For the low-rank cases, imposing $Z_i(n)\in\Z$ at
the generator of the genuine symmetry gives
\begin{equation}
\vcenter{\hbox{%
\begin{tabular}{lcc}
\toprule
Theory & genuine 1-form & $B_{m_0}$ \\
\midrule
$\SU(2)_0$        & $\Z_2^{(1)}$                                  & $\Z$ \\
$\SU(2)_\pi$      & trivial \ ($\theta=\pi$ breaks $\Z_2^{(1)}$)  & $\Z+\tfrac12$ \\
$\SU(3)_{\kCS}$   & $\Z_{\gcd(3,\kCS)}^{(1)}$                     & $\Z$ ($\kCS$ odd),\ \ $\Z+\tfrac12$ ($\kCS$ even) \\
$\SU(4)_{\kCS}$   & $\Z_{\gcd(4,\kCS)}^{(1)}$                     & $\Z$ \\
\bottomrule
\end{tabular}}}
\label{eq:Bm0table}
\end{equation}

Let us illustrate this in the following examples. For $\SU(2)_0$, the gauge flux $n \in \BZ+\frac{1}{2}$ is in a non-trivial coset allowed by the $\Z_2^{(1)}$ 1-form symmetry;\footnote{Note that $n \in \Z$ is also allowed, but this corresponds to the trivial coset} from \eqref{eq:exp-VSU2pure}, this gives $Z_1(n=1/2) = -(B_{m_0}+1)$, whose integrality requires $B_{m_0} \in \BZ$.  On the other hand, for $\SU(2)_\pi$, the $\Z_2^{(1)}$ 1-form symmetry is broken by the theta angle $\theta=\pi$, so the allowed gauge flux $n$ must be integral; from \eqref{eq:exp-VSU2pure}, we see that the integrality of $Z_1(n)=-2n(B_{m_0}+2n)$ implies that $B_{m_0}$ can be half-integral. In this case, $B_{m_0} \in \BZ+\frac{1}{2}$ gives a distinct solution from $\SU(2)_0$. This analysis gives the same result as the geometric method presented in \cite[(3.35)]{Kim:2020hhh}.

This argument can also be applied to $\SU(3)_{\kCS}$ as follows. If $\mathrm{gcd}(3,\kCS)\neq 1$, then the magnetic fluxes $n=(1/3,2/3)$ is allowed by the $\Z_3^{(1)}$ 1-form symmetry and is in a non-trivial coset. Let us now use \eqref{eq:exp-VSU3pure} to determine the quantisation rules. With $n=(1/3,2/3)$, we see that $Z_1(n)=0$ and $Z_2(n) = \frac{1}{2}\left(-3-2B_{m_0}+\frac{1}{3}\kCS \right)$, whose integrality implies that $B_{m_0} \in \BZ$. On the other hand, with $n=(1,0)$, we find that $Z_1(n) =-2B_{m_0}-4$ and $Z_2(n) = \frac{1}{2}(1+2B_{m_0} -\kCS)$. Thus, if $\kCS$ is odd and $\mathrm{gcd}(3,\kCS)=1$, then the integrality of $Z_2(n)$ requires that $B_{m_0} \in \BZ$.  On the other hand, if $\kCS$ is even, then $B_{m_0}$ must be half-integral for $Z_2(n)$ to be integral.  This is again in agreement with the geometric analysis in \cite[(3.59)]{Kim:2020hhh}.

\paragraph{Remark: reading $B_{m_0}$ on the wrong coset.} If $Z_i(n)\in\Z$ is instead imposed at a
coset that is \emph{not} a background of the genuine 1-form symmetry, one obtains the \emph{opposite}
quantisation of $B_{m_0}$ --- and this very mismatch is the signal that the coset is not allowed.
For $\SU(4)_{\kCS}$ with $\kCS\equiv2\pmod4$ (genuine $\Z_2^{(1)}$, generated by $\omega_2$) the
screened cosets $\omega_1,\omega_3$ return $B_{m_0}\in\Z+\tfrac12$ rather than the $B_{m_0} \in \Z$ found at
$\omega_0,\omega_2$: the would-be $\Z_4^{(1)}$ is explicitly broken to $\Z_2^{(1)}$ by the
Chern--Simons level. Similarly, for $\SU(2)_\pi$ the would-be $\omega_1$ background gives
$B_{m_0}\in\Z$, opposite to the $\Z+\tfrac12$ of the genuine sector, signalling that $\theta=\pi$
breaks the $\Z_2^{(1)}$ outright. The quantisation of $B_{m_0}$ is therefore always to be read off
the allowed cosets alone; the value at a wrong coset merely diagnoses that the coset has been
screened or obstructed.

\subsection{General gauge group and matter content}
\label{sec:expVgeneralsub}

The structure of \eqref{eq:expVmaster} extends to a general gauge group with hypermultiplet matter. Consider a gauge theory with gauge fluxes $n$ (in the coroot basis), instanton flux $B_{m_0}$, and hypermultiplets labelled by $h$, each in a representation $\mathbf{R}_h$ with mass $m_h$, flavour fugacity $w_h = e^{-m_h}$, and background flux $B_h$. In a given chamber of the extended Coulomb branch, each weight $\lambda \in \mathbf{R}_h$ is assigned the sign $\sigma_\lambda = \mathrm{sign}\bigl( \langle \lambda, f \rangle + m_h \bigr)$, and we define the chamber-resolved flux pairings
\begin{equation}
    x_\lambda \;=\; \sigma_\lambda \bigl( \langle \lambda, n \rangle + B_h \bigr)\,.
\end{equation}
Then
\begin{equation}
    \exp(-V_n) \;=\; q^{\,Q(n)}\, (p_1 p_2)^{\,P(n)} \prod_{i} z_i^{\,Z_i(n)} \prod_{h} w_h^{\,W_h(n)}\,,
\end{equation}
with
\begin{subequations}
\label{eq:expVgeneral}
\begin{align}
    Q(n) \;=&\; \frac{1}{2} \Tr_F\, n^2\,, \\
    Z_i(n) \;=&\; -\frac{1}{2} \sum_{\alpha \in \Delta^+} \langle n, \alpha \rangle^2 \langle \alpha, \alpha_i^\vee \rangle \;-\; B_{m_0} \Tr_F \bigl( n\, \alpha_i^\vee \bigr) \nonumber\\
    +&\; \frac{1}{4} \sum_{h} \sum_{\lambda \in R_h} \Bigl( x_\lambda^2 - \frac{1}{4} \Bigr)\, \sigma_\lambda \langle \lambda, \alpha_i^\vee \rangle\,, \label{eq:Zgeneral} \\
    W_h(n) \;=&\; -\frac{1}{4} \sum_{\lambda \in R_h} \sigma_\lambda\, x_\lambda^2\,, \\
    P(n) \;=&\; B_{m_0}\, Q(n) \;+\; \frac{1}{6} \sum_{\alpha \in \Delta^+} \Bigl( \langle n, \alpha \rangle^3 - \langle n, \alpha \rangle \Bigr) \nonumber\\
    -&\; \frac{1}{12} \sum_{h} \sum_{\lambda \in R_h} \Bigl( x_\lambda^3 - c_\lambda\, x_\lambda \Bigr)\,, \label{eq:Pgeneral2}
\end{align}
\end{subequations}
where $F$ denotes the fundamental (defining) representation, the $z_i$ are associated with the coroot directions $\alpha_i^\vee$, and the sums over $\lambda$ run over all the \emph{gauge-charged} weights of $\mathbf{R}_h$ --- gauge-neutral weights drop out entirely, as verified for the zero weight of the $\Lambda^2$ of $\USp(4)$ and of the $\mathbf{7}$ of $\Spin(7)$ --- whence the relative factor of $\frac{1}{2}$ between the hypermultiplet and vector contributions, the latter being restricted to the positive roots. 

Two structural features are worth noting. First, the hypermultiplet enters the $z$-exponents \eqref{eq:Zgeneral} through the combination
\begin{equation}
    x_\lambda^2 - \tfrac{1}{4} \;=\; \bigl( x_\lambda - \tfrac{1}{2} \bigr)\bigl( x_\lambda + \tfrac{1}{2} \bigr)\,,
\end{equation}
which vanishes when the flux takes its \emph{canonical} (unobstructed) half-integer value $x_\lambda = \pm\frac{1}{2}$. This is the hypermultiplet counterpart of the W-boson combination
\begin{equation}
    \langle n, \alpha \rangle^3 - \langle n, \alpha \rangle \;=\; \langle n, \alpha \rangle \bigl( \langle n, \alpha \rangle - 1 \bigr)\bigl( \langle n, \alpha \rangle + 1 \bigr)
\end{equation}
in $P(n)$, which vanishes at the canonical integer pairings $\langle n, \alpha \rangle \in \{0, \pm 1\}$. In both sectors, the exponent is built to vanish at canonical quantisation, so a non-zero value directly measures the deviation --- i.e.\ it signals a fractional 't~Hooft flux background. Second, the linear coefficients $c_\lambda$ in \eqref{eq:Pgeneral2} are chamber-dependent, where for the theories concerned in this paper, we will see below that $c_\lambda$ is either $0$ or $\frac{1}{4}$.

To make \eqref{eq:expVgeneral} concrete, the table~\eqref{eq:letters} lists the chamber-resolved letters $x_\lambda$ for the gauge-charged weights $\lambda$ of each matter representation used in this paper, in the orthonormal pairings of the corresponding subsection. Writing $\ell\equiv\langle\lambda, n\rangle$ for the gauge pairing shown in the third column, the two letters of a $\pm\lambda$ pair are $x_{\pm\lambda} = \sigma_{\pm\lambda}\,\bigl(\langle\pm\lambda, n\rangle + B_h\bigr)$ with signs $\sigma_{\pm\lambda} = \operatorname{sign}\bigl(\langle\pm\lambda, f\rangle + m_h\bigr)$; whether the two signs come out opposite or equal is decided by comparing the gauge holonomy $\langle\lambda,f\rangle$ with the mass $m_h$. The pair is \emph{gauge-dominated} when the holonomy wins, $\langle\lambda,f\rangle > m_h > 0$: then $\sigma_{+\lambda} = +$ and $\sigma_{-\lambda} = -$ are opposite, and the letters are $(\ell + B_h,\ \ell - B_h)$ with $c_\lambda = \frac{1}{4}$. It is \emph{mass-dominated} when the mass wins, $m_h > \langle\lambda,f\rangle > 0$: then $\sigma_{+\lambda} = \sigma_{-\lambda} = +$ are equal, and the letters are $(\ell + B_h,\ -\ell + B_h)$ with $c_\lambda = 0$. Requiring every pair to be gauge-dominated thus amounts to taking $m_h$ below all of the gauge holonomies $\langle\lambda,f\rangle$; once a chamber fixes the sign of each pairing, this is simply a finite list of linear inequalities on $m_h$, written out in each subsection below.
\begin{equation}
\scalebox{0.94}{$
\vcenter{\hbox{%
\begin{tabular}{llll}
\toprule
Theory \& matter & weight pair $\pm\lambda$ & $\ell=\langle\lambda, n\rangle$ & letters $(x_\lambda, x_{-\lambda})$ \\
\midrule
$\SU(4)$ with $\mathbf{4}\oplus\overline{\mathbf{4}}$ & $\pm\mu_1$ & $n_1$            & $(\ell + B_h,\ \ell - B_h)$ \\
 & $\pm\mu_2$ & $n_2 - n_1$       & $(\ell + B_h,\ \ell - B_h)$ \\
 & $\pm\mu_3$ & $n_3 - n_2$       & $(\ell + B_h,\ \ell - B_h)$ \\
 & $\pm\mu_4$ & $-n_3$            & $(\ell + B_h,\ \ell - B_h)$ \\
\midrule
$\SU(4)$ with $\Lambda^2 = \mathbf{6}$ & $\pm(\mu_1{+}\mu_2)$ & $n_2$               & $(\ell + B_h,\ \ell - B_h)$ \\
 & $\pm(\mu_1{+}\mu_3)$ & $n_1 - n_2 + n_3$ & $(\ell + B_h,\ \ell - B_h)$ \\
 & $\pm(\mu_2{+}\mu_3)$ & $n_3 - n_1$       & $(\ell + B_h,\ \ell - B_h)$ \\
\midrule
$\USp(4)$ with $\Lambda^2 = \mathbf{5}\oplus\mathbf{1}$ & $\pm(e_1{+}e_2)$ & $n_2$           & $(\ell + B_h,\ \ell - B_h)$ \\
 & $\pm(e_1{-}e_2)$ & $2n_1 - n_2$  & $(\ell + B_h,\ -\ell + B_h)$ \\
\midrule
$\Spin(7)$ with $\mathbf{7}$ & $\pm e_1$ & $n_1$            & $(\ell + B_h,\ \ell - B_h)$ \\
 & $\pm e_2$ & $n_2 - n_1$       & $(\ell + B_h,\ \ell - B_h)$ \\
 & $\pm e_3$ & $2n_3 - n_2$      & $(\ell + B_h,\ \ell - B_h)$ \\
\midrule
$\Spin(8)$ with $\mathbf{8}_v$ & $\pm e_1$ & $n_1$            & $(\ell + B_h,\ \ell - B_h)$ \\
 & $\pm e_2$ & $n_2 - n_1$       & $(\ell + B_h,\ \ell - B_h)$ \\
 & $\pm e_3$ & $n_3 + n_4 - n_2$ & $(\ell + B_h,\ \ell - B_h)$ \\
 & $\pm e_4$ & $n_4 - n_3$       & $(\ell + B_h,\ \ell - B_h)$ \\
\bottomrule
\end{tabular}}}
$}
\label{eq:letters}
\end{equation}
For $\SU(4)$, $\Spin(7)$ and $\Spin(8)$, all listed pairs are gauge-dominated in the chambers of \S\ref{sec:expVSU4AS}, \S\ref{sec:expVSO7} and \S\ref{sec:expVSO8}; for $\USp(4)$ the $\pm(e_1{-}e_2)$ pair is mass-dominated in the chamber of \S\ref{sec:expVUSp4} (whence its distinct letter structure). The zero weights of $\Lambda^2$ and of $\mathbf{7}$ are gauge-neutral and omitted. The remaining $\Lambda^2$ pairs of $\SU(4)$ ($\pm(\mu_1{+}\mu_4)$, etc.) coincide with those listed modulo the gauge conditions.

As first checks of \eqref{eq:expVgeneral}: for pure $\SU(N)_{\kCS}$, it reduces to \eqref{eq:expVexponents} (upon including the Chern--Simons term). We will demonstrate this in the case of $\SU(2)_{\theta=0} + 1\mathbf{F}$ below. The $\SU(4)_{\kCS}$, $\USp(4)$, $\Spin(7)$ and $\Spin(8)$ cases with matter are worked out explicitly in \S\ref{sec:expVSU4AS}--\S\ref{sec:expVSO8}.

\subsubsection{1-form symmetry and allowed gauge fluxes} For pure gauge theories, the 1-form symmetry is determined by the coweight lattice quotiented by the coroot lattice of the gauge group. The allowed gauge fluxes for the various gauge groups are given explicitly in eqs.~(3.13)--(3.20) of~\cite{Kim:2020hhh}. 

In the presence of matter fields that transform non-trivially under the centre of the gauge group, the 1-form symmetry is explicitly broken to the subgroup of the centre that does not act on the matter fields. We shall henceforth refer to this as the {\it genuine 1-form symmetry}. As an example, the $\SU(4)_4$ pure gauge theory has a $\Z_4^{(1)}$ 1-form symmetry given by $\{\omega_0, \omega_1, \omega_2, \omega_3\}$. However, upon adding matter in the representation $\Lambda^2$ of $\SU(4)$, the genuine 1-form symmetry becomes $\Z_2^{(1)} = \{\omega_0, \omega_2\}$. In this example, the cosets $\omega_1, \omega_3$ are screened by the matter fields. 

In general, the valid gauge fluxes are the 't~Hooft fluxes of the \emph{genuine} 1-form symmetry: the unscreened cosets of $\Lambda_{\cw}/\Lambda_{\crt}$, which form the sublattice generated by the genuine generator. In particular, for the theories of this paper:
\begin{equation}
\scalebox{0.97}{$
\vcenter{\hbox{%
\begin{tabular}{lll}
\toprule
Theory & Genuine 1-form & Generators of allowed gauge fluxes $n$ \\
\midrule
$\SU(4)_{\kCS} + N_{\Lambda^2}\Lambda^2$ & $\Z_2^{(1)}$ & $\omega_0=(0,0,0),\ \omega_2 = \bigl(\tfrac12,0,\tfrac12\bigr)$ \\
$\USp(4)_0 + N_{\Lambda^2}\Lambda^2$ & $\Z_2^{(1)}$ & $0,\ \bigl(\tfrac12, 0\bigr)$ \\
$\Spin(7) + N_f\,\mathbf{F}$ & $\Z_2^{(1)}$ & $0,\ \bigl(1,1,\tfrac12\bigr)$ \\
$\Spin(8) + N_f\,\mathbf{F}$ & $\Z_2^{(1)}$ & $0,\ \bigl(1,1,\tfrac12,\tfrac12\bigr)$ \\
\bottomrule
\end{tabular}}}
$}
\label{eq:genuine1form}
\end{equation}
The smallest non-trivial allowed flux generates $\Z_m^{(1)}$: $\omega_{N/m}$ (with $n_i = i/m$) for pure $\SU(N)_k$, $\omega_2$ for $\SU(4)+\Lambda^2$, and the unique $\Z_2$ flux for $\USp(4)_0 + N_{\Lambda^2}\Lambda^2$, $\Spin(7)+N_f\mathbf{F}$ and $\Spin(8) + N_f \mathbf{F}$. Fluxes outside this set --- e.g.\ the screened $\omega_1,\omega_3$ of $\SU(4)+\Lambda^2$, which are finer than the $\Z_2$ spacing --- are not backgrounds of a genuine 1-form symmetry and are excluded from the probe.

\subsubsection{Quantisation of the background fluxes in the presence of matter}  \label{sec:quantisationwithmatter}
The general formula \eqref{eq:expVgeneral} fixes the quantisation of all background fluxes by a two-step recipe.
\begin{enumerate}[leftmargin=*]
    \item \emph{Flavour fluxes $B_h$.} At vanishing gauge flux, \eqref{eq:Zgeneral} gives
    \begin{equation}
        Z_i(n)\big|_{n = 0} \;=\; \frac{1}{4} \sum_h \Bigl( B_h^2 - \frac{1}{4} \Bigr) \sum_{\lambda \in R_h} \sigma_\lambda \langle \lambda, \alpha_i^\vee \rangle\,.
    \end{equation}
    Requiring $\exp(-V_n)\big|_{n=0}$ to be independent of the gauge fugacities $z_i$ therefore fixes
    \begin{equation}
        B_h \;=\; \pm \frac{1}{2}
    \end{equation}
    for every hypermultiplet.
    \item \emph{Instanton flux $B_{m_0}$.} With the $B_h$ fixed as above, evaluate $\exp(-V_n)$ at the \emph{gauge flux $n$ in a non-trivial coset allowed by the 1-form symmetry} and impose that the gauge fugacities appear with integer powers, $Z_i(n) \in \Z$. This two-step recipe is carried out for the $\SU(2)$ gauge group with one fundamental hypermultiplet below, and for $\SU(4)_{\kCS}$ with antisymmetric matter, $\USp(4)$, $\Spin(7)$ and $\Spin(8)$ in the following subsections.
\end{enumerate}

\subsubsection*{Example: $\SU(2) + 1\mathbf{F}$}  The blow-up function is given by \eqref{eq:expVsu2f}. This is reproduced exactly by the general formula \eqref{eq:expVgeneral}, with $c_\lambda = \frac{1}{4}$ for both weights of the fundamental. Since this theory has a trivial 1-form symmetry, the allowed gauge flux is in the trivial coset of $\SU(2)$, namely $n \in\Z$.
We require that $\exp(-V_n)|_{n=0} = 1$, so $z_1^{\frac{1}{8} \left(-1+4 B_{m_1}^2\right)} = 1$, \ie~ $B_{m_1} = \pm \frac{1}{2}$. The quantisation condition of the background fluxes $B_{m_0}$ is then fixed by imposing that the gauge fugacity $z_1$ appears with integer power, which forces to take $B_{m_0} \in \Z \pm \frac{1}{4}$. This is, again, in agreement with \cite[(4.3)]{Kim:2025qaf}. The one- and two-instanton partitions function and the superconformal index of this theory is discussed in detail in \S\ref{sec:exSU2w1F}.

\subsection{$\SU(4)_{\kCS}$ with antisymmetric hypermultiplets}
\label{sec:expVSU4AS}

For $\SU(4)_{\kCS}$, we use the conventions of \S\ref{sec:expVSUN}, with $\tilde{n}_a = n_a - n_{a-1}$ ($n_0 = n_4 \equiv 0$). The antisymmetric representation $\Lambda^2 = \mathbf{6}$ has weights $\lambda_a + \lambda_b$ ($a < b$), pairing with the fluxes as $\tilde{n}_a + \tilde{n}_b$; since $\sum_a \tilde{n}_a = 0$, they organise into the three pairs $\pm n_2$, $\pm (n_1 - n_2 + n_3)$, $\pm (n_3 - n_1)$. We consider $N_{\Lambda^2}$ antisymmetric hypermultiplets, $h = 1, \ldots, N_{\Lambda^2}$, in the chamber
\begin{equation}
    f_3 \geq f_2 \geq f_1 > 0\,, \qquad 0 < m_h < f_2\,,\quad m_h < f_3 - f_1\,,\quad m_h < f_1 + f_3 - f_2\,,
\end{equation}
in which all three weight pairs resolve with opposite signs ($c_\lambda = \frac{1}{4}$ throughout), with chamber-positive pairings $\hat{w} = \bigl( n_2,\; n_1 - n_2 + n_3,\; n_3 - n_1 \bigr)$. In this chamber the cubic prepotential and the effective prepotential on the $\Omega$-background are
\begin{subequations}
\label{eq:prepotSU4AS}
\begin{align}
    \displaystyle \mathcal{F} \;=&\; m_0\bigl( f_1^2 - f_1 f_2 + f_2^2 - f_2 f_3 + f_3^2 \bigr) + \frac{\kCS}{2}\bigl( f_1^2 f_2 - f_1 f_2^2 + f_2^2 f_3 - f_2 f_3^2 \bigr)\nonumber\\
    +&\; \scalebox{0.95}{$ \displaystyle\frac{4}{3}\bigl( f_1^3 + f_2^3 + f_3^3 \bigr) - f_1 f_2^2 - f_2^2 f_3 - \frac{1}{12}\sum_{h=1}^{N_{\Lambda^2}}\sum_{a=1}^{3}\Bigl[ (\hat{w}_a + m_h)^3 + (\hat{w}_a - m_h)^3 \Bigr]\,, $}\\
    \mathcal{E} \;=&\; \scalebox{0.87}{$ \displaystyle+\frac{1}{\epsilon_1 \epsilon_2}\biggl[ \mathcal{F} + \frac{1}{4}(f_1+f_2+f_3)(\epsilon_1+\epsilon_2)^2 + \Bigl( \frac{N_{\Lambda^2}}{12} f_3 - \frac{1}{12}(f_1+f_2+f_3) \Bigr)(\epsilon_1^2+\epsilon_2^2) \biggr]$}\,,
\end{align}
\end{subequations}
where, in the matter sum, $\hat{w} = (f_2,\, f_1 - f_2 + f_3,\, f_3 - f_1)$ now denotes the chamber-positive $\Lambda^2$ pairings evaluated on the VEVs (all gauge-dominated), and the pure theory is recovered at $N_{\Lambda^2} = 0$. The general formula \eqref{eq:expVgeneral} then gives
\begin{subequations}
\label{eq:expVSU4AS}
\begin{align}
    Q(n) \;=&\; n_1^2 - n_1 n_2 + n_2^2 - n_2 n_3 + n_3^2\,, \\
    W_h(n) \;=&\; -2 B_{m_h} n_3\,, \\
    Z_1(n) \;=&\; -\frac{1}{2} (2 n_1 - n_2) \Bigl[ 2 B_{m_0} + 4 n_1 + \bigl( 2 + N_{\Lambda^2} + \kCS \bigr) n_2 - 2 N_{\Lambda^2}\, n_3 \Bigr]\,, \\
    Z_2(n) \;=&\; \scalebox{0.9}{$ \displaystyle-\frac{1}{2} (n_1 - 2 n_2 + n_3) \Bigl[ -2 B_{m_0} + \bigl( N_{\Lambda^2} + \kCS \bigr) n_1 - 4 n_2 + \bigl( N_{\Lambda^2} - \kCS \bigr) n_3 \Bigr]$}\,, \\
    Z_3(n) \;=&\; \sum_h \Bigl( B_{m_h}^2 - \frac{1}{4} \Bigr) + N_{\Lambda^2} \bigl( n_1^2 - n_1 n_2 \bigr) + \frac{N_{\Lambda^2} + 2 - \kCS}{2}\, n_2^2 \nonumber \\
    +&\; B_{m_0} (n_2 - 2 n_3) + \bigl( \kCS - N_{\Lambda^2} \bigr) n_2 n_3 - \bigl( 4 - N_{\Lambda^2} \bigr) n_3^2\,, \\
    P(n) \;=&\; B_{m_0}\, Q(n) + \frac{1}{6} \sum_{\alpha \in \Delta^+} \Bigl( \langle n, \alpha \rangle^3 - \langle n, \alpha \rangle \Bigr) + \frac{\kCS}{6} \sum_{a=1}^{4} \tilde{n}_a^3 \nonumber \\
    -&\; \frac{N_{\Lambda^2}}{6} \bigl( \hat{w}_1^3 + \hat{w}_2^3 + \hat{w}_3^3 \bigr) + \frac{N_{\Lambda^2}}{12}\, n_3 - n_3 \sum_h B_{m_h}^2\,,
\end{align}
\end{subequations}
in exact agreement with the explicit blow-up computation for $N_{\Lambda^2} = 1, 2$, including all the $\kCS$-dependence.

\paragraph{Quantisation.} At $n = 0$, the only dependence on the gauge fugacities resides in $Z_3(0) = \sum_h \bigl( B_{m_h}^2 - \frac{1}{4} \bigr)$, so step 1 of the recipe fixes the flavour flux $B_{m_h} = \pm\frac{1}{2}$ \emph{independently for each of the $N_{\Lambda^2}$ antisymmetric hypermultiplets} ($h = 1, \ldots, N_{\Lambda^2}$). The antisymmetric matter screens $\Z_4^{(1)}$ down to $\Z_2^{(1)}$, generated by $\omega_2 = \bigl( \frac{1}{2}, 0, \frac{1}{2} \bigr)$. With $B_{m_h} = \pm\frac{1}{2}$, step 2 applied to the cosets $\omega_0, \omega_2$ gives, for any even $\kCS$,
\begin{equation}
    B_{m_0} \in \Z + \frac{N_{\Lambda^2}}{2} \pmod{1}\,,
    \label{eq:Bm0SU4}
\end{equation}
i.e.\ half-integral for one antisymmetric hyper and integral for two.

\subsection{$\USp(4)$ with antisymmetric hypermultiplets}
\label{sec:expVUSp4}

For $\USp(4) = C_2$ gauge algebra, we use the orthonormal basis $e_1, e_2$ of the Cartan subalgebra, with simple roots $\alpha_1 = e_1 - e_2$ (short) and $\alpha_2 = 2e_2$ (long), and coroots $\alpha_1^\vee = e_1 - e_2$, $\alpha_2^\vee = e_2$. The fluxes $n = n_1 \alpha_1^\vee + n_2 \alpha_2^\vee$ have orthonormal components $\tilde{n} = (n_1, n_2 - n_1)$, and the positive roots $2e_1, 2e_2, e_1 + e_2, e_1 - e_2$ pair with $n$ as $2n_1$, $2(n_2 - n_1)$, $n_2$, $2n_1 - n_2$. The antisymmetric representation $\Lambda^2 = \mathbf{5} \oplus \mathbf{1}$ has gauge-charged weights $\pm(e_1 + e_2)$ and $\pm(e_1 - e_2)$, pairing with $n$ as $\pm n_2$ and $\pm(2n_1 - n_2)$. We consider $N_{\Lambda^2}$ antisymmetric hypermultiplets, $h = 1, \ldots, N_{\Lambda^2}$, in the chamber
\begin{equation}
    f_2 > 0\,, \qquad \frac{f_2}{2} < f_1 < \frac{2 f_2}{3}\,, \qquad 2 f_1 - f_2 < m_h < f_2 - f_1\,,
\end{equation}
in which the weight pair $\pm(e_1 + e_2)$ resolves with opposite signs ($c_\lambda = \frac{1}{4}$), while $\pm(e_1 - e_2)$ resolves with equal signs ($c_\lambda = 0$). In this chamber, the cubic prepotential and the effective prepotential on the $\Omega$-background are\footnote{Note that, upon setting $(a_1, a_2) = (f_1, f_2 -
f_1)$ and $m_0 = 1/g_0^2$, our classical term in the prepotential is $m_0 (2f_1^2-2 f_1f_2
+f_2^2)$; this is twice the corresponding term in \cite[(4.13)]{Jefferson:2017ahm}. However, our convention agrees with \cite[(5.3)]{Kim:2020hhh}.}
\begin{subequations}
\label{eq:prepotUSp4}
\begin{align}
    \mathcal{F} \;=&\; m_0 \sum_{k=1}^{2} \tilde{\phi}_k^2 + \frac{1}{6}\Bigl[ (2\tilde{\phi}_1)^3 + (2\tilde{\phi}_2)^3 + (\tilde{\phi}_1+\tilde{\phi}_2)^3 + (\tilde{\phi}_1-\tilde{\phi}_2)^3 \Bigr] \nonumber\\
    -&\; \frac{1}{12}\sum_{h=1}^{N_{\Lambda^2}}\Bigl[ (u+m_h)^3 + (u-m_h)^3 + (v+m_h)^3 + (m_h-v)^3 \Bigr]\,, \\
    \mathcal{E} \;=&\; \frac{1}{\epsilon_1 \epsilon_2}\biggl[ \mathcal{F} + \frac{1}{4}(f_1+f_2)(\epsilon_1+\epsilon_2)^2 + \Bigl( \frac{N_{\Lambda^2}}{24} f_2 - \frac{1}{12}(f_1+f_2) \Bigr)(\epsilon_1^2+\epsilon_2^2) \biggr]\,,
\end{align}
\end{subequations}
where $\tilde{\phi} = (f_1,\, f_2 - f_1)$ are the orthonormal Coulomb-branch VEVs and $u = \tilde{\phi}_1 + \tilde{\phi}_2 = f_2$, $v = \tilde{\phi}_1 - \tilde{\phi}_2 = 2 f_1 - f_2$ (the gauge-dominated pair $\pm(e_1{+}e_2)$ and the mass-dominated pair $\pm(e_1{-}e_2)$). The general formula \eqref{eq:expVgeneral} then gives
\begin{subequations}
\label{eq:expVUSp4}
\begin{align}
    Q(n) \;=&\; \frac{1}{2} \Tr_{\mathbf{4}}\, n^2 \;=\; 2 n_1^2 - 2 n_1 n_2 + n_2^2\,, \\
    W_h(n) \;=&\; -\frac{B_{m_h}^2}{2} - B_{m_h} n_2 - \frac{1}{2} (2 n_1 - n_2)^2\,, \\
    Z_1(n) \;=&\; -(2 n_1 - n_2) \Bigl( 2 n_1 + 3 n_2 + 2 B_{m_0} - 2 \textstyle\sum_h B_{m_h} \Bigr)\,, \\
    Z_2(n) \;=&\; -4 (n_2 - n_1)^2 - 2 n_1 (n_2 - n_1) - 2 B_{m_0} (n_2 - n_1) \nonumber\\
   +&\; \sum_h \biggl[ \frac{n_2^2 + B_{m_h}^2}{2} - B_{m_h} (2 n_1 - n_2) - \frac{1}{8} \biggr]\,, \\
    P(n) \;=&\; B_{m_0}\, Q(n) + \frac{4}{3} \Bigl( n_1^3 + (n_2 - n_1)^3 \Bigr) + \frac{n_2^3 + (2 n_1 - n_2)^3}{6} - \frac{n_1 + n_2}{3} \nonumber \\
    -&\; \sum_h \biggl[ \frac{n_2^3 + 3 n_2 B_{m_h}^2 + B_{m_h}^3 + 3 B_{m_h} (2 n_1 - n_2)^2}{6} - \frac{n_2}{24} \biggr]\,,
\end{align}
\end{subequations}
in exact agreement with the explicit blow-up computation for $N_{\Lambda^2} = 1, 2$.

\paragraph{Quantisation rules for $\USp(4)_{\theta=0} + N_{\Lambda^2} \, \Lambda^2$.} At $n = 0$, the only dependence on the gauge fugacities resides in $Z_2(0) = \sum_h \bigl( \frac{1}{2} B_{m_h}^2 - \frac{1}{8} \bigr)$, so step 1 of the recipe of \S\ref{sec:expVgeneralsub} fixes the flavour flux $B_{m_h} = \pm\frac{1}{2}$ \emph{independently for each of the $N_{\Lambda^2}$ antisymmetric hypermultiplets} ($h = 1, \ldots, N_{\Lambda^2}$). The antisymmetric matter does not screen the $\Z_2^{(1)}$ 1-form symmetry, so the generating flux $n = \bigl( \frac{1}{2}, 0 \bigr)$ is allowed. Evaluating $Z_2$ on this flux with $B_{m_h} = \pm\frac{1}{2}$ gives $Z_2 = B_{m_0} - \frac{1}{2} - \frac{N_{\Lambda^2}}{2}$, so that integrality $Z_2 \in \Z$ requires
\begin{equation}
    B_{m_0} \;\in\; \Z + \frac{N_{\Lambda^2} + 1}{2} \pmod{1}\,,
    \label{eq:Bm0USp4}
\end{equation}
i.e.\ integral for odd $N_{\Lambda^2}$ and half-integral for even $N_{\Lambda^2}$; in particular $B_{m_0} \in \Z$ for $N_{\Lambda^2} = 1$ and $B_{m_0} \in \Z + \frac{1}{2}$ for $N_{\Lambda^2} = 2$. This is the opposite parity to the $\SU(4)$ condition \eqref{eq:Bm0SU4}, $B_{m_0} \in \Z + \frac{N_{\Lambda^2}}{2}$.

\paragraph{Quantisation rules for $\USp(4)_{\theta=\pi} + N_{\Lambda^2} \, \Lambda^2$.} As in the above discussion, $B_{m_h} = \pm \frac{1}{2}$. However, in this case, the $\Z_2^{(1)}$ 1-form symmetry is broken by the theta angle, and so the generating flux $n = \bigl( \frac{1}{2}, 0 \bigr)$ is {\it not} allowed. With integral values of $n_1$ and $n_2$ and $B_h =\pm \frac{1}{2}$, the integrality of $Z_1$ and $Z_2$ implies ``opposite'' quantisation rule to \eqref{eq:Bm0USp4}, namely
\begin{equation}
    B_{m_0} \;\in\; \Z + \frac{N_{\Lambda^2} }{2} \pmod{1}\,,
    \label{eq:Bm0USp4thetapi}
\end{equation}
gives a consistent superconformal index for various $N_{\Lambda^2}$. In particular, for $N_{\Lambda^2}=0$ (\ie~ the $\USp(4)_\pi$ pure gauge theory), the quantisation rule agrees with \cite[(3.71)]{Kim:2020hhh}, and this choice of quantisation gives the index that is equal to that of $\SU(3)_5$ pure gauge theory, as expected.

\subsection{$\Spin(7)$ with vector hypermultiplets}
\label{sec:expVSO7}

For $\Spin(7) = B_3$ gauge algebra, we use the orthonormal basis $e_1, e_2, e_3$, with simple roots $\alpha_1 = e_1 - e_2$, $\alpha_2 = e_2 - e_3$ (long) and $\alpha_3 = e_3$ (short), and coroots $\alpha_1^\vee = e_1 - e_2$, $\alpha_2^\vee = e_2 - e_3$, $\alpha_3^\vee = 2 e_3$. The fluxes $n = \sum_i n_i \alpha_i^\vee$ have orthonormal components $\tilde{n} = (n_1, n_2 - n_1, 2 n_3 - n_2)$, and the positive roots $e_i \pm e_j$ ($i < j$) and $e_k$ pair with $n$ as $\tilde{n}_i \pm \tilde{n}_j$ and $\tilde{n}_k$. The vector representation $\mathbf{7}$ has gauge-charged weights $\pm e_k$, pairing with $n$ as $\pm \tilde{n}_k$ (the zero weight drops out). We consider three vector hypermultiplets, $h = 1, 2, 3$, in the chamber
\begin{equation}
    0 < f_1 < f_2 < 2 f_1\,, \qquad f_2 < 2 f_3 < 2 f_2 - f_1\,, \qquad f_3 < f_1\,,
\end{equation}
with $0 < m_h$ taken below each of $f_2 - f_1$, $\,2 f_1 - f_2$, $\,2 f_3 - f_2$, $\,2 f_2 - f_1 - 2 f_3$ and $\,f_1 - f_3$, in which every weight pair $\pm e_k$ resolves with opposite signs ($c_\lambda = \frac{1}{4}$ throughout). In this chamber, the cubic prepotential and the effective prepotential on the $\Omega$-background for $N_f$ vector hypermultiplets are\footnote{This prepotential is in agreement with \cite[(4.21)]{Jefferson:2017ahm}, up to a mass-only constant, upon setting $(a_1, a_2, a_3) =
(f_1,f_2-f_1,2 f_3-f_2)$ and $m_0 = 1/g_0^2$.}
\begin{subequations}
\label{eq:prepotSO7}
\begin{align}
    \mathcal{F} \;&=\; \scalebox{0.9}{$ \displaystyle m_0 \sum_{k=1}^{3} \tilde{\phi}_k^2 + \frac{1}{6}\sum_{1 \le i < j \le 3}\Bigl[ (\tilde{\phi}_i+\tilde{\phi}_j)^3 + (\tilde{\phi}_i-\tilde{\phi}_j)^3 \Bigr] + \frac{1 - N_f}{6}\sum_{k=1}^{3}\tilde{\phi}_k^3 - f_3\sum_{h=1}^{N_f} m_h^2$}\,, \\
    \mathcal{E} \;&=\; \scalebox{0.9}{$ \displaystyle\frac{1}{\epsilon_1 \epsilon_2}\biggl[ \mathcal{F} + \frac{1}{4}(f_1+f_2+f_3)(\epsilon_1+\epsilon_2)^2 + \Bigl( \frac{N_f}{12} f_3 - \frac{1}{12}(f_1+f_2+f_3) \Bigr)(\epsilon_1^2+\epsilon_2^2) \biggr]$}\,,
\end{align}
\end{subequations}
with $\tilde{\phi} = (f_1,\, f_2 - f_1,\, 2 f_3 - f_2)$ the orthonormal Coulomb-branch VEVs; $\mathcal{F}$ is given up to the $\phi$-independent constant $-\tfrac{1}{12}\sum_h m_h^3$ from the zero weight of $\mathbf{7}$. The general formula \eqref{eq:expVgeneral} then gives\footnote{We remark that the gauge flux $n=(1,1,\frac{1}{2})$, \ie $\tilde{n}=(1,0,0) = e_1$, corresponds to a short root of $B_3$ and gives $Q(n)=1$.  On the other hand, the flux $n= (1,2,1)$, \ie $\tilde{n}=(1,1,0) = e_1+e_2$ corresponds to a long root of $B_3$ and gives $Q(n)=2$. It was pointed out in \cite[(2.33), (2.39)]{Kim:2019uqw} that only the fluxes corresponding to the long roots contribute at the one instanton level, where their normalisation is such that the half-squared length of any long root $\vec{k}$ is 1, \ie~$\frac{1}{2} \vec{k} \cdot \vec{k} =1$. Their convention, however, corresponds to our $Q(n)=2$. Thus, their one-instanton partition function (which agrees with that obtained from the one-instanton ADHM construction) corresponds to our two-instanton partition function obtained from the blow-up formula using the convention of \cite{Kim:2020hhh}. \label{foot:shortlongB3}}
\begin{subequations}
\label{eq:expVSO7}
\begin{align}
    Q(n) \;=&\; \frac{1}{2} \Tr_{\mathbf{7}}\, n^2 \;=\; 2 \bigl( n_1^2 - n_1 n_2 + n_2^2 - 2 n_2 n_3 + 2 n_3^2 \bigr)\,, \\
    W_h(n) \;=&\; -2 B_{m_h} n_3\,, \\
    Z_1(n) \;=&\; -(2 n_1 - n_2) \bigl( 2 n_1 - n_2 + 2 B_{m_0} \bigr)\,, \\
    Z_2(n) \;=&\; 2 \bigl( B_{m_0} + n_1 + n_2 - 2 n_3 \bigr) \bigl( n_1 - 2 n_2 + 2 n_3 \bigr)\,, \\
    Z_3(n) \;=&\; -\frac{3}{4} + \sum_h B_{m_h}^2 + 2 (n_2 - 2 n_3) \bigl( 2 B_{m_0} + 3 n_2 - 2 n_3 \bigr)\,, \\
    P(n) \;=&\; B_{m_0}\, Q(n) + \frac{4}{3} n_1^3 - 2 n_1^2 n_2 + n_1 n_2^2 + \frac{2}{3} (n_2 - 2 n_3)^2 (2 n_2 - n_3) \nonumber \\
    -&\; \frac{n_1 + n_2}{3} - \frac{n_3}{12} - n_3 \sum_h B_{m_h}^2\,,
\end{align}
\end{subequations}
in exact agreement with the explicit blow-up computation. Note that the hypermultiplet contributions to $Z_1$ and $Z_2$ cancel against the short-root contributions --- a cancellation specific to three flavours.

\paragraph{Quantisation rules for the $\Spin(7)$ pure gauge theory.} The integrality of the exponents of $z_i$ (with $i=1,2,3$) evaluated at the fluxes $n=(0,0,1/2)$ and $(1,1,1/2)$ fixes $B_{m_0} \in \Z \pm \frac{1}{4}$.

\paragraph{Quantisation rules for $\Spin(7) + 3\mathbf{F}$.} At $n = 0$, the only dependence on the gauge fugacities resides in $Z_3(0) = \sum_h B_{m_h}^2 - \frac{3}{4}$, so step 1 fixes $B_{m_h} = \pm\frac{1}{2}$. The vector matter does not screen the 1-form symmetry, and the 't~Hooft flux $\tilde{n} = (1, 0, 0)$, i.e.\ $n = \bigl( 1, 1, \frac{1}{2} \bigr)$, corresponding to a bundle that does not lift to $\mathrm{Spin}(7)$, is allowed.\footnote{Numerically $n=(1,1,\frac{1}{2})$, \ie $\tilde{n}=(1,0,0)=e_1=\omega_1$, is the highest weight of the vector $\mathbf{7}$, but it enters here as a \emph{magnetic} flux, i.e.\ a cocharacter. It generates $\Z_2^{(1)}$ because the cocharacter lattice of $\SO(7)$ is the coweight lattice $P^\vee=\Z^3$ of $B_3$, whereas genuine $\Spin(7)$ fluxes form the coroot lattice $Q^\vee$ (the integer vectors of even coordinate sum); since $P^\vee/Q^\vee=\Z_2$, there is a single non-trivial class, represented by $\tilde{n}=(1,0,0)$. The spinor weight $\tilde{n} =\bigl(\tfrac12,\tfrac12,\tfrac12\bigr) = \omega_{\mathbf{8}}$, i.e.\ $n=\bigl(\tfrac12,1,\tfrac34\bigr)$, is by contrast \emph{not} an allowed flux: it lies outside $P^\vee$, pairing half-integrally with the short root $\alpha_3=e_3$. The spinor is instead the electric line that \emph{detects} this flux, $\langle\tilde{n}=(1,0,0),\omega_{\mathbf{8}}\rangle=\tfrac12$ (Aharonov--Bohm phase $-1$), so the $\Z_2^{(1)}$ background $n=(1,1,\frac{1}{2})$ already \emph{is} ``the spinor flux''.} Step 2, evaluated on this flux with $B_{m_h} = \pm\frac{1}{2}$, gives $Z_2 = Z_3 = 0$ and $Z_1 = -\bigl( 2 B_{m_0} + 1 \bigr)$, so $z$-integrality fixes only $2 B_{m_0} \in \Z$, \ie~ $B_{m_0}$ can be either integral or half-integral. Unfortunately, this cannot be decided by using $\exp(-V_n)$ alone. As we will see in \S\ref{sec:indexSO7}, the one-instanton partition function must be invariant under $x \to 1/x$ --- required by the fact that $x$ is an $\SU(2)_R$ fugacity --- and only $B_{m_0} \in \BZ$ leads to a partition function with this property. This also turns out to be compatible with the faithful $\SO(3)_I$ instanton symmetry of $\Spin(7) + 3\mathbf{F}$, discussed in \S\ref{sec:indexSO7}. Hence $B_{m_0} \in \Z$ for $\Spin(7) + 3\mathbf{F}$. 

\subsection{$\Spin(8)$ with vector hypermultiplets}
\label{sec:expVSO8}

For $\Spin(8) = D_4$ gauge algebra, we use the orthonormal basis $e_1, \ldots, e_4$, with simple roots $\alpha_1 = e_1 - e_2$, $\alpha_2 = e_2 - e_3$, $\alpha_3 = e_3 - e_4$ and $\alpha_4 = e_3 + e_4$; as $D_4$ is simply laced, the coroots coincide with the roots, $\alpha_i^\vee = \alpha_i$. The fluxes $n = \sum_i n_i \alpha_i^\vee$ have orthonormal components
\begin{equation}
    \tilde{n} = \bigl( n_1,\; n_2 - n_1,\; n_3 + n_4 - n_2,\; n_4 - n_3 \bigr)\,,
\end{equation}
the positive roots $e_i \pm e_j$ ($i < j$) pair with $n$ as $\tilde{n}_i \pm \tilde{n}_j$, and the vector representation $\mathbf{8}_v$ has gauge-charged weights $\pm e_k$, pairing with $n$ as $\pm \tilde{n}_k$. We consider four vector hypermultiplets, $h = 1, \ldots, 4$, in the chamber
\bes{
    &2 f_1 - f_2 > 0\,,\quad 2 f_2 - f_1 - f_3 - f_4 > 0\,,\quad 2 f_3 - f_2 > 0\,,\quad \\& f_4 - f_3 > 0\,,\quad 0 < m_h < f_4 - f_3\,,
}
which orders the orthonormal Coulomb-branch holonomies as $\tilde{\phi}_1 > \tilde{\phi}_2 > \tilde{\phi}_3 > \tilde{\phi}_4 > 0$ (with the smallest being $\tilde{\phi}_4 = f_4 - f_3$) and places every mass below it; all eight weight pairs $\pm e_k$ are then gauge-dominated ($c_\lambda = \frac{1}{4}$ throughout). In this chamber, the cubic prepotential and the effective prepotential on the $\Omega$-background for $N_f$ vector hypermultiplets are\footnote{This prepotential is in agreement with \cite[(4.27)]{Jefferson:2017ahm} upon setting $(a_1, a_2, a_3, a_4) = (f_1,f_2-f_1,f_3+f_4-f_2,f_4-f_3)$ and $m_0 = 1/g_0^2$.}
\begin{subequations}
\label{eq:prepotSO8}
\begin{align}
    \mathcal{F} \;=&\; m_0 \sum_{k=1}^{4} \tilde{\phi}_k^2 + \frac{1}{6}\sum_{1 \le i < j \le 4}\Bigl[ (\tilde{\phi}_i+\tilde{\phi}_j)^3 + (\tilde{\phi}_i-\tilde{\phi}_j)^3 \Bigr] \nonumber\\
   -&\; \frac{N_f}{6}\sum_{k=1}^{4}\tilde{\phi}_k^3 - f_4\sum_{h=1}^{N_f} m_h^2\,, \\
    \mathcal{E} \;=&\; \frac{1}{\epsilon_1 \epsilon_2}\biggl[ \mathcal{F} + \frac{1}{4}(f_1+f_2+f_3+f_4)(\epsilon_1+\epsilon_2)^2 \nonumber\\
    +&\; \Bigl( \frac{N_f}{12} f_4 - \frac{1}{12}(f_1+f_2+f_3+f_4) \Bigr)(\epsilon_1^2+\epsilon_2^2) \biggr]\,,
\end{align}
\end{subequations}
with $\tilde{\phi} = (f_1,\, f_2 - f_1,\, f_3 + f_4 - f_2,\, f_4 - f_3)$ the orthonormal Coulomb-branch VEVs. The general formula \eqref{eq:expVgeneral} then gives
\begin{subequations}
\label{eq:expVSO8}
\begin{align}
    Q(n) \;&=\; \frac{1}{2} \Tr_{\mathbf{8}_v}\, n^2 \;=\; 2 \bigl( n_1^2 - n_1 n_2 + n_2^2 + n_3^2 + n_4^2 - n_2 (n_3 + n_4) \bigr)\,, \\
    W_h(n) \;&=\; -2 B_{m_h} n_4\,, \\
    Z_1(n) \;&=\; -(2 n_1 - n_2) \bigl( 2 B_{m_0} + 2 n_1 - n_2 \bigr)\,, \\
    Z_2(n) \;&=\; 2 \bigl( B_{m_0} + n_1 + n_2 - n_3 - n_4 \bigr) \bigl( n_1 - 2 n_2 + n_3 + n_4 \bigr)\,, \\
    Z_3(n) \;&=\; (n_2 - 2 n_3) \bigl( 2 B_{m_0} + 3 n_2 + 2 n_3 - 4 n_4 \bigr)\,, \\
    Z_4(n) \;&=\; -1 + \sum_h B_{m_h}^2 + 2 B_{m_0} n_2 + 3 n_2^2 - 4 n_2 n_3 + 4 n_3^2 - 4 \bigl( B_{m_0} + n_2 \bigr) n_4\,, \\
    P(n) \;&=\; \scalebox{0.93}{$ \displaystyle B_{m_0}\, Q(n) + \frac{1}{6} \sum_{\alpha \in \Delta^+} \Bigl( \langle n, \alpha \rangle^3 - \langle n, \alpha \rangle \Bigr) - \frac{2}{3} \sum_{k=1}^4 \tilde{n}_k^3 + \frac{n_4}{3} - n_4 \sum_h B_{m_h}^2 $}\,,
\end{align}
\end{subequations}
in exact agreement with the explicit blow-up computation. As for $\Spin(7)$, the hypermultiplet contributions to $Z_1, Z_2, Z_3$ cancel against the root contributions, leaving the matter only in $Z_4$ (and in $W_h, P$) --- a cancellation special to four vectors.
\paragraph{Quantisation rules.} At $n = 0$, the only dependence on the gauge fugacities resides in $Z_4(0) = \sum_h B_{m_h}^2 - 1$, so step~1 of the recipe of \S\ref{sec:expVgeneralsub} fixes $B_{m_h} = \pm\frac{1}{2}$ for each of the four vectors. The vector matter does not screen the 1-form symmetry, and the smallest allowed 't~Hooft flux is the vector class $\tilde{n} = (1, 0, 0, 0)$, i.e.\ $n = \bigl( 1, 1, \tfrac{1}{2}, \tfrac{1}{2} \bigr)$. Step~2, evaluated on this flux with $B_{m_h} = \pm\frac{1}{2}$, gives $Z_2 = Z_3 = Z_4 = 0$ and $Z_1 = -\bigl( 2 B_{m_0} + 1 \bigr)$, so $z$-integrality fixes only $2 B_{m_0} \in \Z$, \ie~ $B_{m_0}$ can be either integral or half-integral. As for $\Spin(7) + 3\mathbf{F}$, this cannot be decided by using $\exp(-V_n)$ alone. As we will see in \S\ref{sec:indexSO8}, the one-instanton partition function must be invariant under $x \to 1/x$ --- required by the fact that $x$ is an $\SU(2)_R$ fugacity --- and only $B_{m_0} \in \BZ$ leads to a partition function with this property. This is also compatible with the faithful $\SO(3)_I$ instanton symmetry of $\Spin(8) + 4\mathbf{F}$, discussed in \S\ref{sec:indexSO8}. Hence $B_{m_0} \in \Z$ for $\Spin(8) + 4\mathbf{F}$. The same conclusion holds for the pure $\Spin(8)$ theory: the integrality of the exponents of $z_i$ (with $i = 1, \ldots, 4$) at the three non-trivial 't~Hooft fluxes --- the vector $\tilde{n} = (1, 0, 0, 0)$, i.e.\ $n = \bigl( 1, 1, \tfrac{1}{2}, \tfrac{1}{2} \bigr)$, the spinor $\tilde{n} = \bigl( \tfrac{1}{2}, \tfrac{1}{2}, \tfrac{1}{2}, \tfrac{1}{2} \bigr)$, i.e.\ $n = \bigl( \tfrac{1}{2}, 1, \tfrac{1}{2}, 1 \bigr)$, and the cospinor $\tilde{n} = \bigl( \tfrac{1}{2}, \tfrac{1}{2}, \tfrac{1}{2}, -\tfrac{1}{2} \bigr)$, i.e.\ $n = \bigl( \tfrac{1}{2}, 1, 1, \tfrac{1}{2} \bigr)$ --- likewise fixes only $2 B_{m_0} \in \Z$. 

For the $\Spin(8)$ pure gauge theory, however, we find that the integer and half-integer options are \emph{equivalent}: $B_{m_0} \in \Z$ and $B_{m_0} \in \Z + \tfrac{1}{2}$ yield the same one-instanton partition function, equal to $x^{6}$ times the Hilbert series of one $\Spin(8)$ instanton (\S\ref{sec:indexSO8}), so either may be used. This is no longer true for the theory with matter: for $\Spin(8) + 4\mathbf{F}$ only $B_{m_0} \in \Z$ survives, as established above. In contrast with the two branches in the quantisation rule $B_{m_0} \in \Z \pm \frac{1}{4}$ of the $\Spin(7)$ pure gauge theory (\S\ref{sec:expVSO7}) originating from the short roots of $B_3$ (which contribute half-integer constants to the $z_i$ exponents), the simply-laced $D_4$ algebra has no short roots, so the pure $\Spin(8)$ exponents carry integer constants and admit no such branch.

\subsection{Non-Lagrangian theories}
\label{sec:expVnonlag}

The construction extends to the non-Lagrangian SCFTs of appendices~\ref{app:BN} and
\ref{app:PF}: the $B_N$ family of~\cite{Eckhard:2020jyr,Morrison:2020ool} (whose rank-one member
$B_3$ is the local $\mathbb{P}^2$, i.e.\ the $E_0$ theory of~\cite{Morrison:1996xf}) and the
rank-two theories $\mathbb{P}^2\cup\mathbb{F}_3$ and $\mathbb{P}^2\cup\mathbb{F}_6$
of~\cite{Jefferson:2018irk,Apruzzi:2019opn}, whose partition functions were computed
in~\cite{Kim:2020hhh}. These and related non-Lagrangian SCFTs, and the geometric extraction of
their 1-form symmetry, were studied in~\cite{Morrison:2020ool}.

There is no gauge Lagrangian, but the blow-up function is still controlled
by the Coulomb branch prepotential \eqref{eq:BNprepot}, assembled from the geometric intersection
data of \eqref{eq:BNcouplings}: the cubic couplings $c_{ijk}=S_i\!\cdot\!S_j\!\cdot\!S_k$, the
gravitational couplings $C^G_i=c_2(X)\cdot S_i$ and the $\SU(2)_R$ couplings $C^R_i=2$. The
magnetic fluxes $n_i$ now thread the compact divisors $S_i$, and $z_i=e^{\phi_i}$ are the
Coulomb fugacities. 

As for the Lagrangian theories, the \emph{allowed} (dynamical) fluxes are
those for which the exponents $Z_i(n)$ of the Coulomb fugacities are integers; we find that this
single condition reproduces the magnetic flux quantisation of these theories, spin$^c$ shifts
included. Explicitly, the blow-up function reads
\begin{equation}
    \exp(-V_n)=(p_1p_2)^{P(n)}\,\prod_i z_i^{Z_i(n)}\,,\qquad
    Z_i(n)=\kappa_i-\tfrac12\,c_{ijk}\,n_jn_k\,,
    \label{eq:expVnonlagform}
\end{equation}
with $p_1p_2=x^2$ the $\SU(2)_R$ fugacity, $P(n)$ the $(p_1p_2)$-exponent \eqref{eq:Pnonlag}, whose
fractional part is the cubic anomaly of \S\ref{sec:anomnonlag}, and $\kappa_i$ a constant spin$^c$
zero-point (the $\tfrac18$ below for the non-spin $\mathbb{P}^2$, zero for a spin divisor). A flavour
$\U(1)$, where present, carries its own magnetic flux $B_h$ --- the flux through the flavour divisor
$D_F$ --- so the total (gauge and flavour) flux is $\mathcal{N}=\sum_i n_iS_i+B_h\,D_F$, and the Coulomb exponents pick up a
cross-term, namely
\begin{equation}
    Z_i=-\tfrac12\,S_i\!\cdot\!\mathcal{N}^2=\kappa_i-\tfrac12 c_{ijk}n_jn_k-(S_iS_jD_F)\,n_j\,B_h-\tfrac12(S_iD_F^2)\,B_h^2\,,
    \label{eq:expVflavflux}
\end{equation}
and a flavour factor $w^{W}$ appears, with exponent $W=\tfrac12\,D_F\cdot\mathcal{N}^2$ (the sign
appropriate to $w=e^{-m}$); the $(p_1p_2)$-exponent $P$ likewise depends on the full flux $\mathcal N$
through its cubic term, $P=\tfrac16 c_{ijk}\mathcal N^i\mathcal N^j\mathcal N^k+\bigl(\tfrac{C^G_i}{48}-\tfrac{C^R_i}{8}\bigr)n_i$,
the gravitational and $\SU(2)_R$ couplings running over the compact fluxes only. For the matter theories of
\S\ref{sec:expVSU4AS}--\S\ref{sec:expVSO7}, the allowed $B_h$ follow from demanding that
$\exp(-V_n)$ be independent of the $z_i$ at the minimal gauge flux $n=0$, where the hypermultiplet
zero-point pins $B_{m_h}=\pm\tfrac12$. The non-Lagrangian theories below have no such zero-point ---
$\exp(-V_n)|_{n=0}$ is $z_i$-independent for \emph{any} $B_h$, so $n=0$ imposes no condition --- and the
flavour flux is instead quantised by $z_i$-integrality of $\exp(-V_{\mathfrak{n}})$ on the
non-trivial $1$-form cosets $\mathfrak{n}$, discussed in more detail below. We collect the effective prepotentials (many of which are new) in the following table, and display $\exp(-V_n)$ for each theory in turn.
\bes{
\begin{tabular}{cc}
\hline
Theory & Effective prepotential $\mathcal{E}$\\
\hline
$B_3=B_3^{(2)}=E_0$ & \cite[(4.20)]{Kim:2021gyj} \\
$B_4$ & \eqref{eq:B4prepot} \\
$B_5$ &  \eqref{eq:B5prepot} \\
$\mathbb{P}^2 \cup \mathbb{F}_{3}$ &  \cite[(5.63)]{Kim:2020hhh} \\
$\mathbb{P}^2 \cup \mathbb{F}_{6}$ &  \cite[(5.67)]{Kim:2020hhh} \\
$B_3^{(1)}$ & \eqref{eq:B31prepot} \\
$B_4^{(1)}$ & \eqref{eq:B41prepot} \\
$B_4^{(2)}$ & \eqref{eq:B42prepot} \\
$B_3^{(3)}$ & \eqref{eq:B33prepot} \\
$B_4^{(3)}$ & \eqref{eq:B43prepot} \\
\hline
\end{tabular}
}

\paragraph{Dynamical fluxes vs.\ symmetry backgrounds.} In every case, the allowed (dynamical)
fluxes are the central-divisor cosets dressed by the spin$^c$ offset
\begin{equation}
    n=\frac{j\,k}{\fm}+\delta\,,\qquad j=0,1,\ldots,\fm-1\,,
    \label{eq:nonlagflux}
\end{equation}
with $k$ the central vector of \eqref{eq:kdef} ($Z=\sum_i k_i S_i$ the generator of $\Z_\fm^{(1)}$)
and $\delta$ the spin$^c$ offset of \eqref{eq:delta}, non-zero precisely when a compact curve is
odd. This is the flux summed over in the blow-up equation. It is to be distinguished from the
bare 1-form \emph{background} 
\bes{
\mathfrak{n}=jk/\fm \quad  \text{(without $\delta$)} \,,
}
 on which the 't~Hooft
anomaly is evaluated in \S\ref{sec:anomalies} and in appendices~\ref{app:BN}, \ref{app:PF}.
We therefore write $\mathfrak{n}$ for the bare 1-form background and
$n=\mathfrak{n}+\delta$ for the physical flux \eqref{eq:nonlagflux}, the two differing precisely
by the spin$^c$ offset $\delta$. For a Lagrangian gauge theory the distinction is moot: the
genuine 1-form background pairs integrally with every dynamical state --- the W-bosons (whose
one-loop term is an integer, \S\ref{sec:anomalies}) and, for the genuine symmetry, the charged
matter --- so there is no spin$^c$ offset, $\delta=0$, and $\mathfrak{n}=n$.\footnote{This statement holds even for the $\SU(2)$ or $\USp(2N)$ gauge group with theta angle $\pi$. As discussed in \cite[(3.41) and around (5.47)]{Kim:2020hhh}, there always exists a choice where $\delta=0$ in the gauge fluxes, but with the half-integer shifts residing in the flavour background fluxes $B_{m_h}$ or topological flux $B_{m_0}$. This is also the case for the other Lagrangian theories, such as $\Spin(7)+3\mathbf{F}$: $B_{m_h}=\pm\tfrac12$, $B_{m_0} \in \Z$, and $\SU(4)_2+\Lambda^2$: $B_{m_h}=\pm\tfrac12$, $B_{m_0}\in\Z+\tfrac12$.} For convenience, we collect the bare $1$-form background
$\mathfrak{n}=jk/\fm$, the spin$^c$ offset $\delta$ (so that the dynamical flux is $n=\mathfrak{n}+\delta$),
and the flavour symmetry together with its background-flux quantisation for every non-Lagrangian
theory studied below in Table~\ref{tab:nonlagfluxes}. A remark on notation, here and in the rest of the paper: when the flavour
symmetry has a single Cartan direction, we denote its background flux by $B_h$ (or $B_m$); when
there are several flavour divisors $D_{F_a}$, we write $B_{m_a}$ ($a = 1, 2, \ldots$) for the flux
through $D_{F_a}$. The two notations are used interchangeably, with $B_h \equiv B_{m_1}$ whenever
only one flavour flux is present.

\begin{table}[h]
\centering
\renewcommand{\arraystretch}{1.3}
\setlength{\tabcolsep}{4.5pt}
\small
\begin{tabular}{lccccc}
\toprule
Theory & $\Gamma^{(1)}$ & $\mathfrak{n}=\tfrac{j}{\fm}k$ & $\delta$ & $F$ & flavour fluxes \\
\midrule
$B_3=E_0$ & $\Z_3^{(1)}$ & $\tfrac{j}{3}$ & $\tfrac16$ & --- & --- \\
$B_4$ & $\Z_7^{(1)}$ & $\tfrac{j}{7}(1,2,4)$ & $(0,0,0)$ & --- & --- \\
\midrule
$\mathbb{P}^2\cup\mathbb{F}_3$ & $\Z_5^{(1)}$ & $\tfrac{j}{5}(1,3)$ & $(0,\tfrac12)$ & --- & --- \\
$\mathbb{P}^2\cup\mathbb{F}_6$ & $\Z_4^{(1)}$ & $\tfrac{j}{4}(2,1)$ & $(-\tfrac14,-\tfrac18)$ & --- & --- \\
\midrule
$B_3^{(1)}$ & $\Z_2^{(1)}$ & $\tfrac{j}{2}$ & $0$ & $\U(1)$ & $B_h\in\Z$ \\
$B_4^{(1)}$ & $\Z_3^{(1)}$ & $\tfrac{j}{3}(2,2,1)$ & $(\tfrac12,0,0)$ & $\SU(2)\times \U(1)$ & $B_{m_1}\in2\Z,\ \ B_{m_2}\equiv j\!\!\pmod2$ \\
\midrule
$B_3^{(2)}=B_3$ & $\Z_3^{(1)}$ & $\tfrac{j}{3}$ & $\tfrac16$ & --- & --- \\
$B_4^{(2)}$ & $\Z_4^{(1)}$ & $\tfrac{j}{4}(1,1,2)$ & $(\tfrac12,0,0)$ & $\SU(2)$ & $B_h\in\Z$ \\
\midrule
$B_3^{(3)}$ & $\Z_2^{(1)}$ & $\tfrac{j}{2}$ & $0$ & $\SU(2)$ & $B_m\in\Z$ \\
$B_4^{(3)}$ & $\Z_3^{(1)}$ & $\tfrac{j}{3}(2,1,1)$ & $(0,\tfrac12,0)$ & $\SU(3)$ & $B_{m_1},B_{m_2}\in\Z$ \\
\bottomrule
\end{tabular}
\caption{Bare $1$-form background $\mathfrak{n}=jk/\fm$ ($j=0,\dots,\fm-1$, with $k$ the central
vector $Z=\sum_i k_iS_i$ generating $\Gamma^{(1)}=\Z_\fm^{(1)}$), spin$^c$ offset $\delta$, flavour
symmetry $F$ and the quantisation of the flavour fluxes for the non-Lagrangian theories. The
dynamical flux is $n=\mathfrak{n}+\delta$. $B_3^{(2)}$ is local $\mathbb{P}^2$ ($B_3$); the rank-one
$B_3^{(1)}$ and $B_3^{(3)}$ are distinct geometries (compact divisor $\mathbb{F}_0$ versus
$\mathbb{F}_2$) that happen to share the coarse invariant $c_{111}=8$. The flavour fluxes are denoted by $B_h$ (or $B_m$) when $F$ has a single Cartan direction, and by $B_{m_a}$ when there are several; see \S\ref{sec:expVnonlag} and appendix~\ref{app:BN}.}
\label{tab:nonlagfluxes}
\end{table}

\subsubsection{The $B_N$ family} \label{sec:BNfamily}

\subsubsection*{The $B_3=E_0$ theory (local $\mathbb{P}^2$)}
This is, in fact, the rank-one $E_0$ theory studied in \cite{Morrison:1996xf}. With a single flux $n_1$, the blow-up function reads
\begin{equation}
    \exp(-V_n)=(p_1p_2)^{\frac32 n_1^3-\frac38 n_1}\;z_1^{\,\frac18-\frac92 n_1^2}\,,
\end{equation}
so the power of the Coulomb fugacity $z_1=e^{\phi_1}$ --- a magnetic charge --- is
$Z_1(n)=\tfrac18-\tfrac92\,n_1^2$, which must be an integer. Its flux-dependent part $-\tfrac92 n_1^2=-\tfrac12
c_{111}\,n_1^2$ is the classical contribution of the cubic prepotential, while the constant
$\tfrac18$ is a gravitational zero-point. To see what the constant does, note that the flux
through the line $\ell\subset\mathbb{P}^2$ is $\langle n,\ell\rangle=(S\cdot\ell)\,n_1=-3n_1$, so
that
\begin{equation}
    Z_1(n)=\tfrac18-\tfrac12\,\langle n,\ell\rangle^2\,.
\end{equation}
Integrality of $Z_1$ then holds \emph{precisely} when $\langle n,\ell\rangle\in\Z+\tfrac12$ ---
the Freed--Witten/spin$^c$ condition that the flux through the \emph{odd} curve $\ell$ (with
$\ell^2=1$, so that $\mathbb{P}^2$ is non-spin) be half-integral. This is exactly $6n_1$ odd,
$n_1\in\{\tfrac16,\tfrac12,\tfrac56\}\pmod1$. The value of the constant is the spin$^c$
zero-point itself: $\tfrac18=\tfrac12\big(\tfrac12\big)^2$ is $\tfrac12\langle n,\ell\rangle^2$
evaluated at the minimal half-integral flux $\langle n,\ell\rangle=\tfrac12$, so it is exactly
what forbids $n_1=0$ and pushes the allowed fluxes off the integer lattice. Were the relevant
curve even --- a spin divisor, as for the $B_4$ surfaces below, whose $Z_i$ carry no such
constant --- the integer flux $n=0$ would be allowed. The three values are thus the three cosets
of $\Z_3^{(1)}$, each dressed by this half-unit offset: the $6n \in 2\mathbb{Z}+1$ lattice.

\subsubsection*{The $B_4$ theory} 
The rank-three member has
\begin{equation}
\begin{aligned}
    P(n)&=\scalebox{0.99}{$ \displaystyle \tfrac43\bigl(n_1^3+n_2^3+n_3^3\bigr)-n_1^2n_3-n_1n_2^2-n_2n_3^2+n_1n_2n_3
    -\tfrac13(n_1+n_2+n_3)$}\,,\\
    Z_1(n)&=-4n_1^2+2n_1n_3+n_2^2-n_2n_3\,,\\
    Z_2(n)&=2n_1n_2-n_1n_3-4n_2^2+n_3^2\,,\\
    Z_3(n)&=n_1^2-n_1n_2+2n_2n_3-4n_3^2\,.
\end{aligned}
\end{equation}
Every compact curve is even, so there is no spin$^c$ shift, and $Z_i\in\Z$ selects the integral
$\Z_7^{(1)}$ cosets $n=\tfrac{j}{7}(1,2,4)$, with $j=0,\dots,6$.

\subsubsection{$\mathbb{P}^2\cup\mathbb{F}_3$} 
With fluxes $(n_1,n_2)$, blow-up function
$\exp(-V_n)=(p_1p_2)^{P(n)}z_1^{Z_1(n)}z_2^{Z_2(n)}$, where
\begin{equation}
\begin{aligned}
    P(n)&=\tfrac32 n_1^3-\tfrac32 n_1^2n_2+\tfrac12 n_1n_2^2+\tfrac43 n_2^3-\tfrac38 n_1-\tfrac13 n_2\,,\\
    Z_1(n)&=-\tfrac18(6n_1-2n_2-1)(6n_1-2n_2+1)\,,\\
    Z_2(n)&=\tfrac12(n_1-2n_2)(3n_1+4n_2)\,,
\end{aligned}
\end{equation}
and $Z_1,Z_2\in\Z$ selects the five cosets
\begin{equation}
    (n_1,n_2)\in\Big\{(0,\tfrac12),\,(\tfrac15,\tfrac1{10}),\,(\tfrac25,\tfrac7{10}),\,
    (\tfrac35,\tfrac3{10}),\,(\tfrac45,\tfrac9{10})\Big\}\pmod1\,,
\end{equation}
i.e.\ $\Z_5^{(1)}$, in agreement with $n_1\in\Z+\tfrac15$, $n_2\in\Z+\tfrac1{10}$ of
\cite[(5.64)]{Kim:2020hhh}; the denominator $10=2\fm$ in $n_2$ is the spin$^c$ half-shift.

\subsubsection{$\mathbb{P}^2\cup\mathbb{F}_6$} 
Here $\exp(-V_n)=(p_1p_2)^{P(n)}z_1^{Z_1(n)}z_2^{Z_2(n)}$, with
\begin{equation}
\begin{aligned}
    P(n)&=\tfrac32 n_1^3-3 n_1^2n_2+2 n_1n_2^2+\tfrac43 n_2^3-\tfrac38 n_1-\tfrac13 n_2\,,\\
    Z_1(n)&=\tfrac18-\tfrac12(3n_1-2n_2)^2\,,\qquad
    Z_2(n)=(n_1-2n_2)(3n_1+2n_2)\,,
\end{aligned}
\end{equation}
and integrality of $Z_1$ and $Z_2$ selects the $\Z_4^{(1)}$ cosets $n_1\in\Z+\tfrac14$, $n_2\in\Z+\tfrac18$ of
\cite[(5.68)]{Kim:2020hhh} (denominator $8=2\fm$).

\subsubsection{The $B_N^{(1)}$ family} 
The sibling family $B_N^{(1)}$ of \cite{Eckhard:2020jyr,Morrison:2020ool} (whose geometry is described in
appendix~\ref{app:BN1}), with 1-form symmetry $\Z_{N-1}^{(1)}$ and flavour symmetry
$\SU(N-2)\times \U(1)$ of rank $N-2$ (Fig.~13 of~\cite{Eckhard:2020jyr})
with fugacities $w_a=e^{-m_a}$, is treated the same way, its Coulomb exponents
$Z_i(n)=-\tfrac12 c_{ijk}n_jn_k$ now carrying no gravitational constant. For $B_3^{(1)}$
($\Z_2^{(1)}$ 1-form symmetry, a single divisor with $c_{111}=8$ and $S_1^2D_F=-2$), we have
\begin{equation}
    \exp(-V_n)=(p_1p_2)^{\frac43 n_1^3-\frac13 n_1}\;w^{\,n_1^2}\;z_1^{-4n_1^2+2n_1 B_h}\,.
\end{equation}
Integrality of $Z_1=-4n_1^2+2n_1B_h$ is the only condition, and it selects the $\Z_2^{(1)}$ cosets
$n_1\in\{0,\tfrac12\}$. At $n_1=0$ one has $Z_1=0$ for \emph{any} $B_h$, so the minimal flux leaves
$B_h$ unconstrained; it is the generating coset $n_1=\tfrac12$ --- where $Z_1=-1+B_h$ --- that forces
the flavour flux to be integral, $B_h\in\Z$ (there being no hypermultiplet zero-point, the $B_{m_h}=\pm\tfrac12$ of the matter
theories has no analogue here).
For $B_4^{(1)}$ ($\Z_3^{(1)}$ 1-form symmetry, rank three), the flavour symmetry has rank two (five boundary points), so
there are \emph{two} masses; with $\mathcal N=\sum_i n_iS_i+B_{m_1}D_{F_1}+B_{m_2}D_{F_2}$
(with $D_{F_1}=A$, $D_{F_2}=E$ in appendix~\ref{app:BN1}), we have
$\exp(-V_n)=(p_1p_2)^{P}\,w_1^{W_1}w_2^{W_2}\,z_1^{Z_1}z_2^{Z_2}z_3^{Z_3}$, with
\begin{equation}
\begin{aligned}
    W_1&=\tfrac12 D_{F_1}\!\cdot\!\mathcal N^2=-n_1^2+n_1n_3-n_3^2\,,\qquad
    W_2=\tfrac12 D_{F_2}\!\cdot\!\mathcal N^2=B_{m_2}n_2-\tfrac32 n_2^2\,,\\
    Z_1&=-\tfrac18-\tfrac72 n_1^2+n_1n_2+\tfrac12 n_2^2-n_2n_3+n_3^2+(2n_1-n_3)B_{m_1}\,,\\
    Z_2&=-\tfrac18+\tfrac12 n_1^2+n_1n_2-n_1n_3-\tfrac72 n_2^2+n_3^2+3n_2 B_{m_2}-\tfrac12 B_{m_2}^2\,,\\
    Z_3&=-n_1n_2+2n_1n_3+2n_2n_3-4n_3^2+(2n_3-n_1)B_{m_1}\,,
\end{aligned}
\end{equation}
and the $(p_1p_2)$-exponent $P$ of \eqref{eq:Pnonlag}. The spin$^c$ zero-points $\kappa=(-\tfrac18,-\tfrac18,0)$ exclude $n=0$; the
single compact--compact odd curve $S_1S_2$ fixes the gauge offset $\delta=(\tfrac12,0,0)$, and
$Z_i\in\Z$ on the central cosets $n=\tfrac{j}{3}(2,2,1)+\delta$ then locks the two flavour fluxes,
\bes{
B_{m_1}\in2\Z~, \quad B_{m_2}\equiv j\pmod2\,. 
}

\subsubsection{The $B_N^{(2)}$ family} 
The third family $B_N^{(2)}$ of~\cite{Eckhard:2020jyr,Morrison:2020ool} (appendix~\ref{app:BN2}, with the 1-form symmetry $\Z_N^{(1)}$) follows
the same pattern.  The $B_3^{(2)}$ theory is simply local $\mathbb{P}^2$ ($B_3$) discussed in \sref{sec:BNfamily}.
Its flavour symmetry is $\SU(N-2)$ (Fig.~13 of~\cite{Eckhard:2020jyr}), present for $N\ge4$; for $B_4^{(2)}$, it
is $\SU(2)$, whose Cartan supplies the single flavour fugacity $w$.

For $B_4^{(2)}$ ($\Z_4^{(1)}$ 1-form symmetry, rank three), the blow-up function
is $\exp(-V_n)=(p_1p_2)^{P}\,w^{W}\,\prod_i z_i^{Z_i}$ with total flux
$\mathcal N=n_1S_1+n_2S_2+n_3S_3+B_hD_F$, the $(p_1p_2)$-exponent $P$ of \eqref{eq:Pnonlag}
(prepotential in appendix~\ref{app:BN2}), and
\begin{equation}
\begin{aligned}
    W&=\tfrac12\,D_F\!\cdot\!\mathcal N^2=-n_1^2-n_2^2-\tfrac32 n_3^2+(B_h+n_1+n_2)\,n_3\,,\\
    Z_1&=-\tfrac18-\tfrac72 n_1^2+n_1n_2+\tfrac12 n_2^2+\tfrac32 n_3^2-(B_h+n_1+n_2)\,n_3+2n_1B_h\,,\\
    Z_2&=-\tfrac18+\tfrac12 n_1^2+n_1n_2-\tfrac72 n_2^2+\tfrac32 n_3^2-(B_h+n_1+n_2)\,n_3+2n_2B_h\,,\\
    Z_3&=\tfrac18-\tfrac12\bigl(B_h+n_1+n_2-3n_3\bigr)^2\,,
\end{aligned}
\end{equation}
including the spin$^c$ zero-points $\kappa=(-\tfrac18,-\tfrac18,\tfrac18)$. All three divisors are
non-spin ($S_{1,2}^3=7$, $S_3^3=9$), so $\kappa\neq0$ and the trivial flux is \emph{excluded}:
$Z_i(0)=\kappa_i\notin\Z$. Integrality of the $Z_i$ is the spin$^c$ flux quantisation; from
$Z_3=\tfrac18-\tfrac12(n_1+n_2-3n_3+B_h)^2$, one has
\begin{equation}
    Z_3\in\Z \iff n_1+n_2-3n_3+B_h\in\Z+\tfrac12\,,
    \label{eq:BN2spinc}
\end{equation}
the exact analogue of $3n_1\in\Z+\tfrac12$ (``$6n_1$ odd") for local $\mathbb{P}^2$, with $Z_{1,2}\in\Z$
imposing the corresponding shifts on $S_1,S_2$. With $B_h\in\Z$ (integer charges $D_F\cdot C_a$),
this forces the \emph{gauge} combination $n_1+n_2-3n_3$ to be half-integral: the spin$^c$ half-unit
sits in the gauge flux, not in $B_h$ (unlike the matter theories, where $B_{m_h}=\pm\tfrac12$ carries
it). The spin$^c$ offset can be chosen to be $\delta=(\tfrac12,0,0)$ and the bare $\Z_4^{(1)}$ backgrounds are
$\mathfrak n=\tfrac{j}{4}(1,1,2)$. The dynamical fluxes are therefore 
\bes{
n = \tfrac{j}{4}(1,1,2) +(\tfrac12,0,0)\,.
}
Note that, as before, the fluxes $n=0$ are excluded. 

\subsubsection{The $B_N^{(3)}$ family}  
The fourth family $B_N^{(3)}$ of~\cite{Eckhard:2020jyr,Morrison:2020ool} (discussed in appendix~\ref{app:BN3}, with $\Z_{N-1}^{(1)}$ 1-form symmetry,
flavour symmetry $\SU(N-1)$ of rank $N-2$; Fig.~13 of~\cite{Eckhard:2020jyr}) follows the same blow-up pattern. For $B_4^{(3)}$ ($\Z_3^{(1)}$ 1-form symmetry,
$Z=2S_1+S_2+S_3$, two masses $D_{F_1}=(4,0)$, $D_{F_2}=(1,3)$), the total flux is
$\mathcal N=\sum_i n_iS_i+B_{m_1}D_{F_1}+B_{m_2}D_{F_2}$, the blow-up function reads
$\exp(-V_n)=(p_1p_2)^{P}\,w_1^{W_1}w_2^{W_2}\,z_1^{Z_1}z_2^{Z_2}z_3^{Z_3}$, with
\begin{equation}
\begin{aligned}
    W_1&=-n_1^2+n_1n_3-n_3^2\,,\qquad W_2=-\tfrac32 n_2^2+n_2 B_{m_2}\,,\\
    Z_1&=-4n_1^2+2n_1n_2-n_1n_3-n_2n_3+\tfrac32 n_3^2+(2n_1-n_3)B_{m_1}\,,\\
    Z_2&=-\tfrac18+n_1^2-n_1n_3-\tfrac72 n_2^2+n_2n_3+\tfrac12 n_3^2+3n_2 B_{m_2}-\tfrac12 B_{m_2}^2\,,\\
    Z_3&=-\tfrac18-\tfrac12 n_1^2-n_1n_2+3n_1n_3+\tfrac12 n_2^2+n_2n_3-\tfrac72 n_3^2+(2n_3-n_1)B_{m_1}\,,
\end{aligned}
\end{equation}
and the $(p_1p_2)$-exponent $P$ of \eqref{eq:Pnonlag}. With
spin$^c$ zero-points $\kappa=(0,-\tfrac18,-\tfrac18)$ ($S_1^3=8$ is even, so $Z_1$ carries none),
$Z_i\in\Z$ excludes $n=0$ and places the dynamical fluxes at 
\bes{
n=\tfrac{j}{3}(2,1,1)+\delta \, , \quad   \delta=(0,\tfrac12,0)\,,
}
where the gauge spin$^c$ offset $\delta$ is obtained from the odd compact--compact curves $S_1S_3,S_2S_3$. The background flavour fluxes are such that
\bes{
B_{m_1},B_{m_2}\in\Z \,. 
} 

\section{'t Hooft anomalies involving the 1-form symmetry}
\label{sec:anomalies}
In this section, we extract the 't~Hooft anomalies involving the 1-form symmetry directly from the
blow-up function. The quantity to evaluate is \emph{not} the dynamical gauge flux but the
\emph{background for the 1-form symmetry}: we turn on a non-trivial $\Z_\fm^{(1)}$ background by
setting the flux to a non-trivial coset of the centre --- the bare 1-form background $\mathfrak{n}$
of \S\ref{sec:expVnonlag} --- \emph{without} the spin$^c$ offset $\delta$ that dresses the
dynamical flux $n=\mathfrak{n}+\delta$ in \eqref{eq:nonlagflux} summed over in the blow-up equation.
The anomalies, including the mixed ones, are then read from the \emph{fractional parts} of the
exponents of $\exp(-V_{\mathfrak{n}})$: the instanton-fugacity exponent $Q(\mathfrak{n})$ gives the
mixed 1-form--$\U(1)_I$ anomaly (more precisely its fractional part rescaled by the Dynkin index, $[Q(\mathfrak{n})/I_2(F)]_{\mod1}$; see \S\ref{sec:kvsQ}), the constant part of the $(p_1p_2)$-exponent $P(\mathfrak{n})$ gives
the cubic self-anomaly together with the mixed 1-form--gravitational and 1-form--$\SU(2)_R$ couplings,
and the flavour-fugacity exponents $W_a(\mathfrak{n})$, where present, give the mixed
1-form--flavour-Cartan anomalies.

Evaluating on the symmetry background $\mathfrak{n}$ rather than on the dynamical flux $n$ is the
conceptually correct prescription: an 't~Hooft anomaly is a property of the background alone, while
the spin$^c$ offset $\delta$ --- a half-integral dressing tied to odd compact curves --- belongs to
the dynamical bundle summed in the partition function and would otherwise shift the very fractional
parts that encode the anomaly. For pure gauge theories, the distinction is immaterial, since
$\delta=0$ and $\mathfrak{n}=n$ (explained in \S\ref{sec:AnomSUNpure}); it becomes essential for the non-Lagrangian
theories of \S\ref{sec:anomnonlag}, where $\delta\neq0$. The method applies equally to pure gauge
theories, to theories with matter in non-trivial gauge representations, and to non-Lagrangian
theories.

\subsection{The instanton charge $k$ versus the blow-up exponent $Q$}
\label{sec:kvsQ}
The mixed 1-form--$\U(1)_I$ anomaly requires one point of normalisation. The topological symmetry $\U(1)_I$ has conserved current $\propto\Tr(F\wedge F)$, and its charge is the \emph{canonical instanton number}
\begin{equation}
    k(n)\;=\;\frac{1}{8\pi^2}\int\Tr(F\wedge F)\Big|_{n}\;=\;\tfrac12\langle n,n\rangle\,,
    \label{eq:kinst}
\end{equation}
with $\langle\cdot,\cdot\rangle$ normalised so that the long roots have length squared equal to $2$; then $k\in\Z$ on the coroot lattice, with the minimal (long-root) instanton at $k=1$. The exponent of the instanton fugacity in $\exp(-V_n)$ is instead the defining-representation trace
\begin{equation}
    Q(n)\;=\;\tfrac12\Tr_F\,n^2\;=\;I_2(F)\,k(n)\,,
    \label{eq:QvsK}
\end{equation}
where $I_2(F)$ is the quadratic Dynkin index of the defining representation $F$ in the normalisation $I_2(\mathrm{adj})=2h^\vee$: $I_2(F)=1$ for the fundamentals of $\SU(N)$ and $\USp(2N)$, and $I_2(F)=2$ for the vector of $\Spin(N)$. Hence $Q=k$ for $\SU(N)$ and $\USp(2N)$, while, for $\Spin(N)$, the blow-up grades by $Q=2k$: genuine bundles carry even $Q$, and the minimal long-root instanton sits at $Q=2$ (consistently with footnote \ref{foot:shortlongB3}).

The mixed $\Z_\fm^{(1)}$--$\U(1)_I$ anomaly is the fractional instanton number carried by the 1-form background --- the part of $k(\mathfrak{n}_{\rm gen})$ that cannot be removed by a properly quantised $\U(1)_I$ counterterm, the charge lattice being $\Z$. It is therefore the fractional part of the \emph{physical} charge,
\begin{equation}
    \Omega_{ii,\U(1)_I}\;=\;\Bigl[\,\frac{Q(\mathfrak{n}_{\rm gen})}{I_2(F)}\,\Bigr]_{\mod\,1}\;=\;\bigl[\,k(\mathfrak{n}_{\rm gen})\,\bigr]_{\mod\,1}\,.
    \label{eq:Ianomaly}
\end{equation}
For $\SU(N)$ and $\USp(2N)$, this is just $[Q(\mathfrak{n}_{\rm gen})]_{\mod 1}$; for $\Spin(N)$, the division by $I_2(F)=2$ is essential, since, on a vector-class background, $Q$ is an odd integer --- so $[Q]_{\mod1}=0$ --- whereas the genuine fractional instanton number is $k=\tfrac12$.

\subsection{Anomalies in $\SU(N)_{\kCS}$ pure gauge theory}\label{sec:AnomSUNpure}
Let us first consider the $\SU(N)_{\kCS}$ pure gauge theory. The 1-form symmetry is $\Z_{\fm}^{(1)}$ with $\fm = \gcd(N, \kCS)$. The bare $1$-form background $\mathfrak{n}=jk/\fm$ and the dynamical
($z$-integral) flux $n=\mathfrak{n}+\delta$ of \eqref{eq:nonlagflux} in fact coincide. The only
charged dynamical states are the W-bosons, whose one-loop contribution
$\tfrac16\sum_{\alpha\in\Delta^+}\bigl(\langle n,\alpha\rangle^3-\langle n,\alpha\rangle\bigr)$ is
integral on \emph{every} coset of $\Lambda_{\cw}/\Lambda_{\crt}$ (proven at the end of this
subsection), so the spin$^c$ offset vanishes, $\delta=0$; we have verified this for all $N$ and
$\kCS$. Any half-integer or fractional shifts reside instead in the instanton background $B_{m_0}$
and in the anomalous exponents $Q$ and $P|_{\rm const}$, never in the gauge flux. We may therefore
take $n=\mathfrak{n}$ and use the two interchangeably throughout this subsection.

For the gauge fluxes, a convenient representative of the generating coset $\omega_{N/\fm}$ is $n = (N/\fm)\, \omega_1$, i.e.
\begin{equation}
    n_i = \frac{i}{\fm}\,, \qquad i = 1, \ldots, N-1\,.
\end{equation}
We evaluate the exponents of \eqref{eq:expVmaster} on such fluxes (note that the exponents are well defined modulo integers under a change of representative within the coset).

We first consider the mixed 't~Hooft anomaly between the 1-form symmetry and the instanton symmetry $\U(1)_I$, which is read off from the exponent $Q(n)$ of the instanton fugacity $q$. Evaluating it on the generating flux $n_i = i/\fm$, one finds
\begin{equation}
    Q(n)\Big|_{n_i = i/\fm} \;=\; \frac{N(N-1)}{2 \fm^2}\,.
    \label{eq:mixedanomaly}
\end{equation}
This is precisely the coefficient of the mixed 't~Hooft anomaly between $\Z_\fm^{(1)}$ and $\U(1)_I$ (see \cite{BenettiGenolini:2020doj}, \cite[(B.22)]{Apruzzi:2021vcu}, \cite[(4.6)]{Apruzzi:2021nmk}, \cite[(4.25)]{Apruzzi:2022dlm}); see also \cite{Bertolini:2025wyj} for a recent analysis of this anomaly and of its counterpart in other global variants of the theory, where it is traded for a symmetry extension. 

Next, we consider the cubic self-anomaly of the 1-form symmetry. It originates from the Chern--Simons coupling $\kCS\, \mathrm{CS}_5(A) = 2\pi\, \frac{\kCS}{6} \int_{M_6} \Tr(F^3)$, where $M_6$ is a six-dimensional manifold whose boundary is the five-dimensional spacetime on which our theory lives, and $F$ is taken to be the background field for the 1-form symmetry. We therefore isolate the term proportional to $\kCS$ that is cubic in the gauge fluxes. This is precisely the fractional part of the $(p_1 p_2)$-exponent $P(n)$ evaluated with the instanton background switched off, $B_{m_0} = 0$:
\begin{equation}
    P(n)\big|_{B_{m_0} = 0} \;\equiv\; \frac{\kCS}{6} \sum_{a=1}^{N} \tilde{n}_a^3 \pmod{1}\,,
    \label{eq:Pmod1}
\end{equation}
where the one-loop W-boson contribution $\frac{1}{6} \sum_{\alpha \in \Delta^+} \bigl( \langle n, \alpha \rangle^3 - \langle n, \alpha \rangle \bigr)$ is an \emph{integer} for $n$ in any coset of $\Lambda_{\cw} / \Lambda_{\crt}$ (proven below), so that only the Chern--Simons term contributes to the fractional part. Using \eqref{eq:CStermalt} and evaluating on the generating flux, we find
\begin{equation}
    \frac{\kCS}{2} \sum_{i=1}^{N-2} \Bigl( n_i^2\, n_{i+1} - n_i\, n_{i+1}^2 \Bigr) \Big|_{n_i = i/\fm} \;=\; -\frac{N(N-1)(N-2)\, \kCS}{6\, \fm^3}\,,
    \label{eq:selfanomaly}
\end{equation}
so that
\begin{equation}
    P(n)\big|_{B_{m_0} = 0} \;\equiv\; -\frac{N(N-1)(N-2)\, \kCS}{6\, \fm^3} \pmod 1\,,
\end{equation}
in agreement with the cubic self 't~Hooft anomaly of the 1-form symmetry (\cite[(5.100), (5.106)]{Gukov:2020btk}, \cite[(B.28)]{Apruzzi:2021vcu}, \cite[(4.5)]{Apruzzi:2021nmk}, \cite[(4.25)]{Apruzzi:2022dlm}).

To prove the integrality, take first the representative $n^{(j)} = j\, \omega_1$ of the coset $\omega_j$, i.e.\ $n_i = \frac{i j}{N}$: then $\tilde{n}_a = \frac{j}{N}$ for $a \leq N-1$ and $\tilde{n}_N = \frac{j}{N} - j$, so that $\langle n^{(j)}, \alpha \rangle = 0$ for the $\frac{(N-1)(N-2)}{2}$ positive roots $e_a - e_b$ with $b \leq N - 1$, and $\langle n^{(j)}, \alpha \rangle = j$ for the $N-1$ roots $e_a - e_N$. Hence
\begin{equation}
    \frac{1}{6} \sum_{\alpha \in \Delta^+} \Bigl( \langle n^{(j)}, \alpha \rangle^3 - \langle n^{(j)}, \alpha \rangle \Bigr) \;=\; \frac{(N-1)\, (j^3 - j)}{6} \;\in\; \Z\,,
\end{equation}
since $j^3 - j \equiv 0 \pmod 6$. Any other element of the coset is $n^{(j)} + \mu$ with $\mu \in \Lambda_{\crt}$, under which every pairing shifts by an integer, $y_\alpha \to y_\alpha + c_\alpha$ with $y_\alpha = \langle n^{(j)}, \alpha \rangle \in \Z$ and $c_\alpha = \langle \mu, \alpha \rangle \in \Z$. The variation of each summand, namely
\begin{equation}
    (y + c)^3 - (y + c) - \bigl( y^3 - y \bigr) \;=\; 3\, y c\, (y + c) + \bigl( c^3 - c \bigr)\,,
\end{equation}
is divisible by $6$: indeed $c^3 - c \equiv 0 \pmod 6$, and $y c (y + c)$ is always even for integers $y, c$ (if $y$ and $c$ are both odd, $y + c$ is even). This proves \eqref{eq:Pmod1} on every coset $\omega_j$ --- not only on those generating the genuine 1-form symmetry $\Z_{\fm}^{(1)}$.


\subsection{Anomalies in the $B_N$, $B^{(1,2,3)}_N$, and $\mathbb{P}^2\cup\mathbb{F}_{3,6}$ non-Lagrangian theories}
\label{sec:anomnonlag}

For the non-Lagrangian theories of \S\ref{sec:expVnonlag}, the diagnostic works in the same way,
with one difference: there are no W-bosons, so the integer combination
$\tfrac16\sum_{\alpha}(\langle n,\alpha\rangle^3-\langle n,\alpha\rangle)$ is absent and the
gravitational coupling it would package must be reinstated by hand. Directly from the prepotential
\eqref{eq:BNprepot}, the $(p_1 p_2)$-exponent evaluated on the bare 1-form background
$\mathfrak{n}$ (in the notation of \S\ref{sec:expVnonlag}) reads
\begin{equation}
    P(\mathfrak{n})=\underbrace{\tfrac16\,c_{ijk}\,\mathfrak{n}^i \mathfrak{n}^j \mathfrak{n}^k}_{\text{cubic CS}}
    \;+\;\underbrace{\sum_i\frac{C^G_i}{48}\,\mathfrak{n}_i}_{\text{gravitational}}
    \;+\;\underbrace{\sum_i\Big(-\frac{C^R_i}{8}\Big)\mathfrak{n}_i}_{\SU(2)_R}\,,
    \label{eq:Pnonlag}
\end{equation}
the cubic part being the Chern--Simons prepotential, the linear part splitting into a
gravitational piece ($C^G_i/48$) and an $\SU(2)_R$ piece ($-C^R_i/8$).

\emph{Which flux to take.} The cubic self-anomaly is read off on the bare 1-form
\emph{background} $\mathfrak{n}=jk/\fm$ --- the central-divisor coset of \eqref{eq:kdef},
\emph{without} the spin$^c$ offset $\delta$. This is exactly the flux on which $\Omega$ is
computed in appendices~\ref{app:BN} and \ref{app:PF}; it should not be confused with the
dynamical $z$-integral flux $n=\mathfrak{n}+\delta$ of \eqref{eq:nonlagflux}, from which it
differs by $\delta$. On this background the cubic and gravitational pieces of \eqref{eq:Pnonlag}
evaluate to
\begin{equation}
\begin{split}
 \tfrac16 \sum_{i,j,k}c_{ijk}\Big(\tfrac{jk^i}{\fm}\Big)\Big(\tfrac{jk^j}{\fm}\Big)\Big(\tfrac{jk^k}{\fm}\Big)
    =\frac{Z^3}{6\fm^3}\,j^3\,, \qquad \sum_i\frac{C^G_i}{48}\frac{j\,k_i}{\fm}=\frac{c_2(X)\cdot Z}{48\,\fm}\,j\,,
\end{split}    
\end{equation}
using $Z^3=\sum_{i,j,k} c_{ijk}k^ik^jk^k$ and $\sum_i C^G_i k_i=c_2(X)\cdot Z$.

Let us compare this with \eqref{eq:BNanomaly}. The cubic piece of
$P$ reproduces the Chern--Simons term of the anomaly outright --- both carry the coefficient
$\tfrac16$, up to the overall sign of the $\exp(-V_n)$ convention --- while its gravitational piece
identifies the intersection number $c_2(X)\cdot Z$ that governs the gravitational term of
\eqref{eq:BNanomaly}. The two encode the \emph{same} mixed 1-form--gravitational coupling,
proportional to $c_2(X)\cdot Z$, but in different normalisations: $P$ carries it as
$\tfrac1{48}(\epsilon_1^2+\epsilon_2^2)\,C^G_i$, with $\epsilon_1^2+\epsilon_2^2$ identified with (minus) the Pontryagin class of the tangent bundle of the 5d spacetime, $-p_1(T)$, see below \eqref{CiGexpression}, and $C^G_i=c_2(X)\!\cdot\!S_i$, whereas \eqref{eq:BNanomaly} writes it as
$\tfrac1{24}\,Z\cdot p_1(TX)=-\tfrac1{12}\,c_2(X)\cdot Z$, using the internal $p_1(TX)=-2c_2(X)$.
Both are fixed by the single intersection number $c_2(X)\cdot Z$; assembling the cubic and
gravitational pieces in the normalisation of \eqref{eq:BNanomaly} gives the cubic self 't~Hooft
anomaly
\begin{equation}
    \Omega(j)=-\frac{Z^3}{6\fm^3}\,j^3-\frac{c_2(X)\cdot Z}{12\,\fm}\,j\pmod1\,,
    \label{eq:anomfromP}
\end{equation}
which is precisely \eqref{eq:BNanomaly}. The $\SU(2)_R$ linear term $-C^R_i \mathfrak{n}_i/8$ does
not enter: it is a mixed 1-form--$\SU(2)_R$ coupling --- the $p_1p_2$ analogue of the mixed
1-form--$\U(1)_I$ anomaly read from $Q(n)$ in \eqref{eq:mixedanomaly} --- distinct from the cubic
self-anomaly; we quantify it at the end of this subsection. Evaluating \eqref{eq:anomfromP} on the cosets reproduces the following values:
\begin{equation} \label{oneformanomnonL}
    \Omega(j)=
    \begin{cases}
    \tfrac19\,j^3 & B_3 = E_0\,\\[2pt]
    \tfrac17\,j^3 & B_4\,\\[2pt]
    -\tfrac3{10}(j^3-j)\equiv\tfrac15(j^3-j) & \mathbb{P}^2\cup\mathbb{F}_3\,\\[2pt]
    \tfrac14\,j^3 & \mathbb{P}^2\cup\mathbb{F}_6\,
    \end{cases}\,,
\end{equation}
and, in particular, exhibits, in the gravitational coefficient $c_2(X)\cdot Z/(48\fm)$ of
\eqref{eq:Pnonlag}, the very datum whose divisibility $2\fm\mid c_2(X)\cdot Z$ separates the
pure-cubic anomalies of $B_N$ and $\mathbb{P}^2\cup\mathbb{F}_6$ from the mixed
1-form--gravitational anomaly of $\mathbb{P}^2\cup\mathbb{F}_3$ (appendix~\ref{app:PF}).  We emphasise that, for $\mathbb{P}^2\cup\mathbb{F}_3$, we can exploit the $\bmod \, 24$ ambiguity of the index congruence to rewrite $\Omega(j) = -\tfrac3{10}(j^3-j)$ as $\tfrac{1}{5}j^3$, which is purely cubic; we discuss this in detail around \eqref{mod24shift} and \eqref{Omegatilde}.

The family $B_N^{(1)}$ of \cite{Eckhard:2020jyr,Morrison:2020ool} (geometry in appendix~\ref{app:BN1})
adds a new ingredient: besides the $\mathbb{Z}_{N-1}^{(1)}$ 1-form symmetry, it has the non-Abelian
flavour symmetry $\SU(N-2)\times \U(1)$ of rank $N-2$ (Fig.~13 of~\cite{Eckhard:2020jyr}; the rank
is the number of boundary lattice points minus three), so that $\exp(-V_{\mathfrak{n}})$ carries,
besides $z_i$ and $p_1p_2$, flavour fugacities $w_a=e^{-m_a}$ ($a=1,\dots,N-2$) for the $N-2$ Cartan
directions. Its cubic self-anomaly is read from $P(\mathfrak{n})$ exactly as above,
\bes{
\Omega^{(1)}=\frac{(N-2)(N-3)}{6(N-1)}\,,
}
which vanishes for $B_3^{(1)}$, and is equal to $\tfrac19$ for $B_4^{(1)}$.

Because the 1-form symmetry is discrete, its mixed 't~Hooft anomaly with a non-Abelian flavour
group can only be carried by the \emph{Cartan} of the latter: a background for the full $\SU(N-2)\times
\U(1)$ enters the anomaly through the Abelian flux that commutes with the 1-form background. The
non-compact divisors $D_{F_a}$ furnish exactly these Cartan $\U(1)$ directions, and each carries a
mixed $\Z_{N-1}^{(1)}$--$\U(1)_a$ anomaly given by the fractional part of $-W_a(\mathfrak{n})$, with
$W_a=\tfrac12 D_{F_a}\!\cdot\!\mathfrak{n}^2=\tfrac12(S_iS_jD_{F_a})\,\mathfrak{n}^i\mathfrak{n}^j$ the
($w_a=e^{-m_a}$) flavour exponent of \S\ref{sec:expVnonlag},
\begin{equation}
    \Omega_{ii,a}=\bigl[\,-W_a(\mathfrak{n})\,\bigr]_{\rm frac}
    =-\frac12\,\frac{Z^2\!\cdot\!D_{F_a}}{\fm^2}\,j^2 \pmod 1\,.
    \label{eq:BN1mixed}
\end{equation}
For $B_3^{(1)}$, the flavour symmetry is a single $\U(1)$ ($D_F=(N-1,0)$): $-W=n_1^2\to\tfrac14$. For
$B_4^{(1)}$ it is $\SU(2)\times \U(1)$ (rank two), with the two Cartan directions $D_{F_1}=(3,0)$ and
$D_{F_2}=(0,3)$: the first gives $-W_1=n_1^2-n_1n_3+n_3^2\to\tfrac13$ (matching
\cite[(4.38)]{Apruzzi:2021nmk}), the second $-W_2=\tfrac32 n_2^2\to\tfrac23$. This is the
non-Lagrangian, flavour-symmetry counterpart of the mixed 1-form--instanton anomaly of
\S\ref{sec:anommixed} below: the flavour $w$-exponent here plays the role of the instanton $q$-exponent
there. 

The same holds for the sibling family $B_N^{(2)}$ (appendix~\ref{app:BN2},
$\mathbb{Z}_N^{(1)}$ 1-form symmetry, flavour symmetry $\SU(N-2)$~\cite{Eckhard:2020jyr}): its cubic self-anomaly is given by \eqref{eq:BN2anom} (see \cite[(4.36)]{Apruzzi:2021nmk}):
\bes{
\Omega^{(2)}=\frac{(N-2)(N-1)}{6N}\,, 
}
and for $N\ge4$, where the $\SU(N-2)$ flavour symmetry appears, the mixed $\Z_N^{(1)}$--Cartan anomaly along the vertex $(N,0)$
is 
\bes{
\Omega_{ii,(N,0)}=\frac{N-2}{2N}\,, 
}
read from the flavour exponent as above; $B_3^{(2)}$ has no flavour symmetry and coincides with local $\mathbb{P}^2$ ($B_3$), while $B_4^{(2)}$ has flavour
$\SU(2)$, whose single Cartan supplies the fugacity $w$. 

The fourth family $B_N^{(3)}$
(appendix~\ref{app:BN3}, $\mathbb{Z}_{N-1}^{(1)}$ 1-form symmetry, flavour symmetry $\SU(N-1)$~\cite{Eckhard:2020jyr})
is the exceptional case: its cubic self-anomaly \emph{vanishes}, $\Omega^{(3)}=0$ --- the genuine
cubic coefficient cancels and the residual gravitational $(j^3-j)$ piece drops mod $1$, see
\eqref{eq:BN3anom} --- yet its $\SU(N-1)$ flavour symmetry still carries non-trivial mixed
$\Z_{N-1}^{(1)}$--Cartan anomalies $\Omega_{ii,a}=-\tfrac12 Z^2\!\cdot\!D_{F_a}/\fm^2$ along its
$N-2$ Cartan directions (e.g.\ $\tfrac13$ and $\tfrac16$ for $B_4^{(3)}$, whose flavour is $\SU(3)$).
It thus realises a 1-form symmetry whose \emph{only} 't~Hooft anomalies are the mixed ones with the
flavour Cartan.

\subsubsection*{The mixed 1-form--$\SU(2)_R$ anomaly} 
Let us now quantify the mixed
1-form--$\SU(2)_R$ coupling carried by the last term of \eqref{eq:Pnonlag}. As discussed in
\S\ref{sec:EffPrep}, the $\epsilon_+^2$ term of the effective prepotential
\eqref{eq:BNprepot} is the equivariant image, under the substitution $c_2(R)\to\epsilon_+^2$, of
the mixed gauge/$\SU(2)_R^2$ Chern--Simons coupling \eref{MixedCSCR}, with levels $C^R_i/2$.
Evaluating this coupling on the bare 1-form background $\mathfrak{n}_i=j k_i/\fm$ produces, on a
background carrying $\SU(2)_R$ instanton number $\int c_2(R)$, the anomalous phase
\begin{equation}
    \exp\left[2\pi i\,\mathcal{A}_R(j)\int c_2(R)\right]\,,\quad \text{with} \quad
    \mathcal{A}_R(j)\;=\;\frac{\sum_i C^R_i\,k_i}{2\,\fm}\,j\pmod 1\,,
    \label{eq:ARnonlag}
\end{equation}
i.e.\ a mixed 't~Hooft anomaly between $\Z_\fm^{(1)}$ and the $\SU(2)_R$ symmetry.
Equivalently, $\mathcal{A}_R(j)$ is $(-4)$ times the $\SU(2)_R$ piece $-C^R_i\mathfrak{n}_i/8$ of
\eqref{eq:Pnonlag}; this is precisely the same conversion factor encountered for the
gravitational piece, which maps $C^G_i\mathfrak{n}_i/48$ of \eqref{eq:Pnonlag} to the term
$-c_2(X)\cdot Z\,j/(12\fm)$ of \eqref{eq:anomfromP}, providing a non-trivial consistency check of
the normalisation. Since $C^R_i=2$ for every compact divisor in all of our examples (see the
prepotentials collected in appendices~\ref{app:BN} and~\ref{app:PF}), \eqref{eq:ARnonlag}
reduces to $\mathcal{A}_R(j)=\big(\textstyle\sum_i k_i/\fm\big)\,j$, namely
\begin{equation}
    \mathcal{A}_R(j)=
    \begin{cases}
    \tfrac13\,j & B_3 = E_0\,\\[2pt]
    0 & B_4\,\\[2pt]
    \tfrac45\,j & \mathbb{P}^2\cup\mathbb{F}_3\,\\[2pt]
    \tfrac34\,j & \mathbb{P}^2\cup\mathbb{F}_6\,
    \end{cases}\,,
    \label{eq:ARvalues}
\end{equation}
using $k=(1)$, $\fm=3$ for $B_3$; $k=(1,2,4)$, $\fm=7$ for $B_4$; $k=(1,3)$, $\fm=5$ for
$\mathbb{P}^2\cup\mathbb{F}_3$; and $k=(2,1)$, $\fm=4$ for $\mathbb{P}^2\cup\mathbb{F}_6$.
Interestingly, the anomaly vanishes for $B_4$.

The value for $B_3=E_0$ provides an independent check against the literature. In
\cite[(1.1)]{Sacchi:2023omn}, it was proposed that the $E_0$ theory possesses a mixed anomaly
\bes{
\frac{2\pi i}{3}\int B_2\,\left(c_2(R)\bmod 3\right)
}
between its $\Z_3^{(1)}$ 1-form symmetry
and the $\SU(2)_R$ symmetry, where $B_2$ is the background field for the 1-form symmetry. In the
parametrisation of \cite[(2.24)]{Sacchi:2023omn}, namely
$\frac{\pi i \alpha}{9}\int B_2^3+\frac{2\pi i\beta}{3}\int B_2\left(c_2(R)\bmod 3\right)$, the
cubic coefficient $\alpha=2$ reproduces $\Omega(j)=\tfrac19\,j^3$ of \eref{oneformanomnonL}. On the other hand, the value
$\beta=1$ was determined indirectly in that reference, by matching discrete anomalies of the 3d
theories obtained by compactifying the $E_0$ theory on Riemann surfaces, a direct 5d computation
being unavailable at the time. Our result $\mathcal{A}_R(j)=\tfrac13\,j$ corresponds precisely to
$\beta=1$: the blow-up function thus supplies an intrinsically five-dimensional derivation of
this mixed anomaly. Note that the reduction of $c_2(R)$ modulo $3$ in the expressions above
reflects the fact that $\mathcal{A}_R$ is well defined modulo integer shifts by local
counterterms, in the same way as the anomalies discussed previously.

For completeness, we also evaluate \eqref{eq:ARnonlag} at general $N$ for the $B_N$ family and
its siblings $B_N^{(1,2,3)}$ (for all of which $C^R_i=2$; see the prepotentials in
appendix~\ref{app:BN}). For $B_N$, we use the generator $k_v$ of \eqref{eq:BNcentral}; for
$B_N^{(2)}$ and $B_N^{(3)}$, the generators are $k_v\equiv x \pmod N$ and
$k_v\equiv N-x-y \pmod{N-1}$ respectively, for the interior point $v=(x,y)$, matching the
explicit central vectors given in appendices~\ref{app:BN2} and~\ref{app:BN3}; for $B_N^{(1)}$,
we take $k_v\equiv x\pmod{N-1}$, the generator for which the cubic anomaly $\Omega^{(1)}$
assumes the value quoted in table~\ref{tab:BN1}. To be explicit: this is the generator $-k$
relative to the central vector $k=(2,2,1)$ of $B_4^{(1)}$ given in appendix~\ref{app:BN1}, with
which \eqref{eq:anomfromP} evaluates instead to $\Omega^{(1)}(1)=-\tfrac19\equiv\tfrac89$; the
two conventions are related by the orientation reversal $j\to-j$, under which
$\mathcal{A}_R(j)\to-\mathcal{A}_R(j)$. With these conventions, one finds
\begin{equation}
    \mathcal{A}_R(j)=
    \begin{cases}
    \tfrac13\,j\ \ \text{if } N\equiv 0\ (\mathrm{mod}\ 3)\,,\quad 0\ \text{ otherwise} & B_N\,,\\[5pt]
    \tfrac{N(N-2)}{6}\,j\pmod 1 & B_N^{(1)}\ \text{and}\ B_N^{(3)}\,,\\[5pt]
    \tfrac{(N-1)(N-2)}{6}\,j \pmod 1 & B_N^{(2)}\,,
    \end{cases}
    \label{eq:ARfamilies}
\end{equation}
which we have verified for $3\le N\le 12$. For $B_N$, the pattern is consistent with the fact
that $\fm=N(N-3)+3$ is divisible by $3$ precisely when $3\mid N$. For $B_N^{(2)}$, the
coefficient $\tfrac{(N-1)(N-2)}{6}$ equals $\tfrac13$ modulo $1$ when $3\mid N$ and vanishes
otherwise, exactly as for $B_N$ (consistently with $B_3^{(2)}=B_3$). For $B_N^{(1)}$ and
$B_N^{(3)}$, the coefficient $\tfrac{N(N-2)}{6}$ follows the period-six pattern
$\tfrac12,\tfrac13,\tfrac12,0,\tfrac56,0$ for $N\equiv 3,4,5,6,7,8 \pmod 6$. The origin of these
closed forms is the amusing identity that, with the stated generators, the sum of the entries of
the central vector is the \emph{same} binomial coefficient for all three sibling families:
\begin{equation}
    \sum_i k_i \;=\; \sum_{c=1}^{N-2} c\,(N-1-c) \;=\; \binom{N}{3}\,,
\end{equation}
so that $\mathcal{A}_R(j)=\binom{N}{3}\,j/\fm \bmod 1$, with $\fm=N-1$ for $B_N^{(1,3)}$ and
$\fm=N$ for $B_N^{(2)}$. In particular, the coincidence of the mixed 1-form--$\SU(2)_R$
anomalies of $B_N^{(1)}$ and $B_N^{(3)}$ extends, to all $N$, the algebraic similarity of these
two families already observed for $N=3$ in appendix~\ref{app:BN3}.

\subsection{Mixed anomaly with the instanton symmetry in theories with matter}
\label{sec:anommixed}

The mixed 't~Hooft anomaly between the 1-form symmetry and the instanton symmetry $\U(1)_I$ is read
in the presence of matter exactly as in the pure case of \eqref{eq:mixedanomaly}: it is the
fractional part of the canonical instanton number $k=Q/I_2(F)$ \eqref{eq:Ianomaly}, evaluated on the generating 1-form
background $\mathfrak{n}_{\rm gen}$:
\begin{equation}
    \Omega_{ii,\U(1)_I}\;=\;\Bigl[\,Q(\mathfrak{n}_{\rm gen})/I_2(F)\,\Bigr]_{\mod\,1}\,,
    \qquad
    \Omega_{ii\alpha}\;=\;-\frac{1}{2}\,\frac{Z_{(i)}^2\cdot D_{(\alpha)}}{n_{(i)}^2}\,,
    \label{eq:mixedabstract}
\end{equation}
the second equality being the geometric form of \cite[(4.21)]{Apruzzi:2021nmk}, with $D_{(\alpha)}$
the non-compact divisor dual to $\U(1)_I$. The key point is that $Q(n)=\tfrac12\Tr n^2$ is the
\emph{pure-gauge} instanton density: it carries no dependence on the matter content. The matter
enters only through which coset generates the (possibly screened) 1-form symmetry and through the
quantisation of the instanton flux $B_{m_0}$, but not through $Q$ itself.

Evaluating $Q$ on the generators of \S\ref{sec:expVSU4AS}--\S\ref{sec:expVSO8} and dividing by
$I_2(F)$ gives Table~\ref{tab:mixed}. As both $Q$ and $I_2(F)$ are matter-blind, the anomaly is
independent of the number $N_{\Lambda^2}$ of antisymmetric hypermultiplets (and of $\kCS$), and
equals $\tfrac12$ in all four cases: for the two antisymmetric families, $Q(\mathfrak{n}_{\rm gen})=\tfrac12$
with $I_2(F)=1$, while, for $\Spin(7)+3\mathbf{F}$ and $\Spin(8)+4\mathbf{F}$, the integer $Q(\mathfrak{n}_{\rm gen})=1$
divided by $I_2=2$ likewise gives $k(\mathfrak{n}_{\rm gen})=\tfrac12$ --- the same value as the pure
$\Spin(7)$ and $\Spin(8)$ theories, since the vector matter does not screen the $\Z_2^{(1)}$.

\begin{table}[h]
\centering
\begin{tabular}{lccccc}
\toprule
theory & 1-form & $\mathfrak{n}_{\rm gen}$ & $Q(\mathfrak{n}_{\rm gen})$ & $I_2$ & $[Q/I_2]_{\mod\,1}$ \\
\midrule
$\SU(4)_{\kCS}+N_{\Lambda^2}\Lambda^2$ \ ($\kCS$ even) & $\Z_2$ & $(\tfrac12,0,\tfrac12)$ & $\tfrac12$ & $1$ & $\tfrac12$ \\
$\USp(4)+N_{\Lambda^2}\Lambda^2$                       & $\Z_2$ & $(\tfrac12,0)$          & $\tfrac12$ & $1$ & $\tfrac12$ \\
$\Spin(7)+3\mathbf{F}$                                   & $\Z_2$ & $(1,1,\tfrac12)$        & $1$ & $2$ & $\tfrac12$ \\
$\Spin(8)+4\mathbf{F}$                                   & $\Z_2$ & $(1,1,\tfrac12,\tfrac12)$ & $1$ & $2$ & $\tfrac12$ \\
\bottomrule
\end{tabular}
\caption{Mixed 't~Hooft anomaly between the $\Z_2^{(1)}$ 1-form symmetry and the instanton symmetry
$\U(1)_I$, $\Omega_{ii,\U(1)_I}=[Q(\mathfrak{n}_{\rm gen})/I_2(F)]_{\mod1}$ \eqref{eq:Ianomaly}. The
fractional instanton number is $\tfrac12$ in all four cases (any $N_{\Lambda^2}$, any even $\kCS$);
for $\Spin(7)$ and $\Spin(8)$, the integer $Q=1$ must be divided by $I_2=2$ to give the physical
$k=\tfrac12$.}
\label{tab:mixed}
\end{table}

A caution attends the dependence on the number of hypermultiplets $N_{\Lambda^2}$. Since
$Q(n)=\tfrac12\Tr n^2$ depends only on the gauge flux $n$ --- and not on $B_{m_0}$ or the flavour
fluxes $B_{m_h}$ --- the value $\tfrac12$ is read off unambiguously, and is in fact the anomaly with
the \emph{Abelian} $\U(1)_I$. The instanton symmetry, however, \emph{may} enhance at the fixed point
to a non-Abelian group $G_I$ --- this is not generic, and many theories keep a plain $\U(1)_I$ ---
and whether it does is controlled by $N_{\Lambda^2}$ and $\kCS$: for instance, $\SU(4)_0+2\Lambda^2$
has $\U(1)_I\to\SU(2)_I$, while $\SU(4)_2+2\Lambda^2$ stays Abelian (\S\ref{sec:faithfulsym}). When the
enhancement occurs, the mixed anomaly is the one between $\Z_\fm^{(1)}$ and the \emph{Cartan}
$\U(1)\subset G_I$, with $F$ the field strength of that Cartan current; it is still read from $Q$,
now interpreted as the Cartan charge. Enhancement to a non-Abelian $G_I$ does not by itself switch
the anomaly off: pure $\SU(4)_4$ has $\U(1)_I\to\SO(3)_I$ and yet retains the mixed anomaly
\eqref{eq:mixedanomaly}, $\tfrac{N(N-1)}{2\fm^2}=\tfrac38$. What the global form of $G_I$ controls
is the Cartan normalisation and hence the precise coefficient; pinning it down requires the
superconformal index of \S\ref{sec:faithfulsym}.

\section{Faithful global symmetries from the superconformal index}
\label{sec:faithfulsym}
In this section, we collect the superconformal indices of the various theories of interest in this paper. They provide information about the faithful continuous global symmetries, which will be useful for the analysis of 2-group symmetries in the next section.

\subsection{Pure gauge theories}
\label{sec:indexpure}
For any simple gauge group $G$ and $\kCS=0$, the instanton partition function in the $k$-instanton sector is
\begin{equation}
\begin{aligned}
    Z^{G}_{\text{pure}} \big|_{\text{$k$-inst}} &\;=\; x^{\,k h^\vee_{G}}\, \mathrm{HS}\bigl[\, \mathcal{M}_{k,G}\,\bigr] = \frac{x^{k h^\vee_{G}}}{(1-x y)(1-x y^{-1})} \mathrm{HS}\bigl[\, \tilde{\mathcal{M}}_{k,G}\,\bigr]  \,, 
\end{aligned}
\end{equation}
where $\mathcal{M}_{k,G}$ is the \emph{full} $k$ $G$-instanton moduli space on $\mathbb{C}^2$, of quaternionic dimension $k h^\vee_G$, with $h^\vee_G$ the dual Coxeter number of $G$. It factorises as $\mathcal{M}_{k,G} = \mathbb{C}^2 \times \tilde{\mathcal{M}}_{k,G}$: the centre-of-mass $\mathbb{C}^2$ --- the position of the instanton on $\mathbb{C}^2$, contributing the $1/[(1-xy)(1-x/y)]$ factor --- times the \emph{centred} (reduced) moduli space $\tilde{\mathcal{M}}_{k,G}$ of instantons. For $k=1$, $\tilde{\mathcal{M}}_{1,G} =\overline{\mathcal{O}_{\min}}(\mathfrak{g})$, the closure of the minimal nilpotent orbit of $\mathfrak{g}$, of quaternionic dimension $h^\vee_{\mathfrak{g}} - 1$, whose Hilbert series is given by \cite{Benvenuti:2010pq} (see also \cite{Keller:2011ek}):
\bes{
\mathrm{HS}\bigl[\, \tilde{\mathcal{M}}_{1,G}\,\bigr] = \sum_{p=0}^\infty \chi^G_{p.\mathrm{hw(adj)}} \, x^{2p} \, ,
}
where $p.\mathrm{hw(adj)}$ denotes the irreducible representation whose highest weight is $p$ times the highest weight of the adjoint representation of $G$. The explicit Hilbert series for $k=2$ are given in \cite{Hanany:2012dm}, and the Hilbert series for general $k$ can be computed using the method described in \cite{Cremonesi:2014xha}. Since the instanton partition function is given by \bes{
Z_{\text{inst}} = \sum_{k=0}^\infty Z^{G}_{\text{pure}} \big|_{\text{$k$-inst}} \, q^k
}
and the Hilbert series starts with $1$, we see from \eqref{Indexgeneralformula} that the one-instanton contribution enters the superconformal index at order $x^{h^\vee_G}$. 

In the special case of the $\SU(2)_{\theta=0}$ pure gauge theory ($h^\vee_{\SU(2)}=2$), from \eqref{instcontr}, the one-instanton partition function contributes $q^{\pm 1}$ at order $x^2$. Along with the contribution of $\U(1)_I$ conserved current, the terms at order $x^2$ are $q+1+q^{-1}$; the global symmetry algebra is thus $\su(2)$. The index of this theory (up to four-instanton contributions) is given by \cite[(4.9)]{Kim:2012gu} (see also \cite[(16)]{Bergman:2013ala}  and \cite[Page 55]{Hwang:2014uwa}). Since only even Dynkin labels of the $\su(2)$ representations appear in the index, the $\Z_2$ centre of $\SU(2)$ acts trivially on such representations, and we conclude that the faithful global symmetry of its UV fixed point, namely the $E_1$ theory \cite{Seiberg:1996bd}, is $\SO(3)$; see also \cite{BenettiGenolini:2020doj, Genolini:2022mpi, Apruzzi:2021vcu}. This is also in agreement with the analysis of \cite[Section 2.2.2]{Cremonesi:2015lsa}, where it was found that the Higgs branch at infinite coupling is $\BC^2/\Z_2$, whose isometry is $\SO(3)$.  
On the other hand, for the theory with theta angle $\theta=\pi$, the instanton contribution appears at higher orders than $x^2$; specifically, for $\SU(2)_\pi$, the index is explicitly given by \cite[(17)]{Bergman:2013ala}, from which we see that the faithful global symmetry is $\U(1)_I$.\footnote{We explicitly checked that the blow-up method provides the same answer with the gauge flux $n \in \Z$, and the topological background $B_{m_0} \in \Z$ for $\SU(2)_0$ and $B_{m_0} \in \Z+\frac{1}{2}$ for $\SU(2)_\pi$.  This was discussed in \cite[Section 3.2.1]{Kim:2020hhh}.}

We remark that the $\USp(4)_{\theta=\pi}$ pure gauge theory also has the faithful global symmetry $\U(1)_I$.  Its index is equal to that of $\SU(3)_5$, namely
\bes{
\mathcal{I}_{\SU(3)_5} \;=\; 1 \;+\; x^2 \;+\; 2\chi_{[1]}(y)\, x^3 \;+\; 2\bigl[1+\chi_{[2]}(y)\bigr]\, x^4\;+\; \ldots\,.
}
For $\SU(N)_N$, the faithful global symmetry is also $\SO(3)$. We demonstrate this using the indices of $\SU(3)_3$ and $\SU(4)_4$ below. Moreover, the magnetic quiver for these theories is described by the 3d $\CN=4$ $\U(1)$ gauge theory with $2$ hypermultiplets of charge $1$ \cite[Phase III, $N_f=0$, Table 7]{Cabrera:2018jxt} whose Higgs/Coulomb branch is $\BC^2/\Z_2$ with isometry $\SO(3)$, independent of $N$.\footnote{We thank Julius Grimminger for confirming this point.} For reference, we present the character expansion of the index of $\SU(3)_3$ up to two instantons; this was computed in \cite[(4)]{Kim:2023qwh}, and we report it here for convenience:
\bes{
\scalebox{0.88}{$
\begin{split}
    &\mathcal{I}_{\SU(3)_3} \;=\; 1 \;+\; \chi_{[2]}^{\su(2)_I}\bigl(q^{1/2}\bigr)\, x^2 \;+\; \chi_{[1]}(y)\Bigl[ 1 + \chi_{[2]}^{\su(2)_I}\bigl(q^{1/2}\bigr) \Bigr]\, x^3 \\
    & \quad +\; \Bigl\{ 2 + \chi_{[2]}^{\su(2)_I}\bigl(q^{1/2}\bigr) + \chi_{[4]}^{\su(2)_I}\bigl(q^{1/2}\bigr)  + \bigl[ \chi_{[2]}(y) - 1 \bigr]\bigl[ 1 + \chi_{[2]}^{\su(2)_I}\bigl(q^{1/2}\bigr) \bigr] \Bigr\}\, x^4 \\
    & \quad +\; \Bigl\{ \chi_{[3]}(y)\bigl[ 1 + \chi_{[2]}^{\su(2)_I}\bigl(q^{1/2}\bigr) \bigr] + \chi_{[1]}(y)\bigl[ 1 + 2\chi_{[2]}^{\su(2)_I}\bigl(q^{1/2}\bigr) + \chi_{[4]}^{\su(2)_I}\bigl(q^{1/2}\bigr) \bigr] \Bigr\}\, x^5 \;+\; \cdots\,,
    \end{split}
    $}
    \label{eq:indexSU33}
}
where $\chi_{[2]}^{\su(2)_I}(q^{1/2}) = q + 1 + q^{-1}$ is the adjoint of the enhanced $\su(2)_I$, the three one-instanton currents accompanying the $\U(1)_I$ Cartan. We have also computed the index of $\SU(4)_4$ up to two instantons, and it coincides with \eqref{eq:indexSU33} up to and including order $x^5$. In both cases only even $\su(2)_I$ Dynkin labels occur --- for example the two-instanton states enter at order $x^4$ as $\chi_{[4]}^{\su(2)_I}(q^{1/2})$ --- so the $\Z_2$ centre acts trivially and the faithful continuous global symmetry is $\SO(3)_I$.

Let us discuss another interesting case, namely $\SU(3)_6$. The index of this theory is given by \cite[(7)]{Kim:2023qwh}, confirming that the faithful global symmetry is $\SU(2)$, since $\su(2)_I$ representations with odd Dynkin labels also appear in the index.  It was also pointed out in the aforementioned reference that the instanton partition function of this theory can be computed via the blow-up method, not from the ADHM construction. It is therefore interesting to contrast this with the result from the magnetic quiver, discussed in \cite[Section 5.3]{Cabrera:2018jxt}, whose Coulomb branch is the moduli space of two $\SU(2)$ instantons on $\BC^2$. Discarding the free hypermultiplet associated with the centre-of-mass of the instantons, we see from \cite[(3.14)--(3.15)]{Hanany:2012dm} that the Hilbert series can be written in terms of the characters of representations of $\SU(2)_y \times \SO(3)_I$, where odd Dynkin labels appear for the symmetry $\SU(2)_y$ of $\BC^2$ where the instantons reside, but not for the internal $\SO(3)_I$ symmetry of the instantons.  Since the derivation of the instanton partition function and of the index of $\SU(3)_6$ is rather subtle, and bears a strong resemblance to that of the $\USp(4)_{\theta=0}$ gauge theory with two antisymmetric hypermultiplets, we provide the details of both in appendix~\ref{app:blow-upSU36USp4}. 

\subsection{$\SU(4)_{\kCS}$ with antisymmetric hypermultiplets}
\label{sec:so3form}

Our richest examples are $\SU(4)_{\kCS}$ with one and with two antisymmetric hypermultiplets. In either case, the gauge group, the matter content and the 1-form $\Z_2^{(1)}$ symmetry are held fixed, while the \emph{enhanced} continuous flavour symmetry at the UV fixed point depends on the Chern--Simons level $\kCS$. Let us summarise the results in the table below and derive them in the following subsections.
\bes{
\begin{tabular}{lc}
\toprule
Theory & Faithful continuous global symmetry  \\
\midrule
$\SU(4)_0 + 1\Lambda^2$ & $\bigl[ \SU(2)_w \times \U(1)_I \bigr] / \Z_2$ \\
$\SU(4)_2 + 1\Lambda^2$ & $\SO(3)_w \times \U(1)_I$  \\
$\SU(4)_4 + 1\Lambda^2$ & $\SU(3)/\BZ_3$  \\
\hline \hline
$\SU(4)_0 + 2\Lambda^2$ & $(\USp(4)_w/\BZ_2) \times \SO(3)_I$ \\
$\SU(4)_2 + 2\Lambda^2$ & $[\USp(4)_w \times \U(1)_I]/\BZ_2$ \\
$\SU(4)_4 + 2\Lambda^2$ & $\SO(7)$ \\
\bottomrule
\end{tabular}
\label{tab:faithfulsym}
}
The expression of the superconformal index of the $\SU(4)$ gauge theory with $N_{\Lambda^2}$ antisymmetric hypermultiplets is given by \eref{Indexgeneralformula}--\eref{Indexpert}, with the Haar measure and the characters of the adjoint and antisymmetric representations of $\su(4)$ collected in appendix~\ref{app:grouptheory}.

\subsubsection{One antisymmetric hypermultiplet}
The superconformal index of $\SU(4)_0 + 1\Lambda^2$ (up to one-instanton contribution) reads
\bes{
\scalebox{0.82}{$
\begin{split}
    &\mathcal{I}_{\SU(4)_0 + 1\Lambda^2} \;=\; 1 \;+\; \bigl[ 1 + \chi_{[2]}(w_1) \bigr]\, x^2 \;+\; \chi_{[1]}(y) \bigl[ 2 + \chi_{[2]}(w_1) \bigr]\, x^3 \\
    &\quad+ \Bigl\{ 4 + 2 \chi_{[2]}(w_1) + \chi_{[4]}(w_1) + \bigl( q + q^{-1} \bigr)\, \chi_{[1]}(w_1)  + \bigl[ \chi_{[2]}(y) - 1 \bigr] \bigl[ 2 + \chi_{[2]}(w_1) \bigr] \Bigr\}\, x^4  \\
    &\quad+ \Bigl\{ \chi_{[3]}(y) \bigl[ 2 + \chi_{[2]}(w_1) \bigr] + \chi_{[1]}(y) \Bigl[ 4 + 4 \chi_{[2]}(w_1) + \chi_{[4]}(w_1)  + \bigl( q + q^{-1} \bigr)\, \chi_{[1]}(w_1) \Bigr] \Bigr\}\, x^5 \;+\; \ldots\,,
    \end{split}
    $}
    \label{eq:indexSU40AS}
}
where $y$ is the fugacity of the spacetime $\SU(2)$ rotations and $q$ that of the topological symmetry $\U(1)_I$. We see that the one-instanton states contribute $\bigl( q + q^{-1} \bigr) \chi_{[1]}(w_1)$ at order $x^4$ and $\bigl( q + q^{-1} \bigr) \chi_{[1]}(w_1)\, \chi_{[1]}(y)$ at order $x^5$: representations with \emph{odd} Dynkin label (half-integer spin) of $\SU(2)_w$ are present, and they carry odd charge under $\U(1)_I$. The faithful continuous global symmetry is therefore
\begin{equation}
    \frac{\SU(2)_w \times \U(1)_I}{\Z_2} \qquad (N_{\Lambda^2} = 1, \, \kCS = 0)\,,
\end{equation}
with the $\Z_2$ identifying the centre of $\SU(2)_w$ with the parity of the instanton charge.

On the other hand, at $\kCS = 2$, one finds
\bes{
\scalebox{0.91}{$
\begin{split}
    &\mathcal{I}_{\SU(4)_2 + 1\Lambda^2} \;=\; 1 \;+\; \bigl[ 1 + \chi_{[2]}(w_1) \bigr]\, x^2 \;+\; \Bigl\{ \bigl( q + q^{-1} \bigr) + \chi_{[1]}(y) \bigl[ 2 + \chi_{[2]}(w_1) \bigr] \Bigr\}\, x^3  \\
    & \qquad+ \Bigl\{ 4 + 2 \chi_{[2]}(w_1) + \chi_{[4]}(w_1) + \bigl( q + q^{-1} \bigr)\, \chi_{[1]}(y) + \bigl[ \chi_{[2]}(y) - 1 \bigr] \bigl[ 2 + \chi_{[2]}(w_1) \bigr] \Bigr\}\, x^4  \\
    & \qquad+ \Bigl\{ \chi_{[3]}(y) \bigl[ 2 + \chi_{[2]}(w_1) \bigr] + \chi_{[1]}(y) \Bigl[ 4 + 4 \chi_{[2]}(w_1) + \chi_{[4]}(w_1) \Bigr] \\ &\qquad \qquad \qquad \quad \, \, \, + \bigl( q + q^{-1} \bigr) \Bigl[ 1 + \chi_{[2]}(w_1) + \chi_{[2]}(y) \Bigr] \Bigr\}\, x^5 \;+\; \ldots\,.
    \label{eq:indexSU42AS}
    \end{split}
    $}
}
We see that only \emph{even} Dynkin labels of $\SU(2)_w$ appear at every order of $x$, including in the instanton sectors, and the faithful continuous global symmetry is
\begin{equation}
    \SO(3)_w \times \U(1)_I \qquad (N_{\Lambda^2} = 1, \, \kCS = 2)\,,
\end{equation}
in agreement with \cite{Apruzzi:2021vcu}.

Finally, at $\kCS = 4$, one finds
\bes{
\scalebox{0.97}{$
\begin{split}
    \mathcal{I}_{\SU(4)_4 + 1\Lambda^2} \;=\; 1 \;&+\; \Bigl[ 1 + \chi_{[2]}^{\su(2)_w}(w_1) + \bigl(q + q^{-1}\bigr)\,\chi_{[1]}^{\su(2)_w}(w_1) \Bigr]\, x^2 \\
    &+\; \chi_{[1]}(y)\Bigl[ 2 + \chi_{[2]}^{\su(2)_w}(w_1) + \bigl(q + q^{-1}\bigr)\,\chi_{[1]}^{\su(2)_w}(w_1) \Bigr]\, x^3 \;+\; \ldots\,,
    \end{split}
    $}
    \label{eq:indexSU44_1AS}
}
where $w_1$ is the fugacity of the flavour algebra $\su(2)_w$ rotating the antisymmetric hypermultiplet and $q$ that of the topological symmetry $\U(1)_I$. The eight conserved currents at order $x^2$ --- the $\su(2)_w \oplus \fu(1)_I$ currents $1 + \chi_{[2]}^{\su(2)_w}(w_1)$ together with the one-instanton states $(q+q^{-1})\,\chi_{[1]}^{\su(2)_w}(w_1)$ --- fill the adjoint of $\su(3)$, exhibiting the enhancement of $\su(2)_w \oplus \fu(1)_I$ to $\su(3)$. The index is covariant under this enhanced symmetry,
\begin{equation}
    \mathcal{I}_{\SU(4)_4 + 1\Lambda^2} \;=\; 1 + \chi_{\mathbf{8}}^{\su(3)}\, x^2 + \chi_{[1]}(y)\bigl[ 1 + \chi_{\mathbf{8}}^{\su(3)} \bigr]\, x^3 + \ldots\,,
    \label{eq:indexSU44_1AS_su3}
\end{equation}
where
\begin{equation}
    \chi_{\mathbf{8}}^{\su(3)} \;=\; \chi_{[2]}^{\su(2)_w}(w_1) + 1 + \bigl(q + q^{-1}\bigr)\,\chi_{[1]}^{\su(2)_w}(w_1)
\end{equation}
is the character of the adjoint $\mathbf{8}$ of $\su(3)$ in the branching $\mathbf{8} \to \mathbf{3}_0 \oplus \mathbf{1}_0 \oplus \mathbf{2}_{+1} \oplus \mathbf{2}_{-1}$ under $\su(2)_w \oplus \fu(1)_I \subset \su(3)$. As in $\SU(4)_4 + 2\Lambda^2$ described in \eqref{eq:indexSU44x4}--\eqref{predictionx4SU4kCS42AS}, the order-$x^4$ coefficient provides a sharper test, since $\SU(3)$ covariance requires a two-instanton contribution:
\begin{equation} \label{predictionx4ofSU4kCS41AS}
    \mathcal{I}_{\SU(4)_4 + 1\Lambda^2}\big|_{x^4} \;=\; 1 + \chi_{\mathbf{8}}^{\su(3)} + \mathrm{Sym}^2\chi_{\mathbf{8}}^{\su(3)} + \bigl[ \chi_{[2]}(y) - 1 \bigr]\bigl[ 1 + \chi_{\mathbf{8}}^{\su(3)} \bigr]\,,
\end{equation}
with $\mathrm{Sym}^2\chi_{\mathbf{8}}^{\su(3)} = \chi_{\mathbf{1}}^{\su(3)} + \chi_{\mathbf{8}}^{\su(3)} + \chi_{\mathbf{27}}^{\su(3)}$, the $\mathbf{27}$ predicting the two-instanton states $(q^2 + q^{-2})\,\chi_{[2]}^{\su(2)_w}(w_1)$ at this order.  We have verified that the two-instanton partition function indeed reads $\chi_{[2]}^{\su(2)_w}(w_1) x^4 + \ldots$, and thus confirms this result.  Since only $\SU(3)$ representations that transform trivially under the $\Z_3$ centre appear in the index (the $\mathbf{1}$, $\mathbf{8}$, $\mathbf{27}$, \ldots, with, for example, the fundamental representation being absent), we conclude that the faithful continuous global symmetry is
\begin{equation}
    \SU(3)/\Z_3 \qquad (N_{\Lambda^2} = 1, \, \kCS = 4)\,,
\end{equation}
in agreement with \cite[(4.70)]{Apruzzi:2021vcu}.

\subsubsection{Two antisymmetric hypermultiplets}
\label{sec:indexSU42AS}
The one-instanton contribution to the conserved-current spectrum determines the enhancement pattern summarised in the table above. Applying the analysis of~\cite[Table 1, Eq. (22)]{Zafrir:2015rga} (see \cite{Tachikawa:2015mha} for the original exposition and also \cite{Yonekura:2015ksa} for generalisations) to count the gauge-invariant one-instanton states built from the fermionic raising operators $A$ (from antisymmetric zero modes) and $B$ (from gaugino zero modes), one finds:
\begin{itemize}[leftmargin=*]
    \item \textbf{$\kCS = 4$}: three gauge-invariant states $\lvert 0 \rangle$, $\epsilon A^2 \lvert 0\rangle$, $(\epsilon A^2)^2 \lvert 0\rangle$, organising into $(\mathbf{1})_{+2} \oplus (\mathbf{3})_0 \oplus (\mathbf{1})_{-2}$ of $\SU(2)_{\text{AS}} \times \U(1)_{\text{AS}} \subset \USp(4)$. This is exactly the branching of the $\mathbf{5}$ (vector) of $\USp(4) \cong \mathrm{Spin}(5)$. The five new conserved currents at $q = +1$, together with five conjugate currents at $q = -1$, the ten currents of the $\USp(4)$ adjoint, and the one $\U(1)_I$ current, sum to $5 + 5 + 10 + 1 = 21 = \dim \SO(7)$. This indicates that the enhanced symmetry algebra is $\so(7) \supset \so(5) \oplus \so(2) \cong \usp(4) \oplus \fu(1)_I$.
    \item \textbf{$\kCS = 0$}: one gauge-invariant state $(\epsilon AB)^2 \lvert 0\rangle$, which (after the Bose-symmetry and $\SU(2)_R$-singlet contraction enforced by the broken-current supermultiplet condition) is a singlet of $\USp(4)$. This single new current at $q = +1$, plus its conjugate at $q = -1$, combines with the existing $\U(1)_I$ current to fill out the $\SU(2)_I$ adjoint, leaving $\USp(4)$ untouched. The global symmetry algebra in this case is therefore $\usp(4) \oplus \su(2)_I$.
    \item \textbf{$\kCS = 2$}: no global symmetry enhancement (see also \cite[Table 4]{Zafrir:2015rga}); there is no $\SU(2)_R$-triplet conserved current at $q = \pm 1$ beyond what $\U(1)_I$ already provides.
\end{itemize}
The above method determines the global symmetry algebras. Below, we will use the superconformal index to determine the faithful symmetry group associated to such symmetry algebras. 

\subsubsection*{$N_{\Lambda^2}=2$ and $\kCS=4$}
For $\kCS = 4$, one finds
\begin{equation}
    \mathcal{I}_{\SU(4)_4 + 2\Lambda^2} \;=\; 1 \;+\; \chi_{\mathrm{adj}}^{\so(7)}\, x^2 \;+\; \chi_{[1]}(y) \Bigl[ 1 + \chi_{\mathrm{adj}}^{\so(7)} \Bigr]\, x^3 \;+\; \ldots\,,
    \label{eq:indexSU44_2AS}
\end{equation}
where
\begin{equation}
    \chi_{\mathrm{adj}}^{\so(7)} \;=\; \chi_{[2,0]}^{\usp(4)_w}(w_1, w_2) \;+\; 1 \;+\; \bigl( q + q^{-1} \bigr)\, \chi_{[0,1]}^{\usp(4)_w}(w_1, w_2)
    \label{eq:SO7branching}
\end{equation}
is precisely the character of the adjoint $\mathbf{21} = [0,1,0]$ of $\so(7)$ in the branching $\mathbf{21} \to \mathbf{10}_0 \oplus \mathbf{1}_0 \oplus \mathbf{5}_{+1} \oplus \mathbf{5}_{-1}$ under $\usp(4)_w \oplus \fu(1)_q \subset \so(7)$, with $w_1, w_2$ the orthonormal fugacities of $\usp(4)_w$ and $\mathbf{10} = [2,0]$, $\mathbf{5} = [0,1]$ in $\usp(4)$ Dynkin labels. The conserved currents at order $x^2$ thus assemble into the adjoint of $\so(7)$: the symmetry algebra of $\SU(4)_4 + 2\Lambda^2$ is enhanced to $\so(7)$, into which $\usp(4)_w$ and $\fu(1)_q$ combine.

The order-$x^4$ coefficient provides a sharper test, since $\SO(7)$ covariance requires a \emph{two-instanton} contribution at this order. The $q^{0}$ and $q^{\pm 1}$ sectors of the index at order $x^4$ are reproduced, uniquely, by the $\SO(7)$-covariant expression
\begin{equation}
    \mathcal{I}_{\SU(4)_4 + 2\Lambda^2} \Big|_{x^4} \;=\; 1 \;+\; \chi_{\mathrm{adj}}^{\so(7)} \;+\; \mathrm{Sym}^2 \chi_{\mathrm{adj}}^{\so(7)} \;+\; \bigl[ \chi_{[2]}(y) - 1 \bigr] \Bigl[ 1 + \chi_{\mathrm{adj}}^{\so(7)} \Bigr]\,,
    \label{eq:indexSU44x4}
\end{equation}
where $\mathrm{Sym}^2 \chi_{\mathrm{adj}}^{\so(7)} = \chi_{\mathbf{1}} + \chi_{\mathbf{27}} + \chi_{\mathbf{35}} + \chi_{\mathbf{168'}}$ is the symmetric square of the adjoint (the conserved-current bilinears), with $\mathbf{27} = [2,0,0]$, $\mathbf{35} = [0,0,2]$ and $\mathbf{168'} = [0,2,0]$. Under $\usp(4)_w \oplus \fu(1)_q$, the $\mathbf{27}$ and $\mathbf{168'}$ contain the charge-2 components $\mathbf{1}_{\pm 2}$ and $\mathbf{14}_{\pm 2}$, so \eqref{eq:indexSU44x4} \emph{predicts} the two-instanton contribution at order $x^4$ to be
\begin{equation} \label{predictionx4SU4kCS42AS}
    \bigl( q^2 + q^{-2} \bigr) \Bigl[ 1 + \chi_{[0,2]}^{\usp(4)_w}(w_1, w_2) \Bigr]\,,
\end{equation}
with $\chi_{[0,2]}^{\usp(4)_w}$ the character of the $\mathbf{14} = [0,2]$. We have checked explicitly that the two-instanton partition function indeed reads $[ 1 + \chi_{[0,2]}^{\usp(4)_w}(w_1, w_2) ]x^4 + \ldots\,,$ which confirms the aforementioned prediction. Note that this is very similar to the discussion around \eqref{eq:indexSU44_1AS_su3}--\eqref{predictionx4ofSU4kCS41AS}. In conclusion, the faithful continuous global symmetry of this theory is $\SO(7)$, in agreement with \cite[(4.77)]{Apruzzi:2021vcu}.

\subsubsection*{$N_{\Lambda^2}=2$ and $\kCS=0$}
At $\kCS = 0$, the superconformal index (up to two-instanton contribution) reads
\begin{equation}
\scalebox{0.83}{$
\begin{split}
    &\mathcal{I}_{\SU(4)_0 + 2\Lambda^2} \;=\; 1 \;+\; \Bigl[ \chi_{[2]}^{\su(2)_I}\bigl(q^{1/2}\bigr) + \chi_{[2,0]}^{\usp(4)_w} \Bigr]\, x^2  +\; \chi_{[1]}(y) \Bigl[ 1 + \chi_{[2]}^{\su(2)_I}\bigl(q^{1/2}\bigr) + \chi_{[2,0]}^{\usp(4)_w} \Bigr]\, x^3 \\&\quad+\; \Bigg\{3 + \chi_{[4,0]}^{\usp(4)_w} + \chi_{[0,2]}^{\usp(4)_w} + \chi_{[2,0]}^{\usp(4)_w} + \chi_{[4]}^{\su(2)_I}\bigl(q^{1/2}\bigr) +  \chi_{[2]}^{\su(2)_I}\bigl(q^{1/2}\bigr) \bigl[ 1 + \chi_{[2,0]}^{\usp(4)_w} + \chi_{[0,1]}^{\usp(4)_w} \bigr] \\ & \qquad \quad \,\,\,\,+ \bigl[ \chi_{[2]}(y) - 1 \bigr] \Bigl[ 1 + \chi_{[2]}^{\su(2)_I}\bigl(q^{1/2}\bigr) + \chi_{[2,0]}^{\usp(4)_w} \Bigr]\Bigg\} x^4 +\ldots\,,
\end{split}
$}
    \label{eq:indexSU40_2AS}
\end{equation}
with $\chi_{[2]}^{\su(2)_I}\bigl(q^{1/2}\bigr) = q + 1 + q^{-1}$: the $13$ conserved currents at order $x^2$ fill the adjoint of $\su(2)_I \oplus \usp(4)_w$, exhibiting the enhancement $\fu(1)_I \to \su(2)_I$. The $\usp(4)_w$ characters appearing in the expression above correspond to the representations $\mathbf{10} = [2,0]$, $\mathbf{5} = [0,1]$, $\mathbf{14} = [0,2]$, $\mathbf{35}^{\prime} = [4,0]$.
Note that only integer powers of $q$ (integer $\SU(2)_I$ spins) and only $\Z_2$-centre-invariant representations of $\USp(4)_w$ appear, consistently with the faithful global form $\SO(3)_I \times \bigl( \USp(4)_w / \Z_2 \bigr) = \SO(3)_I \times \SO(5)_w$, in agreement with \cite{Apruzzi:2021vcu}.

\subsubsection*{$N_{\Lambda^2}=2$ and $\kCS=2$}
Finally, at $\kCS = 2$, one finds
\bes{ \label{eq:indexSU42_2AS}
\scalebox{0.97}{$
\begin{split}
    &\mathcal{I}_{\SU(4)_2 + 2\Lambda^2} \;=\; 1 \;+\; \Bigl[ 1 + \chi_{[2,0]}^{\usp(4)_w} \Bigr]\, x^2 \\
    &\qquad +\; \Bigl\{ \bigl( q + q^{-1} \bigr)\, \chi_{[1,0]}^{\usp(4)_w} + \chi_{[1]}(y) \Bigl[ 2 + \chi_{[2,0]}^{\usp(4)_w} \Bigr] \Bigr\}\, x^3  \\
    &\qquad+\; \Bigg\{4 + \chi_{[4,0]}^{\usp(4)_w} + \chi_{[0,2]}^{\usp(4)_w} + 2 \chi_{[2,0]}^{\usp(4)_w} + \chi_{[0,1]}^{\usp(4)_w}  \\
    &\qquad \qquad \; \; \; + \bigl( q + q^{-1} \bigr)\, \chi_{[1,0]}^{\usp(4)_w}\, \chi_{[1]}(y) \;+ \bigl[ \chi_{[2]}(y) - 1 \bigr] \Bigl[ 2 + \chi_{[2,0]}^{\usp(4)_w} \Bigr] \Bigg\}\, x^4 + \ldots\,.
    \end{split}
    $}
}
The $x^2$ coefficient contains only the $11$ currents of $\fu(1)_I \oplus \usp(4)_w$: there is no symmetry enhancement, in agreement with the table above. The $x^4$ coefficient at $q^0$ is again $1 + \chi_{\mathrm{adj}} + \mathrm{Sym}^2 \chi_{\mathrm{adj}}$, using $\mathrm{Sym}^2 \mathbf{10} = \mathbf{1} \oplus \mathbf{5} \oplus \mathbf{14} \oplus \mathbf{35}'$, where, in terms of the $\usp(4)_w$ characters, $\mathbf{10} = [2,0]$, $\mathbf{5} = [0,1]$, $\mathbf{14} = [0,2]$, $\mathbf{35}^{\prime} = [4,0]$. The instanton states, however, now appear in the \emph{fundamental} $\mathbf{4} = [1,0]$ of $\usp(4)_w$, which is charged under the $\Z_2$ centre: states with non-trivial $\USp(4)_w$ centre charge carry odd $\U(1)_I$ charge, so the faithful continuous global symmetry at $\kCS = 2$ is
\begin{equation}
    (\USp(4)_w \times \U(1)_I)/\Z_2\,.
\end{equation}
Moreover, the two-instanton partition function reads $\chi_{[2,0]}^{\usp(4)_w} x^6 + \ldots\,,$ in agreement with the aforementioned faithful continuous global symmetry.

\subsection{$\USp(4)_0$ with antisymmetric hypermultiplets}
The superconformal index of the $\USp(4)$ gauge theory with $N_{\Lambda^2}$ antisymmetric hypermultiplets can be derived from the general formula \eref{Indexgeneralformula}--\eref{Indexpert}, with the Haar measure and the characters of the adjoint and antisymmetric representations of $\usp(4)$ collected in appendix~\ref{app:grouptheory}.

\subsubsection*{The case of $N_{\Lambda^2}=1$}
The superconformal index of the $\USp(4)_{\theta=0} + 1\Lambda^2$ theory (up to two-instanton contributions) reads
\bes{
\scalebox{0.88}{$
\begin{split}
    &\mathcal{I}_{\USp(4)_0 + 1\Lambda^2} \;=\; 1 \;+\; \Bigl[ \chi_{[2]}^{\su(2)_w}(w_1) + \chi_{[2]}^{\su(2)_I}\bigl(q^{1/2}\bigr) \Bigr]\, x^2  \\
    & \quad +\; \Bigl\{ \chi_{[1]}^{\su(2)_w}(w_1)\, \chi_{[2]}^{\su(2)_I}\bigl(q^{1/2}\bigr)  + \chi_{[1]}(y) \Bigl[ 1 + \chi_{[2]}^{\su(2)_w}(w_1) + \chi_{[2]}^{\su(2)_I}\bigl(q^{1/2}\bigr) \Bigr] \Bigr\}\, x^3  \\
    & \quad+\; \Bigl\{ 2 + \chi_{[4]}^{\su(2)_w}(w_1) + \chi_{[4]}^{\su(2)_I}\bigl(q^{1/2}\bigr) + \chi_{[1]}(y)\,\chi_{[1]}^{\su(2)_w}(w_1)  + \chi_{[2]}(y)\Bigl[ 1 + \chi_{[2]}^{\su(2)_w}(w_1) \Bigr] \\
    &\quad \qquad \; \; + \chi_{[2]}^{\su(2)_I}\bigl(q^{1/2}\bigr)\Bigl[ \chi_{[2]}^{\su(2)_w}(w_1) + \chi_{[2]}(y) + \chi_{[1]}(y)\,\chi_{[1]}^{\su(2)_w}(w_1) \Bigr] \Bigr\}\, x^4 \\
    & \quad+\; \Bigl\{ \chi_{[1]}^{\su(2)_w}(w_1) + \chi_{[2]}(y)\,\chi_{[1]}^{\su(2)_w}(w_1) + \chi_{[3]}(y)\Bigl[ 1 + \chi_{[2]}^{\su(2)_w}(w_1) \Bigr] \\& \quad \qquad \; \;+ \chi_{[4]}^{\su(2)_I}\bigl(q^{1/2}\bigr)\Bigl[ \chi_{[1]}^{\su(2)_w}(w_1) + \chi_{[1]}(y) \Bigr] + \chi_{[1]}(y)\Bigl[ 2 + 2\chi_{[2]}^{\su(2)_w}(w_1) + \chi_{[4]}^{\su(2)_w}(w_1) \Bigr] \\
    &\quad \qquad \; \;+ \chi_{[2]}^{\su(2)_I}\bigl(q^{1/2}\bigr)\Bigl[ \chi_{[3]}^{\su(2)_w}(w_1) + \chi_{[1]}^{\su(2)_w}(w_1)\bigl( 1 + \chi_{[2]}(y) \bigr) \\ & \quad \qquad \qquad+ 2\chi_{[1]}(y)\bigl( 1 + \chi_{[2]}^{\su(2)_w}(w_1) \bigr) + \chi_{[3]}(y) \Bigr] \Bigr\}\, x^5 \;+\; \ldots\,,
    \label{eq:indexUSp4_1AS}
    \end{split}
    $}
}
with $w_1$ the fugacity of the flavour algebra $\su(2)_w$ rotating the antisymmetric hypermultiplet, $q$ the topological fugacity, and $y$ that of the spacetime $\SU(2)$. The six conserved currents at order $x^2$ fill the adjoint of $\su(2)_w \oplus \su(2)_I$: the classical $\su(2)_w$ is accompanied by the enhancement $\fu(1)_I \to \su(2)_I$ of the topological symmetry, signalled by the one-instanton currents $\chi_{[2]}^{\su(2)_I}(q^{1/2}) = q + 1 + q^{-1}$. The two-instanton contribution enters at order $x^4$. Odd Dynkin labels of $\su(2)_w$ start appearing at $x^4$, but the Dynkin labels of $\su(2)_I$ are always even to all orders we can check (up to $x^5$). The faithful continuous global symmetry is therefore
\bes{
\SU(2)_w \times \SO(3)_I \qquad (N_{\Lambda^2}=1)\,, 
}
in agreement with \cite{Apruzzi:2021vcu}.

\subsubsection*{The case of $N_{\Lambda^2}=2$}
For $\USp(4)_{\theta=0} + 2\Lambda^2$, the classical flavour algebra of the two antisymmetric hypermultiplets is $\usp(4)_w$ (fugacities $w_1, w_2$), and the topological symmetry enhances to $\su(2)_I$ (fugacity $q$, with the doublet at $q^{\pm 1}$). Computing the index up to two instantons, and organising the result in representations $(\mathbf{R}_{\usp(4)_w}, \mathbf{R}_{\su(2)_I})$ of $\usp(4)_w \oplus \su(2)_I$, we find
\bes{
\scalebox{0.96}{$
\begin{split}
    &\mathcal{I}_{\USp(4)_0 + 2\Lambda^2} \;=\; 1 \;+\; \Bigl[ (\mathbf{10}, \mathbf{1}) \oplus (\mathbf{1}, \mathbf{3}) \oplus (\mathbf{4}, \mathbf{2}) \Bigr]\, x^2 \\
    & \qquad +\; \left\{ (\mathbf{5},\mathbf{2}) \oplus (\mathbf{4},\mathbf{1})  +\; \chi_{[1]}(y) \Bigl[ (\mathbf{1}, \mathbf{1}) \oplus (\mathbf{10}, \mathbf{1}) \oplus (\mathbf{1}, \mathbf{3}) \oplus (\mathbf{4}, \mathbf{2}) \Bigr] \right\}\, x^3 \;+\; \ldots \\
    & \qquad= 1 \;+\; \chi_{\mathrm{adj}}^{\usp(6)}\, x^2 \;+\; \left( \chi_{[0,0,1]}^{\usp(6)} \; + \; \chi_{[1]}(y) \Bigl[ 1 + \chi_{\mathrm{adj}}^{\usp(6)} \Bigr] \right)\, x^3 \;+\; \ldots\,,
    \end{split}
    $}
    \label{eq:indexUSp4_2AS}
}
which, in characters, translates into
\begin{equation}
\scalebox{0.88}{$
    (\mathbf{10}, \mathbf{1}) \oplus (\mathbf{1}, \mathbf{3}) \oplus (\mathbf{4}, \mathbf{2}) \;=\; \chi_{[2,0]}^{\usp(4)_w}(w_1, w_2) \;+\; \chi_{[2]}^{\su(2)_I}(q) \;+\; \chi_{[1,0]}^{\usp(4)_w}(w_1, w_2)\, \chi_{[1]}^{\su(2)_I}(q)\,,
    $}
\end{equation}
where $\chi_{[1]}^{\su(2)_I}(q) = q + q^{-1}$ and $\chi_{[2]}^{\su(2)_I}(q) = q^2 + 1 + q^{-2}$. The $21$ currents at order $x^2$ are precisely the adjoint representation of the enhanced algebra $\usp(6) \supset \usp(4)_w \oplus \su(2)_I$
\begin{equation}
    \mathbf{21} \;=\; (\mathbf{10}, \mathbf{1}) \oplus (\mathbf{1}, \mathbf{3}) \oplus (\mathbf{4}, \mathbf{2})\,,
\end{equation}
so that $\usp(4)_w$ and $\su(2)_I$ combine into $\usp(6)$. The \emph{two-instanton} states --- the $(\mathbf{1}, \mathbf{3})$ components at $q^{\pm 2}$, i.e.\ the top and bottom of the $\su(2)_I$ adjoint --- are an essential part of the enhanced symmetry: only at two instantons does the full $\usp(6)$ become manifest. Note also that the representation $\mathbf{14'}=[0,0,1]$ of $\USp(6)$, whose branching rule under $\usp(6) \supset \usp(4)_w \oplus \su(2)_I$ is
\bes{
\mathbf{14'} \;=\; (\mathbf{5},\mathbf{2}) \oplus (\mathbf{4},\mathbf{1})~,
}
appears at order $x^3$. This representation transforms non-trivially under the $\Z_2$ centre of $\USp(6)$. This leads us to conclude that the faithful global symmetry is 
\bes{
\USp(6) \qquad (N_{\Lambda^2}=2) \, ,
}
rather than $\USp(6)/\Z_2$, in agreement with \cite[(4.68)]{Apruzzi:2021vcu}. This parallels the case of the pure $\SU(3)_6$ gauge theory discussed in \S\ref{sec:indexpure}, whose faithful global symmetry is $\SU(2)$ rather than $\SO(3) = \SU(2)/\Z_2$. The two theories in fact share several structural features, and their superconformal indices are computed in parallel from the blow-up equations in appendix~\ref{app:blow-upSU36USp4}.

\subsection{$\Spin(7)$ with $3$ hypermultiplets in the vector representation}
\label{sec:indexSO7}
The superconformal index of $\Spin(7) + 3\mathbf{F}$ can be computed from \eref{Indexgeneralformula}--\eref{Indexpert}, with the Haar measure and the characters of the adjoint and vector representations of $\so(7)$ collected in appendix~\ref{app:grouptheory}. Up to the one-instanton contribution, it reads
\begin{align} 
    \mathcal{I}_{\Spin(7) + 3\mathbf{F}} \;=\; 1 \;&+\; \Bigl[ \chi_{\mathrm{adj}}^{\usp(6)} + \chi_{[2]}^{\su(2)_I}\bigl(\tilde{q}^{1/2}\bigr) \Bigr]\, x^2 \nonumber \\
    &+\; \chi_{[1]}(y) \Bigl[ 1 + \chi_{\mathrm{adj}}^{\usp(6)} + \chi_{[2]}^{\su(2)_I}\bigl(\tilde{q}^{1/2}\bigr) \Bigr]\, x^3 \;+\; \ldots\,,
    \label{eq:indexSO7}
\end{align}
where $w_1, w_2, w_3$ are the orthonormal fugacities of the flavour algebra $\usp(6)$ rotating the three vector hypermultiplets, $\tilde{q}$ is the topological fugacity --- related to the blow-up instanton fugacity $q$ of \S\ref{sec:expVSO7} by $\tilde{q} = q^2$, so that the canonical one-instanton states sit at $\tilde{q}^{\pm 1}$ (cf.\ the footnote there) --- and $y$ that of the spacetime $\SU(2)$. The $24$ conserved currents at order $x^2$ fill the adjoint of $\usp(6) \oplus \su(2)_I$: the three one-instanton states $\chi_{[2]}^{\su(2)_I}(\tilde{q}^{1/2}) = \tilde{q} + 1 + \tilde{q}^{-1}$ exhibit the enhancement $\fu(1)_I \to \su(2)_I$ of the topological symmetry. Computing the index to order $x^4$ confirms this structure; in particular the $y^{\pm 2}$ sector at this order is
\begin{equation} \label{indexSO7with3x4}
    \mathcal{I}_{\Spin(7) + 3\mathbf{F}} \big|_{x^4}^{y^{\pm 2}} \;=\; \bigl[ \chi_{[2]}(y) - 1 \bigr] \Bigl[ 1 + \chi_{\mathrm{adj}}^{\usp(6)} + \chi_{[2]}^{\su(2)_I}\bigl(\tilde{q}^{1/2}\bigr) \Bigr]\,.
\end{equation}
Crucially, \emph{only integer powers of $q$} appear at every order --- equivalently, only integer-isospin (even Dynkin label) representations of $\su(2)_I$ occur, the half-integer-isospin doublet $\tilde{q}^{\pm 1/2}$ being absent. The $\su(2)_I$ is therefore realised as $\SO(3)_I = \SU(2)_I/\Z_2$, by the same parity argument used for $\SO(3)_w$ in \S\ref{sec:so3form}. The faithful continuous global symmetry of $\Spin(7) + 3\mathbf{F}$ is (see also \cite[(4.95)]{Apruzzi:2021vcu})
\begin{equation} \label{globalsymmSPin7w3}
    \USp(6) \times \SO(3)_I\,,
\end{equation}
where we emphasise that this is not $\USp(6)/\BZ_2 \times \SO(3)_I$ for the following reason. A more detailed analysis reveals that the $y^{\pm 1}$ sector of the index at order $x^6$ contains the term $2 \chi_{[0,0,1]}^{\usp(6)}$, where the Dynkin label $[0,0,1]$ corresponds to the representation $\mathbf{14'}$ of $\usp(6)$. The latter carries non-trivial charge under the $\BZ_2$ centre of $\USp(6)$, hence confirming that the faithful global symmetry is given by \eref{globalsymmSPin7w3}.

\subsection{$\Spin(8)$ with $4$ hypermultiplets in the vector representation}
\label{sec:indexSO8}
The superconformal index of $\Spin(8) + 4\mathbf{F}$ can be computed analogously, with the $\so(8)$ group theory data collected in appendix~\ref{app:grouptheory}. Up to the one-instanton contribution, it reads
\begin{align}
    \mathcal{I}_{\Spin(8) + 4\mathbf{F}} \;=\; 1 \;&+\; \Bigl[ \chi_{\mathrm{adj}}^{\usp(8)} + \chi_{[2]}^{\su(2)_I}\bigl(\tilde{q}^{1/2}\bigr) \Bigr]\, x^2 \nonumber \\
    &+\; \chi_{[1]}(y) \Bigl[ 1 + \chi_{\mathrm{adj}}^{\usp(8)} + \chi_{[2]}^{\su(2)_I}\bigl(\tilde{q}^{1/2}\bigr) \Bigr]\, x^3 \;+\; \ldots\,,
    \label{eq:indexSO8}
\end{align}
where $w_1, \ldots, w_4$ are the orthonormal fugacities of the flavour algebra $\usp(8)$ rotating the four vector hypermultiplets, $\tilde{q}$ is the topological fugacity, and $y$ that of the spacetime $\SU(2)$. The $39$ conserved currents at order $x^2$ fill the adjoint of $\usp(8) \oplus \su(2)_I$: the $36$ flavour currents assemble into $\chi_{\mathrm{adj}}^{\usp(8)}$ (the $32$ roots realised as $w$-monomials, together with the four Cartan generators), while the three one-instanton states $\chi_{[2]}^{\su(2)_I}(\tilde{q}^{1/2}) = \tilde{q} + 1 + \tilde{q}^{-1}$ exhibit the enhancement $\fu(1)_I \to \su(2)_I$ of the topological symmetry. Since the vector $\mathbf{8}_v$ is a strictly real representation, the flavour symmetry of $N_f$ vectors is enhanced from $\fu(N_f)$ to $\usp(2 N_f)$; for $N_f = 4$ this is the rank-four $\usp(8)$. As in $\Spin(7) + 3\mathbf{F}$, only integer powers of $q$ appear at every order, so the half-integer-isospin doublet $\tilde{q}^{\pm 1/2}$ is absent and the topological symmetry is realised as $\SO(3)_I = \SU(2)_I/\Z_2$. Moreover, upon examining higher orders in $x$, we find that all representations of $\USp(8)$ that appear in the index transform trivially under its $\Z_2$ centre.\footnote{This includes, for example, the representation $[0,0,0,1]$, which is contained in the fourth exterior power $\Lambda^4 [1,0,0,0]$ of the fundamental representation.} The faithful continuous global symmetry of $\Spin(8) + 4\mathbf{F}$ is therefore
\begin{equation}
(\USp(8)/\BZ_2) \times \SO(3)_I\,,
\end{equation}
in agreement with \cite[(4.51)]{Apruzzi:2021vcu}.

\section{Two-group symmetries}
\label{sec:twogroup}
In this section, we consider another generalised global symmetry, known as {\it 2-groups} \cite{Cordova:2018cvg,Benini:2018reh}, which arises as a group extension between the 1-form symmetry and a subgroup of the centre of the continuous global symmetry. Consider a 5d $\CN=1$ SCFT that flows to a gauge theory with gauge group $G$. Let $\mathfrak{f}$ be the global symmetry algebra of the SCFT and $F$ be a simply connected Lie group associated to $\mathfrak{f}$, and let $\mathcal{Z}_F$ be a subgroup of the centre of $F$ such that $\mathcal{Z}_F$ acts trivially on the operators in the SCFT. The faithful zero-form global symmetry is then $F/\mathcal{Z}_F$. Furthermore, let $\BZ_\fm$ denote the subgroup of the centre of $G$ that acts trivially on the matter fields; the associated electric 1-form symmetry of the gauge theory is $\BZ_\fm^{(1)}$. We are interested in the 2-groups that arise from the extension of the 1-form symmetry by $\mathcal{Z}_F$, characterised by
\bes{ \label{nonclosureB2}
\delta B_2 = \mathrm{Bock}(w_2(F))\,,
}
where $B_2$ is the 2-form background field for $\BZ_\fm^{(1)}$, and $w_2(F)$ is the $\mathcal{Z}_F$-valued obstruction class for lifting the background $F/\mathcal{Z}_F$ bundle to an $F$ bundle. $\mathrm{Bock}$ is the Bockstein homomorphism associated with the short exact sequence
\bes{ \label{extension}
0 \, \rightarrow \, \BZ_\fm \, \rightarrow \, \mathbb{G}\, \rightarrow \, \mathcal{Z}_F \, \rightarrow 0~,
}
where $\mathbb{G}$ is the maximal subgroup of the product of the centres of $G$ and $F$ that acts trivially on the matter fields, so that $\BZ_\fm = \mathbb{G} \cap Z(G)$.

Since the matter fields transform in representations of $(G\times F)/\mathbb{G}$, the gauge and flavour bundles combine into a single bundle for the \emph{structure group}
\bes{ \label{structuregroup}
\mathcal{S} = (G \times F)/\mathbb{G}~.
}
Let us emphasise the following two points. First, only the gauge factor of $\mathcal{S}$ is dynamical: in a background $F/\mathcal{Z}_F$ bundle with a given obstruction class $w_2(F)$, the partition function sums over $G/\BZ_\fm$ bundles whose obstruction class $B_2$ is correlated with $w_2(F)$, while the flavour bundle itself is a fixed background. In particular, gauging the 1-form symmetry $\BZ^{(1)}_\fm$ --- i.e.\ summing over the $G/\BZ_\fm$ bundles --- is consistent on its own, whereas activating $w_2(F)$ alone is not. Second, the structure group \eqref{structuregroup} is determined by the matter representations alone, irrespective of whether a 2-group is present. The 2-group structure resides instead in the extension \eqref{extension}: it is present when $\BZ_\fm$ is non-trivial and \eqref{extension} does not split, in which case the correlation between $B_2$ and $w_2(F)$ cannot be removed --- which is precisely the non-closure \eqref{nonclosureB2}.

The purpose of this section is to determine the presence of a 2-group symmetry using the blow-up function $\exp(-V_n)$, along with the information about faithful continuous global symmetry derived in Section \ref{sec:faithfulsym}.

\subsection{A flux criterion for the 2-group}
\label{sec:diag}

The construction above amounts to a sharp and practical recipe, which we state here and then verify theory by theory in the rest of the section. It has two independent ingredients: the blow-up function, which is purely computational, and the faithful global symmetry of \S\ref{sec:faithfulsym}.

\paragraph{Step 1: the forced flux, from the blow-up.}
Activate the 1-form background by setting the gauge flux to the generator $\mathfrak{n}_{\rm gen}=\omega$ of the genuine $\Z_\fm^{(1)}$, i.e.\ the coset of $\Lambda_\cw/\Lambda_\crt$ that acts trivially on the matter. Demanding that the gauge-fugacity exponents $Z_i(\omega)$ be integers fixes the admissible background fluxes $(B_{m_0}, B_{m_h})$ and leaves a definite \emph{fractional} flux for the partner instanton and flavour symmetries: a fractional instanton number $k = Q/I_2$ together with the fractional flavour fluxes $B_{m_h}$. We denote this forced flux by $\mu$; it is a coweight of the faithful symmetry $F/\mathcal{Z}_F$, dragged along by the 1-form background --- precisely the non-closure \eqref{nonclosureB2}.\footnote{Here $I_2$ is the Dynkin index of the defining representation of the gauge group $G$. As discussed previously, for classical gauge groups $I_2=2$ for $\SO(N)$ or $\Spin(N)$, and $I_2=1$ otherwise.}

\paragraph{Step 2: information from the faithful global symmetry.}
Whether $\mu$ signals a 2-group is settled entirely by what $\mu$ is inside $F/\mathcal{Z}_F$, and there are three possibilities:
\begin{itemize}[leftmargin=*]
\item \textbf{Non-trivial 2-group.} $\mu$ is a non-trivial second Stiefel-Whitney class $w_2$ of a \emph{non-Abelian} factor of $F/\mathcal{Z}_F$ with $\pi_1(F/\mathcal{Z}_F) = \Z_2$; for example, such a non-Abelian factor may be $\SO(3)$, $\SO(5)$ or $\SO(7)$. The 1-form background then forces $\delta B_2 = \mathrm{Bock}(w_2)\neq 0$, and $\mathbb{G}$ is a non-trivial extension.
\item \textbf{Trivial or split 2-group.} Suppose that $\pi_1(F/\mathcal{Z}_F)$ is non-trivial, but $\mu$ is trivial in $\pi_1(F/\mathcal{Z}_F)$. This occurs when the order of $\mu$ is coprime to the relevant centre $\mathcal{Z}_F$. For example, if $\mathcal{Z}_F=\BZ_3$ and the forced flux $\mu$ is order 2, say $(B_{m_0}, B_h)$ are such that $2 (B_{m_0}, B_h)\in \BZ^2$, an order‑2 element of $\BZ_3$ must be the identity, so the forced flux is the trivial element of $\mathcal{Z}_F$. Then, $\mu$ lifts to a genuine $F$ bundle, and can be removed by a large $F$ gauge transformation. In this case, the extension \eqref{extension} splits, and the corresponding 2-group is trivial.
\item \textbf{Mixed anomaly.} Suppose that a non-Abelian factor in $F/\mathcal{Z}_F$ is simply connected, \eg, $G=\SU(2)$ or $\USp(4)$, but $w_2(F/\mathcal{Z}_F)$ is non-trivial, say through a quotient $F/\mathcal{Z}_F = [\,G\times \U(1)_X\,]/\Z_2$. The fractional flux $\mu$ is then locked onto the \emph{Abelian} $\U(1)_X$. A 1-form symmetry tied to an Abelian $0$-form flux is then a mixed 't~Hooft anomaly between this 1-form symmetry and $\U(1)_X$, not a 2-group.
\end{itemize}
To summarise, the important message is as follows: \emph{the blow-up function supplies the forced flux $\mu$, while the faithful symmetry $F/\mathcal{Z}_F$ obtained in \S\ref{sec:faithfulsym} supplies the group in which $\mu$ lives; only the latter separates a genuine 2-group from a mixed anomaly or a split extension.} This is most striking for the $\SU(4)_{\kCS=0,2,4}$ gauge theories with one or two antisymmetric hypermultiplets: the blow-up returns the \emph{same} forced flux for all of them, and it is solely the faithful global symmetry that decides their fate. Table~\ref{tab:2groupsummary} collects the outcomes, which the following subsections derive.
\begin{table}[t]
\centering
\footnotesize
\setlength{\tabcolsep}{5pt}
\resizebox{\textwidth}{!}{%
\begin{tabular}{lccll}
\toprule
Theory & $\Z_\fm^{(1)}$ & forced flux $\mu$ & faithful $F/\mathcal{Z}_F$ & outcome \\
\midrule
$\SU(2)_{\theta=0}$ & $\Z_2$ & $q^{1/4}$ & $\SO(3)_I$ & 2-group, non-split $\mathbb{G}=\Z_4$ \\
$\SU(3)_3$ & $\Z_3$ & $q^{1/3}$ & $\SO(3)_I$ & split, $\mathbb{G}=\Z_6 \cong \Z_2 \times \Z_3$ \\
$\SU(4)_4$ & $\Z_4$ & $q^{3/8}$ & $\SO(3)_I$ & 2-group, non-split $\mathbb{G}=\Z_8$ \\
$\SU(4)_6$ & $\Z_2$ & $q^{1/2}$ & $\U(1)_I$ & mixed anomaly \\
\midrule
$\SU(4)_0+1\Lambda^2$ & $\Z_2$ & $q^{1/2}w^{1/2}$ & $[\SU(2)_w\times \U(1)_I]/\Z_2$ & mixed anomaly \\
$\SU(4)_2+1\Lambda^2$ & $\Z_2$ & $q^{1/2}w^{1/2}$ & $\SO(3)_w\times \U(1)_I$ & 2-group, non-split $\mathbb{G}=\Z_4$ \\
$\SU(4)_4+1\Lambda^2$ & $\Z_2$ & $q^{1/2}w^{1/2}$ & $\SU(3)/\Z_3$ & split, $\mathbb{G}=\Z_2 \times \Z_3$ \\
$\SU(4)_0+2\Lambda^2$ & $\Z_2$ & $q^{1/2}w_1^{1/2}w_2^{1/2}$ & $\SO(3)_I\times(\USp(4)_w/\BZ_2)$ & 2-group, non-split $\mathbb{G}=\Z_4\times\Z_2$ \\
$\SU(4)_2+2\Lambda^2$ & $\Z_2$ & $q^{1/2}w_1^{1/2}w_2^{1/2}$ & $[\USp(4)_w\times \U(1)_I]/\Z_2$ & mixed anomaly \\
$\SU(4)_4+2\Lambda^2$ & $\Z_2$ & $q^{1/2}w_1^{1/2}w_2^{1/2}$ & $\SO(7)$ & 2-group, non-split $\mathbb{G}=\Z_4$ \\
\midrule
$\USp(4)_0+1\Lambda^2$ & $\Z_2$ & $q^{1/2}w^{-1/2}$ & $\SO(3)_I \times \SU(2)_w$ & 2-group, non-split $\mathbb{G}=\Z_4$ \\
$\USp(4)_0+2\Lambda^2$ & $\Z_2$ & $q^{1/2}w_1^{-1/2}w_2^{-1/2}$ & $\USp(6)$ & no 2-group \\
\midrule
$\Spin(7)+3\mathbf{F}$ & $\Z_2$ & $\tilde q^{1/2}\,\prod_h w_h^{1/2}$ & $\SO(3)_I \times\,\USp(6)_w$ & 2-group, non-split $\mathbb{G}=\Z_4$\\
$\Spin(8)+4\mathbf{F}$ & $\Z_2$ & $\tilde q^{1/2}\,\prod_h w_h^{1/2}$ & $\SO(3)_I \times\,(\USp(8)_w/\Z_2)$ & 2-group, non-split $\mathbb{G}=\Z_4 \times \Z_2$\\
\bottomrule
\end{tabular}%
}
\caption{The flux criterion of \S\ref{sec:diag} across all theories considered. The blow-up forces the fractional partner flux $\mu$ at the generator of $\Z_\fm^{(1)}$; the faithful symmetry $F/\mathcal{Z}_F$ then determines whether $\mu$ is a non-trivial $w_2$ of a non-Abelian factor (non-trivial 2-group), an Abelian flux (mixed anomaly), or split/trivial 2-group. Note that the six theories with $\SU(4)$ gauge group and antisymmetric hypermultiplets share the \emph{same} $\mu$ and are distinguished only by $F/\mathcal{Z}_F$. For $\Spin(N)$, the blow-up instanton exponent is the \emph{integer} $Q=1$, so the $\SO(3)_I$ flux is the half-unit $k=Q/I_2=\tfrac12$, written as $\tilde q^{1/2}$ in the $\SO(3)_I$ fugacity $\tilde q=q^2$; it is accompanied by the forced half-unit flavour flux $\prod_h w_h^{1/2}$.}
\label{tab:2groupsummary}
\end{table}

\subsection{Pure $\SU(N)_{\kCS}$ gauge theories}
\label{sec:su2warmup}
To illustrate the above discussion, let us consider the pure $\SU(N)_{\kCS}$ gauge theories listed in \eqref{eq:SUNkCS2groups}, whose 1-form symmetry is $\Z_\fm^{(1)}$ with $\fm= \gcd(N, \kCS)$. The continuous 0-form global symmetry is the instanton symmetry. This is the partner symmetry that we will focus on upon turning on a non-trivial background for $\Z_\fm^{(1)}$.

When $N = \kCS$, the $\U(1)_I$ instanton symmetry of the infrared gauge theory gets enhanced to $\SO(3)_I$, with $F = \SU(2)$ and $\mathcal{Z}_F = \BZ_2$, at the superconformal fixed point in the ultraviolet. This can be seen from the superconformal index and from the magnetic quiver \cite[Table 7 (III), $N_f=0$]{Cabrera:2018jxt}, which is the $\U(1)$ gauge theory with $2$ hypermultiplets of charge $1$, showing that the Higgs branch at infinite coupling of the 5d $\SU(N)_N$ is always $\BC^2/\BZ_2$ with isometry $\SO(3)$.  This argument also holds for the $\SU(2)_{\theta=0}$ pure gauge theory whose UV fixed point is the $E_1$ theory \cite{Seiberg:1996bd}. We point out that there has recently been an argument that strongly supports the presence of such a 2-group symmetry in the $E_1$ SCFT based on the symmetries of the theories arising from diagonally gauging the $\SO(3)$ global symmetry of two and four copies of the $E_1$ theory and its dimensional reduction  \cite{Zafrir:2025xca}.

On the other hand, for $\SU(4)_6$, the continuous global symmetry of the ultraviolet fixed point remains $\U(1)_I$, which can also be seen from the magnetic quiver discussed in \cite[(5.48), $N_f=0$]{Cabrera:2018jxt}; here there is no 2-group symmetry since $\mathcal{Z}_F$ is trivial.
\begin{equation}
\scalebox{0.91}{$
\vcenter{\hbox{%
\small
\setlength{\tabcolsep}{5pt}
\begin{tabular}{lccll}
\toprule
Theory & $\fm$ & Generator of $\Z_\fm^{(1)}$ & Partner symmetry & $q$-power: $Q(\omega_1)$ from \eqref{eq:expVmaster}\\
\midrule
$\SU(2)_{\theta=0}$ & $2$ & $\omega_1 = (\tfrac{1}{2})$                & $\SO(3)_{I}$ & $q^{1/4}$: non-split $\mathbb{G} = \Z_4$ \\
$\SU(3)_3$ & $3$ & $\omega_1 = (\tfrac{1}{3}, \tfrac{2}{3})$  & $\SO(3)_{I}$ & $q^{1/3}$: split $\mathbb{G} = \Z_6 \cong \Z_2\times\Z_3$ \\
$\SU(4)_4$ & $4$ & $\omega_1 = (\tfrac{1}{4}, \tfrac{1}{2}, \tfrac{3}{4})$ & $\SO(3)_{I}$ & $q^{3/8}$: non-split $\mathbb{G} = \Z_8$ \\
$\SU(4)_6$ & $2$ & $\omega_2 = (\tfrac{1}{2}, 0, \tfrac{1}{2})$ & $\U(1)_I$        & $q^{1/2}$: mixed $\Z_2^{(1)}$--$\U(1)_I$ anomaly \\
\bottomrule
\end{tabular}}}
$}
\label{eq:SUNkCS2groups}
\end{equation}
We see that, in the cases of $\SU(2)_{\theta=0}$ and $\SU(4)_4$, the denominators of the exponents of $q$ evaluated at $\omega_1$ are $2\fm$, with $\fm = 2$ and $4$, respectively. This means that $\mathbb{G}$ in the group extension \eqref{extension} (with $\mathcal{Z}_F=\BZ_2$) is $\BZ_{2 \fm}$ for these cases. Such a non-trivial extension implies the presence of non-trivial 2-groups in these theories. On the other hand, for $\SU(3)_3$, the denominator of $q$ remains $\fm$ with $\fm=3$; indeed, the extension in this case splits and the corresponding 2-group is trivial. For $\SU(4)_6$, there is no 2-group, and the fractional power of $q$ indicates a mixed anomaly between $\Z_2^{(1)}$ and $\U(1)_I$ symmetry, as discussed in \eqref{eq:mixedabstract}.

\subsection{$\SU(4)_{\kCS}$ with antisymmetric matter}
\label{sec:su4_2as}
We now apply the criterion of \S\ref{sec:diag} to $\SU(4)_{\kCS}$ with $N_{\Lambda^2}=1,2$ antisymmetric hypermultiplets at $\kCS=0,2,4$. The antisymmetric matter is neutral under the order-two element $\omega_2=(\tfrac12,0,\tfrac12)$ of the gauge centre $\Z_4$, which therefore generates the genuine $\Z_2^{(1)}$; the faithful continuous global symmetries $F/\mathcal{Z}_F$ of \S\ref{sec:faithfulsym} are collected in Table~\ref{tab:faithfulsym}.

Evaluating $\exp(-V_n)$ of \eqref{eq:expVSU4AS} at the representative $\omega_2=(\tfrac12,0,\tfrac12)$ of the genuine $\Z_2^{(1)}$ (every $\kCS$-dependent term carries a factor of $n_2$ and drops out), one finds, for any even $\kCS$ and $B_{m_h}=\pm\tfrac12$, the blow-up functions
\begin{equation}
    \exp(-V_{\omega_2}) = q^{\frac12}\,(p_1 p_2)^{\frac{B_{m_0}}{2}-\frac14}\, w_1^{-B_{m_1}}\, z_1^{-B_{m_0}-\frac12}\, z_2^{B_{m_0}-\frac12}\, z_3^{-B_{m_0}-\frac12}\,, \quad B_{m_0}\in\Z+\tfrac12
    \label{eq:expVomega2_1AS}
\end{equation}
for $N_{\Lambda^2}=1$, and
\begin{equation}
    \exp(-V_{\omega_2}) = q^{\frac12}\,(p_1 p_2)^{\frac{B_{m_0}}{2}-\frac12}\, w_1^{-B_{m_1}}\, w_2^{-B_{m_2}}\, z_1^{-B_{m_0}}\, z_2^{B_{m_0}-1}\, z_3^{-B_{m_0}}\,, \quad B_{m_0}\in\Z
    \label{eq:expVomega2_2AS}
\end{equation}
for $N_{\Lambda^2}=2$. The exponents of gauge fugacities $z_i$ are integral, and the fractional content collects into a \emph{single} half-unit,
\begin{equation}
    \left(q\,\textstyle\prod_{h=1}^{N_{\Lambda^2}} w_h \right)^{1/2} \qquad (\text{for }B_{m_h}=-\tfrac12)\,.
    \label{eq:combinedhalfunit}
\end{equation}
In other words, the forced flux of \S\ref{sec:diag} is the single half-unit $\mu=\bigl(q\prod_h w_h\bigr)^{1/2}$ of \eqref{eq:combinedhalfunit}: the $\Z_2^{(1)}$ background locks a half-unit of flavour 't~Hooft flux to a half-integer instanton number.

The fractional content is identical for all even $\kCS$; what changes is the faithful continuous global symmetry of \S\ref{sec:faithfulsym} into which the activated classes embed, and this is what fixes the 2-group. Explicitly, for $\SU(4)_{\kCS} + 2 \Lambda^2$, we have
\bes{
\small
\begin{tabular}{lll}
\toprule
$\kCS$ & Faithful global symmetry & Outcome \\
\midrule
$0$ & $\SO(3)_I \times (\USp(4)_w/\Z_2)$ & 2-group, non-split $\mathbb{G} =\BZ_4 \times\BZ_2$ \\
$2$ & $[\USp(4)_w \times \U(1)_I]/\Z_2$ & mixed $\Z_2^{(1)}$--$\U(1)_I$ anomaly \\
$4$ & $\SO(7)$ & 2-group, non-split $\mathbb{G} =\BZ_4$ \\
\bottomrule
\end{tabular}
}
For convenience, in the following we write $\USp(4)_w/\BZ_2 \cong \SO(5)_w$. For $\kCS=0$, the $\U(1)_I$ instanton symmetry in the gauge theory gets enhanced to $\SO(3)_I$ at the UV fixed point, and the faithful symmetry is $\SO(3)_I \times \SO(5)_w$. The centre $\mathcal{Z}_F$ in this case is $\BZ_2 \times \BZ_2$. The result \eqref{eq:combinedhalfunit} suggests that there is a 2-group between the $\Z_2^{(1)}$ 1-form symmetry and $w_2(\SO(3)_I)+w_2(\SO(5)_w)$; in which case, the group extension is $\mathbb{G} = \BZ_4 \times \BZ_2$. Likewise, for $\kCS=4$, the $\SO(5)_w$ flavour symmetry along with the $\U(1)_I \cong \SO(2)_I$ instanton symmetry gets enhanced to $\SO(7) \supset \SO(5)_w \times \SO(2)_I$. Let us denote the first Chern class of $\SO(2)_I$ by $c_1$. Since both summands of the branching rule $\mathbf{7}=\mathbf{5}\oplus\mathbf{2}$ are oriented ($w_1=0$), the Whitney sum formula gives $w_2(\SO(7))|_{\SO(5)\times\SO(2)}=w_2(\SO(5))+w_2(\SO(2))=w_2(\SO(5))+c_1\bmod2$, using $w_2(\SO(2))=c_1\bmod2$ for the rank-two oriented (complex line) bundle. In both $\kCS =0$ and $4$ cases, the activated class is a non-trivial $w_2$ of a non-Abelian factor, giving a non-trivial 2-group. At $\kCS=2$, by contrast, the non-Abelian flavour $\USp(4)_w$ is simply connected and the half-unit is locked onto the Abelian $\U(1)_I$ through the quotient $[\USp(4)_w \times \U(1)_I]/\Z_2$, giving a mixed $\Z_2^{(1)}$--$\U(1)_I$ anomaly, rather than a 2-group.

It is instructive to examine the case of $\kCS=4$ and $N_{\Lambda^2}=2$ in detail. The 2-group is the extension
\begin{equation}
    0 \;\longrightarrow\; \Z_2^{(1)} \;\longrightarrow\; \mathbb{G} \;\longrightarrow\; \mathcal{Z}_F \;\longrightarrow\; 0\,,
    \label{eq:su4ext}
\end{equation}
of the $0$-form flavour centre $\mathcal{Z}_F=\Z_2$ by the $1$-form $\Z_2^{(1)}$; abstractly, it is either the split $\Z_2\times\Z_2$ or the non-split $\Z_4$, and the centre arithmetic decides which. The antisymmetric $\Lambda^2$ carries gauge-centre charge $2\bmod4$ under the centre of $\SU(4)$, which is $\Z_4=\langle\omega_1\rangle$, with $\omega_1=(\tfrac14,\tfrac12,\tfrac34)$. Hence (i) the order-two element $\omega_2=\omega_1^2=(\tfrac12,0,\tfrac12)$ acts trivially on the matter and generates the genuine $\Z_2^{(1)}$, while the order-four elements $\omega_1,\omega_3$ are screened; and (ii) the matter is left invariant only by the \emph{diagonal} combination $(\omega_1,\zeta_F)$ of the gauge centre and the flavour centre, where $\zeta_F$ is the order-two generator of $\mathcal{Z}_F$ that compensates the charge-$2$ phase $e^{2\pi i\cdot 2/4}=-1$. Turning on the flavour flux $w_2(F/\mathcal{Z}_F)$ — the generator of the quotient $\mathcal{Z}_F$ in \eqref{eq:su4ext} — therefore drags along the gauge-centre partner $\omega_3=\omega_1^3=(\tfrac34,\tfrac12,\tfrac14)$, which is the value used in~\cite[(4.76)--(4.79)]{Apruzzi:2021vcu}. This lift has \emph{order four}: $\omega_3^2=\omega_2$ is precisely the generator of $\Z_2^{(1)}$. A generator of $\mathcal{Z}_F$ that squares to the generator of $\Z_2^{(1)}$ is exactly a non-split extension, so $\mathbb{G}=\Z_4$ rather than $\Z_2\times\Z_2$.

The $N_{\Lambda^2}=1$ theories follow the same logic with the flavour symmetry algebra $\su(2)_w$, instead of $\usp(4)_w$. At $\kCS=2$, the faithful $\SO(3)_w\times \U(1)_I$ yields a 2-group between $\Z_2^{(1)}$ and $w_2(\SO(3)_w)$; at $\kCS=0$, the simply connected $\SU(2)_w$ in $[\SU(2)_w\times \U(1)_I]/\Z_2$ gives only a mixed anomaly; and, at $\kCS=4$, the flavour and instanton symmetries combine into $\SU(3)/\Z_3$, whose centre $\Z_3$ is coprime to $\Z_2^{(1)}$, so the extension splits into a trivial 2-group. All six cases are collected in Table~\ref{tab:2groupsummary}.

\subsection{$\Spin(7)+3\mathbf{F}$, $\Spin(8)+4\mathbf{F}$ and $\USp(4)_0$ with antisymmetric matter}
\label{sec:twogroupSO}

The remaining theories all have gauge centre $\Z_2$ (or $\Z_2\times\Z_2$ for $\Spin(8)$), unscreened by the matter, and can be treated together.

\subsubsection*{$\Spin(7)+3\mathbf{F}$ and $\Spin(8)+4\mathbf{F}$}
The vector matter ($\mathbf{7}$, $\mathbf{8}_v$) is neutral under the $\Z_2$ centre generating $\Z_2^{(1)}$, which is therefore unscreened. The generating 't~Hooft flux is the vector class $\tilde n=(1,0,\dots)$, i.e.\ $n=(1,1,\tfrac12)$ for $\Spin(7)$ and $n=(1,1,\tfrac12,\tfrac12)$ for $\Spin(8)$. Evaluating $\exp(-V_n)$ of \eqref{eq:expVSO7} and \eqref{eq:expVSO8} there, with the flavour fluxes fixed at $B_{m_h}=\pm\tfrac12$ (step~1 at $n=0$) and $B_{m_0}\in\Z$, gives
\begin{align}
    \Spin(7)+3\mathbf{F}:\quad \exp(-V_{\omega}) &= q\,(p_1 p_2)^{B_{m_0}-\frac34}\, z_1^{-(2B_{m_0}+1)}\prod_{h=1}^{3} w_h^{-B_{m_h}}\,, \label{eq:expVomegaSO7}\\
    \Spin(8)+4\mathbf{F}:\quad \exp(-V_{\omega}) &= q\,(p_1 p_2)^{B_{m_0}-1}\, z_1^{-(2B_{m_0}+1)}\prod_{h=1}^{4} w_h^{-B_{m_h}}\,. \label{eq:expVomegaSO8}
\end{align}
The gauge-fugacity exponents are integral (only $z_1$ survives; the exponents of $z_2,z_3$, and $z_4$ for $\Spin(8)$ vanish). The forced flux $\mu$ has \emph{two} fractional pieces. First, the instanton exponent is the \emph{integer} $Q=1$, so the canonical instanton number is the half-unit $k=Q/I_2=\tfrac12$ (Table~\ref{tab:mixed}); there is no fractional power of $q$ itself. Second, each flavour fugacity carries a half-integer flux
\begin{equation}
    W_h = -2B_{m_h}\,n_r = -B_{m_h} = \mp\tfrac12\,, \qquad n_r=\begin{cases} n_3 & \Spin(7)\\ n_4 & \Spin(8)\end{cases}\ =\tfrac12 \ \text{at}\ \omega\,,
\end{equation}
arising from the gauge flux coupling to the flavour background. So the $\Z_2^{(1)}$ background forces \emph{both} the canonical instanton number $k=\tfrac12$ and a flavour flux $\prod_h w_h^{\mp1/2}$.

The two theories need separate discussions.  For $\Spin(8)+4\mathbf{F}$, the faithful continuous global symmetry is $\SO(3)_I \times (\USp(8)_w/\BZ_2)$, and the above analysis suggests that there is a 2-group between the $\Z_2^{(1)}$ 1-form symmetry and $w_2(\SO(3)_I) + w_2(\USp(8)_w/\BZ_2)$, where the corresponding group extension is $\mathbb{G} = \BZ_4 \times \BZ_2$. This is very similar to $\SU(4)_0 + 2 \Lambda^2$, discussed in \S\ref{sec:su4_2as}.  In contrast, for $\Spin(7)+3\mathbf{F}$, the faithful symmetry is $\SO(3)_I \times \USp(6)$, and the 2-group turns out to be between the $\Z_2^{(1)}$ 1-form symmetry and $w_2(\SO(3)_I)$, since $\USp(6)$ is simply connected, and the corresponding group extension is $\mathbb{G} = \BZ_4$.

Note, however, that the mechanism here differs from the order-four gauge-centre lift of \S\ref{sec:su4_2as}: the gauge centre is only $\Z_2$ for $\Spin(7)$ (resp.\ $\Z_2\times\Z_2$ for $\Spin(8)$), so it contains no order-four element. Let us consider the case of $\Spin(7)$ for definiteness. The geometric analysis of~\cite[(4.94)--(4.97)]{Apruzzi:2021vcu} nonetheless presents the same non-split $\Z_4$ through an order-four generator of $\Z_4$ with gauge part $(\tfrac12,\tfrac12,\tfrac14)$, whose square is the $\Z_2^{(1)}$ generator $(1,1,\tfrac12)$. The two descriptions agree, but this order-four element lies in the \emph{maximal torus} of fractional gauge holonomies, not in the gauge centre $\mathcal{Z}(\Spin(7))=\Z_2$ --- in contrast to the $\SU(4)$ case of \S\ref{sec:su4_2as}, where the analogous $\omega_3$ is a genuine centre element. In the blow-up function, the flux $(\tfrac12,\tfrac12,\tfrac14)$, i.e.\ $\tilde{n}=(\tfrac12,0,0)$, is \emph{not} a coweight of $B_3$ --- it pairs half-integrally with the long root $e_1-e_2$, and \eqref{eq:expVSO7} returns the fractional gauge-fugacity exponent $Z_1=-\tfrac14-B_{m_0}\notin\Z$ --- so it is not an admissible $1$-form background on its own; the genuine generator $n=(1,1,\tfrac12)$ instead gives $Q=1$ and $Z_1=-(2B_{m_0}+1)\in\Z$ (the half-unit $k=\tfrac12$). The non-split structure is therefore characterised intrinsically by the topological Postnikov class $\mathrm{Bock}(w_2(\SO(3)_I))=w_3(\SO(3)_I)\neq0$ of the $0$-form partner $\SO(3)_I=\SU(2)_I/\Z_2$. This is, in fact, the same mechanism as for the 5d $\SU(2)$ pure gauge theory of \S\ref{sec:su2warmup}, whose gauge centre is likewise only $\Z_2$, yet whose $\SO(3)_{I}$ partner forces $\mathbb{G}=\Z_4$; as we shall now see, the same mechanism operates in $\USp(4)_0 + 1\Lambda^2$.

\subsubsection*{$\USp(4)_0 + N_{\Lambda^2}\Lambda^2$}
The antisymmetric matter is neutral under the $\Z_2$ centre of $\USp(4)$, so the $\Z_2^{(1)}$ 1-form symmetry is unscreened, with generating 't~Hooft flux $\omega = \bigl(\tfrac12, 0\bigr)$ (\S\ref{sec:expVUSp4}). Evaluating $\exp(-V_n)$ of \eqref{eq:expVUSp4} there, with $B_{m_h}=\pm\tfrac12$ and $B_{m_0}$ quantised as in \eqref{eq:Bm0USp4}, all gauge-fugacity exponents are integers, namely $Z_1 = -\bigl(2B_{m_0} - 2\textstyle\sum_h B_{m_h} + 1\bigr)$ and $Z_2 = B_{m_0} - \sum_h B_{m_h} - \tfrac12$. In contrast with the $\SU(4)$ theories of \S\ref{sec:su4_2as}, here $\exp(-V_n)\big|_{n=0}$ is itself non-trivial\footnote{This indeed leads to the non-zero parameter $\Lambda_{(0)} = (p_1 p_2)^{-\frac{1}{48} N_{\Lambda^2}} \prod_h w_h^{-\frac{1}{8}}$ in the blow-up equation; see appendix~\ref{app:blow-upSU36USp4}.} --- the mass-dominated pair contributes the flux-independent constants $W_h(0) = -\tfrac12 B_{m_h}^2$ and $P(0) = -\tfrac16 \sum_h B_{m_h}^3$ --- so the 1-form data is cleanly exhibited by the ratio
\begin{equation}
    \frac{\exp(-V_{\omega})}{\exp(-V_{0})}
    \;=\; q^{\frac12}\, (p_1 p_2)^{\frac12 \left(B_{m_0} - \sum_h B_{m_h}\right)}\, \prod_{h=1}^{N_{\Lambda^2}} w_h^{-\frac12}\; z_1^{Z_1}\, z_2^{Z_2}\,.
    \label{eq:expVomegaUSp4}
\end{equation}
The $\Z_2^{(1)}$ background therefore forces the half-unit flux $\mu = q^{1/2}\prod_h w_h^{-1/2}$: a half-integer instanton number ($Q=\tfrac12$ with $I_2=1$; Table~\ref{tab:mixed}) locked to a half-unit of flavour flux for each antisymmetric hypermultiplet --- note that the latter arises from the mass-dominated pair and is independent of the sign of $B_{m_h}$.

The fate of this forced flux is again decided by the faithful global symmetry of \S\ref{sec:faithfulsym}. For $N_{\Lambda^2}=1$ the faithful symmetry is $\SU(2)_w \times \SO(3)_I$: the flavour factor $\SU(2)_w$ is simply connected and plays no role, while $\mathcal{Z}_F$ is the $\Z_2$ centre of $\SU(2)_I$, which acts trivially since only integer $\su(2)_I$ isospins appear in the index. The forced half-instanton flux activates $w_2(\SO(3)_I)$, and the corresponding group extension is the non-split $\mathbb{G}=\Z_4$, by the same intrinsic Postnikov argument as for $\SU(2)_{\theta=0}$ and the $\Spin(N)$ theories above: $\mathrm{Bock}(w_2(\SO(3)_I)) = w_3(\SO(3)_I) \neq 0$, the gauge centre $\Z_2$ containing no order-four element. For $N_{\Lambda^2}=2$, by contrast, the faithful symmetry is the simply connected $\USp(6)$, into which both $\usp(4)_w$ and $\su(2)_I$ are absorbed: $\mathcal{Z}_F$ is trivial (the centre-odd representation $\mathbf{14'}$ appears in the index, \S\ref{sec:faithfulsym}), the extension \eqref{extension} is trivial, and there is \emph{no} 2-group. The fractional fluxes in \eqref{eq:expVomegaUSp4} are then interpreted as mixed 't~Hooft anomalies between $\Z_2^{(1)}$ and the Cartan directions of $\usp(6)$ --- in particular, the half-integer instanton number is precisely the mixed $\Z_2^{(1)}$--$\U(1)_I$ anomaly of Table~\ref{tab:mixed}. The pair of theories thus provides a sharp illustration of the criterion of \S\ref{sec:diag}: the blow-up returns the same forced flux for $N_{\Lambda^2}=1,2$, and it is solely the faithful symmetry --- $\SU(2)_w \times \SO(3)_I$ versus $\USp(6)$ --- that decides between a non-split 2-group and no 2-group at all.

\acknowledgments
We are grateful to Fabio Apruzzi, Julius Grimminger, Alessandro Tomasiello and Gabi Zafrir for useful discussions. W.H. also acknowledges the Abdus Salam Centre for Theoretical Physics, Imperial College London, especially Amihay Hanany, for hospitality during the realisation of this project. N.M. would especially like to thank Carlotta Meneghini and Michele Sarzana for their warm hospitality during the completion of this work. The research of W.H. and N.M. is partially supported by the MUR-PRIN grant No.~2022NY2MXY (Finanziato dall’Unione europea~-- Next Generation EU, Missione 4 Componente 1, CUP H53D23001080006 and I53D23001330006).

\appendix

\section{Non-Lagrangian $B_N$ and $B_N^{(1,2,3)}$ theories}
\label{app:BN}

The diagnostic of \S\ref{sec:anomalies} reads the cubic self-anomaly of the 1-form
symmetry off the fractional part of the $(p_1 p_2)$-exponent $P(n)$ of $\exp(-V_n)$, and
relies on the W-boson one-loop term $\frac{1}{6}\sum_{\alpha\in\Delta^+}(\langle n,\alpha\rangle^3-\langle n,\alpha\rangle)$
being an integer on every coset. For genuinely non-Lagrangian theories there are no
W-bosons and this term is absent; the gravitational coupling that it would otherwise
package together with the cubic Chern--Simons term must then be reinstated explicitly.
In this appendix, we compute the cubic anomaly for the non-Lagrangian toric models $B_N$
of \cite{Eckhard:2020jyr,Morrison:2020ool} --- whose rank-one member $B_3$ is the
$E_0$ (local $\mathbb{P}^2$) theory \cite{Morrison:1996xf} --- directly from the toric geometry of the
resolved Calabi--Yau threefold $X$, and we give explicit data for $B_3$ and $B_4$, as
well as closed forms for general $N$. We then treat the sibling families $B_N^{(1)}$, $B_N^{(2)}$ and $B_N^{(3)}$
(\S\ref{app:BN1}--\S\ref{app:BN3}), which, in addition, carry the non-Abelian flavour symmetries
$\SU(N-2)\times \U(1)$, $\SU(N-2)$ and $\SU(N-1)$, respectively~\cite{Eckhard:2020jyr}, and hence mixed
1-form--flavour-Cartan anomalies.

A remark on notation is in order. Whenever a theory possesses a non-trivial flavour symmetry, each Cartan direction is associated with a non-compact flavour divisor $D_{F_a}$ carrying its own background magnetic flux. When there is a single such direction we denote its flux by $B_h$ (or $B_m$), as in \S\ref{sec:expVnonlag}; when there are several, we write $B_{m_a}$ ($a = 1, 2, \ldots$) for the flux through $D_{F_a}$. The two notations are therefore used interchangeably, with $B_h \equiv B_{m_1}$ whenever only one flavour flux is present.

\subsection{Geometric formula for the anomaly}
\label{app:BN-formula}
We follow the analysis described in \cite[(4.21)]{Apruzzi:2021nmk}. Let $X$ be the resolved toric Calabi--Yau threefold, $S_i$ its compact toric divisors, and let the
1-form symmetry be $\mathbb{Z}_\fm^{(1)}$ with
\begin{equation}
    \fm \;=\; N(N-3)+3 \,.
    \label{eq:BNorder}
\end{equation}
Denote by $Z$ the \emph{central divisor} representing the generator of the 1-form
symmetry, i.e.\ the integral combination $Z=\sum_i k_i S_i$ such that $Z\cdot C\equiv 0
\pmod{\fm}$ for every compact curve $C$, so that $Z/\fm$ is an admissible fractional flux.
The cubic self-anomaly is the $\mathbb{R}/\mathbb{Z}$-valued quantity
\bes{
\scalebox{1.13}{$
    \Omega
    \;=\; \left[\, -\frac{1}{6}\,\frac{Z\cdot Z\cdot Z}{\fm^3}
    \;+\; \frac{1}{24}\,\frac{Z\cdot p_1(TX)}{\fm}\, \right]_{\bmod 1}
    \;=\; \left[\, -\frac{Z^3}{6\,\fm^3}
    \;-\; \frac{c_2(X)\cdot Z}{12\,\fm}\, \right]_{\bmod 1}\,,
    \label{eq:BNanomaly}
    $}
}
where, in the second equality, we used $p_1(TX)=c_1(TX)^2-2c_2(TX)=-2c_2(TX)$, valid on a
Calabi--Yau threefold. As emphasised in \cite{Apruzzi:2021nmk}, the two terms in
\eqref{eq:BNanomaly} are not separately well defined; only their sum is, the ambiguity
being fixed by the index-theoretic congruence
$x\,p_1\equiv 4x^3 \pmod{24}$ for $x\in H^2(M_6;\mathbb{Z})$.

The same intersection data fix the Coulomb branch prepotential. In the conventions of
\cite{Jefferson:2017ahm}, the blow-up function is governed by
\begin{equation}
\scalebox{0.99}{$
    \mathcal{E}
    \;=\; \frac{1}{\epsilon_1\epsilon_2}\left[\,
    \frac{1}{6}\,c_{ijk}\,\phi^i\phi^j\phi^k
    \;+\; \frac{\epsilon_1^2+\epsilon_2^2}{48}\,C^G_i\,\phi^i
    \;+\; \frac{\epsilon_+^2}{2}\,C^R_i\,\phi^i \,\right]\,,
    \qquad \epsilon_+ \equiv \tfrac{1}{2}(\epsilon_1+\epsilon_2)\,,
    $}
    \label{eq:BNprepot}
\end{equation}
with the cubic, gravitational and $\SU(2)_R$ couplings given by the intersection numbers
\begin{equation}
    c_{ijk} \;=\; S_i\cdot S_j\cdot S_k\,, \qquad
    C^G_i \;=\; c_2(X)\cdot S_i\,, \qquad
    C^R_i \;=\; 2\,\chi(\mathcal{O}_{S_i})\,.
    \label{eq:BNcouplings}
\end{equation}
The gravitational level $C^G_i$ entering \eqref{eq:BNanomaly} is thus the very same
coefficient that multiplies $\epsilon_1^2+\epsilon_2^2$ in \eqref{eq:BNprepot}; this is
the bridge between the prepotential read off from the blow-up and the topological
anomaly.\footnote{The $\tfrac{\epsilon_1^2+\epsilon_2^2}{48}C^G_i$ term in
\eqref{eq:BNprepot} is the $\Omega$-background form of the mixed gauge/gravitational
Chern--Simons term $-\tfrac1{48}C^G_i\,A^i\wedge p_1(T)$ of \cite[(2.4)]{Kim:2020hhh},
obtained through the equivariant replacement $p_1(T)\to-(\epsilon_1^2+\epsilon_2^2)$ of
\cite[(2.13)]{Kim:2020hhh} (a relation with unit coefficient). Here $p_1(T)$ is the
Pontryagin class of the \emph{five-dimensional spacetime}, and is not to be confused with
the \emph{internal} $p_1(TX)=-2c_2(X)$ appearing in the anomaly \eqref{eq:BNanomaly}; the
level $C^G_i=c_2(X)\cdot S_i$ is the internal datum common to both. In particular, the
$\tfrac1{48}$ here and the $\tfrac1{24}$ of \eqref{eq:BNanomaly} are prefactors of these
two distinct couplings, not two normalisations of the same one.}

\paragraph{Toric intersection rules.}
All numbers defined in \eqref{eq:BNcouplings} are computed on a smooth (crepant) resolution,
i.e.\ a fine star triangulation of the toric polytope with vertices
$(N{-}1,0)$, $(1,N{-}1)$, $(0,1)$; each lattice point $v$ gives a ray $r_v=(v,1)$ and a
toric
divisor $D_v$, the interior points giving the compact $S_i$. Writing $D_uD_vD_w=1$ when
$\{u,v,w\}$ spans a triangle and $0$ otherwise, the self-intersections follow from the
linear-equivalence (wall) relations: for an interior edge $(u,v)$ with the two adjacent
apexes $a,b$, unimodularity gives $r_a+r_b=\alpha\,r_u+\beta\,r_v$, and hence
\begin{equation}
    D_u^2 D_v=-\alpha\,, \qquad D_u D_v^2=-\beta\,, \qquad
    D_u^2 D_v + D_u D_v^2 = -2\,,
    \label{eq:wall}
\end{equation}
the last equality being the statement that every compact curve is a rational curve of
the Calabi--Yau (normal-bundle degrees summing to $-2$). Cubes follow either by iterating
\eqref{eq:wall} or, more directly, from adjunction $N_{S}=K_{S}$: a smooth compact toric
surface with $\rho$ rays has $\chi(S)=\rho$ and $S^3=K_S^2=12-\rho$, while
$c_2(X)\cdot S=\chi(S)-K_S^2=2\rho-12$ and $\chi(\mathcal{O}_S)=1$ (so $C^R_i=2$).

\paragraph{Derivation of the wall relations.}
For readers less familiar with toric geometry, we spell out \eqref{eq:wall}; the argument
combines three standard facts. The curve $C=D_u\cap D_v$ is the $\mathbb{P}^1$ dual to the
wall (two-dimensional cone) $\langle r_u,r_v\rangle$; in a fine triangulation, this wall is
shared by exactly two top-dimensional cones $\langle r_u,r_v,r_a\rangle$ and
$\langle r_u,r_v,r_b\rangle$, so the four rays $r_a,r_u,r_v,r_b$ form the entire star of $C$.

\emph{(i) Linear equivalence.} For any character $m$ in the dual lattice $M$, the principal
divisor of $\chi^m$ is $\mathrm{div}(\chi^m)=\sum_w\langle m,r_w\rangle D_w$, which is
linearly --- hence numerically --- trivial
\begin{equation}
    \sum_w \langle m,r_w\rangle\,D_w \;\sim\; 0\,.
    \label{eq:lineq}
\end{equation}
We are free to choose $m$; take $\langle m,r_u\rangle=-1$ and $\langle m,r_v\rangle=0$.

\emph{(ii) Capping with the curve.} Intersecting \eqref{eq:lineq} with $C=D_uD_v$, a triple
intersection $D_wD_uD_v$ is non-zero --- and equal to $1$ by smoothness --- precisely when
$\{w,u,v\}$ spans a top-dimensional cone, i.e.\ only for $w\in\{a,b\}$, while $w=u,v$ return
$D_u^2D_v$ and $D_uD_v^2$. Hence
\begin{equation}
    0 \;=\; \langle m,r_u\rangle\,D_u^2D_v + \langle m,r_v\rangle\,D_uD_v^2
    + \langle m,r_a\rangle + \langle m,r_b\rangle\,,
\end{equation}
using $D_aD_uD_v=D_bD_uD_v=1$. With the above choice of $m$, this collapses to
$D_u^2D_v=\langle m,\,r_a+r_b\rangle$.

\emph{(iii) The wall relation.} Substituting $r_a+r_b=\alpha\,r_u+\beta\,r_v$ gives
$D_u^2D_v=\alpha\langle m,r_u\rangle+\beta\langle m,r_v\rangle=-\alpha$, and the mirror choice
$\langle m,r_u\rangle=0$, $\langle m,r_v\rangle=-1$ gives $D_uD_v^2=-\beta$. The relation
itself, with unit coefficients on $r_a,r_b$, is forced by smoothness: since $\{r_u,r_v,r_b\}$
is a lattice basis, we may write $r_a=p\,r_u+q\,r_v+s\,r_b$ with $p,q,s\in\mathbb{Z}$, and
$\det(r_u,r_v,r_a)=s\,\det(r_u,r_v,r_b)$, so smoothness of \emph{both} cones forces $|s|=1$;
as $a,b$ lie on opposite sides of the wall, then $s=-1$, i.e.\ $r_a+r_b=p\,r_u+q\,r_v$ with
$(\alpha,\beta)=(p,q)$. Finally, all rays lie on the height-one plane $r_w=(v_w,1)$ --- the
Calabi--Yau condition --- so the last coordinate of the wall relation reads $1+1=\alpha+\beta$,
giving $D_u^2D_v+D_uD_v^2=-2$. This is adjunction: the normal bundle of the rational curve is
$N_{C/X}=\mathcal{O}(-\alpha)\oplus\mathcal{O}(-\beta)$ with $\deg N_{C/X}=2g(C)-2=-2$, so
$(\alpha,\beta)=(1,1)$ is a $(-1,-1)$ curve, $(0,2)$ a $(0,-2)$ curve, and so on. The
worked $B_4$ edge $(S_1,S_2)$ below, with apexes $A,S_3$ and $(\alpha,\beta)=(0,2)$,
illustrates the rule.

\subsection{The intersection matrix $M$, the generator $k$, and the spin$^c$ shift $\delta$}
\label{app:BN-flux}

The order \eqref{eq:BNorder} and the generator $Z$ used above are both read off a single
object: the pairing between the compact divisors and the compact curves. We record here
how that matrix is assembled from the data of \S\ref{app:BN-formula}, how the generator
$k$ is extracted from it, and how it fixes the quantisation of the magnetic flux including
the spin$^c$ shift. All three steps are purely linear-algebraic and require no field-theory
input beyond the geometry already in hand.

\paragraph{The intersection matrix $M$.} The compact curves $C_a$ are the internal walls of
the triangulation, i.e.\ the edges $(u,v)$ shared by two triangles (boundary edges of the
polytope give non-compact curves). Each pairing $M_{ia}=S_i\cdot C_a$ is the triple
intersection $D_{w_i}\!\cdot D_u\!\cdot D_v$, which the rules around \eqref{eq:wall} evaluate
\emph{entry by entry}: given $a,b$ the two apexes of the edge $(u,v)$ and $(\alpha,\beta)$ its
wall data, then
\begin{equation}
    M_{ia}\;=\;D_{w_i}\!\cdot D_u\!\cdot D_v\;=\;
    \begin{cases}
    -\alpha & w_i=u\\
    -\beta & w_i=v\\
    +1 & w_i\in\{a,b\}\\
    0 & \text{otherwise}
    \end{cases}\,.
    \label{eq:Mentry}
\end{equation}
Only the two ``diagonal'' entries $w_i\in\{u,v\}$ carry the wall data; the remaining entries
are pure combinatorics of the triangulation. A useful check is that each column, summed over
\emph{all} divisors (compact and non-compact alike), vanishes: $\sum_w D_w\sim0$ is the
height-one relation \eqref{eq:lineq} with $m=(0,0,1)$, and $-\alpha-\beta+1+1=0$ is precisely
$\alpha+\beta=2$. The matrix \eqref{eq:Mentry} is nothing but the charge matrix of the
dynamical states: its columns are the curve volumes
$\mathrm{vol}(C_a)=-\sum_i(S_i\cdot C_a)\,\phi^i$, so $M$ is equally available from the field
theory whenever the BPS central charges are known.

\paragraph{The generator $k$.} Regard $M$ as the linear map that sends each compact curve to
its vector of pairings with the compact divisors, $C_a\mapsto(S_i\cdot C_a)_i$. Its
\emph{cokernel} is the divisor lattice modulo the sublattice spanned by those image vectors,
and by the structure theorem for finitely generated Abelian groups --- equivalently the
Smith normal form $M=U\,\mathrm{diag}(d_1,d_2,\dots)\,V$, with $U,V$ unimodular and invariant
factors $d_1\mid d_2\mid\cdots$ --- this is given by
\begin{equation}
    \coker M\;=\;\mathbb{Z}^{\#\{S_i\}}\big/ M\,\mathbb{Z}^{\#\{C_a\}}
    \;\cong\;\underbrace{\mathbb{Z}^{\,\#\{S_i\}-\rho}}_{\text{free}}\ \oplus\
    \underbrace{\textstyle\bigoplus_{d_i>1}\mathbb{Z}_{d_i}}_{(\coker M)_{\mathrm{tor}}}\,,
    \qquad \rho=\rank M\,.
    \label{eq:coker}
\end{equation}
The subscript $\mathrm{tor}$ denotes the finite \emph{torsion} subgroup, the second summand
$\bigoplus_{d_i>1}\mathbb{Z}_{d_i}$. The free part $\mathbb{Z}^{\#\{S_i\}-\rho}$ arises only
when the pairing is degenerate --- a combination of compact divisors pairing trivially with
every compact curve --- which signals a continuous (flavour) symmetry and is discarded; the
discrete 1-form symmetry is precisely the torsion,
\begin{equation}
    \mathbb{Z}_\fm^{(1)}\;=\;\big(\coker M\big)_{\mathrm{tor}}\,.
\end{equation}
For the $B_N$ theories, there are no flavour branes, so $M$ has full row rank
($\rho=\#\{S_i\}$, no free part) and exactly one invariant factor exceeds one, namely $\fm$;
hence $(\coker M)_{\mathrm{tor}}=\mathbb{Z}_\fm$. Its
generator is the central divisor $Z=\sum_i k_i S_i$, where $k$ is the left null-vector of $M$
modulo $\fm$
\begin{equation}
    k^{\mathsf T}M\equiv 0 \pmod{\fm}\,, \qquad k\notin\fm\,\mathbb{Z}^{\rank}\,,
    \label{eq:kdef}
\end{equation}
equivalently the minimal fractional combination $k/\fm$ whose pairing with every compact
curve is integral. This is the same vector that enters $Z^3$ and the anomaly
\eqref{eq:BNanomaly}; it is defined only up to multiplication by a unit of $\mathbb{Z}_\fm$.

\paragraph{Flux quantisation and the spin$^c$ shift $\delta$.} A magnetic flux
$n=\sum_i n_i S_i$ is admissible when its pairing with every compact curve obeys the
Freed--Witten (spin$^c$) condition
\begin{equation}
    \langle n,C_a\rangle\;=\;(M^{\mathsf T}n)_a\;\in\;\mathbb{Z}+\tfrac12\,C_a^2\,,
    \qquad C_a^2=D_u^2 D_v\ \ (\equiv D_u D_v^2 \bmod 2)\,,
    \label{eq:fluxquant}
\end{equation}
the half-integer offset being present exactly on the curves of \emph{odd} self-intersection.
Writing $n=k/\fm+\delta$ and using \eqref{eq:kdef}, the offset solves the congruence
\begin{equation}
    M^{\mathsf T}\delta\;\equiv\;s \pmod 1\,, \qquad
    s_a=\tfrac12\,C_a^2\in\{0,\tfrac12\}\,,
    \label{eq:delta}
\end{equation}
solvable by the Smith decomposition $M^{\mathsf T}=U\,D\,V$ (the zero rows of $D$ encode the
consistency of $s$). Thus $\delta=0$ precisely when every compact curve is even, while a
single odd curve forces a non-trivial half-integer offset. The invariant content of $n$ is
the offset $\delta$ --- which curve-fluxes are half-integral --- since the individual $n_i$
depend on the choice of lift and of the generator $k$.

\subsection{$B_3=E_0$ (local $\mathbb{P}^2$)}
\label{app:B3}

Here, the unique compact divisor is $S=\mathbb{P}^2$, the 1-form symmetry is
$\mathbb{Z}_3^{(1)}$ ($\fm=3$), and the generator is $Z=S$. From $K_{\mathbb{P}^2}=
\mathcal{O}(-3)$, one has $\chi(\mathbb{P}^2)=3$, $K^2=9$, hence
\begin{equation}
    S^3=K_S^2=9\,, \qquad
    c_2(X)\cdot S=\chi(S)-K_S^2=3-9=-6\,, \qquad
    C^R=2\,.
\end{equation}
The prepotential \eqref{eq:BNprepot} reduces to the familiar
$\mathcal{E}=\bigl(\tfrac32\phi^3-\tfrac18\phi(\epsilon_1^2+\epsilon_2^2)
+\epsilon_+^2\phi\bigr)/\epsilon_1\epsilon_2$, and \eqref{eq:BNanomaly} gives
\begin{equation}
    \Omega
    \;=\; -\frac{9}{6\cdot 27}-\frac{-6}{12\cdot 3}
    \;=\; -\frac{1}{18}+\frac{1}{6}
    \;=\; \frac{1}{9}\,.
\end{equation}
This $\tfrac19$ is the known cubic $\mathbb{Z}_3^{(1)}$ anomaly of local $\mathbb{P}^2$ (the
$E_0$ theory) \cite{Apruzzi:2021nmk,Apruzzi:2022dlm,Cvetic:2025lat}; only the total is physical, the split
between the cubic and gravitational contributions being scheme-dependent
(cf.\ \S\ref{app:PF-counterterm}); in \cite{Cvetic:2025lat}, where this anomaly is obtained
from the $\eta$-invariant of the boundary $S^5/\Z_3$ (with $B_3=E_0$ realised as $\mathbb{C}^3/\Z_3(1,1,1)$), an analogous scheme dependence appears as a $\Z_6$ redundancy of the two coefficients. Note that the often-quoted ``$c_1=1$'' of
\cite[(4.27)]{Apruzzi:2022dlm} is a rescaled coupling, not the intersection number
$c_2(X)\cdot S=-6$ (which is always even); the two differ by the field-normalisation and
prefactor conventions, and \eqref{eq:BNanomaly} works directly with the geometric $-6$.

\paragraph{Flux lattice.} The three internal edges join the interior point to the polytope
vertices and are all homologous to the hyperplane $\ell\subset\mathbb{P}^2$; \eqref{eq:Mentry}
gives $S\cdot\ell=D_S^2 D_v=-3$ for each, so the pairing is $M=(-3,-3,-3)$, with single
elementary divisor $\fm=3$ and left null-vector $k=1$, $Z=S$. The line is odd,
$\ell^2=1$ (equivalently $S\cdot\ell=-3$ is odd), so $s=\tfrac12$ and \eqref{eq:delta} reads
$-3\delta\equiv\tfrac12\pmod1$, giving $\delta=-\tfrac16$ and
\begin{equation}
    n\;=\;\frac{k}{\fm}+\delta\;=\;\frac13-\frac16\;=\;\frac16\,,
\end{equation}
i.e.\ the spin$^c$-shifted lattice $6n\in 2\mathbb{Z}+1$. The three admissible fluxes
$\{\tfrac16,\tfrac12,\tfrac56\}$ are the $\mathbb{Z}_3$ translates of $\tfrac16$ by the
generator $\tfrac13$; in particular $n=\tfrac12=\tfrac16+\tfrac13$ is the doubly-charged
coset, not an uncharged half-integer flux, because the half-shift is locked onto every coset
by the odd curve $\ell$ --- the would-be integer-charge fluxes $\{0,\tfrac13,\tfrac23\}$
($6n$ even) are simply absent.

\subsection{$B_4$}
\label{app:B4}

The polytope has vertices $A=(3,0)$, $B=(1,3)$, $C=(0,1)$ and three interior points
\begin{equation}
    S_1=(1,1)\,, \qquad S_2=(2,1)\,, \qquad S_3=(1,2)\,,
\end{equation}
so the rank is $\tfrac{(N-1)(N-2)}{2}=3$ and $\fm=N(N-3)+3=7$. We work in the
$\mathbb{Z}_3$-symmetric chamber defined by the triangulation (7 unimodular triangles, as required by $\fm=7$)
\begin{equation}
\scalebox{0.9}{$
    (A,S_2,S_1),\ (A,S_1,C),\ (C,S_1,S_3),\ (B,S_3,C),\ (B,S_2,S_3),\ (B,A,S_2),\ (S_1,S_2,S_3)\,.
    $}
\end{equation}
Applying the rules \eqref{eq:wall}, the
non-vanishing triple intersections are
\begin{equation}
    c_{111}=c_{222}=c_{333}=8\,, \qquad
    c_{113}=c_{122}=c_{233}=-2\,, \qquad
    c_{123}=1\,,
    \label{eq:B4cijk}
\end{equation}
(with $c_{112}=c_{223}=c_{331}=0$), and each $S_i$ is a four-ray surface, so
\begin{equation}
    S_i^3=8\,, \qquad C^G_i=c_2(X)\cdot S_i=-4\,, \qquad C^R_i=2 \qquad (i=1,2,3)\,.
\end{equation}
For example, the edge $(S_1,S_2)$ has adjacent apexes $A,S_3$ with
\begin{equation} \label{rS1rS2}
r_A+r_{S_3}=(3,0,1)+(1,2,1)=(4,2,2)=0\cdot r_{S_1}+2\,r_{S_2} \, , 
\end{equation}
giving $c_{112}=0$ and
$c_{122}=-2$ via \eqref{eq:wall}. The prepotential \eqref{eq:BNprepot} is then
\begin{equation}
\begin{split}
    \mathcal{E}_{B_4}
    \;=\; \frac{1}{\epsilon_1\epsilon_2}\Bigl[\,
    &\tfrac{4}{3}\bigl(\phi_1^3+\phi_2^3+\phi_3^3\bigr)
    -\phi_1^2\phi_3-\phi_2^2\phi_1-\phi_3^2\phi_2+\phi_1\phi_2\phi_3 \\
    &-\tfrac{1}{12}\bigl(\epsilon_1^2+\epsilon_2^2\bigr)(\phi_1+\phi_2+\phi_3)
    +\epsilon_+^2(\phi_1+\phi_2+\phi_3)\,\Bigr]\,.
\end{split}
\label{eq:B4prepot}
\end{equation}
The generating coset of $\mathbb{Z}_7^{(1)}$ is represented by the central divisor
\begin{equation}
    Z=S_1+2S_2+4S_3 \qquad (k=(1,2,4))\,,
\end{equation}
which satisfies $Z\cdot C\equiv 0\pmod 7$ for all nine compact curves. Explicitly, ordering
the nine curves as the internal edges
\begin{equation}
\begin{split}
&(A,S_1),(A,S_2),(S_1,S_2), \\
&(C,S_1),(C,S_3),(S_1,S_3), \\
& (B,S_3),(B,S_2),(S_2,S_3), 
\end{split}
\end{equation}
the rule
\eqref{eq:Mentry} gives the pairing matrix\footnote{To see the rule \eqref{eq:Mentry} at
work, take the third curve, the edge $(S_1,S_2)$, with wall relation \eqref{rS1rS2}. Its two
``diagonal'' entries are self-intersections in which the row divisor is squared, each equal
to minus the coefficient of \emph{its own} ray in \eqref{rS1rS2}:
$M_{13}=D_{S_1}^2\!\cdot\! D_{S_2}=0$ (coefficient of $r_{S_1}$) and
$M_{23}=D_{S_2}^2\!\cdot\! D_{S_1}=-2$ (coefficient of $r_{S_2}$) --- which are just
$c_{112}$ and $c_{122}$ of \eqref{eq:B4cijk}. The remaining nonzero entry of the column,
$M_{33}=D_{S_1}\!\cdot\! D_{S_2}\!\cdot\! D_{S_3}=+1$, arises because $S_3$ is the other apex
of the edge.}
\begin{equation}
    M\;=\;\begin{pmatrix}
    -2 & 1 & 0 & -4 & 1 & -2 & 0 & 0 & 1\\
    1 & -4 & -2 & 0 & 0 & 1 & 1 & -2 & 0\\
    0 & 0 & 1 & 1 & -2 & 0 & -4 & 1 & -2
    \end{pmatrix}\,,
    \label{eq:B4M}
\end{equation}
whose Smith normal form is $\mathrm{diag}(1,1,7)$, confirming $\mathbb{Z}_7^{(1)}$; the
generator \eqref{eq:kdef} is its left null-vector $k=(1,2,4)$. The three columns whose edge
joins two interior points --- $(S_1,S_2),(S_1,S_3),(S_2,S_3)$ --- merely reproduce the
compact triple intersections $c_{ijk}$ of \eqref{eq:B4cijk}, whereas the remaining columns
involve a non-compact (boundary) divisor and are therefore not among the $c_{ijk}$, so $M$
encodes strictly more than the cubic prepotential: it also records how the compact curves
meet the non-compact divisors. Every edge has even wall
coefficient $\alpha$, so all nine curves have even self-intersection; hence $s\equiv0$ and
$\delta=0$ in \eqref{eq:delta}, and the flux lattice is integral with generator
$n=k/\fm=(\tfrac17,\tfrac27,\tfrac47)$. Thus $B_4$ is the spin$^c$-unshifted member of the
family, in contrast to the odd-curve case $B_3$. Expanding
$Z^3=\sum_{ijl}k_ik_jk_l\,c_{ijl}$ multilinearly with \eqref{eq:B4cijk}, we find
\bes{
\scalebox{0.92}{$
\begin{split}
    Z^3 &= \underbrace{8(1^3+2^3+4^3)}_{584}
    + 3\bigl[\underbrace{(1^2)(4)(-2)}_{c_{113}}
    +\underbrace{(2^2)(1)(-2)}_{c_{122}}
    +\underbrace{(4^2)(2)(-2)}_{c_{233}}\bigr]
    + 6\underbrace{(1)(2)(4)(1)}_{c_{123}} \\
    &= 584 - 240 + 48 = 392\,,
\end{split}
$}
}
while $c_2(X)\cdot Z=\sum_i k_i\,C^G_i=(1+2+4)(-4)=-28$. Hence, \eqref{eq:BNanomaly} gives
\begin{equation}
    \Omega
    \;=\; -\frac{392}{6\cdot 7^3}-\frac{-28}{12\cdot 7}
    \;=\; \frac{1}{7}\,,
\end{equation}
in agreement with \cite[(4.34)]{Apruzzi:2021nmk}. Note that, for $B_3$, one has $Z=S$ and
$Z^3=c_{111}$ directly, whereas, for $B_4$, the central divisor is a genuine
multi-divisor combination and the full $c_{ijk}$ tensor is required.

\subsection{General $B_N$}
\label{app:BNgen}

The construction extends to all $N$. In the $\mathbb{Z}_3$-symmetric chamber we find the
following closed forms, verified explicitly for $3\le N\le 8$:
\begin{align}
    \fm &= N(N-3)+3\,, \qquad \mathrm{rank}=\tfrac{(N-1)(N-2)}{2}\,, \qquad C^R_i=2\,, \\
    C^G_i &= 2\chi_i-12 = 12-2\,c_{iii}\,, \qquad \chi_i\in\{4,5,6\}\,, \\
    k_v &\equiv \scalebox{0.95}{$ \displaystyle x+(N-1)y-(N-1) \pmod \fm \quad\text{for the interior point } v=(x,y) $}\,,
    \label{eq:BNcentral}
\end{align}
(the central vector $k$ being defined up to an overall unit mod $\fm$), and the central
divisor evaluates to
\begin{equation}
    Z^3=(N-2)^3\,\fm^2\,, \qquad c_2(X)\cdot Z=-2(N-2)\,\fm\,.
    \label{eq:BNZdata}
\end{equation}
Inserting \eqref{eq:BNZdata} into \eqref{eq:BNanomaly} and using $\fm-(N-2)^2=N-1$,
\begin{equation}
    \Omega
    \;=\; -\frac{(N-2)^3}{6\,\fm}+\frac{N-2}{6}
    \;=\; \frac{(N-1)(N-2)}{6\,\fm}\,,
    \label{eq:BNOmega}
\end{equation}
reproducing \cite[(4.34)]{Apruzzi:2021nmk}. Table~\ref{tab:BN} collects the data.
\begin{table}[h]
\centering
\begin{tabular}{cccccc}
\toprule
$N$ & $\fm=N(N-3)+3$ & rank & $Z^3$ & $c_2(X)\cdot Z$ & $\Omega$ \\
\midrule
$3$ & $3$  & $1$  & $9$    & $-6$   & $1/9$    \\
$4$ & $7$  & $3$  & $392$  & $-28$  & $1/7$    \\
$5$ & $13$ & $6$  & $4563$ & $-78$  & $2/13$   \\
$6$ & $21$ & $10$ & $28224$& $-168$ & $10/63$  \\
$7$ & $31$ & $15$ & $120125$&$-310$ & $5/31$   \\
$8$ & $43$ & $21$ & $399384$&$-516$ & $7/43$   \\
\bottomrule
\end{tabular}
\caption{Toric data and cubic 1-form anomaly $\Omega$ of the non-Lagrangian $B_N$
theories, in the $\mathbb{Z}_3$-symmetric chamber. The values of $\Omega$ agree with
\cite[(4.34)]{Apruzzi:2021nmk}; the closed forms \eqref{eq:BNZdata}--\eqref{eq:BNOmega}
hold for all $N$ checked.}
\label{tab:BN}
\end{table}

\paragraph{Flux lattice.} The central vector \eqref{eq:BNcentral} is the left null-vector of
the matrix \eqref{eq:Mentry} for the chamber at hand, and the physical flux is
$n=k/\fm+\delta$ with $\delta$ fixed by the spin$^c$ congruence \eqref{eq:delta}. Whether the
offset is non-trivial is a property of the chamber: every compact curve of $B_4$ is even, so
$\delta=0$ and its flux lattice is integral, whereas $B_3$ has odd curves and a genuine
half-integer offset $\delta=-\tfrac16$. The offset never affects the anomaly, however: by
\eqref{eq:BNZdata}, $c_2(X)\cdot Z=-2(N-2)\,\fm$ is a multiple of $2\fm$ for every $N$, so the
gravitational term in \eqref{eq:BNanomaly} is insensitive to the half-integer shift and
$\Omega$ retains the value \eqref{eq:BNOmega}. This is the same index-theoretic integrality
that the (absent) W-boson one-loop term of \S\ref{sec:anomalies} would supply in a Lagrangian
description.

The triple intersections $c_{ijk}$ (hence the cubic part of $\mathcal{E}$) are
chamber-dependent, related across chambers by flops; the anomaly $\Omega$ and the closed
forms \eqref{eq:BNZdata}--\eqref{eq:BNOmega} are not. The $B_N$ theories carry no flavour
symmetry (no boundary lattice points beyond the vertices), so $\mathcal{E}$ has no mass
parameters. As an illustration, in the chamber specified above, the rank-six prepotential
of $B_5$ ($\fm=13$, interior points
$S_1{=}(1,1),S_2{=}(1,2),S_3{=}(1,3),S_4{=}(2,1),S_5{=}(2,2),S_6{=}(3,1)$, central
$k=(9,6,3,5,2,1)$, $C^G=(-4,-2,-4,-2,-2,-4)$, $C^R_i=2$) has cubic part
\bes{
\scalebox{0.97}{$
\begin{split}
    \mathcal{F}_{B_5}
    \;=\; & \tfrac43\bigl(\phi_1^3+\phi_3^3+\phi_6^3\bigr)
    +\tfrac76\bigl(\phi_2^3+\phi_4^3+\phi_5^3\bigr)
    -\phi_1^2\phi_2+\tfrac12\phi_1^2\phi_4+\phi_1\phi_2\phi_4
    -\tfrac32\phi_1\phi_4^2 \\
    -\;&\tfrac32\phi_2^2\phi_3-\tfrac12\phi_2^2\phi_4-\tfrac12\phi_2^2\phi_5
    +\tfrac12\phi_2\phi_3^2+\phi_2\phi_3\phi_5-\tfrac12\phi_2\phi_4^2
    +\phi_2\phi_4\phi_5-\tfrac12\phi_2\phi_5^2 \\
    -\; &\phi_3^2\phi_5-\tfrac12\phi_4^2\phi_5-\tfrac12\phi_4\phi_5^2
    +\phi_4\phi_5\phi_6-\phi_4\phi_6^2-\tfrac32\phi_5^2\phi_6
    +\tfrac12\phi_5\phi_6^2\,.
\end{split}
$}
}
The effective prepotential of this theory reads
\begin{equation}
    \mathcal{E}_{B_5}=\frac{1}{\epsilon_1\epsilon_2}\Bigl[\,\mathcal{F}_{B_5}
    -\frac{\epsilon_1^2+\epsilon_2^2}{24}\bigl(2\phi_1+\phi_2+2\phi_3+\phi_4+\phi_5+2\phi_6\bigr)
    +\epsilon_+^2 \sum_{j=1}^6 \phi_j \,\Bigr]\,,
    \label{eq:B5prepot}
\end{equation}
with $\mathcal{F}_{B_5}$ the cubic part above. The flux lattice carries a non-trivial spin$^c$
offset: $9$ of the $18$ compact curves are odd, so $\delta\neq0$ in \eqref{eq:delta}, and the
generating flux of $\mathbb{Z}_{13}^{(1)}$ is
\begin{equation}
    n=\frac{k}{13}+\delta=\Bigl(\tfrac{9}{26},\,\tfrac{3}{13},\,\tfrac{3}{26},\,
    \tfrac{9}{13},\,\tfrac{1}{13},\,\tfrac{1}{26}\Bigr)\,,\qquad
    13\,n=\bigl(\tfrac92,3,\tfrac32,9,1,\tfrac12\bigr)\,,
    \label{eq:B5flux}
\end{equation}
with the half-integer numerators of $13\,n$ being the spin$^c$ half-shift on the odd curves (cf.\ $B_3$,
$n=\tfrac16$, against the spin$^c$-unshifted $B_4$, $n=k/7$). The offset does not affect the
anomaly, since $c_2(X)\cdot Z=-2(N-2)\fm$ is a multiple of $2\fm$ \eqref{eq:BNZdata}.

The relations \eqref{eq:BNZdata} are clean enough to suggest a direct geometric proof,
which we leave for future work.

\subsection{The non-Lagrangian $B_N^{(1)}$ family}
\label{app:BN1}

The same toolkit applies to the sibling family $B_N^{(1)}$ of \cite{Eckhard:2020jyr,Morrison:2020ool},
which is the first non-Lagrangian example in our list carrying \emph{both} a
$\mathbb{Z}_{N-1}^{(1)}$ 1-form symmetry and the non-Abelian flavour symmetry $\SU(N-2)\times \U(1)$
of rank $N-2$ --- so that the mixed 1-form--flavour-Cartan anomaly \eqref{eq:mixedabstract} can be
evaluated directly from the geometry.
The toric polytope is the convex hull of the height-one rays $r_v=(v,1)$ over
\begin{equation}
    (N-1,0)\,,\quad (N-1,1)\,,\quad (0,j)\,, \quad \text{with} \quad j=1,\dots,N-1\,,
\end{equation}
its interior lattice points being the compact divisors $S_i$, of which there are
$\tfrac{(N-1)(N-2)}{2}$. The pairing matrix \eqref{eq:Mentry} has Smith normal form
$\mathrm{diag}(1,\dots,1,N-1)$, so
\begin{equation}
    \fm=N-1\,,\qquad \mathbb{Z}_\fm^{(1)}=\mathbb{Z}_{N-1}^{(1)}\,,
\end{equation}
with central divisor $Z=\sum_i k_i S_i$ its left null-vector \eqref{eq:kdef}. Building a fine star
triangulation and applying the rules of \S\ref{app:BN-formula} --- we have done so explicitly for
$3\le N\le 9$ --- gives the closed forms
\begin{equation}
    Z^3=2(N-2)^2\,\fm^2\,,\qquad c_2(X)\cdot Z=-2(N-2)\,\fm\,,\qquad Z^2\cdot D_F=-(N-2)\,\fm\,,
    \label{eq:BN1data}
\end{equation}
where $D_F=(N-1,0)$ is the boundary vertex generating one of the $N-2$ flavour $\U(1)$ factors (the one used
in \cite{Apruzzi:2021nmk}); the closed form $Z^2\cdot D_F=-(N-2)\fm$ gives its mixed anomaly
\eqref{eq:BN1mixed}, the others being read off the remaining flavour divisors.

\paragraph{Cubic self-anomaly.} Since $c_2(X)\cdot Z=-2(N-2)\,\fm$ is a multiple of $2\fm$, the
gravitational $(j^3-j)$ piece of \eqref{eq:BNanomaly} drops mod $1$ (\S\ref{app:PF-criterion}) and
the anomaly is purely cubic. Inserting \eqref{eq:BN1data} into \eqref{eq:BNanomaly} gives
$-\tfrac{(N-2)^2}{3\fm}j^3+\tfrac{N-2}{6}j$; using the index congruence $j^3\equiv j\pmod 6$ to
absorb the linear (gravitational) term into the cubic and $\fm-2(N-2)=-(N-3)$, then
\begin{equation}
    \Omega^{(1)}(j)\;=\;\frac{(N-2)(N-3)}{6\,(N-1)}\,j^3 \pmod 1\,,
    \label{eq:BN1cubic}
\end{equation}
(up to the overall sign of the $\exp(-V_n)$ convention and the choice of generator $k$ modulo a
unit of $\mathbb{Z}_\fm$), reproducing \cite[(4.35)]{Apruzzi:2021nmk}. It vanishes for $N=3$: the
rank-one member $B_3^{(1)}$ ($\mathbb{Z}_2^{(1)}$) has no cubic 1-form anomaly. Table~\ref{tab:BN1} collects the
data.

\paragraph{Mixed 1-form--flavour anomalies.} Each of the $N-2$ flavour $\U(1)$ factors, with non-compact
divisor $D_{F_a}$, yields a mixed 1-form--flavour anomaly
$\Omega_{ii,a}=-\tfrac12\,Z^2\!\cdot\!D_{F_a}/\fm^2$ \eqref{eq:BN1mixed}, computed directly from the
intersection ring. For $B_4^{(1)}$ ($\fm=3$, $Z=2S_1+2S_2+S_3$), the two directions $D_{F_1}=(3,0)$,
$D_{F_2}=(0,3)$ give
\begin{equation}
    Z^2\!\cdot\!D_{F_1}=-6\ \Rightarrow\ \Omega_{ii,1}=\tfrac13\,,\qquad
    Z^2\!\cdot\!D_{F_2}=-12\ \Rightarrow\ \Omega_{ii,2}=\tfrac23\,,
    \label{eq:BN1mixedexplicit}
\end{equation}
the first reproducing \cite[(4.38)]{Apruzzi:2021nmk} (the general closed form
$Z^2\!\cdot\!(N-1,0)=-(N-2)\fm$ of \eqref{eq:BN1data}). These match the $\exp(-V_{\mathfrak n})$
computation of \S\ref{sec:expVnonlag} term by term: at the bare background $\mathfrak n=jk/\fm$, the
flavour exponent is $W_a(\mathfrak n)=\tfrac12(j/\fm)^2\,Z^2\!\cdot\!D_{F_a}$, so
$\Omega_{ii,a}=\bigl[-W_a(\mathfrak n)\bigr]_{\rm frac}$ --- explicitly
$-W_1=n_1^2-n_1n_3+n_3^2\to\tfrac13$ and $-W_2=\tfrac32 n_2^2\to\tfrac23$ at the generator $j=1$
(both evaluated at zero flavour flux $B_{m_a}=0$; the full second exponent is
$W_2=B_{m_2}n_2-\tfrac32 n_2^2$, the $B_{m_2}$-term not entering the anomaly). The
anomalies thus reside in the fractional \emph{charges} (the $w_a$-exponents), the dynamical flavour
fluxes being quantised as in \eqref{eq:BN1flavflux}.

\begin{table}[h]
\centering
\begin{tabular}{ccccccc}
\toprule
$N$ & $\fm=N-1$ & rank & $Z^3$ & $c_2(X)\cdot Z$ & $\Omega^{(1)}$ & $\Omega^{(1)}_{ii,\U(1)_F}$ \\
\midrule
$3$ & $2$ & $1$  & $8$    & $-4$  & $0$    & $1/4$  \\
$4$ & $3$ & $3$  & $72$   & $-12$ & $1/9$  & $1/3$  \\
$5$ & $4$ & $6$  & $288$  & $-24$ & $1/4$  & $3/8$  \\
$6$ & $5$ & $10$ & $800$  & $-40$ & $2/5$  & $2/5$  \\
$7$ & $6$ & $15$ & $1800$ & $-60$ & $5/9$  & $5/12$ \\
$8$ & $7$ & $21$ & $3528$ & $-84$ & $5/7$  & $3/7$  \\
\bottomrule
\end{tabular}
\caption{Toric data and anomalies of the non-Lagrangian $B_N^{(1)}$ theories: the cubic 1-form
self-anomaly $\Omega^{(1)}$ \eqref{eq:BN1cubic} and the mixed 1-form--$\U(1)$-flavour anomaly
$\Omega^{(1)}_{ii,\U(1)_F}$ \eqref{eq:BN1mixed}, both quoted as the coefficient at the generating
coset $j=1$, reproducing \cite[(4.35),(4.38)]{Apruzzi:2021nmk}. The closed forms
\eqref{eq:BN1data} hold for all $N$ checked ($3\le N\le 8$).}
\label{tab:BN1}
\end{table}

Whereas every compact curve of the $B_N$-family member $B_4$ was even, the $B_N^{(1)}$ theories with
$N\ge4$ have odd compact curves and hence a non-trivial spin$^c$ offset $\delta$: the dynamical flux
is $n=jk/\fm+\delta$ (given explicitly below), so that $n=0$ is excluded, the half-integral flux
residing in the gauge directions. The bare 1-form background $\mathfrak{n}=jk/\fm$ pairs integrally
with every compact curve and is what enters \eqref{eq:BN1cubic} and \eqref{eq:BN1mixed}; in any case,
$c_2(X)\cdot Z$ is a multiple of $2\fm$, so the offset does not affect the cubic anomaly.

\paragraph{Explicit data: $B_3^{(1)}$ and $B_4^{(1)}$.}
We label the boundary (non-compact) divisors $A,B,C,\dots$ and the compact ones $S_i$. The flavour
symmetry has rank equal to the number of boundary lattice points minus three: \emph{one} mass for
$B_3^{(1)}$ (4 boundary points) but \emph{two} for $B_4^{(1)}$ (5 boundary points). Turning on a mass
$m_a$ along a non-compact divisor $D_{F_a}$ extends the prepotential: the well-defined couplings
$S_i\!\cdot\!S_j\!\cdot\!D_{F_a}$ add $\tfrac12(S_iS_jD_{F_a})\,\phi^i\phi^j\,m_a$ to the cubic part,
and $S_i \cdot D_{F_a}D_{F_b}$ adds $\tfrac12(S_iD_{F_a}D_{F_b})\,\phi^i\,m_a m_b$.

For $B_3^{(1)}$ ($\mathbb{Z}_2^{(1)}$ 1-form symmetry, rank one), the polytope is the convex hull of $A=(2,0)$,
$B=(2,1)$, $C=(0,1)$, $D=(0,2)$ with the single compact divisor $S_1=(1,1)$ and flavour divisor
$D_F=A$. The four compact curves $S_1A,S_1B,S_1C,S_1D$ give the row matrix $M=(-2,-2,-2,-2)$, with
$c_{111}=S_1^3=8$, $C^G_1=-4$, $C^R_1=2$ and mass coupling $S_1^2D_F=-2$. Hence $k=(1)$, $Z=S_1$, and
\begin{equation}
    \mathcal{E}_{B_3^{(1)}}=\frac{1}{\epsilon_1\epsilon_2}\Bigl[\,\tfrac43\phi_1^3-\phi_1^2\,m
    -\frac{\epsilon_1^2+\epsilon_2^2}{12}\,\phi_1+\epsilon_+^2\,\phi_1\,\Bigr]\,.
    \label{eq:B31prepot}
\end{equation}
All four curves are even ($C^2\in\{0,-2\}$), so there is no spin$^c$ shift and the generating flux
of $\mathbb{Z}_2^{(1)}$ is $n=k/2=\tfrac12$.

For $B_4^{(1)}$ ($\mathbb{Z}_3^{(1)}$ 1-form symmetry, rank three), the polytope is the hull of $A=(3,0)$, $B=(3,1)$,
$C=(0,1)$, $D=(0,2)$, $E=(0,3)$ with compact divisors $S_1=(1,1)$, $S_2=(1,2)$, $S_3=(2,1)$ and
\emph{two} flavour divisors $D_{F_1}=A$ and $D_{F_2}=E$.\footnote{The flavour lattice is the group of
non-compact divisor classes modulo the compact $S_i$ and the three toric linear relations
$\sum_\rho\langle m,v_\rho\rangle D_\rho=0$, of rank $N-2$; any $N-2$ non-compact divisors
\emph{independent} in this quotient may be taken as the masses. We choose the two vertices
$A=(N-1,0)$ and $E=(0,N-1)$: they are independent, their mass couplings $(S_iS_jD_{F_a})$ are
diagonal in the gauge node, and $A$ reproduces the mixed anomaly of \cite[(4.38)]{Apruzzi:2021nmk}.
A different admissible basis rescales the anomaly values but carries the same flavour-lattice data.}
In the fine triangulation with the nine triangles $CDS_1$, $CS_1A$,
$DES_2$, $DS_1S_2$, $ES_2B$, $S_1S_2S_3$, $S_1S_3A$, $S_2S_3B$, $S_3AB$, the eleven compact curves
in the order
\begin{equation}
    CS_1,\ DS_1,\ S_1A,\ DS_2,\ ES_2,\ S_1S_2,\ S_2B,\ S_1S_3,\ S_2S_3,\ S_3A,\ S_3B
\end{equation}
give the $3\times11$ pairing matrix $M$ (rows $S_1,S_2,S_3$; columns in the above order)
{\setlength{\arraycolsep}{3.5pt}
\begin{equation}
    M=\left(\begin{array}{*{11}{r}}
    -3&-1&-2&1&0&-1&0&0&1&1&0\\
    0&1&0&-1&-3&-1&-2&1&0&0&1\\
    0&0&1&0&0&1&1&-2&-2&-2&-2
    \end{array}\right)\,,
    \label{eq:B41M}
\end{equation}}
of Smith normal form $\mathrm{diag}(1,1,3)$ with central vector $k=(2,2,1)$, i.e.\ $Z=2S_1+2S_2+S_3$.
The non-zero triple intersections are $c_{111}=c_{222}=7$, $c_{333}=8$, $c_{112}=c_{122}=-1$,
$c_{133}=c_{233}=-2$, $c_{123}=1$, with $C^G=(-2,-2,-4)$, $C^R_i=2$. The two masses couple as
$(S_iS_jD_{F_1})=\left(\begin{smallmatrix}-2&0&1\\0&0&0\\1&0&-2\end{smallmatrix}\right)$,
$S_iD_{F_1}^2=0$, and
$(S_iS_jD_{F_2})=\left(\begin{smallmatrix}0&0&0\\0&-3&0\\0&0&0\end{smallmatrix}\right)$,
$S_2D_{F_2}^2=1$ (with $S_iD_{F_1}D_{F_2}=0$), so
\begin{equation}
\begin{split}
    \mathcal{E}_{B_4^{(1)}}=\frac{1}{\epsilon_1\epsilon_2}\Bigl[\,
    \tfrac76 &\bigl(\phi_1^3+\phi_2^3\bigr)+\tfrac43\phi_3^3
    -\tfrac12\phi_1^2\phi_2-\tfrac12\phi_1\phi_2^2-\phi_1\phi_3^2-\phi_2\phi_3^2+\phi_1\phi_2\phi_3\\
    +&\bigl(-\phi_1^2+\phi_1\phi_3-\phi_3^2\bigr)m_1
    +\bigl(-\tfrac32\phi_2^2\bigr)m_2+\tfrac12\phi_2\,m_2^2\\
    -&\frac{\epsilon_1^2+\epsilon_2^2}{24}\bigl(\phi_1+\phi_2+2\phi_3\bigr)
    +\epsilon_+^2\bigl(\phi_1+\phi_2+\phi_3\bigr)\,\Bigr]\,.
    \label{eq:B41prepot}
\end{split}
\end{equation}
The spin$^c$ flux quantisation is read off the $z_i$-exponents of $\exp(-V_n)$
(\S\ref{sec:expVnonlag}), with zero-points $\kappa=(-\tfrac18,-\tfrac18,0)$ ($S_{1,2}^3=7$ odd,
$S_3^3=8$ even) so $Z_i(0)=\kappa_i\notin\Z$ and the trivial flux is excluded. Of the five odd curves
$CS_1,DS_1,DS_2,ES_2,S_1S_2$, only $S_1S_2$ is compact--compact: it fixes the \emph{gauge} spin$^c$
offset
\begin{equation}
    \delta=\bigl(\tfrac12,0,0\bigr)\,,\qquad
    n=\frac{k}{\fm}+\delta=\frac{j}{3}(2,2,1)+\bigl(\tfrac12,0,0\bigr)\,,
    \label{eq:BN1delta}
\end{equation}
while the half-shifts on the four curves touching non-compact divisors are carried by the two
\emph{flavour} fluxes $B_{m_1}$ (through $D_{F_1}$) and $B_{m_2}$ (through $D_{F_2}$). The total flux is
$\mathcal N=\sum_i n_iS_i+B_{m_1}D_{F_1}+B_{m_2}D_{F_2}$, and $Z_i\in\Z$ on every central coset then
requires
\begin{equation}
    B_{m_1}\in 2\Z\,,\qquad B_{m_2}\equiv j \pmod 2\,,
    \label{eq:BN1flavflux}
\end{equation}
with $n=0$ excluded (it is the absence of the second flavour flux that obstructed a purely-gauge
spin$^c$ offset: $z_i$-integrality of $Z_2$, which is $B_{m_1}$-independent, can only be met once
$B_{m_2}$ is turned on, $S_2D_{F_2}^2=1$). The bare $\mathfrak n=jk/\fm$ pairs integrally with all
eleven curves and is the 1-form background entering \eqref{eq:BN1cubic} and \eqref{eq:BN1mixed}.

\subsection{The non-Lagrangian $B_N^{(2)}$ family}
\label{app:BN2}

The third sibling $B_N^{(2)}$ of \cite{Eckhard:2020jyr,Morrison:2020ool} carries a $\mathbb{Z}_N^{(1)}$
1-form symmetry. Its toric polytope is the convex hull of the height-one rays over
\begin{equation}
    (N,0)\,,\qquad (0,j)\, \quad \text{with} \quad j=1,\dots,N-1\,,
\end{equation}
a triangle with $\tfrac{(N-1)(N-2)}{2}$ interior points (the compact divisors $S_i$) and $N$
boundary points, so the flavour symmetry is $\SU(N-2)$ (rank $N-3$~\cite{Eckhard:2020jyr}): none
for $B_3^{(2)}$ --- which is simply the local $\mathbb{P}^2$, i.e.\ $B_3$ --- and $\SU(2)$ for
$B_4^{(2)}$, whose Cartan is the single $\U(1)_F$ used below. The pairing
matrix \eqref{eq:Mentry} has Smith normal form $\mathrm{diag}(1,\dots,1,N)$, and a fine
triangulation gives the closed forms (verified for $3\le N\le 7$)
\begin{equation}
    Z^3=(N-2)\,N^2\,,\qquad c_2(X)\cdot Z=-2(N-2)\,N\,,\qquad Z^2\cdot D_F=-(N-2)\,N\,,
    \label{eq:BN2data}
\end{equation}
with $D_F$ the boundary vertex $(N,0)$. As $c_2(X)\cdot Z$ is a multiple of $2\fm=2N$, the cubic
anomaly is pure cubic; inserting \eqref{eq:BN2data} into \eqref{eq:BNanomaly}, and
\eqref{eq:mixedabstract} with $D_F$, one obtains
\begin{equation}
    \Omega^{(2)}=\frac{(N-2)(N-1)}{6N}\,,\qquad
    \Omega_{ii,\U(1)_F}=-\frac12\,\frac{Z^2\cdot D_F}{N^2}=\frac{N-2}{2N}\quad(N\ge4)\,,
    \label{eq:BN2anom}
\end{equation}
the first reproducing \cite[(4.36)]{Apruzzi:2021nmk}, the second the mixed 1-form--flavour anomaly
(obtained as for $B_N^{(1)}$; not given in \cite{Apruzzi:2021nmk}). Table~\ref{tab:BN2} collects the
data.
\begin{table}[h]
\centering
\begin{tabular}{ccccccc}
\toprule
$N$ & $\fm=N$ & rank & $Z^3$ & $c_2(X)\cdot Z$ & $\Omega^{(2)}$ & $\Omega_{ii,\U(1)_F}$ \\
\midrule
$3$ & $3$ & $1$  & $9$   & $-6$  & $1/9$  & --- \ ($\cong B_3$) \\
$4$ & $4$ & $3$  & $32$  & $-16$ & $1/4$  & $1/4$  \\
$5$ & $5$ & $6$  & $75$  & $-30$ & $2/5$  & $3/10$ \\
$6$ & $6$ & $10$ & $144$ & $-48$ & $5/9$  & $1/3$  \\
$7$ & $7$ & $15$ & $245$ & $-70$ & $5/7$  & $5/14$ \\
\bottomrule
\end{tabular}
\caption{Toric data and anomalies of the non-Lagrangian $B_N^{(2)}$ theories: the cubic 1-form
anomaly $\Omega^{(2)}$ (reproducing \cite[(4.36)]{Apruzzi:2021nmk}) and the mixed
1-form--$\U(1)_F$ anomaly $\Omega_{ii,\U(1)_F}$ (present for $N\ge4$). $B_3^{(2)}$ has no flavour
symmetry and coincides with the local $\mathbb{P}^2$.}
\label{tab:BN2}
\end{table}

\paragraph{Explicit data: $B_4^{(2)}$.} With boundary divisors $A=(4,0)$ ($=D_F$), $C=(0,1)$,
$D=(0,2)$, $E=(0,3)$ and compact divisors $S_1=(1,1)$, $S_2=(1,2)$, $S_3=(2,1)$, the fine
triangulation $CDS_1$, $CS_1A$, $DES_2$, $DS_1S_2$, $ES_2A$, $S_1S_2S_3$, $S_1S_3A$, $S_2S_3A$ gives
the ten compact curves $CS_1$, $DS_1$, $S_1A$, $DS_2$, $ES_2$, $S_1S_2$, $S_2A$, $S_1S_3$,
$S_2S_3$, $S_3A$ and the pairing matrix
{\setlength{\arraycolsep}{3.5pt}
\begin{equation}
    M=\left(\begin{array}{*{10}{r}}
    -4&-1&-2&1&0&-1&0&1&1&1\\
    0&1&0&-1&-4&-1&-2&1&1&1\\
    0&0&1&0&0&1&1&-3&-3&-3
    \end{array}\right)\,,
\end{equation}}
of Smith normal form $\mathrm{diag}(1,1,4)$ with central vector $k=(1,1,2)$, $Z=S_1+S_2+2S_3$. The
non-zero triple intersections are $c_{111}=c_{222}=7$, $c_{333}=9$, $c_{112}=c_{122}=-1$,
$c_{113}=c_{223}=1$, $c_{123}=1$, $c_{133}=c_{233}=-3$, with $C^G=(-2,-2,-6)$, $C^R_i=2$ and mass
couplings $(S_iS_jD_F)=\left(\begin{smallmatrix}-2&0&1\\0&-2&1\\1&1&-3\end{smallmatrix}\right)$,
$S_iD_F^2=(0,0,1)$, so
\bes{ \label{eq:B42prepot}
\scalebox{0.98}{$
\begin{split}
    \mathcal{E}_{B_4^{(2)}}=\frac{1}{\epsilon_1\epsilon_2}\Bigl[\,
    \tfrac76 & (\phi_1^3+\phi_2^3)+\tfrac32\phi_3^3-\tfrac12\phi_1^2\phi_2+\tfrac12\phi_1^2\phi_3
    -\tfrac12\phi_1\phi_2^2+\phi_1\phi_2\phi_3  -\tfrac32\phi_1\phi_3^2 \\+ & \tfrac12\phi_2^2\phi_3-\tfrac32\phi_2\phi_3^2
    +\bigl(-\phi_1^2-\phi_2^2-\tfrac32\phi_3^2+\phi_1\phi_3+\phi_2\phi_3\bigr)m+\tfrac12\phi_3\,m^2\\
    -&\frac{\epsilon_1^2+\epsilon_2^2}{24}(\phi_1+\phi_2+3\phi_3)+\epsilon_+^2(\phi_1+\phi_2+\phi_3)\,\Bigr]\,.
\end{split}
$}
}
As for $B_N^{(1)}$, the BPS curves carry integer charges $D_F\cdot C_a\in\Z$, so the flavour flux is
integral, $B_h\in\Z$. In contrast to $B_N^{(1)}$, however, the spin$^c$ congruence \eqref{eq:delta}
is now \emph{solvable} over the compact divisors. The six odd curves $DS_1,DS_2,S_1S_2,S_1S_3,S_2S_3,
AS_3$ ($C^2$ odd) impose $\langle\delta,C_a\rangle\equiv\tfrac12$, and \eqref{eq:delta} has the
solution
\begin{equation}
    \delta=\bigl(\tfrac12,0,0\bigr)\,,\qquad
    n=\frac{k}{\fm}+\delta=\frac{j}{4}(1,1,2)+\bigl(\tfrac12,0,0\bigr)\,,
    \label{eq:BN2delta}
\end{equation}
so the dynamical fluxes are the central cosets shifted by $\delta$ --- e.g.\ $n=(\tfrac12,0,0)$ at
$j=0$ --- and the trivial flux $n=0$ is \emph{excluded}, exactly as for local $\mathbb{P}^2$.
Equivalently, with $B_h\in\Z$, the gauge combination $n_1+n_2-3n_3$ is forced half-integral
(the $z_i$-integrality condition $Z_3\in\Z\Leftrightarrow n_1+n_2-3n_3+B_h\in\Z+\tfrac12$ of
\S\ref{sec:expVnonlag}); the spin$^c$ half-unit sits in the gauge flux, not in $B_h$. The bare
$\mathfrak{n}=jk/N$ remains the 1-form background entering \eqref{eq:BN2anom}.

\subsection{The non-Lagrangian $B_N^{(3)}$ family}
\label{app:BN3}

The last sibling $B_N^{(3)}$ of \cite{Eckhard:2020jyr,Morrison:2020ool} has toric fan generated by the
height-one rays over $(0,1)$ and $(N-k,k)$, $k=0,\dots,N-1$ (the latter $N$ points collinear on
$x+y=N$). It carries a $\mathbb{Z}_{N-1}^{(1)}$ 1-form symmetry, rank $\tfrac{(N-1)(N-2)}{2}$
(interior points), and the flavour symmetry $\SU(N-1)$ of rank $N-2$~\cite{Eckhard:2020jyr}. A fine star triangulation gives the closed
forms (verified for $3\le N\le 7$)
\begin{equation}
    Z^3=(N-1)^3(N-2)\,,\qquad c_2(X)\cdot Z=-2(N-1)(N-2)\,,
    \label{eq:BN3data}
\end{equation}
$Z=\sum_i k_i S_i$ generating $\mathbb{Z}_{N-1}^{(1)}$. Strikingly, the cubic 1-form anomaly
\emph{vanishes}: with $\fm=N-1$, this yields
\begin{equation}
    \Omega^{(3)}(j)=-\frac{Z^3}{6\fm^3}\,j^3-\frac{c_2(X)\cdot Z}{12\fm}\,j
    =-\frac{\fm-1}{6}\,(j^3-j)\equiv 0\pmod 1\,,
    \label{eq:BN3anom}
\end{equation}
because the genuine cubic coefficient cancels, $A+B=-\tfrac{\fm-1}{6}+\tfrac{\fm-1}{6}=0$, and the
leftover gravitational $(j^3-j)$ piece is divisible by $6$. This reproduces the vanishing $B_2^3$
anomaly of \cite{Apruzzi:2021nmk}. The mixed 1-form--flavour anomalies are instead non-trivial:
$\Omega_{ii,a}=-\tfrac12\,Z^2\!\cdot\!D_{F_a}/\fm^2$ for each flavour divisor $D_{F_a}$.

\paragraph{Explicit data: $B_3^{(3)}$.} ($\mathbb{Z}_2^{(1)}$ 1-form symmetry, rank one, flavour rank one.) The
polytope is the hull of $(0,1)$ and $(3,0),(2,1),(1,2)$, with a single compact divisor $S_1=(1,1)$
and central vector $k=(1)$, $Z=S_1$. Labelling the boundary divisors $A=(3,0)$, $B=(2,1)$,
$C=(1,2)$, $E=(0,1)$, the fine triangulation has the four triangles $ES_1C$, $ES_1A$, $S_1CB$,
$S_1BA$, hence the four compact curves $ES_1,\ S_1C,\ S_1A,\ S_1B$ and the $1\times4$ pairing matrix
$M=(-4,\,-2,\,-2,\,0)$, of Smith normal form $\mathrm{diag}(2)$, confirming the $\mathbb{Z}_2^{(1)}$
1-form symmetry. The single triple intersection is $c_{111}=S_1^3=8$, with $C^G_1=-4$, $C^R_1=2$.
Taking the mass along $D_F=A=(3,0)$, with couplings $S_1^2D_F=-2$ and $S_1D_F^2=0$, the effective prepotential reads
\begin{equation}
    \mathcal{E}_{B_3^{(3)}}=\frac{1}{\epsilon_1\epsilon_2}\Bigl[\,\tfrac43\phi_1^3-\phi_1^2 m
    -\frac{\epsilon_1^2+\epsilon_2^2}{12}\phi_1+\epsilon_+^2\phi_1\,\Bigr]\,,
    \label{eq:B33prepot}
\end{equation}
of the same algebraic form as $\mathcal{E}_{B_3^{(1)}}$ \eqref{eq:B31prepot}. The two are nevertheless
\emph{distinct} geometries: the $B_3^{(1)}$ toric diagram is the parallelogram $(2,0),(2,1),(0,1),(0,2)$,
whose compact divisor is $\mathbb{F}_0=\mathbb{P}^1\!\times\!\mathbb{P}^1$ (curve self-intersections
$\{0,0,0,0\}$), whereas the $B_3^{(3)}$ diagram above is a triangle with compact divisor $\mathbb{F}_2$
($\{-2,0,0,2\}$); having four versus three convex-hull vertices, they are \emph{not} related by any
$GL(2,\Z)$ transformation. Only the coarse invariants $c_{111}=S_1^3=8$ and $C^G_1=-4$ coincide ---
both compact divisors being degree-eight Hirzebruch surfaces. As $\mathbb{F}_0$ and $\mathbb{F}_2$ are
flop-related, the two are plausibly the same rank-one SCFT in different toric phases, but this is a
birational (flop) statement, not a unimodular equivalence of the diagrams. The spin$^c$ zero-point is
$\kappa_1=0$ ($S_1^3=8$ even), so the trivial flux is \emph{allowed} and there is no offset;
$Z_1\in\Z$ on the $\Z_2^{(1)}$ cosets gives
\begin{equation}
    n=\tfrac{j}{2}+\delta\,,\qquad \delta=0\,,\qquad B_m\in\Z\,.
\end{equation}
The cubic 1-form anomaly vanishes, $\Omega^{(3)}=-\tfrac{\fm-1}{6}(j^3-j)\equiv0$
\eqref{eq:BN3anom}, so the only 't~Hooft anomaly is the mixed 1-form--flavour-Cartan one,
$\Omega_{ii,F}=-\tfrac12 Z^2\!\cdot\!D_F/\fm^2=\tfrac14$ (since $Z^2\!\cdot\!D_F=-2$).

\paragraph{Explicit data: $B_4^{(3)}$.} ($\mathbb{Z}_3^{(1)}$ 1-form symmetry, rank three, flavour rank two.) The
polytope is the hull of $(0,1)$ and $(4,0),(3,1),(2,2),(1,3)$, with compact divisors $S_1=(1,1)$,
$S_2=(1,2)$, $S_3=(2,1)$ and central vector $k=(2,1,1)$, $Z=2S_1+S_2+S_3$. Labelling the boundary
divisors $A=(4,0)$, $B=(3,1)$, $C=(2,2)$, $D=(1,3)$, $E=(0,1)$, the fine triangulation has the nine
triangles $ES_1S_2$, $AES_1$, $EDS_2$, $S_1S_2S_3$, $AS_1S_3$, $DCS_2$, $CS_2S_3$, $BCS_3$, $ABS_3$,
hence the eleven compact curves (in order)
\begin{equation}
    ES_1,\ ES_2,\ S_1S_2,\ S_1A,\ S_2D,\ S_1S_3,\ S_2S_3,\ S_3A,\ S_2C,\ S_3C,\ S_3B
\end{equation}
and the $3\times11$ pairing matrix (rows $S_1,S_2,S_3$; columns in the above order)
{\setlength{\arraycolsep}{3.5pt}
\begin{equation}
    M=\left(\begin{array}{*{11}{r}}
    -5&1&-2&-2&0&1&1&1&0&0&0\\
    1&-2&0&0&-3&1&-1&0&-1&1&0\\
    0&0&1&1&0&-3&-1&-2&1&-1&0
    \end{array}\right)\,,
    \label{eq:B43M}
\end{equation}}
of Smith normal form $\mathrm{diag}(1,1,3)$, confirming the $\mathbb{Z}_3^{(1)}$ 1-form symmetry. The
non-zero triple
intersections are $c_{111}=8$, $c_{222}=c_{333}=7$, $c_{112}=-2$, $c_{113}=c_{123}=1$, $c_{133}=-3$,
$c_{223}=c_{233}=-1$, with $C^G=(-4,-2,-2)$, $C^R_i=2$. Taking the two masses along $D_{F_1}=A=(4,0)$
and $D_{F_2}=D=(1,3)$ --- whose couplings
$(S_iS_jD_{F_1})=\left(\begin{smallmatrix}-2&0&1\\0&0&0\\1&0&-2\end{smallmatrix}\right)$, $S_iD_{F_1}^2=0$
and $(S_iS_jD_{F_2})=\left(\begin{smallmatrix}0&0&0\\0&-3&0\\0&0&0\end{smallmatrix}\right)$,
$S_2D_{F_2}^2=1$ coincide with those of $B_4^{(1)}$ --- one finds
\begin{equation}
\begin{split}
    \mathcal{E}_{B_4^{(3)}}=\frac{1}{\epsilon_1\epsilon_2}\Bigl[\,
    \tfrac43 & \phi_1^3+\tfrac76\phi_2^3+\tfrac76\phi_3^3-\phi_1^2\phi_2+\tfrac12\phi_1^2\phi_3+\phi_1\phi_2\phi_3 -\tfrac32\phi_1\phi_3^2-\tfrac12\phi_2^2\phi_3 \\-& \tfrac12\phi_2\phi_3^2
    +\bigl(-\phi_1^2+\phi_1\phi_3-\phi_3^2\bigr)m_1-\tfrac32\phi_2^2 m_2+\tfrac12\phi_2 m_2^2\\
    -& \frac{\epsilon_1^2+\epsilon_2^2}{24}\bigl(2\phi_1+\phi_2+\phi_3\bigr)
    +\epsilon_+^2\bigl(\phi_1+\phi_2+\phi_3\bigr)\,\Bigr]\,.
    \label{eq:B43prepot}
\end{split}
\end{equation}
The flux quantisation follows the same route as for $B_N^{(1)}$: with spin$^c$ zero-points
$\kappa=(0,-\tfrac18,-\tfrac18)$ ($S_1^3=8$ even, $S_{2,3}^3=7$ odd), $Z_i(0)=\kappa_i\notin\Z$, so the
trivial flux is excluded; the two odd compact--compact curves $S_1S_3,S_2S_3$ fix the gauge spin$^c$
offset, and $Z_i\in\Z$ on the central cosets gives
\begin{equation}
    n=\tfrac{j}{3}(2,1,1)+\delta\,,\qquad \delta=\bigl(0,\tfrac12,0\bigr)\,,\qquad B_{m_1},B_{m_2}\in\Z\,,
    \label{eq:BN3flux}
\end{equation}
with $n=0$ excluded and the half-integral spin$^c$ flux residing in the gauge direction $S_2$. The
cubic 1-form anomaly vanishes \eqref{eq:BN3anom}, so the \emph{only} 't~Hooft anomalies are the mixed
1-form--flavour ones, $\Omega_{ii,a}=-\tfrac12 Z^2\!\cdot\!D_{F_a}/\fm^2$:
\begin{equation}
    Z^2\!\cdot\!D_{F_1}=-6\Rightarrow\Omega_{ii,1}=\tfrac13\,,\qquad
    Z^2\!\cdot\!D_{F_2}=-3\Rightarrow\Omega_{ii,2}=\tfrac16\,.
\end{equation}
The value $\tfrac13$ coincides numerically with the $B_N^{(1)}$ mixed anomaly of
\cite[(4.38)]{Apruzzi:2021nmk}, but it is an independent computation here --- \cite[(4.38)]{Apruzzi:2021nmk}
is stated only for $B_N^{(1)}$, and \cite{Apruzzi:2021nmk} gives no mixed anomaly for $B_N^{(3)}$.

\subsection{Comments on the mixed 1-form--gravitational anomalies} \label{app:PF-criterion}
We emphasise that the cubic anomalies of the $B_N$ theories of appendix~\ref{app:BN} were all \emph{pure
cubic}: by \eqref{eq:BNZdata}, the gravitational coupling $c_2(X)\cdot Z$ is a multiple of
$2\fm$, and --- as we explain below --- the would-be gravitational contribution then drops
modulo~1.

One evaluates the anomaly \eqref{eq:BNanomaly} on the $j$-th element of the 1-form symmetry, i.e.\
on the central divisor scaled to $jZ$, with $j\in\mathbb{Z}_\fm$. Since $Z^3$ scales as $j^3$ and
$c_2(X)\cdot Z$ as $j$, then
\begin{equation}
    \Omega(j)=A\,j^3+B\,j\,,\qquad
    A=-\frac{Z^3}{6\fm^3}\,,\quad B=-\frac{c_2(X)\cdot Z}{12\fm}\,,
    \label{eq:Omegaj}
\end{equation}
and we may split off the genuine cubic monomial,
\begin{equation}
    \Omega(j)=(A+B)\,j^3\;-\;B\,(j^3-j)\,.
    \label{eq:Omegasplit}
\end{equation}
The second term is what we call the \emph{gravitational piece}: it carries the coefficient $B$
of the mixed gauge/gravitational coupling, organised in the W-boson/index combination $j^3-j$.
Because $j^3-j=(j-1)\,j\,(j+1)$ is a product of three consecutive integers, with $6\mid(j^3-j)$,
writing $j^3-j=6\,t_j$ with $t_j\in\mathbb{Z}$, one gets
\begin{equation}
    -B\,(j^3-j)=\frac{c_2(X)\cdot Z}{12\,\fm}\,6\,t_j=\frac{c_2(X)\cdot Z}{2\,\fm}\,t_j\,,
\end{equation}
which is an integer for every $j$ --- so the gravitational piece drops modulo~1 --- if and only
if
\begin{equation}
    2\fm \;\big|\; c_2(X)\cdot Z\,.
    \label{eq:dropcrit}
\end{equation}
The condition is sharp: $t_2=1$, so $j=2$ already detects a failure. Two ingredients enter. The
factor $6$ is the index ($x^3\equiv x\bmod 6$) integrality that, in a Lagrangian theory, is
supplied by the one-loop term
$\tfrac16\sum_{\alpha\in\Delta^+}\!\big(\langle n,\alpha\rangle^3-\langle n,\alpha\rangle\big)$
of the W-bosons (cf.\ the discussion opening appendix~\ref{app:BN}); the extra factor of $2$
--- so that the threshold is $2\fm$ rather than $\fm$ --- is the spin$^c$ shift $\delta$ of
\eqref{eq:delta}, present whenever a relevant compact curve has odd self-intersection. When \eqref{eq:dropcrit} holds, the
anomaly is the pure cubic $\Omega(j)=(A+B)j^3$; when it fails, the $(j^3-j)$ term survives mod~1.
For the $B_N$ theories, \eqref{eq:BNZdata} gives $c_2(X)\cdot Z=-2(N-2)\fm$, manifestly divisible
by $2\fm$, so \eqref{eq:dropcrit} always holds and \eqref{eq:BNOmega} is a pure cubic term.

\section{Non-Lagrangian $\mathbb{P}^2\cup\mathbb{F}_{3}$ and $\mathbb{P}^2\cup\mathbb{F}_{6}$ theories}
\label{app:PF}
 In this appendix we apply the same machinery to two rank-two non-Lagrangian SCFTs
of \cite{Jefferson:2018irk,Apruzzi:2019opn}, the local $\mathbb{P}^2\cup\mathbb{F}_3$ and $\mathbb{P}^2\cup\mathbb{F}_6$
theories, which differ only in the gluing of the two compact surfaces, and find that they sit
on opposite sides of the divisibility criterion \eqref{eq:dropcrit} of appendix~\ref{app:PF-criterion}. $\mathbb{P}^2\cup\mathbb{F}_6$ has a pure cubic anomaly,
exactly like the $B_N$; $\mathbb{P}^2\cup\mathbb{F}_3$ does not --- its anomaly is carried
entirely by a mixed 1-form--gravitational term that survives precisely because the divisibility \eqref{eq:dropcrit}
fails. Nevertheless, we can explicitly show that the 't Hooft anomaly for 
$\mathbb{Z}^{(1)}_5$ is truly cubic due to the $\bmod \, 24$ index ambiguity, which amounts to adding appropriate local counterterms. We note that $\mathbb{P}^2\cup\mathbb{F}_3$ is the first member of the family of isolated orbifolds $\mathbb{C}^3/\Z_{2n+3}(1,1,2n+1)$, whose resolutions are linear chains $\mathbb{P}^2\cup\mathbb{F}_3\cup\mathbb{F}_5\cup\cdots\cup\mathbb{F}_{2n+1}$; the 1-form anomalies of this family were recently computed by means of extra-dimensional $\eta$-invariants in \cite{Cvetic:2025lat}, and from the quiver Hilbert series in \cite{Chakrabhavi:2026iku}, where a $\Z_6$ redundancy of the pair of cubic and mixed 1-form--gravitational anomaly coefficients, rooted in the same index congruence as the $\bmod \, 24$ ambiguity exploited here, is discussed.

\subsection{$\mathbb{P}^2\cup\mathbb{F}_3$}
\label{app:PF3}

The theory \cite[\S5.2.1]{Kim:2020hhh} has compact divisors $S_1=\mathbb{P}^2$ and
$S_2=\mathbb{F}_3$. From the prepotential \cite[(5.63)]{Kim:2020hhh},
$6 \mathcal{F}=9\phi_1^3-9\phi_1^2\phi_2+3\phi_1\phi_2^2+8\phi_2^3$, with
\begin{equation}
    \mathcal{E}_{\mathbb{P}^2\cup\mathbb{F}_3}=\frac{1}{\epsilon_1\epsilon_2}\Bigl[\,\mathcal{F}
    -\frac{\epsilon_1^2+\epsilon_2^2}{24}\, \left(3 \phi_1 + 2 \phi_2\right)+\epsilon_+^2\,\left(\phi_1 + \phi_2\right)\,\Bigr]\,,
    \label{eq:P2F3prepot}
\end{equation}
one reads off, in the conventions of
\eqref{eq:BNcouplings}, the following data:
\begin{equation}
    c_{111}=9,\ \ c_{112}=-3,\ \ c_{122}=1,\ \ c_{222}=8\,,
    \qquad C^G=(-6,-4)\,,\quad C^R=(2,2)\,,
\end{equation}
the two surfaces contributing $S_1^3=9,\ c_2(X)\cdot S_1=-6$ ($\mathbb{P}^2$) and
$S_2^3=8,\ c_2(X)\cdot S_2=-4$ (a Hirzebruch surface). The curve volumes
\cite[(5.62)]{Kim:2020hhh}, $\mathrm{vol}(\ell)=3\phi_1-\phi_2$ and
$\mathrm{vol}(f_2)=-\phi_1+2\phi_2$, give, through $\mathrm{vol}(C)=-\sum_i(S_i\cdot C)\phi^i$, the
pairing matrix \eqref{eq:Mentry}
\begin{equation}
    M=\begin{pmatrix} S_1\cdot\ell \, & \, S_1\cdot f_2\\[2pt] S_2\cdot\ell \, & \, S_2\cdot f_2\end{pmatrix}
    =\begin{pmatrix}-3 & 1\\ 1 & -2\end{pmatrix}\,,\qquad \det M=5\,,
\end{equation}
so the 1-form symmetry is $\mathbb{Z}_5^{(1)}$, consistent with the flux $n_1\in\mathbb{Z}+\tfrac15$
of \cite[(5.64)]{Kim:2020hhh}. The generator \eqref{eq:kdef} is $k=(1,3)$, \ie $Z=S_1+3S_2$, giving
\begin{equation}
    Z^3=225\,,\qquad c_2(X)\cdot Z=-18\,,\qquad \fm=5\,.
\end{equation}
Here $2\fm=10\nmid 18$, so the criterion \eqref{eq:dropcrit} \emph{fails}. From \eqref{eq:Omegaj} and \eqref{eq:Omegasplit}, with $A=-\tfrac3{10}$
and $B=+\tfrac3{10}$ the pure cubic coefficient vanishes, $A+B=\Omega(1)=0$, and the entire
anomaly is the surviving gravitational piece
\begin{equation}
    \Omega(j)=-\tfrac3{10}(j^3-j)\;\equiv\;\tfrac15(j^3-j)\pmod 1
    =\Big\{0,\ \tfrac15,\ \tfrac45,\ 0\Big\}\qquad(j=1,2,3,4)\,,
    \label{eq:F3anom}
\end{equation}
the second equality using $\tfrac12(j^3-j)\in\mathbb{Z}$. The spin$^c$ origin is captured by
the offset $\delta$ of \eqref{eq:delta}. Since $\mathbb{P}^2$ is non-spin, the line $\ell$ has
$\ell^2=1$ odd, while the fibre $f_2$ takes integral flux \cite[(5.62),(5.64)]{Kim:2020hhh}, so
the source is $s=(\tfrac12,0)$ and \eqref{eq:delta} gives $\delta=(0,\tfrac12)$. The physical
flux is then
\begin{equation}
    n=\frac{k}{\fm}+\delta=\Big(\tfrac15,\tfrac35\Big)+\Big(0,\tfrac12\Big)\equiv\Big(\tfrac15,\tfrac1{10}\Big)\pmod 1\,,
\end{equation}
reproducing \cite[(5.64)]{Kim:2020hhh}; the half-shift $n_2\in\mathbb{Z}+\tfrac1{10}$
(denominator $2\fm$) is the imprint of the odd curve.

\subsection*{The $\bmod\,24$ ambiguity}
\label{app:PF-counterterm}

The vanishing $\Omega(1)=0$ is a property of the representative delivered by the geometry, not
of the anomaly class. As noted below \eqref{eq:BNanomaly} (and in \cite{Apruzzi:2021nmk}), the
cubic and gravitational terms are not separately well defined; the ambiguity is the index
congruence $x\,p_1\equiv 4x^3\pmod{24}$, i.e.\ $Z\cdot p_1=-2\,c_2(X)\cdot Z$ is defined only
modulo $24$. We utilise this ambiguity as follows. For
$\mathbb{P}^2\cup\mathbb{F}_3$, one has $Z\cdot p_1=36$; adding a single unit of 24, it follows that
\begin{equation} \label{mod24shift}
    Z\cdot p_1:\ 36\ \longrightarrow\ 60\,,\qquad\text{equivalently}\quad
    c_2(X)\cdot Z:\ -18\ \longrightarrow\ -30\,,
\end{equation}
makes $c_2(X)\cdot Z$ divisible by $2\fm=10$, so the gravitational piece now drops and
\begin{equation} \label{Omegatilde}
    \tilde{\Omega}(j)=-\frac16\frac{Z^3}{\fm^3}j^3+\frac1{24}\frac{Z\cdot p_1}{\fm}j
    =-\tfrac3{10}j^3+\tfrac12 j\;\equiv\;\tfrac15\,j^3\pmod1\,,
\end{equation}
which is the pure cubic $\mathbb{Z}^{(1)}_5$ anomaly with $\tilde\Omega(1)=\tfrac15$.

\subsection{$\mathbb{P}^2\cup\mathbb{F}_6$}
\label{app:PF6}

The sister theory \cite[\S5.2.2]{Kim:2020hhh} replaces $\mathbb{F}_3$ by $\mathbb{F}_6$; the
gluing changes only the mixed cubic couplings. From \cite[(5.67)]{Kim:2020hhh}, the prepotential is $6\mathcal{F}=9\phi_1^3-18\phi_1^2\phi_2+12\phi_1\phi_2^2+8\phi_2^3$, with effective prepotential
\begin{equation}
    \mathcal{E}_{\mathbb{P}^2\cup\mathbb{F}_6}=\frac{1}{\epsilon_1\epsilon_2}\Bigl[\,\mathcal{F}
    -\frac{\epsilon_1^2+\epsilon_2^2}{24}\, \left(3 \phi_1 + 2 \phi_2\right)+\epsilon_+^2\,\left(\phi_1 + \phi_2\right)\,\Bigr]\,,
    \label{eq:P2F6prepot}
\end{equation}
from which
\begin{equation}
    c_{111}=9,\ \ c_{112}=-6,\ \ c_{122}=4,\ \ c_{222}=8\,,\qquad C^G=(-6,-4)\,,
\end{equation}
and the curve volumes \cite[(5.66)]{Kim:2020hhh}, $\mathrm{vol}(\ell)=3\phi_1-2\phi_2$,
$\mathrm{vol}(f_2)=-\phi_1+2\phi_2$, give
\begin{equation}
    M=\begin{pmatrix}-3 & 1\\ 2 & -2\end{pmatrix}\,,\qquad \det M=4\,,
\end{equation}
so the 1-form symmetry is now $\mathbb{Z}_4^{(1)}$ (matching $n_1\in\mathbb{Z}+\tfrac14$ of
\cite[(5.68)]{Kim:2020hhh}). The generator \eqref{eq:kdef} is the order-four element
$k=(2,1)$, \ie $Z=2S_1+S_2$, with
\begin{equation}
    Z^3=32\,,\qquad c_2(X)\cdot Z=-16\,,\qquad \fm=4\,.
\end{equation}
Now $2\fm=8\mid 16$, so \eqref{eq:dropcrit} \emph{holds}: the gravitational piece drops and the
anomaly is the pure cubic
\begin{equation}
    \Omega(j)=(A+B)\,j^3=\tfrac14\,j^3\pmod 1
    =\Big\{\tfrac14,\ 0,\ \tfrac34,\ 0\Big\}\,,\qquad \Omega(1)=\tfrac14\neq0\,.
    \label{eq:F6anom}
\end{equation}
As for $\mathbb{F}_3$, the line $\ell$ ($\ell^2=1$) is the unique odd curve, so $s=(\tfrac12,0)$
and \eqref{eq:delta} now gives $\delta\equiv(-\tfrac14,-\tfrac18)$, hence the physical flux
$n=k/\fm+\delta=(\tfrac14,\tfrac18)$ of \cite[(5.68)]{Kim:2020hhh}. Thus, although $\mathbb{F}_6$
also carries a spin$^c$ shift in its flux lattice ($n_2\in\mathbb{Z}+\tfrac18$, denominator
$2\fm$), the shift does not reach the anomaly: $\mathbb{P}^2\cup\mathbb{F}_6$ behaves exactly like
the $B_N$ theories, with a genuine cubic $\mathbb{Z}_4$ anomaly. Table~\ref{tab:PF} summarises
the contrast.

\begin{table}[h]
\centering
\begin{tabular}{lcccccc}
\toprule
theory & 1-form & $Z^3$ & $c_2(X)\!\cdot\!Z$ & $2\fm\mid c_2(X)\!\cdot\!Z$ & $\Omega(j)$ \\
\midrule
$\mathbb{P}^2\cup\mathbb{F}_3$ & $\mathbb{Z}_5$ & $225$ & $-18$ & no  & $-\tfrac3{10}(j^3-j)\,  \rightarrow \, \frac{1}{5}j^3$\\
$\mathbb{P}^2\cup\mathbb{F}_6$ & $\mathbb{Z}_4$ & $32$  & $-16$ & yes & $\tfrac14\,j^3$\\
\bottomrule
\end{tabular}
\caption{The two rank-two SCFTs of \cite{Kim:2020hhh}, differing only in the gluing of the
$\mathbb{P}^2$ and Hirzebruch surfaces. The divisibility \eqref{eq:dropcrit} separates a pure
cubic anomaly ($\mathbb{F}_6$, like the $B_N$) from one carried by the mixed
1-form--gravitational term ($\mathbb{F}_3$). However, in the latter case, taking into account the $\bmod \, 24$ ambiguity, we recover the cubic anomaly.}
\label{tab:PF}
\end{table}

\section{Blow-up equations for $\SU(3)_{\kCS=6}$ and $\USp(4)_{\theta=0} + 2 \Lambda^2$}
\label{app:blow-upSU36USp4}
In this appendix, we provide detailed analyses of the blow-up equations for the pure $\SU(3)_{\kCS=6}$ gauge theory and for the $\USp(4)_{\theta=0}$ gauge theory with two antisymmetric hypermultiplets. The two theories share the following features:
\begin{enumerate}
    \item \label{pt1} The one-instanton partition function starts at order $x$, and the two-instanton partition function starts at order $x^2$; in both cases, the leading term contains no gauge fugacities. The presence of the order-$x$ term in the one-instanton partition function means that, in computing the superconformal index, the contribution of a free hypermultiplet has to be taken into account appropriately.
    \item \label{pt2} The computation of the parameters $\Lambda_{(k)}$ entering the $k$-instanton blow-up equation \eqref{blow-upEqPowerq} is subtle and deserves explanation.
\end{enumerate}
For point \ref{pt1}, the index of $\SU(3)_6$ was given in \cite[(7)]{Kim:2023qwh}, and for point \ref{pt2}, many of the steps for $\SU(3)_6$ were already presented in \cite[(3.72)--(3.78)]{Kim:2020hhh}. In the following, we first present the details for $\SU(3)_6$ and subsequently discuss $\USp(4)_0 + 2 \Lambda^2$.

\subsubsection*{The $\SU(3)_6$ pure gauge theory}
The blow-up function is given by \eqref{eq:exp-VSU3pure}. It follows from \eqref{eq:defLambda} that $\Lambda_{(0)}=0$, while at the one-instanton level
\bes{
\Lambda_{(1)}=0 \quad \text{for} \quad B_{m_0} = \pm \frac{1}{2}~, \qquad \Lambda_{(1)}=-(p_1 p_2)^{1/2} \quad \text{for} \quad B_{m_0} = \frac{3}{2}~,
}
as stated in \cite[(3.73)]{Kim:2020hhh}. Using \eqref{blow-upEqPowerq}, we obtain the one-instanton partition function
\bes{ \label{eq:Zinst1SU36}
\hat{Z}_{\text{inst}, (1)} = x &+ \chi_{[1]}(y)\, x^2 + \Bigl[ \chi_{[2]}(y) + \chi^{\su(3)}_{\text{adj}}(z_i) \Bigr]\, x^3 \\
&+ \Bigl[ \chi_{[3]}(y) + \chi_{[1]}(y)\, \chi^{\su(3)}_{\text{adj}}(z_i) \Bigr]\, x^4 + \ldots~,
}
where $\chi^{\su(3)}_{\text{adj}}(z_i)$ is the character of the adjoint representation of $\su(3)$, whose explicit form is collected in appendix~\ref{app:grouptheory}. At the two-instanton level, \eqref{eq:defLambda} gives
\bes{ \label{eq:Lambda2SU36}
\Lambda_{(2)}= \begin{cases}
-\dfrac{2\, p_1 p_2}{(1-p_1)(1-p_2)(p_1 - p_2)^2} &\qquad B_{m_0} = -\dfrac{1}{2} \\[10pt]
-\dfrac{2\, p_1^2 p_2^2}{(1-p_1)(1-p_2)(p_1 - p_2)^2} &\qquad B_{m_0} = +\dfrac{1}{2} \\[10pt]
-\dfrac{F(p_1, p_2)}{(1-p_1)(1-p_2)(p_1 - p_2)^2} &\qquad B_{m_0} = \dfrac{3}{2}
\end{cases}\,,
}
with
\bes{ \label{eq:FSU36}
F(p_1, p_2) = p_1 p_2 \bigl( & 2 p_1 p_2 + 2 p_1^2 p_2^2 - p_1^2 - p_2^2 - p_1^4 - p_2^4 \\
& + p_1^3 p_2 + p_1 p_2^3 + p_1^4 p_2 + p_1 p_2^4 - p_1^3 p_2^2 - p_1^2 p_2^3 \bigr)~,
}
in agreement with \cite[(3.75)--(3.76)]{Kim:2020hhh}. Using this in the blow-up equation \eqref{blow-upEqPowerq}, we obtain
\bes{
\hat{Z}_{\text{inst}, (2)} = \Bigl[ \chi^{\su(3)}_{\text{adj}}(z_i) - \chi_{[2]}(y) \Bigr]\, x^4 + \ldots~.
}
However, this is not the desired result for the two-instanton partition function, owing to the observation that the plethystic logarithm contains, at order $q^2$, the $z_i$-independent term
\bes{ \label{eq:PLXSU36}
\PL\Bigl[1 + q\, \hat{Z}_{\text{inst}, (1)} + q^2\, \hat{Z}_{\text{inst}, (2)}\Bigr] \Big\lvert_{q^2} = -\frac{p_1 p_2}{(1-p_1)^2(1-p_2)^2} \equiv X(p_1, p_2)~,
}
cf.~\cite[(3.77)]{Kim:2020hhh}, where $\PL\left[f(a)\right] = \sum_{n=1}^{\infty} \frac{\mu(n)}{n} \log{f(a^n)}$. The correct parameters $\Lambda_{(2)}$ can then be computed as follows. We consider the following terms on the right-hand side of \eqref{blow-upEqPowerq} at the two-instanton level (order $q^2$):
\bes{
p_1^{2B_{m_0}}\, \hat{Z}^{(N)}_{\text{inst}, (2)} + p_2^{2B_{m_0}}\, \hat{Z}^{(S)}_{\text{inst}, (2)}~.
}
Substituting $X(p_1, p_2/p_1)$ for $\hat{Z}^{(N)}_{\text{inst}, (2)}$ and $X(p_1/p_2, p_2)$ for $\hat{Z}^{(S)}_{\text{inst}, (2)}$, we obtain the quantity
\bes{ \label{eq:defY}
Y(p_1, p_2) =  \begin{cases} \Lambda_{(2)}\big\lvert_{B_{m_0} =\pm \frac{1}{2}} &\qquad B_{m_0} =\pm \dfrac{1}{2} \\[6pt]
\Lambda_{(2)}\big\lvert_{B_{m_0} =\frac{3}{2}} + p_1 p_2 (p_1 + p_2) &\qquad B_{m_0} = \dfrac{3}{2}
\end{cases}\,,
}
where $\Lambda_{(2)}$ for each $B_{m_0}$ is given in \eqref{eq:Lambda2SU36}. Subtracting \eqref{eq:defY} from $\Lambda_{(2)}$ in \eqref{eq:Lambda2SU36} yields the appropriate parameters $\Lambda'_{(2)}$:
\bes{ \label{eq:Lambda2primeSU36}
\Lambda'_{(2)} = \Lambda_{(2)}-Y(p_1, p_2) = 
\begin{cases} 
0 &\qquad  B_{m_0} = \pm \dfrac{1}{2} \\[6pt]
-p_1 p_2 (p_1+p_2) &\qquad B_{m_0} = \dfrac{3}{2}
\end{cases}\,,
}
in agreement with \cite[(3.78)]{Kim:2020hhh}. The two-instanton partition function obtained from the parameters $\Lambda'_{(2)}$ is
\bes{
\hat{Z}'_{\text{inst}, (2)} = x^2 + 2\, \chi_{[1]}(y)\, x^3 + \Bigl[ 1 + 2\, \chi_{[2]}(y) + \chi^{\su(3)}_{\text{adj}}(z_i) \Bigr]\, x^4 + \ldots~,
}
which is the desired result. In order to compute the index, we have to turn the hatted partition functions into the unhatted ones according to \eqref{ZtoZhatshift}, which amounts to changing the signs $x \rightarrow -x$ and $y \rightarrow -y$ simultaneously. The index computed by using $Z_{\text{inst}} = 1 + Z_{\text{inst}, (1)}\, q + Z'_{\text{inst}, (2)}\, q^2$ in \eqref{Indexgeneralformula} is
\bes{ \label{eq:idxSU36nofree}
1 &- \chi_{[1]}^{\su(2)_I}(q)\, x + \Bigl[ 1 + \chi_{[2]}^{\su(2)_I}(q) - \chi_{[1]}^{\su(2)_I}(q)\, \chi_{[1]}(y) \Bigr]\, x^2 \\
&+ \Bigl[ 2\, \chi_{[1]}(y) \Bigl( 1 + \chi_{[2]}^{\su(2)_I}(q) \Bigr) - \chi_{[1]}^{\su(2)_I}(q) \Bigl( 2 + \chi_{[2]}(y) \Bigr) \Bigr]\, x^3 + \ldots~,
}
where $\chi_{[1]}^{\su(2)_I}(q) = q + q^{-1}$ and $\chi_{[2]}^{\su(2)_I}(q) = q^2 + 1 + q^{-2}$. The negative coefficient at order $x$ stems from going from $\hat{Z}_{\text{inst}, (1)}$ to ${Z}_{\text{inst}, (1)}$, and implies that we need to multiply this result by the index of a free hypermultiplet,
\bes{ \label{eq:Ifree}
\CI_{\text{free}} = \PE \left[ \frac{x}{(1- x y)(1-\frac{x}{y})}\, \chi_{[1]}^{\su(2)_I}(q)\right]~.
}
Upon doing so and keeping the terms in the power series in $x$ up to $q^2$ (two-instanton contributions), we obtain
\bes{
1 + \chi_{[2]}^{\su(2)_I}(q)\, x^2 + \chi_{[1]}(y) \Bigl[ 1 + \chi_{[2]}^{\su(2)_I}(q) \Bigr]\, x^3 + \ldots~,
}
in agreement with \cite[(7)]{Kim:2023qwh}.

\subsubsection*{The $\USp(4)_0 + 2\Lambda^2$ gauge theory}
The blow-up function is given in \S\ref{sec:expVUSp4}, with the flavour fluxes fixed to $B_{m_1} = B_{m_2} = \frac{1}{2}$ and $B_{m_0} \in \Z + \frac{1}{2}$, in accordance with \eqref{eq:Bm0USp4}. A novel feature, compared with the $\SU(3)_6$ case, is that $\Lambda_{(0)}$ is non-trivial:
\bes{ \label{eq:Lambda0USp4AS}
\Lambda_{(0)} = (p_1 p_2)^{-\frac{1}{24}}\, (w_1 w_2)^{-\frac{1}{8}}~.
}
This is precisely the blow-up function at vanishing gauge flux, $\exp(-V_n)\big\lvert_{n=0}$, which is non-trivial for this theory owing to the mass-dominated weight pair of the antisymmetric representation, as pointed out in \S\ref{sec:twogroupSO}. At the one-instanton level, we find
\bes{ \label{eq:Lambda1USp4AS}
\Lambda_{(1)}=0 \quad \text{for} \quad B_{m_0} = \pm \frac{1}{2}~, 
}
whereas, for $B_{m_0} = \frac{3}{2}$, we have
\bes{ \label{eq:Lambda1USp4AS32}
\Lambda_{(1)} = -\Lambda_{(0)}\, (p_1 p_2)^{\frac{1}{2}} \Bigl[ 1 + (p_1 p_2)^{\frac{1}{2}} \left(w_1 + w_2\right) - p_1 p_2\, w_1 w_2 \Bigr]~,
}
whose leading term, $-\Lambda_{(0)}\,(p_1 p_2)^{1/2}$, parallels the value $\Lambda_{(1)} = -(p_1 p_2)^{1/2}$ of the $\SU(3)_6$ theory. Solving \eqref{blow-upEqPowerq}, we obtain the one-instanton partition function
\bes{ \label{eq:Zinst1USp4AS}
\scalebox{0.94}{$
\begin{split}
\hat{Z}_{\text{inst}, (1)} = x &+ \Bigl[ \chi^{\usp(4)_w}_{[1,0]}(w_i) + \chi_{[1]}(y) \Bigr]\, x^2 \\
&+ \Bigl[ \chi_{[2]}(y) + \chi^{\usp(4)_w}_{[1,0]}(w_i)\, \chi_{[1]}(y) + \chi^{\usp(4)}_{\Lambda^2}(z_i) - \chi^{\usp(4)_w}_{[0,1]}(w_i) \Bigr]\, x^3 + \ldots~,
\end{split}
$}
}
where $\chi^{\usp(4)}_{\Lambda^2}(z_i)$ is the character \eqref{eq:chiUSp4AS} of the antisymmetric representation of the \emph{gauge} algebra, while
\bes{
\chi^{\usp(4)_w}_{[1,0]}(w_i) &= w_1 + w_1^{-1} + w_2 + w_2^{-1}~, \\
\chi^{\usp(4)_w}_{[0,1]}(w_i) &= 1 + w_1 w_2 + \frac{w_1}{w_2} + \frac{w_2}{w_1} + \frac{1}{w_1 w_2}
}
are the characters of the representations $\mathbf{4} = [1,0]$ and $\mathbf{5} = [0,1]$ of the \emph{flavour} algebra $\usp(4)_w$.
At the two-instanton level, \eqref{eq:defLambda} yields the naive parameters
\bes{ \label{eq:Lambda2USp4AS}
\Lambda_{(2)}= \Lambda_{(0)} \times \begin{cases}
-\dfrac{2\, p_1 p_2}{(1-p_1)(1-p_2)(p_1 - p_2)^2} &\qquad B_{m_0} = -\dfrac{1}{2} \\[10pt]
-\dfrac{2\, p_1^2 p_2^2}{(1-p_1)(1-p_2)(p_1 - p_2)^2} &\qquad B_{m_0} = +\dfrac{1}{2}
\end{cases}\,,
}
whereas for $B_{m_0} = \frac{3}{2}$,
\bes{ \label{eq:Lambda2USp4AS32}
\Lambda_{(2)} = \Lambda_{(0)} \biggl[ &-\frac{F(p_1, p_2)}{(1-p_1)(1-p_2)(p_1 - p_2)^2} + (p_1 p_2)^{\frac{3}{2}} \left(1 - p_1 p_2\right) \left(w_1 + w_2\right) \\
&- (p_1 p_2)^{\frac{5}{2}} \left(w_1 + w_2\right) w_1 w_2 - (p_1 p_2)^{3}\, w_1 w_2 \biggr]~,
}
with $F(p_1, p_2)$ the \emph{same} function \eqref{eq:FSU36} as in the $\SU(3)_6$ theory. The parallel with \eqref{eq:Lambda2SU36} is remarkable: up to the overall factor of $\Lambda_{(0)}$, the values at $B_{m_0} = \pm\frac{1}{2}$ coincide with those of $\SU(3)_6$, and the $w_h$-independent part at $B_{m_0} = \frac{3}{2}$ is governed by the same function $F(p_1, p_2)$; the flavour fugacities enter only through the additional terms in \eqref{eq:Lambda2USp4AS32}. With these parameters, the blow-up equation \eqref{blow-upEqPowerq} gives the two-instanton partition function
\bes{ \label{eq:Zinst2USp4AS}
\hat{Z}_{\text{inst}, (2)} = \chi^{\usp(4)_w}_{[1,0]}(w_i)\, x^3 + \Bigl[ & \chi^{\usp(4)_w}_{[2,0]}(w_i) - \chi^{\usp(4)_w}_{[0,1]}(w_i) - \chi_{[2]}(y) \\
& + 2\, \chi^{\usp(4)_w}_{[1,0]}(w_i)\, \chi_{[1]}(y) + \chi^{\usp(4)}_{\Lambda^2}(z_i) \Bigr]\, x^4 + \ldots~,
}
where $\chi^{\usp(4)_w}_{[2,0]}(w_i)$ is the character of the adjoint representation $\mathbf{10} = [2,0]$ of the flavour algebra $\usp(4)_w$. As in the $\SU(3)_6$ case, this is not the desired result for the two-instanton partition function: the plethystic logarithm contains, at order $q^2$, precisely the same $z_i$- and $w_h$-independent term $X(p_1, p_2)$ as in \eqref{eq:PLXSU36}, namely
\bes{
\PL\Bigl[1 + q\, \hat{Z}_{\text{inst}, (1)} + q^2\, \hat{Z}_{\text{inst}, (2)}\Bigr] \Big\lvert_{q^2} = -\frac{p_1 p_2}{(1-p_1)^2(1-p_2)^2} = X(p_1, p_2)~.
}
The correction proceeds exactly as in the $\SU(3)_6$ case: substituting $X(p_1, p_2/p_1)$ and $X(p_1/p_2, p_2)$ into $\hat{Z}^{(N)}_{\text{inst}, (2)}$ and $\hat{Z}^{(S)}_{\text{inst}, (2)}$ in the following terms of the right-hand side of \eqref{blow-upEqPowerq} at order $q^2$, that is
\bes{
\Lambda_{(0)} \left[ p_1^{2B_{m_0}}\, \hat{Z}^{(N)}_{\text{inst}, (2)} + p_2^{2B_{m_0}}\, \hat{Z}^{(S)}_{\text{inst}, (2)} \right]~,
}
yields the analogue of \eqref{eq:defY}, and subtracting it from \eqref{eq:Lambda2USp4AS}--\eqref{eq:Lambda2USp4AS32} gives the appropriate parameters
\bes{
\Lambda'_{(2)} = 0 \quad \text{for} \quad B_{m_0} = \pm \frac{1}{2}~,
}
whereas, for $B_{m_0} = \frac{3}{2}$, we find
\bes{ \label{eq:Lambda2primeUSp4AS}
\Lambda'_{(2)} = \Lambda_{(0)} \biggl[ &-p_1 p_2 \left(p_1 + p_2\right) + (p_1 p_2)^{\frac{3}{2}} \left(1 - p_1 p_2\right) \left(w_1 + w_2\right) \\
&- (p_1 p_2)^{\frac{5}{2}} \left(w_1 + w_2\right) w_1 w_2 - (p_1 p_2)^{3}\, w_1 w_2 \biggr]~.
}
Remarkably, \eqref{eq:Lambda2primeUSp4AS} is obtained from the naive parameter \eqref{eq:Lambda2USp4AS32} simply by replacing $-F(p_1,p_2)/\bigl[(1-p_1)(1-p_2)(p_1-p_2)^2\bigr]$ with $-p_1 p_2 (p_1 + p_2)$, leaving the $w_h$-dependent terms untouched --- in exact parallel with the passage from \eqref{eq:Lambda2SU36} to \eqref{eq:Lambda2primeSU36} in the $\SU(3)_6$ theory. The two-instanton partition function obtained from the parameters $\Lambda'_{(2)}$ is
\bes{ \label{eq:Zinst2pUSp4AS}
\hat{Z}'_{\text{inst}, (2)} = x^2 &+ \Bigl[ \chi^{\usp(4)_w}_{[1,0]}(w_i) + 2\, \chi_{[1]}(y) \Bigr]\, x^3 \\
&+ \Bigl[ 1 + \chi^{\usp(4)_w}_{[2,0]}(w_i) - \chi^{\usp(4)_w}_{[0,1]}(w_i) + 2\, \chi_{[2]}(y) \\
&\qquad \, + 2\, \chi^{\usp(4)_w}_{[1,0]}(w_i)\, \chi_{[1]}(y) + \chi^{\usp(4)}_{\Lambda^2}(z_i) \Bigr]\, x^4 + \ldots~,
}
which is the desired result.
As in the $\SU(3)_6$ case, we turn the hatted partition functions into the unhatted ones according to \eqref{ZtoZhatshift} and compute the index using $Z_{\text{inst}} = 1 + Z_{\text{inst}, (1)}\, q + Z'_{\text{inst}, (2)}\, q^2$ in \eqref{Indexgeneralformula}. Setting the flavour fugacities $w_h = 1$ for brevity, the result reads
\bes{ \label{eq:idxUSp4ASnofree}
1 - \chi_{[1]}^{\su(2)_I}(q)\, x &+ \Bigl[ 11 + \chi_{[2]}^{\su(2)_I}(q) + 4\, \chi_{[1]}^{\su(2)_I}(q) - \chi_{[1]}^{\su(2)_I}(q)\, \chi_{[1]}(y) \Bigr]\, x^2 \\
&+ \Bigl[ \chi_{[1]}(y) \Bigl( 12 + 2\, \chi_{[2]}^{\su(2)_I}(q) + 4\, \chi_{[1]}^{\su(2)_I}(q) \Bigr) \\
&\qquad - 7\, \chi_{[1]}^{\su(2)_I}(q) - 4\, \chi_{[2]}^{\su(2)_I}(q) - \chi_{[1]}^{\su(2)_I}(q)\, \chi_{[2]}(y) \Bigr]\, x^3 + \ldots~.
}
The negative coefficient at order $x$ again signals the presence of a free hypermultiplet carrying instanton charges $q^{\pm 1}$, and we multiply the result by the index $\CI_{\text{free}}$ of \eqref{eq:Ifree}. Upon doing so and keeping the terms up to $q^2$ (two-instanton contributions), the index organises into characters of the enhanced algebra $\usp(6) \supset \usp(4)_w \oplus \su(2)_I$:
\bes{ \label{eq:idxUSp4ASfinal}
1 + \chi_{\mathrm{adj}}^{\usp(6)}\, x^2 + \Bigl( \chi_{[0,0,1]}^{\usp(6)} + \chi_{[1]}(y) \Bigl[ 1 + \chi_{\mathrm{adj}}^{\usp(6)} \Bigr] \Bigr)\, x^3 + \ldots~,
}
in exact agreement with the index \eqref{eq:indexUSp4_2AS} of \S\ref{sec:faithfulsym}, with the branchings of $\chi_{\mathrm{adj}}^{\usp(6)}$ and $\chi_{[0,0,1]}^{\usp(6)}$ as given there. In arriving at \eqref{eq:idxUSp4ASfinal}, we have discarded the terms $\bigl(q^3 + q^{-3}\bigr) x^3$: these pair the two-instanton states with the instanton charge of the free hypermultiplet and can only be completed into $\su(2)_I$ characters by the three-instanton sector, which lies beyond the order computed here.\footnote{More precisely, the terms $\bigl(q^3 + q^{-3}\bigr) x^3$ are expected to be cancelled by the leading contribution of the three-instanton partition function: in analogy with \eqref{eq:Zinst1USp4AS} and \eqref{eq:Zinst2pUSp4AS}, $\hat{Z}_{\text{inst}, (3)}$ is expected to start at order $x^3$, and this leading term becomes $-x^3$ in the unhatted partition function upon applying the shift \eqref{ZtoZhatshift}.}

\section{Group theory conventions}
\label{app:grouptheory}

In this appendix, we collect the explicit expressions for the Haar measures and the characters used in the computation of the superconformal indices throughout this paper. The Haar measure for a gauge group $G$ is given by \eqref{Haarmeasure}, with the product running over the positive roots of the Lie algebra $\fg$; below, we list the resulting expressions, together with the characters of the adjoint representation and of the relevant matter representations, in the fugacity conventions of the main text.

\subsection*{$\su(2)$}
\bes{
\oint \left[d\mu_{\SU(2)}(z_1)\right] =& \oint \frac{d z_1}{2 \pi i z_1} \left(1-\frac{1}{z^2_1}\right)\,, \\
\chi^{\su(2)}_{\text{adj}} =& z_1^2 + 1 + \frac{1}{z_1^2}\,, \qquad
\chi^{\su(2)}_{\mathbf{F}} = z_1 + \frac{1}{z_1}\,.
}

\subsection*{$\su(3)$}
\bes{
&\oint \left[d\mu_{\SU(3)}(z_j)\right] = \oint \frac{d z_1}{2 \pi i z_1} \frac{d z_2}{2 \pi i z_2} \left(1-\frac{z_2}{z^2_1}\right) \left(1-\frac{z_1}{z^2_2}\right) \left(1-\frac{1}{z_1 z_2}\right)\,, \\
&\chi^{\su(3)}_{\text{adj}} = 2 + z_1 z_2 + \frac{z_1^2}{z_2} + \frac{z_2^2}{z_1} + \frac{1}{z_1 z_2} + \frac{z_2}{z_1^2} + \frac{z_1}{z_2^2}\,, \qquad
\chi^{\su(3)}_{\mathbf{F}} = z_1 + \frac{z_2}{z_1} + \frac{1}{z_2}\,.
}

\subsection*{$\su(4)$}
\bes{
\oint \left[d\mu_{\SU(4)}(z_j)\right] &= \oint \frac{d z_1}{2 \pi i z_1} \frac{d z_2}{2 \pi i z_2} \frac{d z_3}{2 \pi i z_3} \prod_{j=1}^6 \left(1-\tilde{z}^+_j\right)\,, \\
\chi^{\su(4)}_{\text{adj}} &= 3 + \frac{z^2_1}{z_2}+\frac{z_2}{z^2_1}+\frac{z_1 z_2}{z_3}+\frac{z_3}{z_1 z_2}+z_1 z_3+\frac{1}{z_1 z_3} \\
&\quad +\frac{z^2_2}{z_1 z_3}+\frac{z_1 z_3}{z^2_2}+\frac{z_2 z_3}{z_1}+\frac{z_1}{z_2 z_3}+\frac{z^2_3}{z_2}+\frac{z_2}{z^2_3}\,, \\
\chi^{\su(4)}_{\Lambda^2} &= z_2 + \frac{z_1 z_3}{z_2}+\frac{z_1}{z_3}+\frac{z_3}{z_1}+\frac{z_2}{z_1 z_3}+\frac{1}{z_2}\,,
}
where $\tilde{z}^+_j = \left\{\frac{z_2}{z^2_1}, \frac{z_2}{z^2_3}, \frac{1}{z_1 z_3}, \frac{z_1}{z_2 z_3}, \frac{z_1 z_3}{z^2_2}, \frac{z_3}{z_1 z_2}\right\}$.

\subsection*{$\usp(4)$}
\bes{ \scalebox{0.99}{$
\begin{split}
\oint \left[d\mu_{\USp(4)}(z_j)\right] &= \oint \frac{d z_1}{2 \pi i z_1} \frac{d z_2}{2 \pi i z_2} \left(1-\frac{1}{z^2_1}\right) \left(1-\frac{z^2_1}{z^2_2}\right) \left(1-\frac{1}{z_2}\right) \left(1-\frac{z_2}{z^2_1}\right)\,, \\
\chi^{\usp(4)}_{\text{adj}} &= 2 + \frac{1}{z^2_1}+z^2_1+\frac{z^2_1}{z^2_2}+\frac{1}{z_2}+\frac{z^2_1}{z_2}+z_2+\frac{z_2}{z^2_1}+\frac{z^2_2}{z^2_1}\,, \\
\chi^{\usp(4)}_{\Lambda^2} &= 1 + \frac{1}{z_2}+\frac{z^2_1}{z_2}+z_2+\frac{z_2}{z^2_1}\,. \label{eq:chiUSp4AS}
\end{split}
$}
}

\subsection*{$\so(7)$}
\bes{
\oint \left[d\mu_{\Spin(7)}(z_j)\right] &= \oint \prod_{j=1}^{3} \frac{d z_j}{2 \pi i z_j} \prod_{j=1}^{9} \left(1-\tilde{z}^+_j\right)\,, \\
\chi^{\so(7)}_{\text{adj}} &= 3 + \sum_{j=1}^{9} \left( \tilde{z}^+_j + \frac{1}{\tilde{z}^+_j} \right)\,, \\
\chi^{\so(7)}_{\mathbf{F}} &= 1 + z_1 + \frac{1}{z_1} + \frac{z_2}{z_1} + \frac{z_1}{z_2} + \frac{z_3^2}{z_2} + \frac{z_2}{z_3^2}\,,
}
where
\bes{
\tilde{z}^+_j = \left\{ \frac{z_2}{z_1^2},\ \frac{z_1 z_3^2}{z_2^2},\ \frac{z_2}{z_3^2},\ \frac{z_3^2}{z_1 z_2},\ \frac{z_1}{z_2},\ \frac{1}{z_1},\ \frac{z_1}{z_3^2},\ \frac{z_2}{z_1 z_3^2},\ \frac{1}{z_2} \right\}\,.
}

\subsection*{$\so(8)$}
\bes{
\oint \left[d\mu_{\Spin(8)}(z_j)\right] &= \oint \prod_{j=1}^{4} \frac{d z_j}{2 \pi i z_j} \prod_{j=1}^{12} \left(1-\tilde{z}^+_j\right)\,, \\
\chi^{\so(8)}_{\text{adj}} &= 4 + \sum_{j=1}^{12} \left( \tilde{z}^+_j + \frac{1}{\tilde{z}^+_j} \right)\,, \\
\chi^{\so(8)}_{\mathbf{8}_v} &= z_1 + \frac{1}{z_1} + \frac{z_2}{z_1} + \frac{z_1}{z_2} + \frac{z_3 z_4}{z_2} + \frac{z_2}{z_3 z_4} + \frac{z_4}{z_3} + \frac{z_3}{z_4}\,,
}
where
\bes{ \scalebox{1.08}{$
\tilde{z}^+_j = \left\{ \frac{z_2}{z_1^2},\ \frac{z_1 z_3 z_4}{z_2^2},\ \frac{z_2}{z_3^2},\ \frac{z_2}{z_4^2},\ \frac{z_3 z_4}{z_1 z_2},\ \frac{z_1 z_4}{z_2 z_3},\ \frac{z_1 z_3}{z_2 z_4},\ \frac{z_4}{z_1 z_3},\ \frac{z_3}{z_1 z_4},\ \frac{z_1}{z_3 z_4},\ \frac{z_2}{z_1 z_3 z_4},\ \frac{1}{z_2} \right\}\,.
$}
}

\bibliographystyle{JHEP}
\bibliography{bibli}

\end{document}